\documentclass[12pt,preprint]{aastex}
\begin{document}

\title{A Correlated Study of Optical and X-ray Afterglows of GRBs}
\author{Liang Li\altaffilmark{1,2,3,4}, Xue-Feng Wu\altaffilmark{5,6,7}, Yong-Feng Huang\altaffilmark{8,9}, Xiang-Gao Wang\altaffilmark{10,11}, Qing-Wen Tang\altaffilmark{8,9}, Yun-Feng Liang\altaffilmark{12}, Bin-Bin Zhang\altaffilmark{13}, Yu Wang\altaffilmark{14,15}, Jin-Jun Geng\altaffilmark{8,9}, En-Wei Liang\altaffilmark{10}, Jian-Yan Wei\altaffilmark{16}, Bing Zhang\altaffilmark{11} and Felix Ryde\altaffilmark{1,2}}
\altaffiltext{1}{Department of Physics, KTH Royal Institute of Technology, SE-106 91 Stockholm, Sweden; fryde@kth.se}
\altaffiltext{2}{The Oskar Klein Centre for Cosmoparticle Physics}
\altaffiltext{3}{Department of Physics, Stockholm University, SE-106 91 Stockholm, Sweden; liang.li@fysik.su.se}
\altaffiltext{4}{Erasmus Mundus Joint Doctorate in Relativistic Astrophysics}
\altaffiltext{5}{Chinese Center for Antarctic Astronomy, Chinese Academy of Sciences, Nanjing 210008, China}
\altaffiltext{6}{Joint Center for Particle Nuclear Physics and Cosmology of Purple Mountain Observatory and Nanjing University, Nanjing 210008, China}
\altaffiltext{7}{Key laboratory of Dark Matter and Space Astronomy, Purple Mountain Observatory, Chinese Academy of Sciences, Nanjing 210008, China.}
\altaffiltext{8}{Department of Astronomy, Nanjing University, Nanjing 210093, Jiangsu, China}
\altaffiltext{9}{Key laboratory of Modern Astronomy and Astrophysics (Nanjing University), Ministry of Education, Nanjing 210093, China}
\altaffiltext{10}{GXU-NAOC Center for Astrophysics and Space Sciences, Department of Physics, Guangxi University, Nanning 530004,China}
\altaffiltext{11}{Department of Physics and Astronomy, University of Nevada, Las Vegas, NV 89154, USA}
\altaffiltext{12}{Purple Mountain Observatory, Chinese Academy of Sciences, Nanjing 210008, China}
\altaffiltext{13}{Center for Space Plasma and Aeronomic Research (CSPAR), University of Alabama in Huntsville, Huntsville, AL 35899, USA}
\altaffiltext{14}{Dip. di Fisica and ICRA, Sapienza Universit¨¤ di Roma, Piazzale Aldo Moro 5, I-00185 Rome, Italy.}
\altaffiltext{15}{ICRANet, Piazza della Repubblica 10, I-65122 Pescara, Italy.}
\altaffiltext{16}{National Astronomical Observatories, Chinese Academy of Sciences, Beijing, 100012, China}

\begin{abstract}
We study an extensive sample of 87 GRBs for which there are well sampled and simultaneous optical and X-ray light-curves.
We extract the cleanest possible signal of the afterglow component, and compare the temporal behaviors of the X-ray light-curve, observed by Swift XRT, and optical data, observed by UVOT and ground-based telescopes for each individual burst.
Overall we find 62\% GRBs that are consistent with the standard afterglow model. When more advanced modeling is invoked, up to 91\% of the bursts in our sample may be consistent with the external shock model.
A large fraction of these bursts are consistent with occurring in a constant interstellar density medium (ISM) (61\%) while only 39\% of them occur in a wind-like medium.
Only 9 cases have afterglow light-curves that exactly match the standard fireball model prediction, having a single power law decay in both energy bands which are observed during their entire duration.  In particular, for the bursts with chromatic behavior additional model assumptions must be made over limited segments of the light-curves in order for these bursts to fully agree with the external shock model.
Interestingly, for 54\% of the X-ray and 40\% of the optical band observations the end of the shallow decay ($t^{\sim-0.5}$) period coincides with the jet break ($t^{\sim-p}$) time, causing an abrupt change in decay slope.
The fraction of the burst that consistent with the external shock model is independent of the observational epochs in the rest frame of GRBs.
Moreover, no cases can be explained by the cooling frequency crossing the X-ray or optical band.

\end{abstract}

\section{Introduction}

While the prompt, gamma-ray emission of gamma-ray bursts (GRBs) generally is interpreted as being due to energy  dissipation processes internally in the relativistic outflow (e.g. Usov 1992; Rees \& M\'esz\'aros 1994; Thompson 1994; Kobayashi et al.1997; Daigne \& Mochkovitch 1998; Drenkhahn \& Spruit 2004; Giannios \& Spruit 2006; Zhang \& Pe'er 2009; Zhang \& Yan 2011), the afterglow emission is interpreted  as synchrotron emission generated in a forward shock as the blastwave interacts with the circum-burst medium (e.g., M\'esz\'aros \& Rees 1997; Sari et al. 1998; Panaitescu et al. 1998; Huang et al. 2000; Panaitescu \& Kumar 2001;  see Zhang \& M\'{e}szaros 2004 for review). Sari et al. (1998) calculated the broadband spectrum and corresponding light-curves and reported quantitative relationships (the standard afterglow model) with a constant interstellar medium (ISM), which can generally describe the X-ray and optical afterglow emission to be from the same synchrotron emission process. Prior to the launch of the Swift satellite observational studies largely confirmed the standard model. By comparing the light-curve decay at different wavelengths, Wijers, Rees, \& Meszaros (1997) showed that the standard model fitted the afterglow data for GRB970228 and 970402. Likewise, Panaitescu et al. (2005a, b) concluded that the X-ray and optical afterglow emission of a sample of 10 bursts were consistent with a single component origin.  Furthermore, Nardini et al. (2006) studied the spectral energy distribution over the optical and X-ray bands of 24 bursts and concluded that the majority of the bursts were consistent with a single emission component. Furthermore, Li et al. (2015 in preparation) recently studied the evolution of color indices for a large GRB sample which have well multi-band, optical afterglow observations, and also find that most of these GRBs can be explained by the standard synchrotron afterglow model.

Although the standard afterglow model is successful in explaining the overall features of the observed afterglow behavior, the increasing amount of  observational data, as well as its gain in detail, clearly call for additional assumptions to be made, e.g., on the ambient density profile, prolonged energy injection, jet, accretion around central engine, and viewing angle effects (e.g. M\'esz\'aros et al. 1998; Dai \& Lu 1998, 2002; Rees \& M\'esz\'aros 1998; Rhoads 1999; Sari et al. 1999; Chevalier \& Li 1999, 2000; Huang et al. 2000; Zhang \& M\'esz\'aros 2001; Pisani et al. 2013; Zhang et al. 2014; Ryan et al. 2014; Ruffini et al. 2014). Findings based on these assumptions indicate that the afterglow emission is more complicated than that predicted by the simplest standard afterglow model, and further developments studying ``post-standard effects'' were made by various authors. For example, Chevalier \& Li (1999) took a wind-like medium into account, that is, a circum-burst medium density that decreases as $r^{-2}$, where $\textit{r}$ is the distance from the center, due to the wind from the progenitor. This gave rise to new, quantitative ``closure relations'' of the external shock model. Detailed modeling of early and late afterglow properties in a wind medium has been carried out by Panaitescu \& Kumar (2000), Wu et al. (2003), Kobayashi \& Zhang (2003), and Kobayashi et al. (2004). The energy injection models have been studied by Dai \& Lu (1998), Rees \& M\'esz\'aros (1998), Sari \& M\'esz\'aros (2000), and Zhang \& M\'esz\'aros (2001).
The jet models are studied in detail by Rhoads (1999), Sari et al. (1999), Huang et al. (2000) and Wu et al. (2004), and more recently, simulations have been carried out by Eerten \& MacFadyen (2013) and Duffell \& MacFadyen (2014). Zhang et al. (2006) derived  the ``closure relations'' of relevant external shock models and collected them in their Table 2. A more complete review of closure relations can be found in Gao et al. (2013).

The Swift satellite (Gehrels et al. 2004) has provided an unprecedented sample of long time series of broad energy band observations of bursts. Simultaneous observations with the Swift instruments BAT, XRT, UVOT, and  ground-based optical telescopes have revolutionized our understanding but have also revealed some critical problems with the standard model (e.g., Zhang 2007, 2011). In contrast to the expected results from the standard model, many bursts showed a chromatic behavior between that of the X-ray and optical band, that is, showing different behaviors at different frequencies. For instance, Panaitescu et al. (2006) studied 6 Swift bursts that exhibited distinctly different behaviors of the X-ray and optical light-curves, prompting them to argue for a different origin of the emission in the two bands, at least for the bursts in the study.
Moreover, Oates et al. (2009) studied a sample of 26 Swift bursts and showed that the XRT X-ray and UVOT optical light-curves have similar statistical properties. However, they found that they are remarkably different during the first 500s after the BAT trigger. The late time light-curves have a better resemblence. Zhang et al. (2007) and Liang et al. (2007, 2008, 2009) in a series of papers, who studied X-ray afterglow and their consistency with the external shock model, also suggest that the X-ray and the optical band have a different origin within the emission. Systematic studies of Swift X-ray afterglows and their consistency with external shock models have been also carried out by O'Brien et al. (2006), Willingale et al. (2007), Evans et al. (2007, 2009) and Zaninoni et al. (2013).
In (Ruffini et al 2014), GRBs originating from binary system is investigated, one family of GRBs is classified as that a standard black hole is formed by a neutron star accreting to the critical mass of gravitational collapse, this family holds typical scaling laws and overlapping features in the X-ray afterglow (Pisani et al 2013).

In this paper, we analyze a sample of 87 bursts, focussing on the clean afterglow emission after omitting the components which are related to internal shock origin, afterglow reverse shock emission, and the late-time associated supernova component. These bursts are observed by XRT and simultaneously by UVOT and/or ground-based telescopes and are able to make a large statistical study. We examine whether, and how often, the  X-ray and optical afterglow light-curves are consistent with the external shock model. This paper is organised as follows. We define the data sample, the fitting results are presented and analysed in section 2. We carry out a detailed statistical analysis in section 3. In section 4, we perform several tests to the external shock models for all our afterglow samples. A more detailed investigation regarding how well the models can interpret the data can be found in Wang et al. (2015). We discuss the properties of afterglow break behaviors in section 5, and Conclusions and Discussion in section 6.

\section{Sample Selection and Light-Curve Fitting\label{sec:data}}

We compiled a data base of 270 GRBs with optical afterglow detections made from February 1997 until December 2013 which have been presented in GCN Circulars and in the literature. We included observations made by  Swift/UVOT, as well as  other ground based telescopes, such as the Gamma-Ray burst Optical/NIR Detector (GROND; Greiner et al. 2007, 2008), ROTSE-III (Akerlof et al. 2003), TAROT (Klotz et al. 2009), RAPTOR (Vestrand et al. 2004), and REM (Zerbi et al. 2001). For all these bursts we also compile the full sample of XRT data  from the Swift data archive.

In order to study the afterglow emission expected from the external shock model, we need to define a subsample of these bursts, since multiple emission components can exist.  For instance, Zhang et al. (2006) summarized the X-ray afterglow behavior with a synthetic cartoon of an X-ray light-curve based on the observational data from the Swift archive, showing that X-ray afterglow included five emission components that have distinctly different physical origins or processes.  Likewise, Li et al. (2012) summarized the optical afterglow behavior with a synthetic cartoon optical light-curves, based on an extensive analysis of the optical data statistics, morphology and theory. In contrast to the X-ray afterglow, they identified eight clear emission components in the optical afterglow light-curves, which may have different physical origins or processes (Li et al. 2012).
To extract the cleanest possible signal of the afterglow component we therefore apply the following 4 rules:

{\it  Rule 1:}  We exclude the following components (i) the steep decay in early X-ray afterglow light-curves, that is involved in
 the tail of prompt emission related to internal energy dissipation (Zhang et al 2006) (ii) the flare behavior that may be related to the late-time activity of the central engine (Burrows et al. 2005; Fan \& Wei 2005; Zhang et al. 2006; Dai et al. 2006; Proga \& Zhang. 2006; Perna et al. 2006) (iii) Using the relation $\alpha\geq2+\beta$, we exclude the 'internal plateau' samples (Troja et al. 2007; Liang et al. 2007) (iv) the early optical reverse shock emission (v) the GRBs-SNe bumps that are typically detected in the optical afterglow light-curves at late time.

{\it Rule 2:}  We restrict our sample to include only bursts for which there are good, simultaneous observations in both the X-ray and optical bands, since we want to compare the optical and X-ray light-curve behaviors.

{\it Rule 3:} All the optical light-curves correspond to  R-band observations. For bursts which only have, or have better, observations in other bands, we corrected the observations into the R-band, by assuming a spectrum $ F_\nu = F_{\nu,0} \nu^{-\beta_{\rm o}}$, ${\beta_{\rm o}}$ is the optical spectral index. This is particularly the case when a longer time series exists in other bands.

{\it Rule 4:}  We fit the broadest possible time interval with either a single power-law or a broken power-law. Such a fit will ignore wiggles on short time scales.  For instance, GRB 051109A is fitted with a broken power-law,  which ignores the flare behavior (maybe a late re-brightening bump behavior) that is embedded in the light-curve in the narrow time range $~[10^{4} \rm{s},10^{5}\rm {s}]$.

\begin{figure*}
\includegraphics[angle=0,scale=0.350,width=0.24\textwidth,height=0.2\textheight]{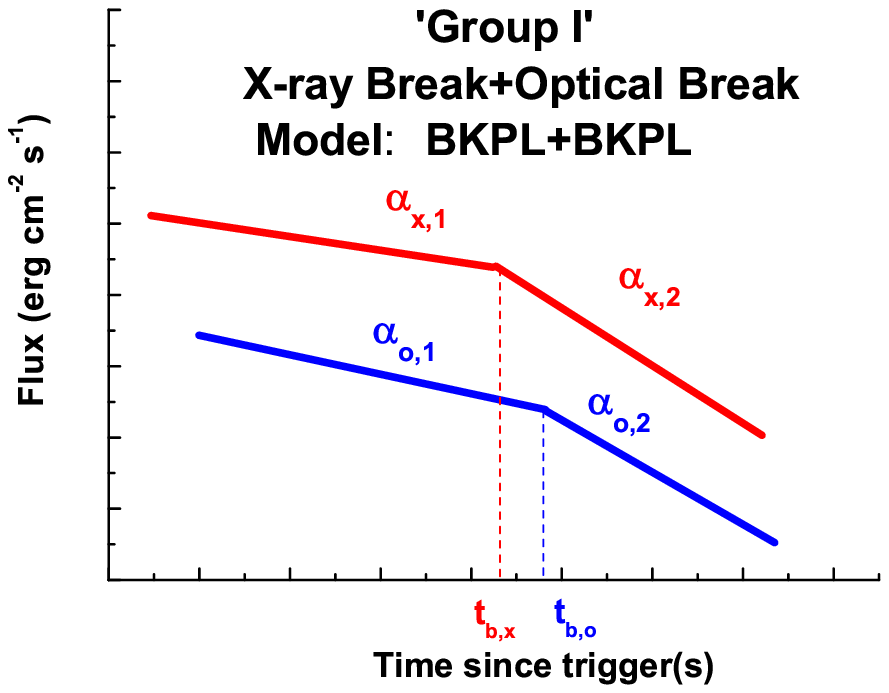}
\includegraphics[angle=0,scale=0.350,width=0.24\textwidth,height=0.2\textheight]{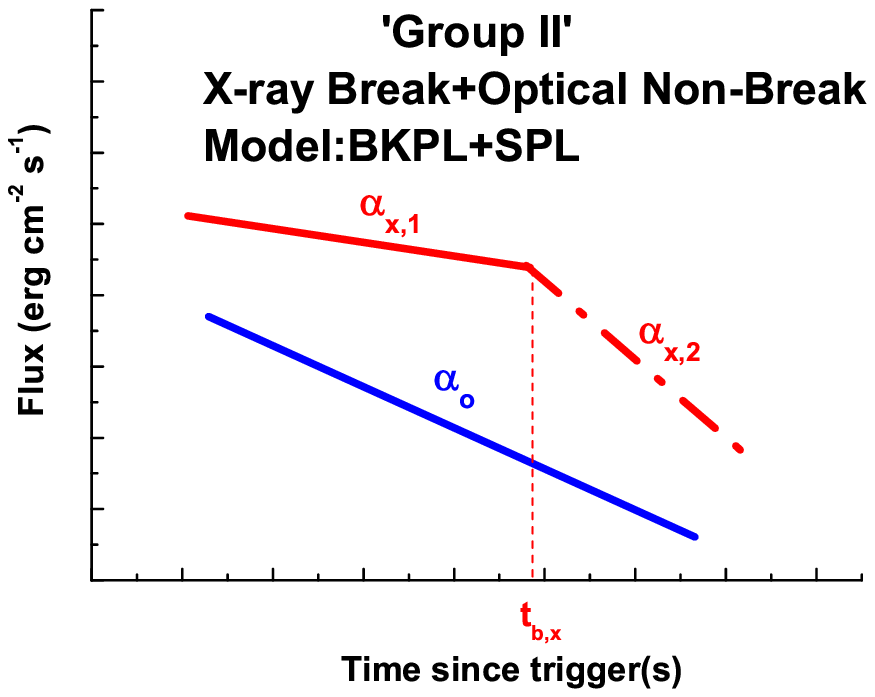}
\includegraphics[angle=0,scale=0.350,width=0.24\textwidth,height=0.2\textheight]{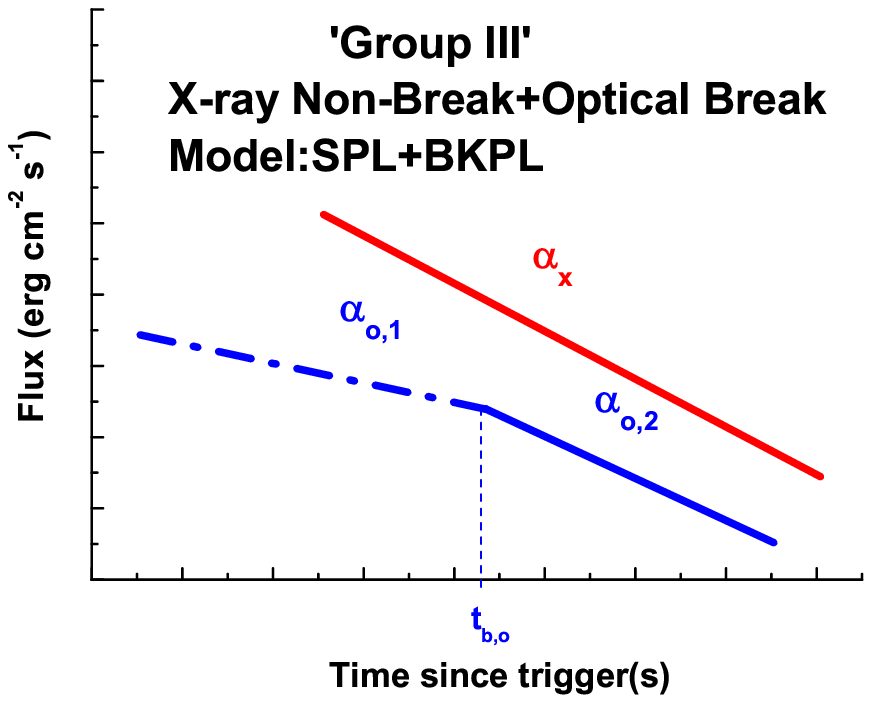}
\includegraphics[angle=0,scale=0.350,width=0.24\textwidth,height=0.2\textheight]{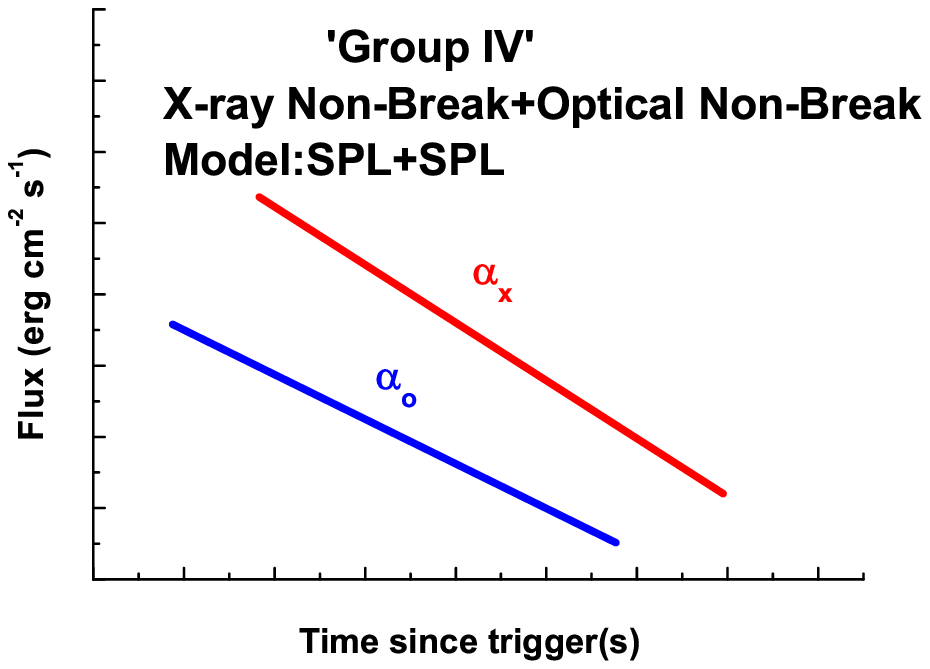}
\caption{The cartoons showing our four samples which, with different shapes with or without breaks, comparing X-ray and optical afterglow light-curves and  the models used. Note that Group II GRBs do not always point towards a chromatic behavior. This is because in many GRBs there is no optical data around the X-ray break time, so that the existence of an X-ray break around the same time is not ruled out. Again some GRBs in Group III do not have X-ray data around the optical break time, so that they are not necessarily chromatic. We therefore plot the post-break segment for Group II and the pre-break segment for Group III by the dot-dashed lines.}\label{Cartoon}
\end{figure*}

These rules give us a total of 87 bursts which we classify as four samples depending on their light-curves in the X-ray and optical bands. We denote these samples as Group I, Group II, Group III, and Group IV which are defined in the following way:

\begin{itemize}
\item Group I:
All the bursts for which there is a clear light-curve break detected in both the X-ray and optical
afterglows, as shown in the left-hand panel in the cartoon of light curves in Figure \ref{Cartoon}. Of 87 GRBs, our full sample consists of 24 such bursts, and the broken power-law fits to the light-curves are shown in Figure \ref{Both Break}.

\item Group II:
All bursts for which there is a  break detected only in the X-rays, while the optical band generally shows a single power-law decay. Note that a few bursts have a detected afterglow onset which is followed by the characteristic decay of the sample. A cartoon of a typical light-curve in this sample is shown in the centre-left panel in Figure \ref{Cartoon}. Of 87 GRBs, our full sample consists of  24 such bursts, and the fits to the light-curves are shown in Figure \ref{Xray Break}. We note that Group II GRBs do not always point towards a chromatic behavior. This is because in many GRBs there is no optical data around the X-ray break time, so that the existence of an X-ray break around the same time is not ruled out. This issue is fully addressed in Wang et al. (2015).

\item Group III:
All bursts for which the break is detected only in the optical band, while the X-ray light-curve in general is described by a single power-law. Note that a few bursts have a detected afterglow onset. A cartoon example is shown in the centre-right panel in Figure \ref{Cartoon}. Of 87 GRBs, This sample comprises 9 bursts and fitted afterglows are depicted in Figure \ref{Optical Break}. Again some GRBs in this group do not have X-ray data around the optical break time, so that they are not necessarily chromatic (Wang et al. 2015).

\item Group IV:
All bursts which do not have any obvious breaks, either in the X-ray or in the optical bands.
They are fitted with  single power-law functions.  A few of these bursts have afterglow onsets.  Of 87 GRBs, this sample includes 30 bursts and the best fitted light-curves are shown in Figure \ref{Non Break}.
\end{itemize}

\begin{figure*}
\includegraphics[angle=0,scale=0.350,width=0.24\textwidth,height=0.2\textheight]{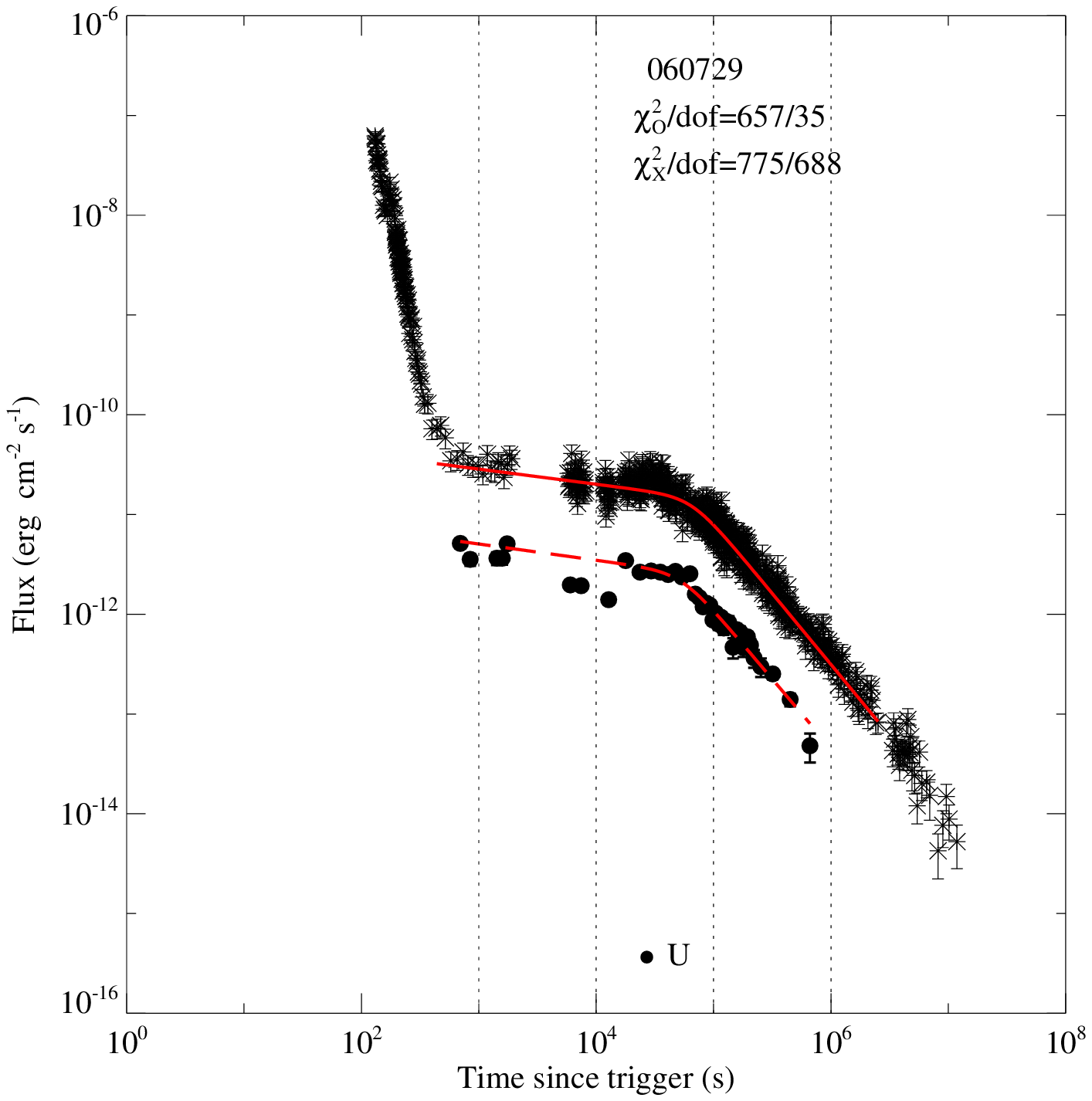}
\includegraphics[angle=0,scale=0.350,width=0.24\textwidth,height=0.2\textheight]{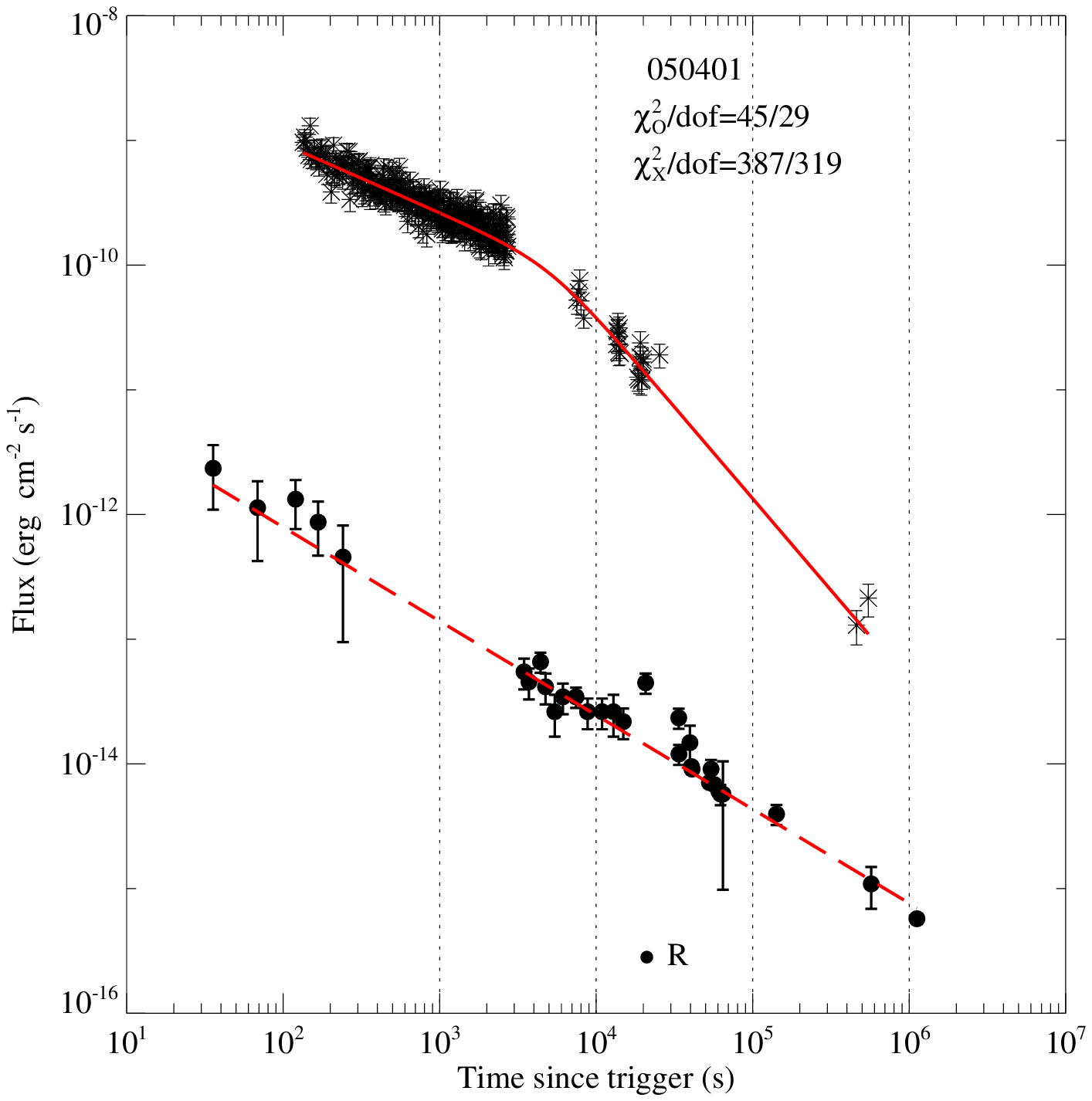}
\includegraphics[angle=0,scale=0.350,width=0.24\textwidth,height=0.2\textheight]{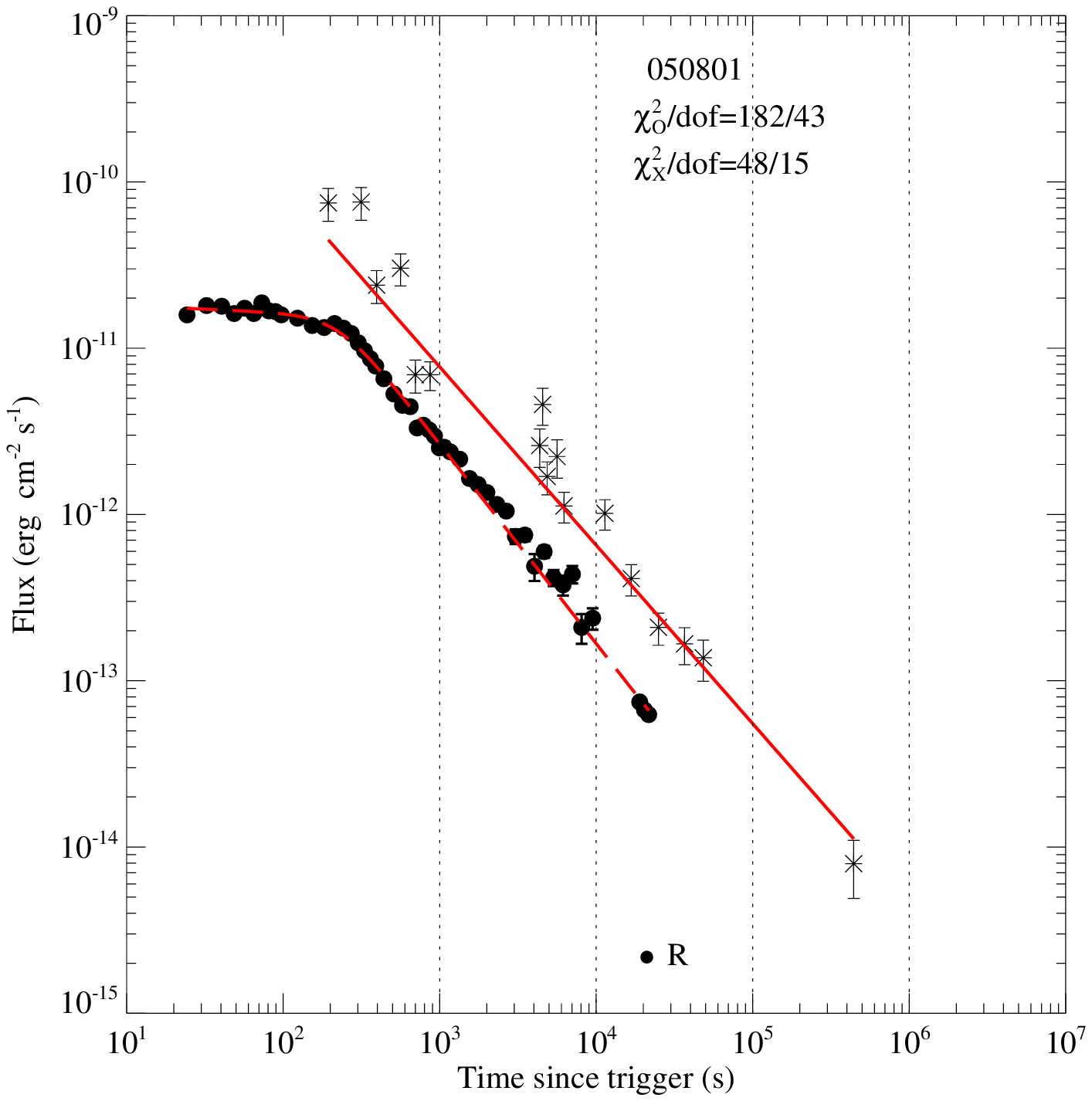}
\includegraphics[angle=0,scale=0.350,width=0.24\textwidth,height=0.2\textheight]{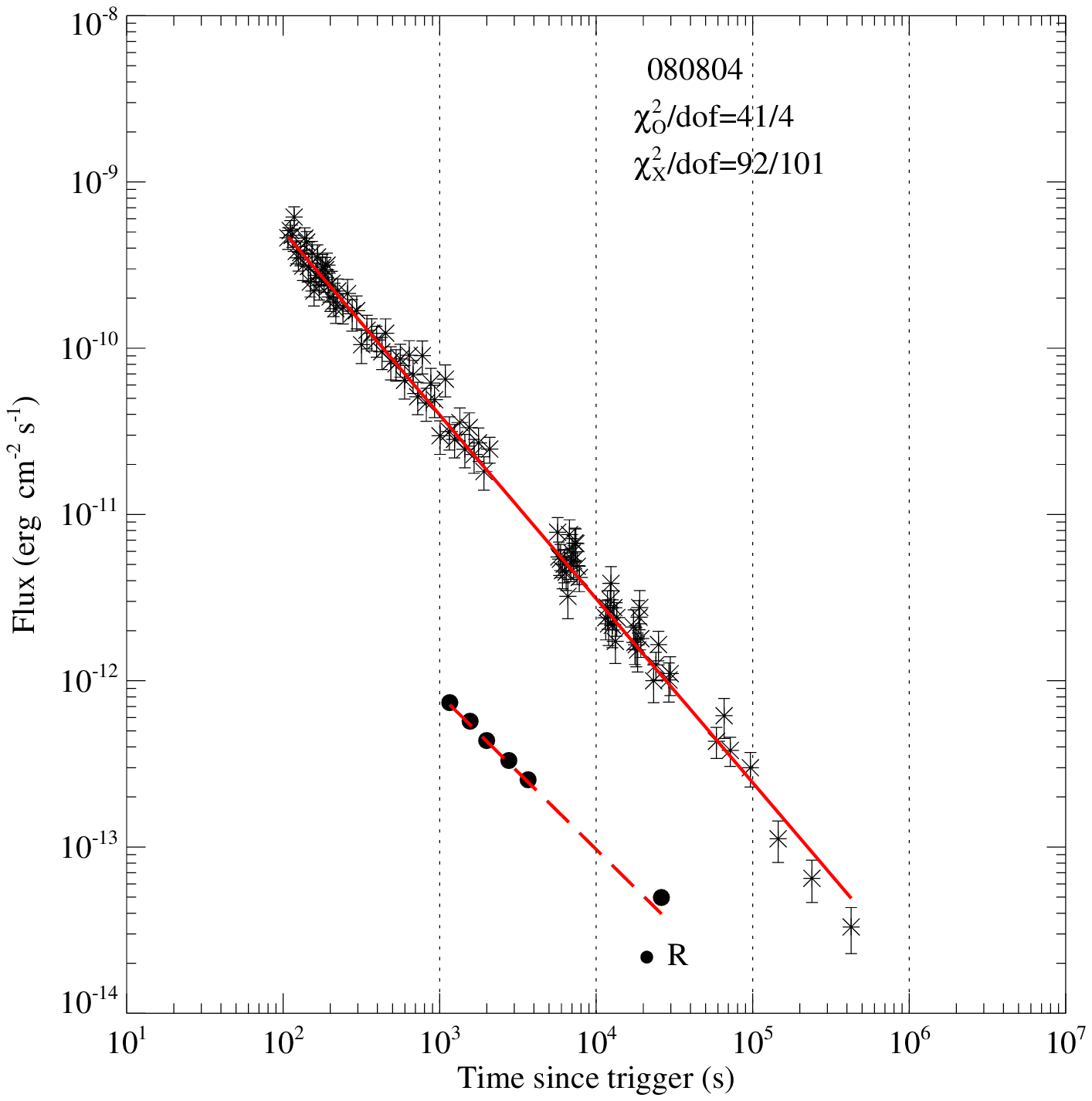}
\caption{Examples of the best afterglow light-curve fittings, that distinguish four shapes with and without breaks for our samples.}\label{Examples}
\end{figure*}

\begin{figure*}
\includegraphics[angle=0,scale=0.350,width=0.3\textwidth,height=0.25\textheight]{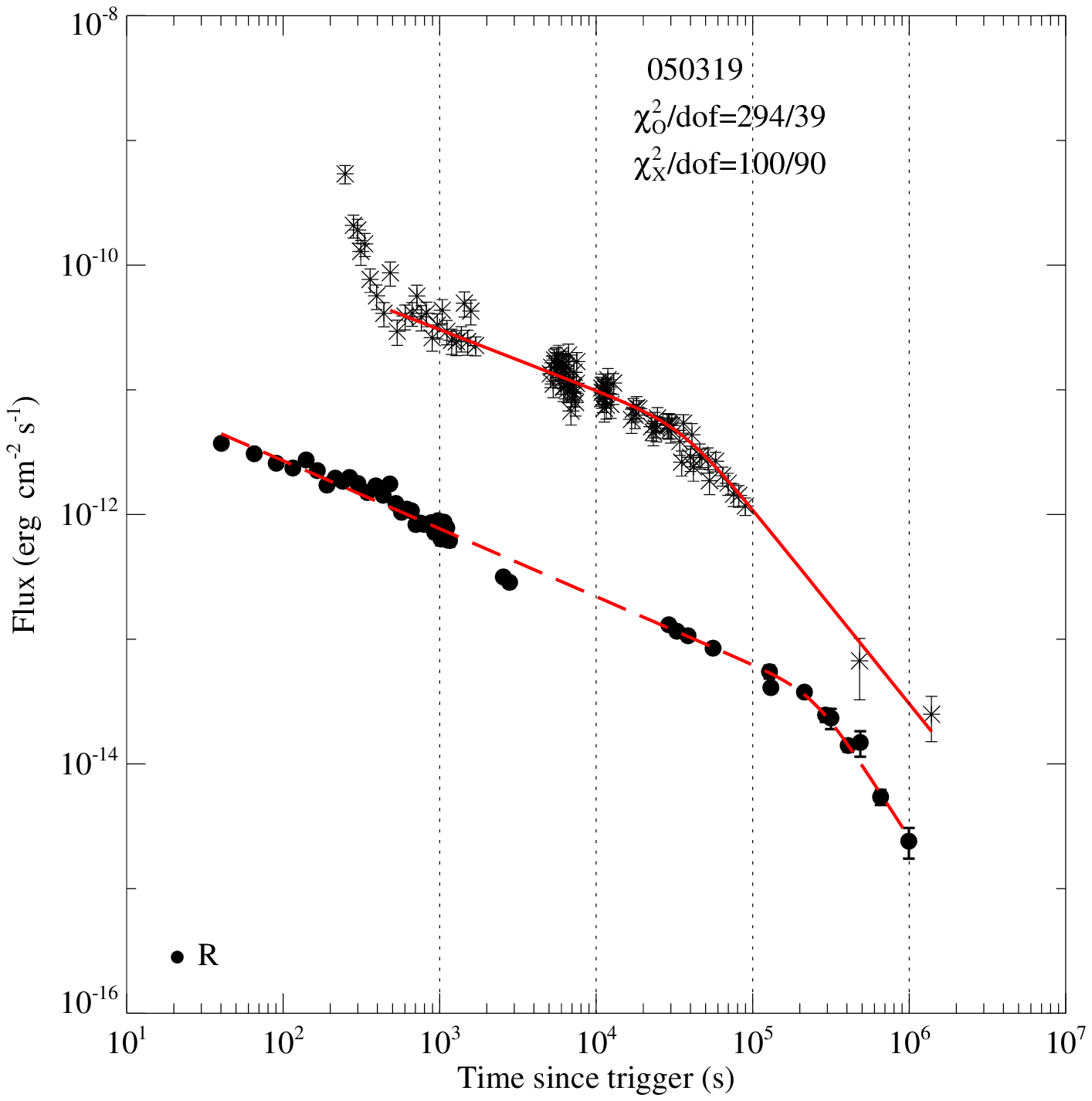}
\includegraphics[angle=0,scale=0.350,width=0.3\textwidth,height=0.25\textheight]{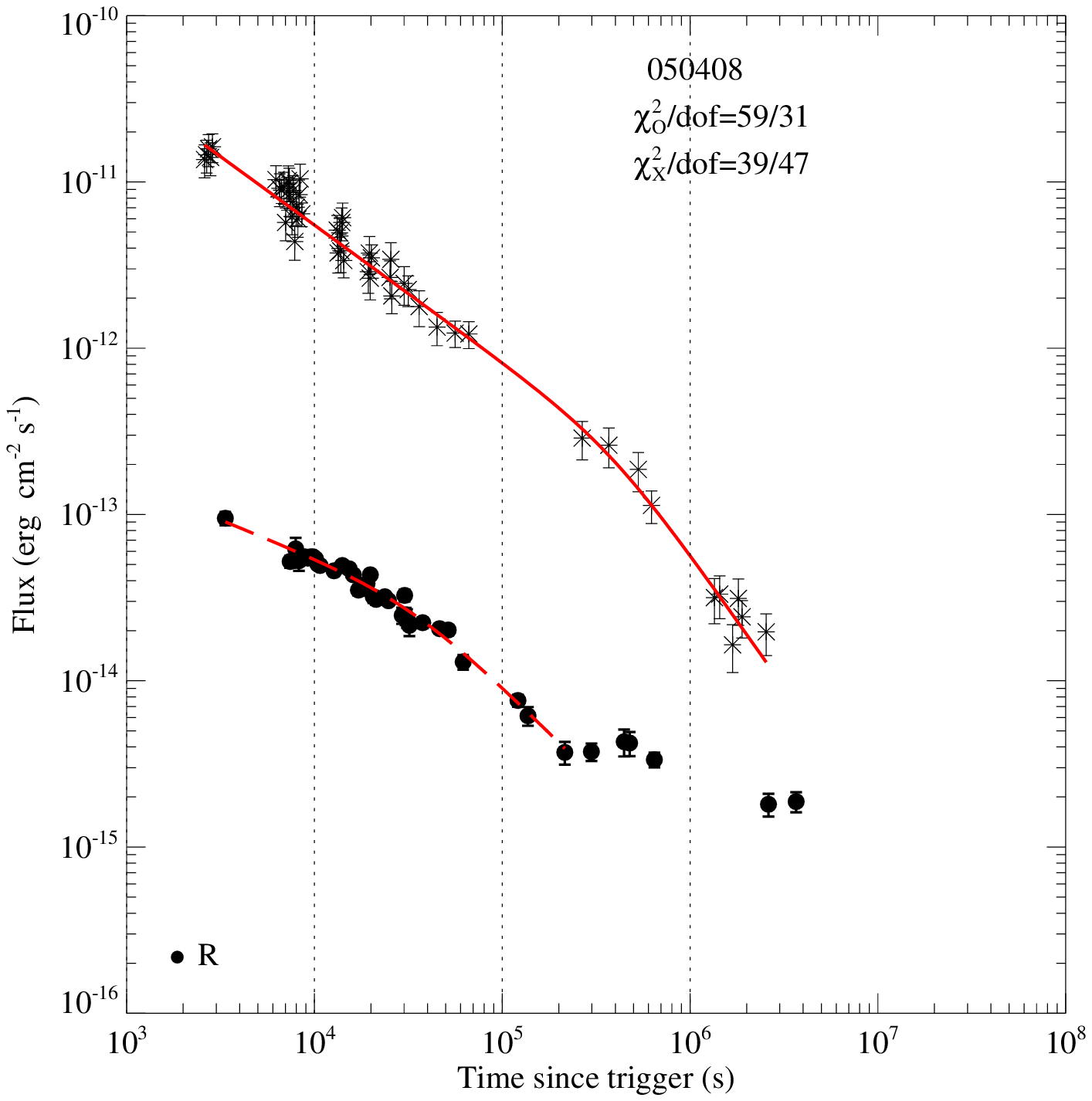}
\includegraphics[angle=0,scale=0.350,width=0.3\textwidth,height=0.25\textheight]{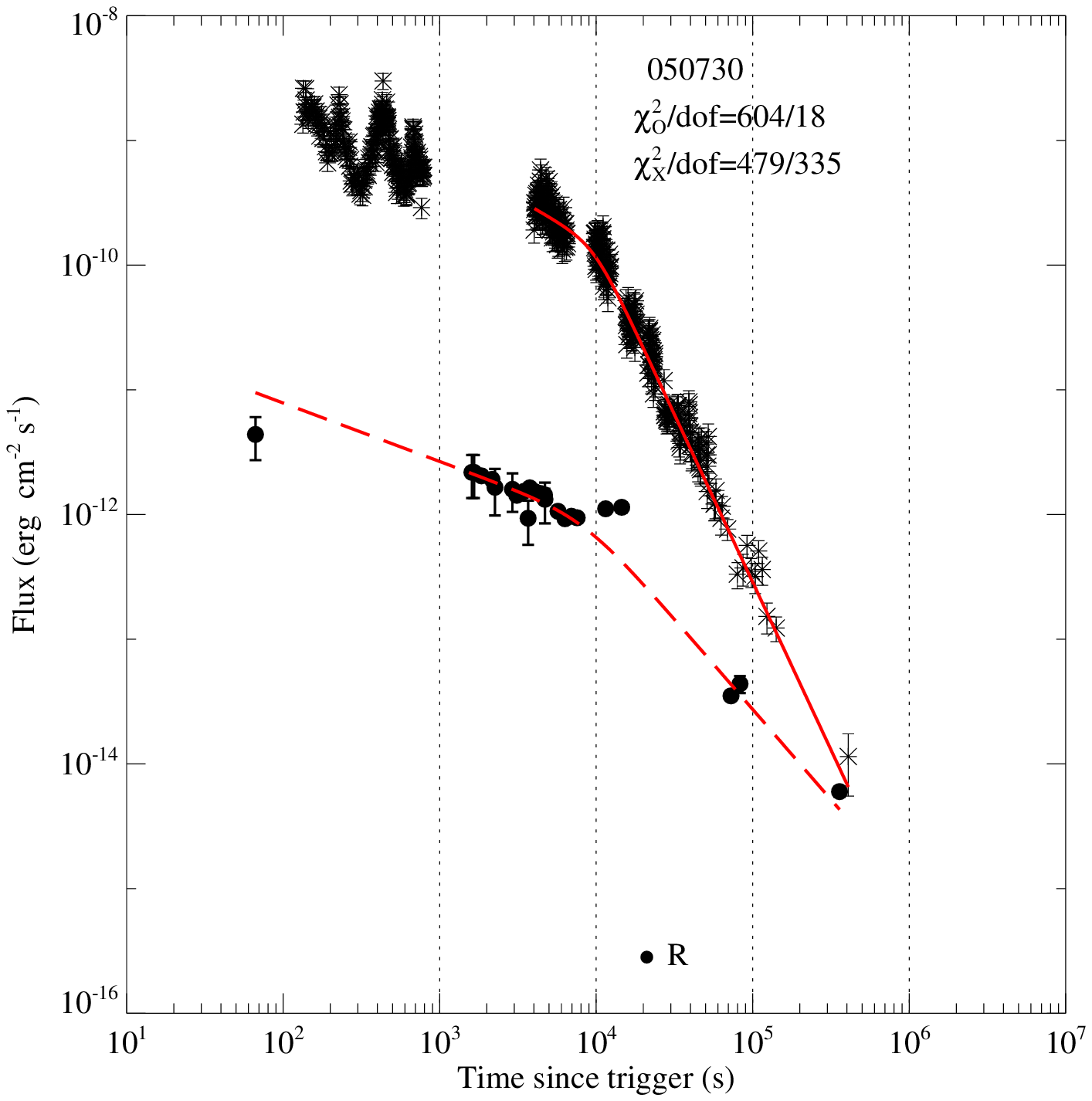}
\includegraphics[angle=0,scale=0.350,width=0.3\textwidth,height=0.25\textheight]{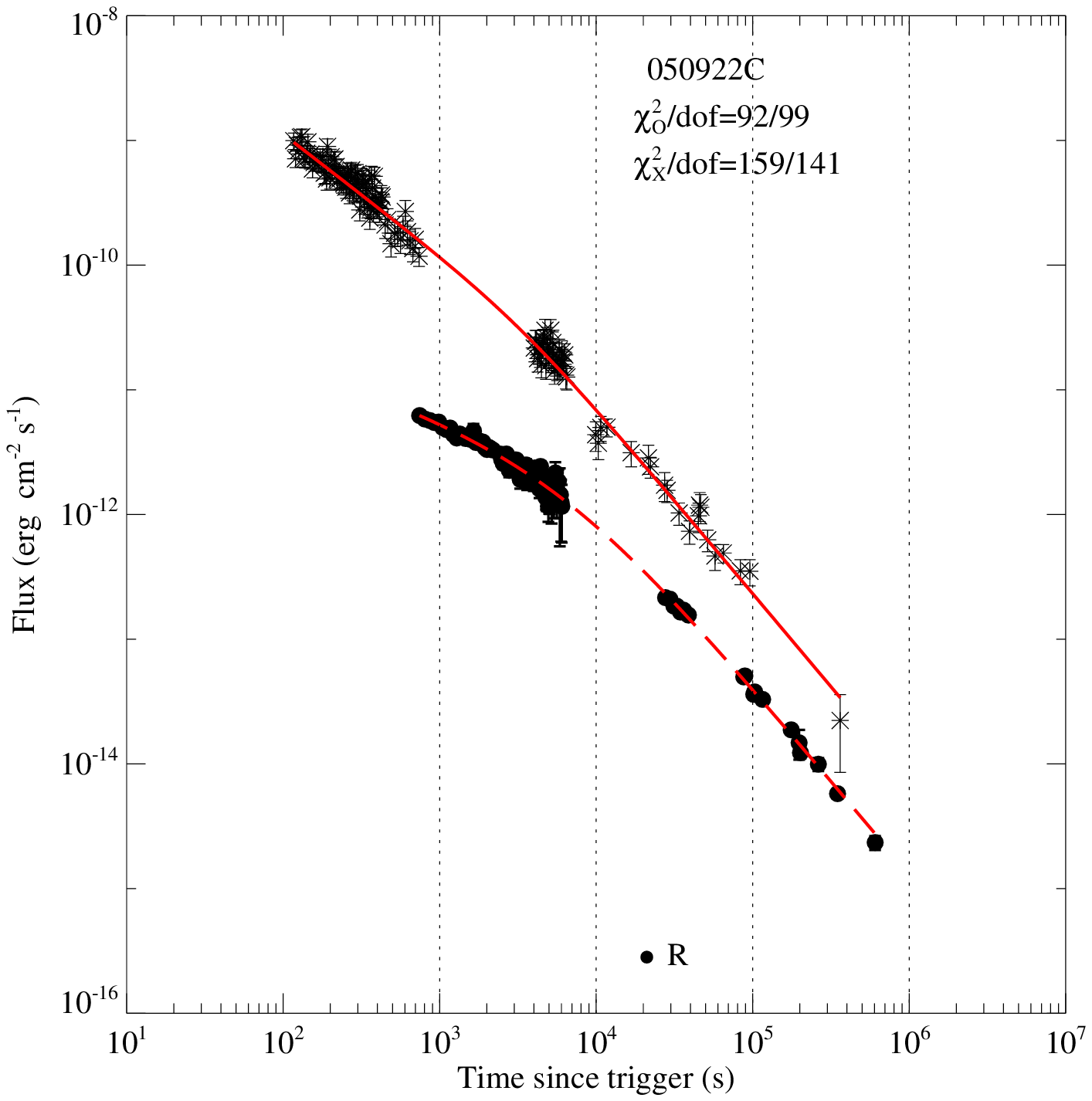}
\includegraphics[angle=0,scale=0.350,width=0.3\textwidth,height=0.25\textheight]{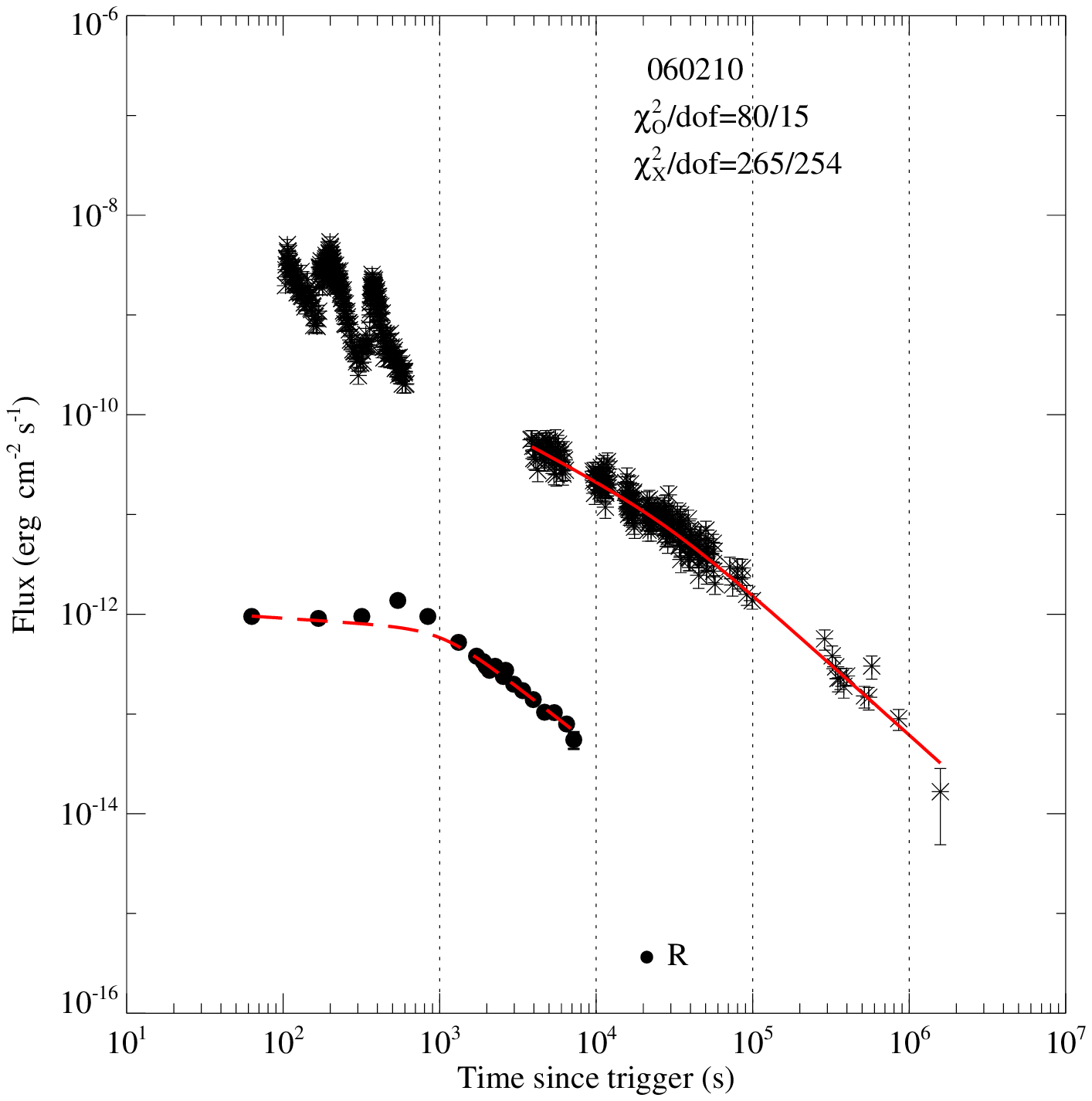}
\includegraphics[angle=0,scale=0.350,width=0.3\textwidth,height=0.25\textheight]{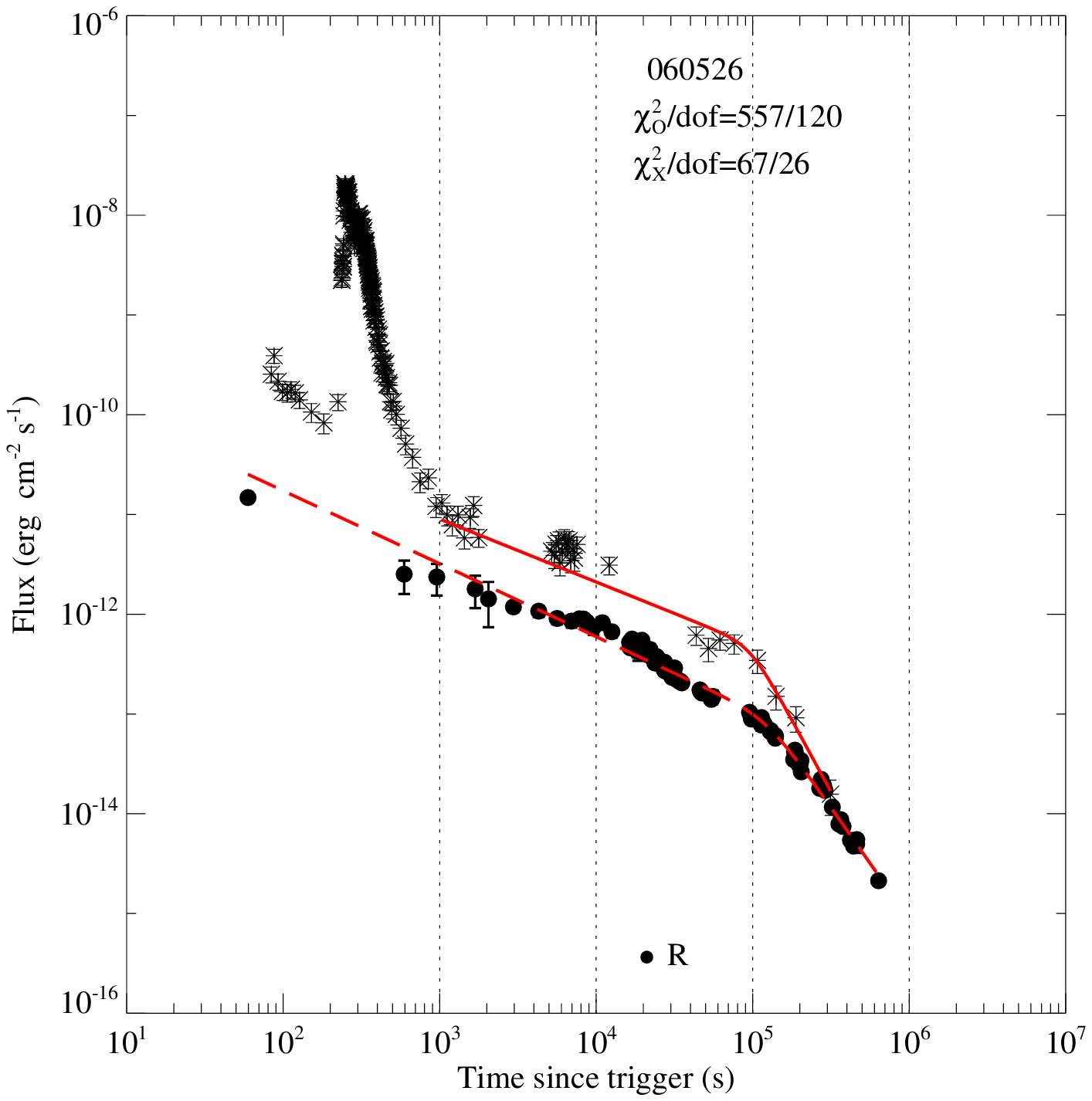}
\includegraphics[angle=0,scale=0.350,width=0.3\textwidth,height=0.25\textheight]{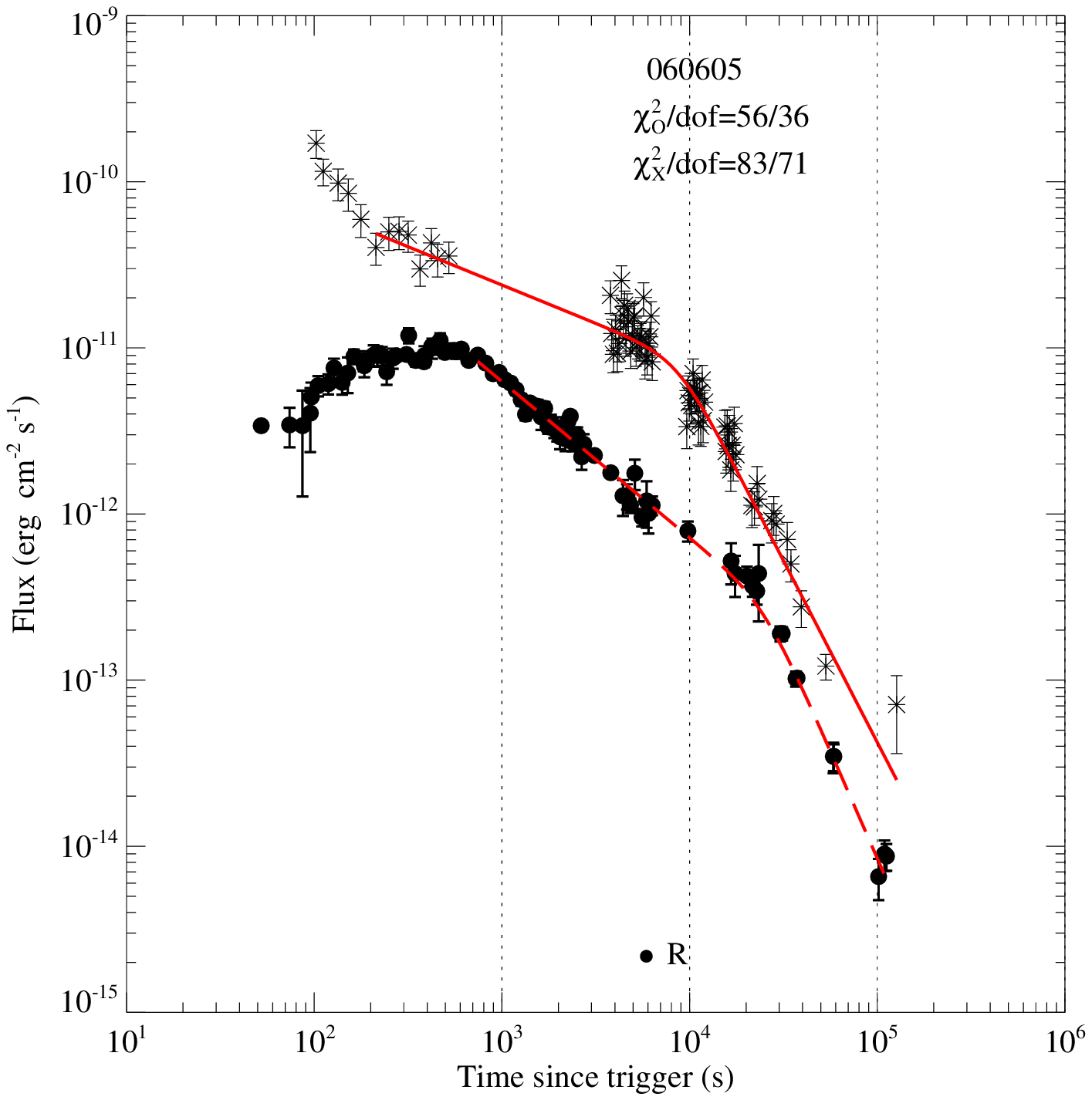}
\includegraphics[angle=0,scale=0.350,width=0.3\textwidth,height=0.25\textheight]{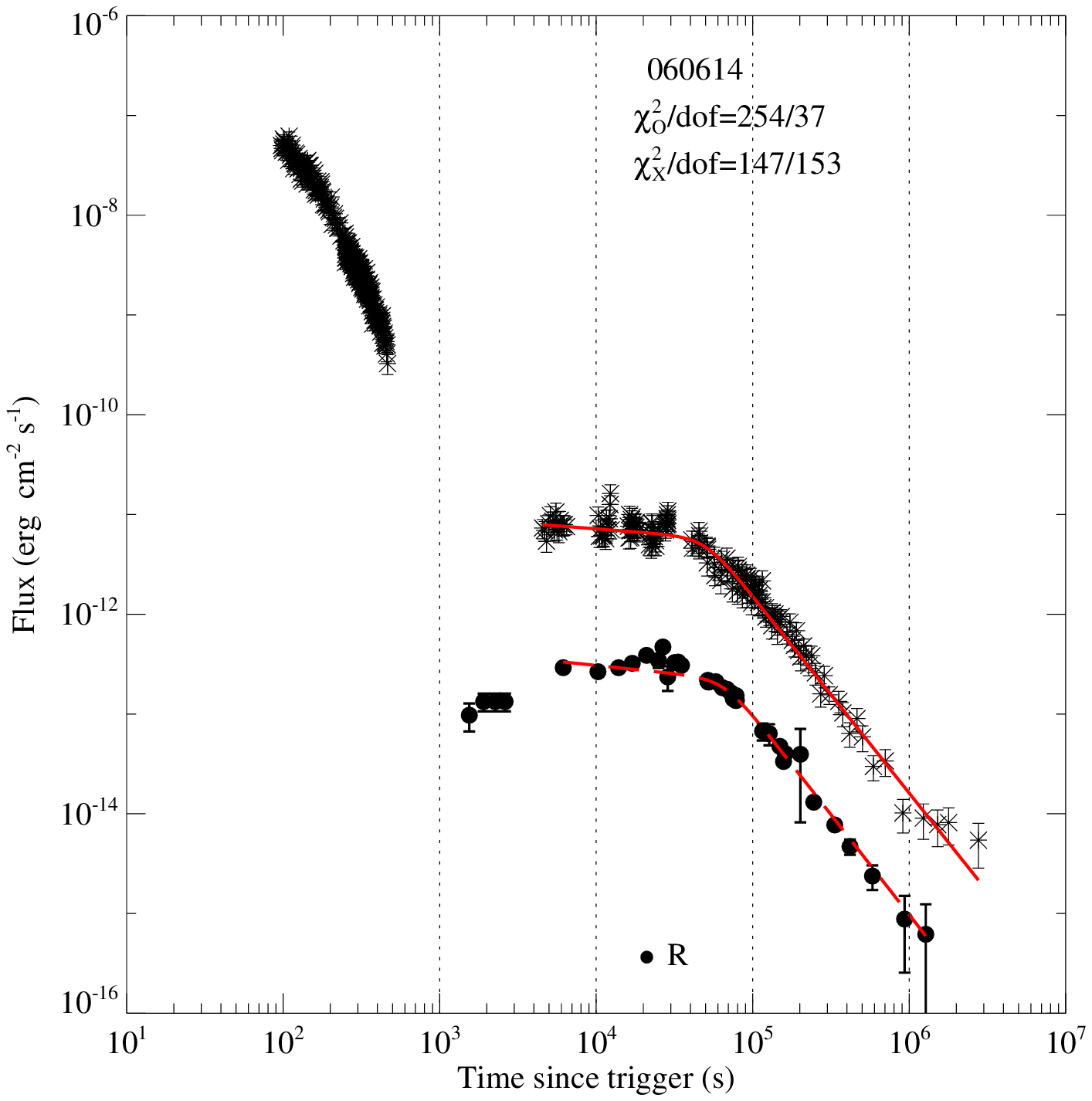}
\includegraphics[angle=0,scale=0.350,width=0.3\textwidth,height=0.25\textheight]{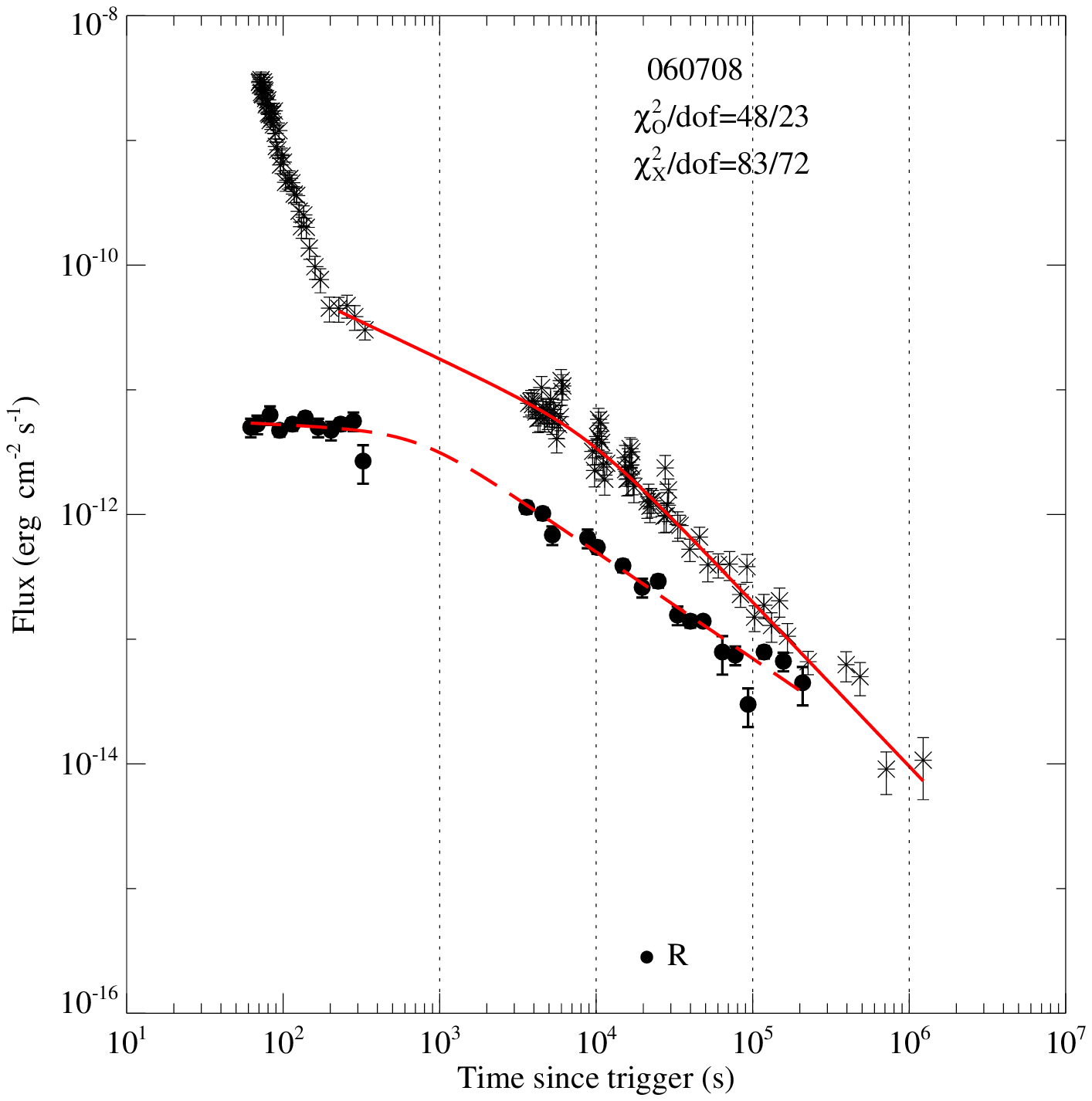}
\includegraphics[angle=0,scale=0.350,width=0.3\textwidth,height=0.25\textheight]{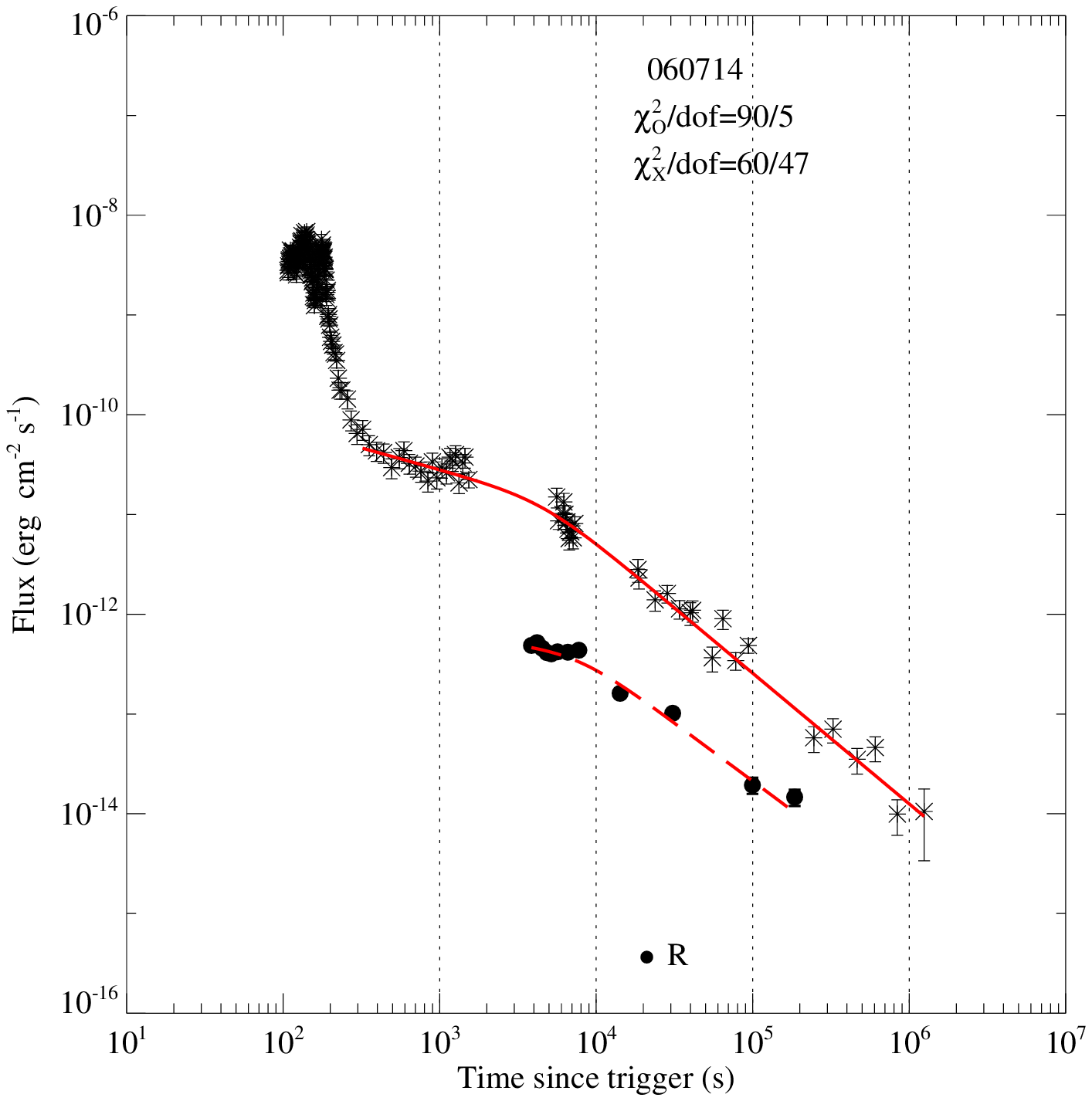}
\includegraphics[angle=0,scale=0.350,width=0.3\textwidth,height=0.25\textheight]{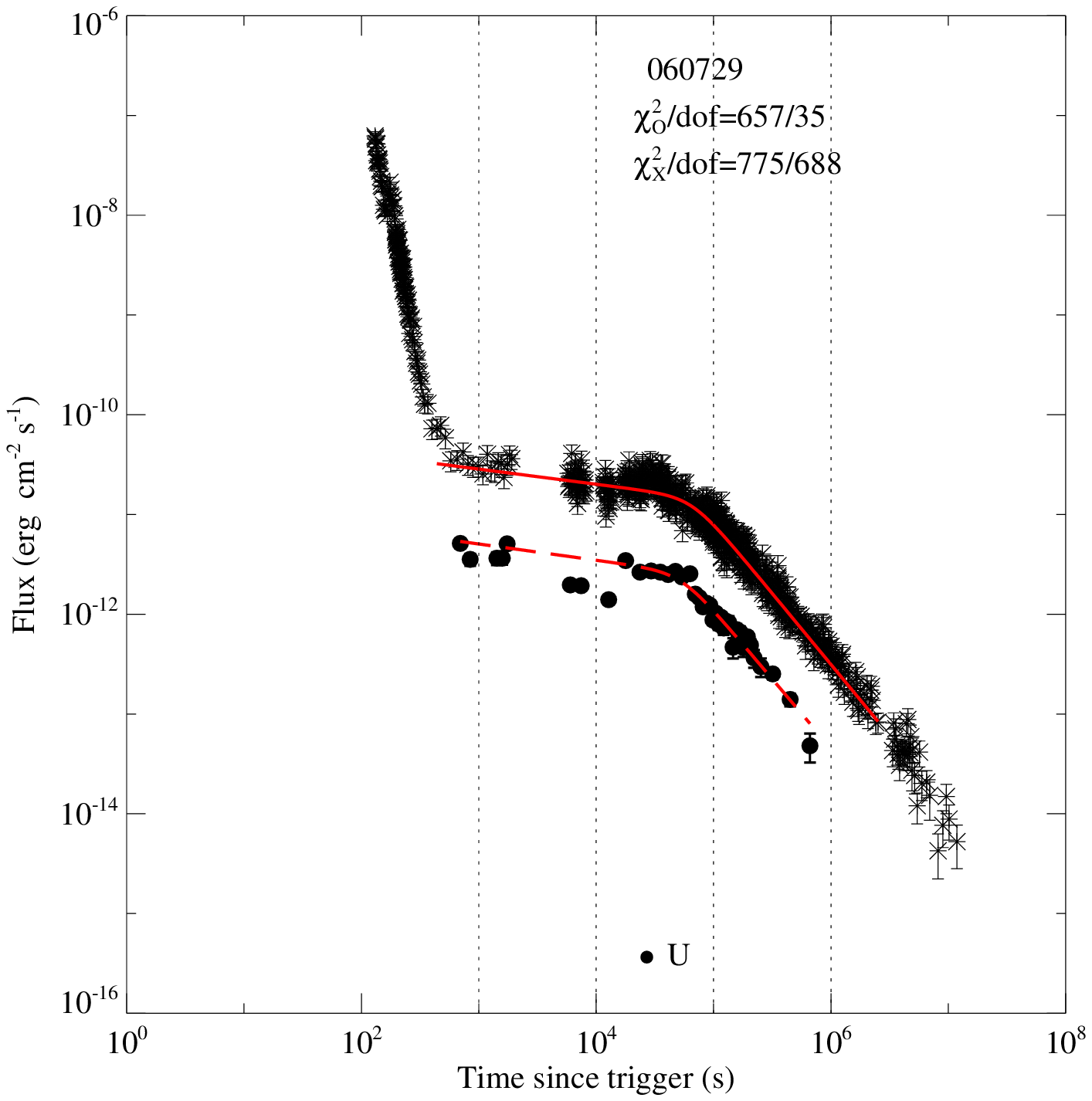}
\includegraphics[angle=0,scale=0.350,width=0.3\textwidth,height=0.25\textheight]{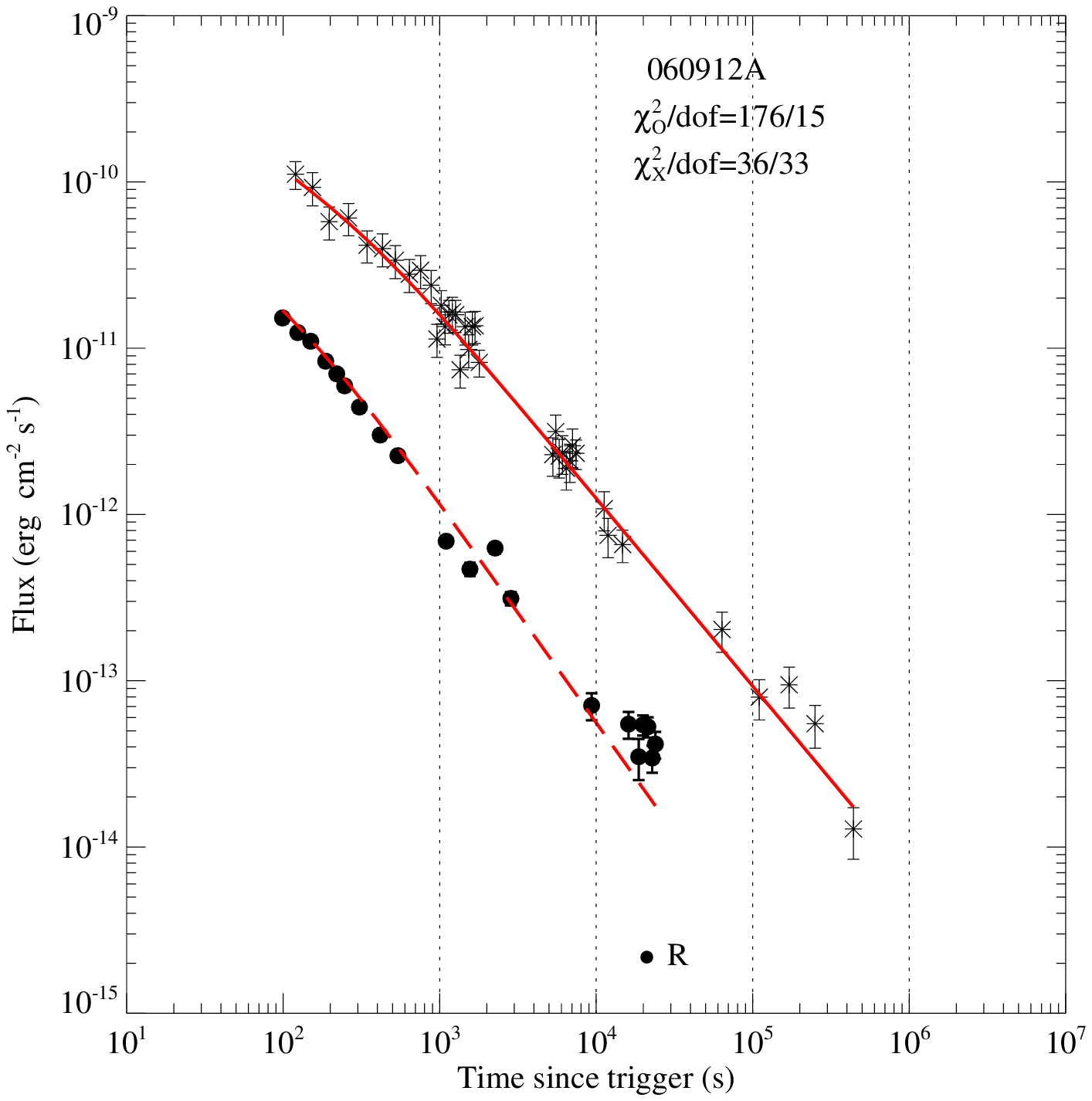}
\caption{The cases in which breaks were detected in their afterglow light-curves both in the X-ray and the optical band.}\label{Both Break}

\end{figure*}

One example of the light-curve fittings is given in Figure \ref{Examples}.

These light-curves were fitted by either a single power-law (SPL)
\begin{equation}
 F = F_0\, t^{-\alpha},
\end{equation}
or a smoothly broken power-law (BKPL)
\begin{equation}
F =  F_0 \, [(t/t_{\rm b})^{\alpha_1\omega}+(t/t_{\rm b})^{\alpha_2\omega}]^{-1/\omega}.
\label{eq:bkpl}
\end{equation}
Here $\alpha$, $\alpha_1$, $\alpha_2$ are the temporal slopes, $t_{\rm b}$ is the break time, $F_{b} = F_0\, 2^{-1/\omega}$ is the flux of the break time, $\omega$ describes the sharpness of a break; the smaller the $\omega$, the smoother the break. We often fixed this parameter to 3 (for a few cases which have a smoother break we fixed it to 1 which improves the fit).
Equation (\ref{eq:bkpl}) is used both for light-curves which have a break and light-curves which exhibit the afterglow onset.
These functions are fitted to the data with a non-linear, least-square fitting using the Levenberg-Marquardt algorithm. We use an IDL routing called mpfitfun.pro\footnote{http://purl.com/net/mpfit} (Markwardt  2009).
Errors in the parameters are calculated either through error propagation or through the bootstrap method, for which we produce 1000 realisations which give a Gaussian distribution, whose standard deviation is an estimate of the parameter error.

The fitting results of the X-ray afterglow, including the break time ($T_{\rm x,b}$), the power-law decay slopes before ($\alpha_{\rm x,1}$) and after ($\alpha_{\rm x,2}$) the break\footnote{Notice that for the light-curves with a single power-law decay we use subscript "2". For example, the X-ray and optical temporal indices are denoted as $\alpha_{\rm x,2}$ and $\alpha_{\rm o,2}$. Similarly, the X-ray photon spectral indices in Group III and IV samples are denoted as $\Gamma_{\rm x,2}$, since the X-ray light curves in these samples are single power laws.}, and time intervals during which model fittings are performed\footnote{The time intervals are selected according to the four rules listed above. We use the subscripts "x" and "o" for parameters in the X-ray and optical bands, respectively.},  [$T_{\rm x,1}, \,T_{\rm x,2}$], are summarized in Table 1.
Based on the fitting parameters, we also calculated the fluence of the X-ray band ($S_{\rm x}$), which is the integral of the observed flux over the time internals. We also calculate the properties in the rest frame for bursts that have a measured red shift (80 of 87 bursts), including the isotropic luminosity at the break ($L_{\rm x,b}$) and the total isotropic energy release in the X-ray band ($E_{\rm x,iso}$).
The photon spectral index before ($\Gamma_{\rm x,1}$) and after ($\Gamma_{\rm x,2}$) the break time, are also given and were retrieved from the Swift website\footnote{$http://www.swift.ac.uk/xrt\_curves/allcurves.php$.}. Most of the spectral indices were derived using the PC mode. Our results are reported in Table 1.

Similarly for the X-rays, in Table 2, we report the fitting results of the optical afterglow over the time interval $[T_{\rm o,1},T_{\rm o,2}]$,
the break time, $T_{\rm o,b}$, the power-law slope before (after) the break, $\alpha_{\rm o,1}$ ($\alpha_{\rm o,2}$), the fluence in the optical band, $S_{o}$, the break luminosity, $L_{\rm o,b}$, and the isotropic energy release in the optical band, $E_{\rm R,iso}$.
We also include in the table the optical afterglow spectral index, $\beta_{\rm o}$,  and the redshift, $z$, found in the literature.
We converted the observed magnitudes to energy fluxes and calculated the magnitude of the $K$-correction by $k=-2.5(\beta_{\rm o}-1)\log(1+z)$ while the Galactic extinction correlation was calculated by a reddening map presented in Schlegel et al. (1998). Any flux contribution in the very late epoch ($\sim 10^6$ s after the GRB trigger) that possibly comes from the host galaxy  is also subtracted.

A comparasion of the relation properties between the X-ray and the optical afterglow is summarized in Table 3.
We compare the X-ray and optical data set and, in particular, the fit results of the light-curves.
We present the ratio of break times ($\frac{T_{\rm x,b}}{T_{\rm o,b}}$).
Next, we calculate the differences in power-law slopes in different temporal segments and/or different energy bands
as well as differences in the spectral slopes in the two bands. We define
\begin{eqnarray}
 \Delta\alpha & \equiv & \alpha_{x}-\alpha_{o}, \\
 \Delta\alpha_{1} & \equiv & \alpha_{x,1}-\alpha_{o,1}, \\
 \Delta\alpha_{2} & \equiv & \alpha_{x,2}-\alpha_{o,2}, \\
 \Delta\alpha_{x} & \equiv & \alpha_{x,2}-\alpha_{x,1}, \\
 \Delta\alpha_{o} & \equiv & \alpha_{o,2}-\alpha_{o,1}, \\
 \Delta\beta & \equiv & \beta_{x}-\beta_{o}, \\
 \Delta\beta_{x} & \equiv & \beta_{x,2}-\beta_{x,1},
\end{eqnarray}
and present their values in Table 3.
We also calculate the X-ray-to-optical ratio of the break luminosity ($\frac{L_{\rm X,b}}{L_{\rm R,b}}$), isotropic energy release ($\frac{E_{\rm X,iso}}{E_{\rm R,iso}}$), as well as the flux at 1 hour ($\frac{F_{\rm X}}{F_{\rm o}}$).

\section{Comparison of the Statistical Properties of the X-ray and Optical Bands}

Totally 833 Swift-GRBs were detected by the Swift Satellite until the end of 2013 (from December 18, 2004 to December 31, 2013), of which only one-tenth (87/833) had good, simultaneously observed X-ray and optical afterglow data.
The fraction is far smaller than of the totally 691/833 Swift-GRBs with good XRT data observations.

The distributions of statistical properties comparing the X-ray with the optical band are shown in Figure \ref{Distributions}.
We compare the distribution of decay slope of the X-ray and the optical bands in Figure \ref{Distributions} (a). We gather all information on X-ray and optical decay slopes. Totally we had 135 decay slope indices for the X-ray ($\alpha_{\rm x}$) and 120 for the optical bands ($\alpha_{\rm o}$)\footnote{Including the pre- and post-break segment of all Group samples. X-ray statistics including temporal slopes: 48 $\alpha_{\rm x,1}$ (Group I+Group II) and 87 $\alpha_{\rm x,2}$: (Group I+Group II+Group III+Group IV). Optical statistics including: 37 $\alpha_{\rm o,1}$ (Group I+Group III) and 87 $\alpha_{\rm o,2}$: (Group I+Group II+Group III+Group IV).}. The distribution of $\alpha_{\rm x}$ ranges within [-0.01, 2.89] and of $\alpha_{\rm o}$ within [-0.08, 2.59]. The X-ray emission usually decays faster than the optical emission\footnote{As summarized in Table 3, overall the $\Delta\alpha$ value of 75 out of 111 (68\%) GRBs are greater than 0 (13 of 24 cases for $\Delta\alpha_{1}$ and 62 of 87 cases for $\Delta\alpha_{2}$), which indicates that the X-rays usually decay faster than the optical.}. This is consistent with previous findings (e.g. Oates et al. 2011; Zaninoni et al. 2013), and is consistent with theoretical predictions of the afterglow model (e.g. Sari et al. 1998; Gao et al. 2013).
In Figure \ref{Distributions} (b), (c), (d), (e), we show the distributions of the break times and the break luminosity for the bursts which have a break behavior in their afterglow light-curves ($t_{\rm b}$, $L_{\rm b}$; X-ray: Group I and Group II; Optical: Group I and Group III), of the fluences ($S_{\rm x}, S_{\rm o}$), and finally the distribution of the isotropic energy ($E_{\rm X,iso}, E_{\rm R,iso}$). The break time ranges from $\sim$ hundreds to $\sim 10^{6}$s with a typical value $\sim 10^{4}$s, and the X-ray and the optical distributions are similar.
The peak distribution of X-ray fluence is $\sim10^{-6} {\rm erg \, cm^{-2}}$ and in the optical band it is $\sim10^{-8} {\rm erg \, cm^{-2}}$. The isotropic break luminosity ($L_{\rm X,b}$ or $L_{\rm R,b}$) with a wider distribution, ranges from $\sim10^{43}$erg s$^{-1}$ to $\sim10^{49}$erg s$^{-1}$ for the X-rays with a typical value of $10^{47}$ erg s$^{-1}$ and from $\sim10^{42}$ erg s$^{-1}$ to $\sim10^{47}$ erg s$^{-1}$ for the optical band with a typical value of $10^{46}$ erg s$^{-1}$.  The isotropic energy release in the X-ray band ($E_{\rm X,iso}$) generally ranges from $\sim10^{49}$ to $\sim10^{53}$ erg, while for the optical band ($E_{\rm R,iso}$) it is from $10^{48} $ to $10^{51}$ erg. The peak distribution value of the X-ray band is $\sim10^{52} {\rm erg}$ and of the optical band $\sim10^{50} {\rm erg}$.

Figure \ref{Correlations} shows the correlation between the X-ray and the optical properties for Group I.
We examine the pair correlations for $\Delta\alpha_{1}-\Delta\alpha_{2}$ and $\Delta\alpha_{\rm x}-\Delta\alpha_{\rm o}$, and obtain $r$=0.40 with $p$=0.42 for $\Delta\alpha_{1}-\Delta\alpha_{2}$ and $r$=0.40 with $p$=0.55 for $\Delta\alpha_{\rm x}-\Delta\alpha_{\rm o}$, which indicates that these are only weak correlations. Notices that the peak value of the distribution for $T_{\rm x,b}/T_{\rm o,b}$ is around 0 in the logarithmic space. This indicates that a good fraction of Group I GRBs have the same break time in both the X-ray and the optical band. This is generally related to the achromatic break behavior. A detailed analysis of the chromaticity of afterglow behavior is presented in Wang et al. (2015).

We compare the start ($T_{\rm x,1}, T_{\rm o,1}$) and end times ($T_{\rm x,2}, T_{\rm o,2}$) of the X-ray and optical band observations in Figure \ref{Times}.
The $T_{\rm x,1}-T_{\rm o,1}$ panel (Figure \ref{Times} (a)) shows that the data points generally cluster around the equality line in the logarithmic space, and the X-rays range from $\sim10^{2}$ seconds to $\sim10^{5}$ seconds and the optical band from $\sim 10$ seconds to $\sim10^{5}$ seconds.
Most of the data points are below the equality line in the logarithmic space for the $T_{\rm x,2}-T_{\rm o,2}$ panel (Figure \ref{Times} (b)), and the distribution is from $\sim10^{4}$ to $\sim10^{7}$ seconds for the X-ray band and from $\sim 10^{2}$ to $\sim10^{7}$ seconds for the optical band.
This indicates that the clear afterglow component was observed for the optical band earlier than for the X-ray band (it is likely that the early X-ray afterglow is usually polluted by the steep decay), and also for the end time of the observation.

\section{Comparison of the X-ray and Optical Temporal and Spectral Indices}

Our discussion concerns the slow cooling case, since most GRBs in our sample have the starting time ($T_{x,1}$) hundreds of seconds after the trigger.
If the cooling frequency ($\nu_{c}$) lies between that of the X-ray ($\nu_{x}$) and the optical ($\nu_{o}$) bands ($\nu_{m}<\nu_{o}<\nu_{c}<\nu_{x}$), then the differences in the temporal indices between the X-ray and the optical bands are the following:
(i) for the cases with no energy injection ($q=1$), one has $\Delta\alpha=\frac{1}{4}$ for the interstellar (ISM) medium scenario (Sari et al. 1998) and $\Delta\alpha=-\frac{1}{4}$ for the wind scenario (Chevaler \& Li 2000);
(ii) for the energy injection case with $q=0$ due to pulsar spin-down (Dai \& Lu 1998; Zhang \& M\'{e}szaros 2001), one has $\Delta\alpha=\frac{1}{2}$ for the ISM case and $\Delta\alpha=-\frac{1}{2}$ for the wind case, which have the maximum difference e between the X-ray and optical bands (e.g. Oates et al. 2009);
(iii) The difference in the spectral indices between the X-ray and the optical bands is at most $\Delta\beta=\frac{1}{2}$ for both the ISM and the wind models (the cooling break).
If the cooling frequency is above the X-rays band ($\nu_{m}<\nu_{o}<\nu_{x}<\nu_{c}$), one has the same spectral and temporal indices in both the X-ray and optical bands, regardless of the medium type and whether or not there is energy injection.  The ranges of relations in the temporal and spectral indices are summarized in Table 5 (see also Table 2 of Zhang et al. 2006). The expressions of $\Delta\alpha$ and $\Delta\beta$ can be summarized in the following two formulae.
\begin{eqnarray}
\label{n} \Delta\alpha= \left\{ \begin{array}{ll}
\frac{1}{2}           , & \nu_{m}<\nu_{o}<\nu_{c}<\nu_{x},  \textrm{injection (q=0)}, (\textrm{ISM})\\
\frac{1}{4}           , & \nu_{m}<\nu_{o}<\nu_{c}<\nu_{x},  \textrm{no injection (q=1)}, (\textrm{ISM})\\
0                     , & \nu_{m}<\nu_{o}<\nu_{x}<\nu_{c},  \textrm{injection (q=0) and no injection (q=1)}, (\textrm{ISM and wind})\\
-\frac{1}{4}          , & \nu_{m}<\nu_{o}<\nu_{c}<\nu_{x}, \textrm{no injection (q=1)}, (\textrm{wind})\\
-\frac{1}{2}          , & \nu_{m}<\nu_{o}<\nu_{c}<\nu_{x}, \textrm{injection (q=0)}, (\textrm{wind})\\
\end{array} \right.
\end{eqnarray}

\begin{eqnarray}
\label{n} \Delta\beta= \left\{ \begin{array}{ll}
0           , & \nu_{m}<\nu_{o}<\nu_{x}<\nu_{c} ,  \textrm{injection (q=0) and no injection (q=1)}, (\textrm{ISM and wind})\\
\frac{1}{2} , & \nu_{m}<\nu_{o}<\nu_{c}<\nu_{x} ,  \textrm{injection (q=0) and no injection (q=1)}, (\textrm{ISM and wind})\\
\end{array} \right.
\end{eqnarray}

\subsection{Temporal and spectral indices for the 4 groups}

In Figure \ref{Temporal Group}, we present the correlations between the temporal decay indices in the optical band versus those in the X-ray band ($\alpha_{x}-\alpha_{o}$ panel). We present the correlation separately for our 4 groups. In the figure we mark the range deviating by -0.25 and 0.25 from the equality line, which is the range expected from the standard forward shock model for different circum-stellar media (ISM and Wind-like), as well as the range of -0.5, 0.5 (dashed lines) which is related to the energy injection model for different circum-stellar media (ISM and Wind-like).
Note that if the points lying above the equality line decay more quickly in the optical than in the X-ray band, then the data are consistent with the wind-like model; if the points below the equality line decay more quickly in the X-ray than in the optical band, then the data are consistent with the ISM model. In the Figure \ref{Spectral Group}, we similarly present the correlation between the spectral indices in the optical band versus those in the X-ray band, for each of the 4 groups. We indicate the line corresponding to $\beta_{x}-\beta_{o}$ lying within the range [0.0, 0.5] which is expected for different spectral regimes of the external shock model.

The difference in temporal indices between the X-ray and the optical bands are compared to the empirical relations for various external shock models and are summarized in Table 4.
For Group I (Figure \ref{Temporal Group} (a)), 18 of 24 and 16 of 24 data points fall within the external shock model ranges (-0.5,0.5) for $\alpha_{\rm x,1}-\alpha_{\rm o,1}$ and $\alpha_{\rm x,2}-\alpha_{\rm o,2}$, respectively. Out of these, 13 of 24  and 10 of 24 data points fall within the standard afterglow model ranges (-0.25,0.25).
Moving to Group IV (Figure \ref{Temporal Group}(d)), there are 27 of 30 data points which fall within (-0.5,0.5), which includes 13 of 30 data points within (-0.25,0.25) for $\alpha_{\rm x,2}-\alpha_{\rm o,2}$.
This indicates that the data for Group I and Group IV satisfy the external shock model for the global behavior.
However, for Group II, although there are 17 of 24 data points which fall within [-0.5,0.5] and includes 5 of 24 points within [-0.25,0.25] for $\alpha_{\rm x,1}-\alpha_{\rm o,2}$, only 8 of 24 data points fall within (-0.5,0.5) and only rare points (only one case)  within the ranges of (-0.25, 0,25) for $\alpha_{\rm x,2}-\alpha_{\rm o,2}$.
Similar to Group II, for Group III, 3 of 9 data points fall within [-0.5,0.5] and no point within [-0.25,0.25] for $\alpha_{\rm x,2}-\alpha_{\rm o,1}$. However there are 8 of 9 data points which fall within [-0.5,0.5] and 6 of 9 points within [-0.25,0.25] for $\alpha_{\rm x,2}-\alpha_{\rm o,2}$.
These observations indicate that the data are generally inconsistent with the external shock model for the global behavior for Group II and Group III. Since these two groups include GRBs that potentially have a chromatic behavior (which is not predicted in the simplest afterglow model), this result is understandable: The GRBs that do not comply with the models may require two emission components to interpret the X-ray and optical emission, respectively. One of the two components should still come from the external shock. The data indeed show that for a certain section of the afterglow the data are still be consistent with the external shock model. A full investigation of this issue is presented in Wang et al. (2015).

Based on the statistics of $\alpha_{\rm x}-\alpha_{\rm o}$ relation with the definitions of 4 Group samples, within the error, totally
we find 54 of 87 (62\%) GRBs that are consistent with the standard afterglow model ($\Delta\alpha=[-0.25,0.25]$).
However, when more advanced modeling is invoked (e.g. long lasting central engine emission, reverse shock emission, or structured jets), 79 of 87 (91\%) GRBs may also satisfy the external shock models ($\Delta\alpha=[-0.5,0.5]$).
We also find that 53 of 87 (61\%) GRBs correspond to the ISM model, and 34 of 87 (39\%) GRBs correspond to the wind model.

The spectral indices for the GRBs in each Group are shown in the $\beta_{\rm x}-\beta_{\rm o}$ panels of Figure \ref{Spectral Group}, within error, the data generally fall well within the range of [0, 0.5], which is expected for the external shock models (the data of Group II have a little scatter around the external shock model ranges). Notice that we have collected the optical spectral indices from the literature. No time-resolved optical spectral analysis was available for most GRBs. A detailed time-resolved joint analysis of afterglow power density spectrum using X-ray and optical data will be presented in a separate paper (Li et al. 2015, in preparation).

In order to find the properties of the mean values of the temporal and spectral indices,
in Figure \ref{Mean}, we show the distributions of the pre-, and post-break decay slopes, and the spectral indices for the X-ray and the optical bands and compare them\footnote{The pre-break decay slope indices include 48 $\alpha_{\rm x,1}$ (Group I+Group II) and 37 $\alpha_{\rm o,1}$ (Group I+Group III). The post-break decay slope indices include 87 $\alpha_{\rm x,2}$ and 87 $\alpha_{\rm o,2}$ (for all Groups). The spectral indices include 87 $\beta_{\rm x,2}$ and 80 $\beta_{\rm o}$ (for all Groups)}. All the distributions can be reasonably fitted with a Gaussian function in order to find the mean values. The decay slope of the pre-break slopes are $0.46 \pm 0.01$ (optical) and $0.68 \pm 0.03$ (X-rays). This indicates that a shallow decay segment is commonly detected in both the X-ray and optical band at the early time. The corresponding values of the  post-break slopes are $1.23 \pm 0.03$ (optical) and $1.48 \pm 0.02$ (X-rays), for which the value of $\alpha_{\rm x,2}-\alpha_{\rm o,2}$ is 0.25, and fit very well with the case of an ISM medium, and slow cooling with spectrum regime ($\nu_{m}<\nu_{o}<\nu_{c}<\nu_{x}$). The spectral indices are $0.99\pm 0.23$ (X-ray) and $0.69\pm 0.25$ (optical), also well-consistent with the external shock model.

\subsection{Temporal indices for different epochs}

In order to study evolution in the relations, we compare the X-ray and the optical temporal indices at different epochs, and divide the data into 4 epochs and test the data against the external shock model.

We study a sample of GRBs that is selected according to the following: (i) We select the GRBs with redshift measurements (80 out of 87 bursts) so that their observed times can be corrected to the burst rest-frames; (ii) We chose four epochs in the rest frames: epoch 1: $< 10^{3} {\mathrm s}$, epoch 2: $10^{3}{\mathrm s}-10^4 {\mathrm s}$, epoch 3: $10^4 {\mathrm s} - 10^5 {\mathrm s}$, and epoch 4: $>10^{5}s$. The reason to choose these different epochs is to investigate how the joint X-ray/optical emission properties are consistent or not consistent with each other in different epochs.
(iii) We selected the cases for which there was at least one epoch during which both X-ray and optical data are available. (iv) If a light-curve break time (X-ray or optical) lies within one epoch, we include both the pre- and post-decay indices in this epoch. Using these selection criteria we finally get 50, 97, 80, 35 cases in the four epochs, respectively. For the different samples the number of cases are (19, 45, 41, 11) for Group I, (14, 26, 15, 7) for Group II, (5, 7, 5, 2) for Group III, and (12, 19, 19, 15) for Group IV.

As shown in Figure \ref{Temporal Bins}, in the $\alpha_{x}-\alpha_{o}$ panels for our epoch samples, consider the error, 42 of 50 (84\%), 76 of 97 (78\%), 60 of 80 (75\%) and 28 of 35 (80\%) data points for $\Delta\alpha$ fall within our external shock model ranges (-0.5,0.5) for epochs 1, 2, 3 and 4, respectively. Out of these, 23 of 50 (46\%), 45 of 97 (46\%), 37 of 80 (46\%) and 19 of 35 (54\%) data points are within the standard afterglow model ranges (-0.25,0.25). This indicates that the data are generally consistent with the external shock models and there is no significant evolutionary effect in our epoch samples. The results can be also seen in Table 3.

In order to consolidate these results, and in particular to show that the data are not randomly distributed, we compare the observational data with a simulation using the Monte Carlo (MC) method. We simulated a random set of decay slopes with a Gaussian distribution in both the X-ray and optical bands, lying within the observed range of [0,3]. For each epoch the simulated set contains the same number of points as the observed set. The simulated points (red open circles) are also shown in Fig.13. We evaluate the consistency between the simulated and observed sets by a Kolmogorov-Smirnov (K-S) test and use the probability of the K-S test ($ P_{\rm K-S}$) to measure the consistency.  A larger value of $P_{\rm K-S}$ indicates a better consistency. If a value of $ P_{\rm K-S}$ is less than 0.1, then there is no consistency. Our results are shown in Table 5, which suggests that the simulated and observed data are generally inconsistent with each other. We conclude that the observed data sets are not randomly distributed.

\section{Confronting the Break Behavior with Various External Shock Models}

The canonical X-ray light-curve (Zhang et al 2006) can be described as an initially steep decay phase I (with a temporal power-law index, $\alpha_{\rm x}\sim-3$ and $F_{\rm (t)}\propto t^{\alpha}$), then changing to a shallow decay segment (or plateau) II ($-1.0<\alpha_{\rm x}<0$). This phase is then followed by a normal decay segment III ($-1.5<\alpha_{\rm x}<1.0$) and finally the light-curve changes into a post jet-break decay IV. Compared to the X-rays, the synthetic optical light-curves can be generally summarized as eight emission components with distinct physical origins\footnote{These eight components are: Ia. early prompt optical emission, Ib. early reverse shock emission, II. early shallow decay segment, III. standard afterglow (normal decay), IV. post jet-break phase, V. optical flares, VI. late re-brightening emission, VII. late SN-GRB bump.}, the segments II-IV are usually observed as counterparts in optical afterglow light-curves (see Li et al. 2012, Liang et al. 2013). Thus, break behaviors in afterglow light-curves are commonly observed, the shallow-to-normal breaks at early times and the normal-to-jet break behavior at late times. Another possibility is a spectral break which is due to the cooling frequency crossing the X-ray or optical band. We summarized the properties of afterglow break behaviors, comparing the X-ray with optical bands and their statistic behaviors in Table 6.

The external forward shock model provides well-predicted spectral and temporal properties and allows for a wide variety of behaviors, it
has been successful in explaining the broadband afterglow emission (see, e.g., Rees \& M\'esz\'aros 1992; M\'esz\'aros \& Rees 1993, 1997; Paczy\'nski \& Rhoads 1993; Sari, Piran \& Narayan 1998).
The relations between spectral and temporal properties are denoted "closure relations", which have been widely used to interpret the rich multi-wavelength afterglow observations.
The temporal index $\alpha$ and spectral index $\beta$ in various afterglow models and the typical numerical values of each case are summarized in Table 5 (see also Table 2 in  Zhang et al. 2006; Sari et al. (1998, 1999); Chevalier \& Li (1999)).

The closure relations ($\alpha-\beta$) of the various external shock models in different regimes for all the cases are shown in Figure \ref{Closure Relations}, together with the closure relations for the ISM and wind-like models (Liang et al. 2008). The X-ray data are shown in Figures \ref{Closure Relations} (a) for the ISM model and \ref{Closure Relations} (b) for the wind-like model.
For X-ray break samples (Group I+Group II), the pre-break segments (the black solid) are shallower than the model predictions according to the standard afterglow model, generally due to the energy injection causing a shallow decay segment. The post-break segments (the blue triangles) are well-consistent with the standard afterglow model, or fall well within the jet-break model region. We find the data for the X-ray non-break samples (the pink half-solid points, Group III+Group IV) are well-consistent with the standard afterglow model.
The optical data are shown in Figure \ref{Closure Relations} (c) and (d) for the ISM model and the wind-like model\footnote{Same conventions, but since there are no time-resolved optical spectral data analyses available, we use the same spectral index for both the pre- and post-break segments.}.
It can be seen that optical data in jet-break regions (are clustered in our optical break samples: Group I and Group III) are fewer than X-ray data.  By contrast, the fraction of GRBs which fall in the jet-break region for optical non-break samples is significantly larger than in the X-ray cases.
A few GRBs fall in the jet-break regions with spectral regime I ($\nu_{opt}>max(\nu_{m},\nu_{c})$).

\subsection{Energy injection break}

The shallow decay segments were commonly detected in the early afterglow light-curves, generally believed to be caused by the GRBs outflow catching up with the external forward shock and with a long-lasting energy injection into the external shock which refreshes the afterglow emission.
It can be defined by the afterglow theory, by which the temporal decay is shallower than what is predicted in
the external shock model.

We defined the shallow decay segment by the closure relation: $\alpha_{1}<\frac{3\beta_{1}-1}{2}$ for the spectral regime I ($\nu>max(\nu_{m},\nu_{c})$) for both the ISM and the wind-like models, $\alpha_{1}<\frac{3\beta_{1}}{2}$ for ISM model and $\alpha_{1}<\frac{3\beta_{1}+1}{2}$ for wind-like model for spectral regime II ($\nu_{m}<\nu<\nu_{c}$), respectively.
We find that generally a very high fraction of $\alpha_{1}$ are in a shallow decay segment for both the X-ray and the optical break samples. As summarized in Table 6, 42 of 48 GRBs have a shallow decay segment for X-ray break samples (including: 21 GRBs for Group I and Group II, respectively). 27 of 30 GRBs with a shallow decay segment for optical break samples (including: 18 of 21 GRBs for Group I and 9 of 9 GRBs for Group III).
No significant spectral evolution is observed for the shallow decay segment to the following phase for our 42 X-ray, shallow decay samples (see also in Table 6), only 4 of 42 GRBs showed significant spectral evolution, indicating that the shallow decay should have an external-shock origin (see also Liang et al 2007).

The energy injection behavior can be described as $L(t)\propto t^{-q}$ (see, e.g. , Zhang \& M\'{e}szaros  2001), here \emph{q} is the temporal index.
The difference between the decay slopes before and after the time of the break depends on the spectral regime observed and the type of ambient medium, which can be summarized as in Table 5 (see also Zhang et al.(2006) in Table 2).
This model is called the "energy injection model" (Zhang et al 2006) and produces a break in the light-curves, which are denoted "injection breaks" (Zhang et al. 2004 for review). The pre-break slope $\alpha_{1}$ should give $q=0$ for a constant luminosity,

\begin{eqnarray}
\label{n} \alpha_{1}= \left\{ \begin{array}{ll}
(q-1)+\frac{(2+q)\beta}{2} =\frac{(2p-6)+(p+3)q}{4}, &{\rm \nu_{m}<\nu_{}<\nu_{c}  \textrm{(ISM)}}\\
\frac{q}{2}+\frac{(2+q)\beta}{2}=\frac{(2p-2)+(p+3)q}{4}, &{\rm \nu_{m}<\nu_{}<\nu_{c}  \textrm{(wind)}}\\
\frac{(q-2)}{2}+\frac{(2+q)\beta}{2}=\frac{(2p-4)+(p+3)q}{4}, &{\rm \nu_{}>\nu_{c}  \textrm{(ISM and wind)}},
\end{array} \right.
\end{eqnarray}
while the post-break slope $\alpha_{2}$ should correspond to \textit{q}=1 (for a constant energy fireball, the
scaling law is the same as \textit{q}=1, Zhang \& M\'{e}szaros 2001). Then, according to Zhang et al. (2006), one should have

\begin{eqnarray}
\label{n} \alpha_{2}= \left\{ \begin{array}{ll}
\frac{3\beta}{2}=\frac{3(p-1)}{4}=\frac{3\alpha_{1}+3}{2},&{\rm \nu_{m}<\nu_{}<\nu_{c}  \textrm{(ISM)}},\\
\frac{3\beta+1}{2}=\frac{3p-1}{4}=\frac{3\alpha_{1}+1}{2},&{\rm \nu_{m}<\nu_{}<\nu_{c} \textrm{(wind)}},\\
\frac{3\beta-1}{2}=\frac{3p-2}{4}=\frac{3\alpha_{1}+2}{2},&{\rm \nu_{}>\nu_{c}  \textrm{(ISM and wind)}},\\
\end{array} \right.
\end{eqnarray}

The temporal and spectral properties of the afterglow after the break (the normal decay phase)
should satisfy the 'closure relation' of the external shock model (e.g. Zhang \& M\'{e}szaros 2004), i.e. as described in Table 5.
The relation of the $\alpha_{1}-\alpha_{2}$ panel for all our break samples (Group I+II for the X-ray emission and Group I+III for the optical emission) is shown in Figure \ref{Alpha1Alpha2}.

Since the calculation of the \emph{q} and \emph{p} indices depend on which afterglow model is used,
our first step is to use the observed quantity of the temporal index $\alpha$ and spectral index $\beta$ to determine which spectral regime the burst belongs to. Spectral regime I ($\nu_{obs}>max(\nu_{m},\nu_{c})$) corresponds to the observed band lying above the cooling frequency; and Spectral regime II ($\nu_{\rm m}<\nu_{\rm obs}<\nu_{\rm c}$) corresponds the observed band lying below the cooling frequency. Since different spectral regimes give different predicted decay indices through the ``closure relations'' of the external shock model (e.g., Table 2 of Zhang et al 2006; Gao et al. 2013), we can combine the $\alpha$ and $\beta$ information to determine the spectral regime of a temporal segment in each GRB. Since the normal decay phase is not contaminated by energy injection (Zhang et al. 2007), we use its temporal index $\alpha$ to perform the analysis.
We first use the temporal index $\alpha$ to calculate the theoretically expressed spectral index ($\beta^{'}$) for the different spectral regimes:
\begin{eqnarray}
\label{n} \beta^{'}_{\alpha}=\left\{ \begin{array}{ll}
\frac{2\alpha+1}{3}, &  \nu>max{(\nu_{m},\nu_{c})}  \textrm{(ISM and wind)}\\
\frac{2\alpha}{3},   &  \nu_{m}<\nu<\nu_{c} \textrm{(ISM)}\\
\frac{2\alpha-1}{3}, &  \nu_{m}<\nu<\nu_{c} \textrm{(wind)}\\
\end{array} \right.
\end{eqnarray}
Second, we compare this index ($\beta^{'}_{\alpha}$) to the observed spectral index ($\beta_{\rm obs}$): $\chi^{2}=(\beta^{'}_{\alpha}-\beta_{\rm obs})^{2}/(\sigma_{\beta(\alpha)}^{2}+\sigma_{\beta(\rm obs)}^{2})$.
We thereby can identify which spectral regime the observation belongs to.
We find that for our X-ray observations, 19 of 87 GRBs belong to spectral regime I, and 68 of 87 GRBs belong to spectral regime II.
For optical emission the following was found: 9 of 80 GRBs belong to spectral regime I, and 68 of 80 GRBs belong to spectral regime II.
Note that the optical spectral index of 7 GRBs is not available and therefore we did not judge the spectral regime of the optical band for those GRBs.

After determining the spectral regimes for each individual burst, the electron spectral index $p$ can be determining through,
\begin{eqnarray}
\label{n} p=\left\{ \begin{array}{ll}
\frac{4\alpha_{2}+2}{3}, &  \nu>max{(\nu_{m},\nu_{c})}  \textrm{(ISM and wind)}\\
\frac{4\alpha_{2}+3}{3},   &  \nu_{m}<\nu<\nu_{c} \textrm{(ISM)}\\
\frac{4\alpha_{2}+1}{3}, &  \nu_{m}<\nu<\nu_{c} \textrm{(wind)}\\
\end{array} \right.
\end{eqnarray}

Similarly, the the luminosity injection index \textit{q} (for energy injection cases, calculated by shallow decay segments) can be determined from the temporal index $\alpha$ and the spectral index $\beta$,

\begin{eqnarray}
\label{n} q=\left\{ \begin{array}{ll}
\frac{2\alpha_{1}-2\beta_{1}+2}{1+\beta_{1}}, &  \nu>max{(\nu_{m},\nu_{c})}  \textrm{(ISM and wind)}\\
\frac{2\alpha_{1}-2\beta_{1}+2}{2+\beta_{1}}, &  \nu_{m}<\nu<\nu_{c} \textrm{(ISM)}\\
\frac{2\alpha_{1}-2\beta_{1}}{1+\beta_{1}},   &  \nu_{m}<\nu<\nu_{c}  \textrm{(wind)}\\
\end{array} \right.
\end{eqnarray}

The distributions of the \emph{p} and \emph{q} indices for both the X-ray and optical bands, are shown in Figure \ref{deltaalphaqp} (a) and (b). We calculated the \emph{q} index for our break samples with the temporal index of pre-break in a shallow decay segment.
The \emph{q} value ranges within [-0.90, 0.88] for the X-ray band and within [-0.79, 0.90] for the optical band, and the best Gaussian function fitting with a center value $q_{\rm x(c)}=0.46\pm0.03$ and $q_{\rm o(c)}=0.13\pm0.11$ for the X-ray and the optical band, respectively. The \emph{p} index is derived from the observed spectral index of the post-break decay slope for all the cases. The \emph{p} value ranges within [1.78, 4.18] and 6 of 87 cases had $p<2.0$ for the X-ray band and within [1.28, 4.42] and 17 of 87 cases has $p<2.0$ for the optical band. The best Gaussian fitting with a center value $p_{\rm x(c)}=2.63\pm0.02$ and $p_{\rm o(c)}=2.58\pm0.04$ for the X-ray and the optical band, respectively.

The $\Delta \alpha$ for each case is summarized in Table 4. The functional dependence of $\Delta \alpha - p, q$ for the different external shock model cases can be summarized as:
\begin{eqnarray}
\label{n} \Delta\alpha= \left\{ \begin{array}{ll}
\frac{1}{4} (p+2)(1-q),&{\rm \nu_{}>\nu_{c} \textrm{(ISM and wind)}},\\
\frac{1}{4} (p+3)(1-q),&{\rm \nu_{m}<\nu_{}<\nu_{c}  \textrm{(ISM)}},\\
\frac{1}{4} (p+1)(1-q),&{\rm \nu_{m}<\nu_{}<\nu_{c}  \textrm{(wind)}},
\end{array} \right.
\end{eqnarray}
The two-dimensional panels between the observed quantity $\Delta\alpha$ and the value of the model parameters \emph{p} and \emph{q} are shown in Figure \ref{deltaalphaqp} (c). For the $\Delta\alpha-p$ panel, the various external shock model lines are well covered by all the data points when \textit{q}=-0.5, 0, 0.5, respectively. For the $\Delta\alpha-q$ panel, we also find the external shock model lines are well covered by the data, when we have \textit{p}=2.0, 3.0, respectively. Compared with the optical band, we find that the X-ray data are gathered in the regions which have higher \emph{p} values.

\subsection{Jet break}

Another common break behavior in the afterglow is a post-jet break.
Physically there are two phenomena that could cause a steepening decay in the late afterglow light-curves: the edge effect and sideways expansion.
$\Delta\alpha=0.75$ (ISM) and $\Delta\alpha=0.5$ (wind-like) are expected from the edge effect. A sharper break (from $\alpha_{normal}\approx-1.2$ to $\alpha_{jet}\approx-p$) is suggested by sideways expansion, the post-jet break decay index is approximately $\sim-p$. Thus the criterion for a good candidate for jet-break can generally be summarized as:
(i) $\alpha_{2}\sim-p$. Therefore, we take $\alpha_{2}>1.5$ as our definition of a jet-break sample (see also Liang et al. 2008), since this value is difficult to interpret by the standard fireball model.
(ii) $\Delta\alpha$ ranges within [0.5, 1.2] for normal-to-jet break\footnote{Here the $\Delta\alpha$=1.2 is supposed to be a typical value of \emph{p}.}.
(iii) The 'Closure relation' should be satisfied. As shown in the Figure \ref{Closure Relations}, in the $\alpha-\beta$ panel, a good fraction of GRBs in our samples (both the X-ray and optical band) are consistent with the expectations of the external shock jet models, especially in the wind-like model.
(iv) No spectral evolution. Since jet-break behavior is purely of dynamical effects, there should be no spectral evolution across the break and achromatic behavior should be observed in the multiple-band afterglow light-curves.

The distribution of the \emph{p} index is shown in Figure \ref{deltaalphaqp}, and the peak distribution of the \emph{p} value is generally 2.5 for both the X-ray and the optical bands.
As summarized in our Table 6, there are no jet-break cases with a spectral evolution.
32 of 48 GRBs for the X-ray break cases (including: 14 of 24 for Group I and 18 of 24 GRBs for Group II, respectively) and 13 of 33 GRBs for optical break cases (including: 11 of 24 GRBs for Group I and 2 of 9 GRBs for Group III) satisfy the criterion of $\alpha_{2}>1.5$.
Based on this criterion, the $\Delta\alpha_{x}$ ranges within [0.56, 2.27] for the X-ray and $\Delta\alpha_{o}$ within [0.75, 1.84] for the optical jet-break cases, respectively.
This indicates that the data are difficult to fully interpret by the criterion for a normal-to-jet break decay segment($\Delta\alpha$ ranges within [0.5, 1.2] expected by the theory), since we find that there are also 16 of 32 GRBs for the X-ray and 10 of 13 GRBs for the optical band, with $\Delta\alpha>1.2$.
We find that the typical time for our jet-break samples is $\sim10^{5}$ seconds, and for our energy injection break samples $\sim10^{4}$ seconds.
However, it is interesting that we find that $\alpha_{1}$ is generally in a shallow decay segment and $\alpha_{2}$ is generally in the post-jet break segment. All these cases have $\Delta\alpha>1.2$, indicating that the energy injection break time appears similar to the jet-break times for those GRBs. This is the reason for $\Delta\alpha$ being large.

\subsection{Spectral break}

Another possibility that can cause a break in the X-ray (or optical) light-curve is due to spectral evolution where the spectral peak of the synchrotron emission moves through the X-ray band. However, since, in general,  there is a  lack of  significant spectral evolution, this possibility  is largely excluded (De Pasquale et al. 2006; Nousek et al. 2006).
If the cooling frequency ($\nu_{\rm c}$) crosses the X-ray band, the change of the  X-ray spectrum should be reflected by $\Delta\beta_{\rm x}=\frac{1}{2}$ for the ISM scenario and $\Delta\beta_{\rm x}=-\frac{1}{2}$ for the wind-like circum-burst medium scenario (Sari et al. (1999); Chevalier \& Li (1999)).
The change in the light-curve should then correspond to $\Delta\alpha_{\rm x}=\frac{1}{4}$ for both the ISM and wind-like scenario.

We show the distributions of $\Delta\alpha$ for all our cases with breaks in Figure \ref{Cooling}.
We have 44 X-ray break cases (Group I+II), and the distribution of $\Delta\alpha_{\rm x}$ is in the range [0.44,2.27], which is larger than $\Delta\alpha=0.25$ as expected by the cooling frequency crossing the X-ray band model.
This is similar for the 33 cases with optical afterglow breaks (Group I+III). The distribution of $\Delta\alpha_{\rm o}$ is in the range of [0.47, 1.84]. All bursts have $\Delta\alpha$  values that are larger than the value expected by the cooling frequency crossing the optical band ($\Delta\alpha=0.25$).

On the other hand, we derived the  photon spectral indices of the pre-break and post-break  portions of the X-ray afterglow for our break samples (Group I+ II) and then calculated the spectral indices using $\beta_{\rm x}=\Gamma_{\rm x}-1$.
The distribution of the spectral indices of the pre-break and post-break portions are shown in Figure \ref{Cooling} (b), and the center values with fitting by Gaussian functions to the distributions are $\beta_{\rm x,1}=0.95\pm0.03$ and $\beta_{\rm x,2}=0.98\pm0.01$. The $\beta_{\rm x,2}-\beta_{\rm x,1}$ panel shows that all the bursts (to within the  error) are in the vicinity of the equality line. This indicates that the pre-break and post-break portions of the light-curves lack the significant spectral evolution in most of the cases. Since no optical time-resolution was available for analysis of all the optical break sample, we are not able to examine the change of the spectral index for the transition from pre-break to post-break.

Both the X-ray and optical afterglow break behavior can thus not be interpreted by the cooling frequency crossing the X-ray or optical band. Indeed, there is also a lack of any significant spectral evolution from pre-break to post-break.

\section{Conclusions and Discussion}

We extracted the cleanest possible signals of the afterglow emission components, and include the observations not only by Swift/UVOT, but also by other ground-based telescopes, and obtain a sample of, in total, 87 bursts that have good quality, simultaneous detections in both the X-ray and the optical bands.
We divided the sample into four groups depending on the existence of breaks in their afterglow light-curves, and examine their behavior in light of various external shock model empirical relations. We find that the data generally satisfy the various external shock empirical relations.
Our main results can be summarized as follows:

\begin{itemize}

\item About 10\% Swift-GRBs (until the end of 2013) had good simultaneous observational data in both the X-ray and optical afterglow.
      Studying this sample in detail, we find that the X-ray and optical afterglow light-curves show similar statistical properties.

\item Characterizing the light-curves in 4 different groups (Fig.1), we investigate how well GRBs in each group satisfy the external shock model predictions. We find that in groups I and IV, a majority of GRBs follow the predictions the external shock model.
    The bursts in groups II and III have, per definition, chromatic light-curves, so that one cannot interpret the full evolution directly with the external shock model. However, a large fraction of these burst  have individual segments in the light-curves that do satisfy the relations of the external shock model. This is consistent with the following picture:
    Four these bursts, one cannot interpret both X-ray and optical data simultaneously in the external shock model. Only one band may be of the external shock origin. Part  of the other band may arises from a different emission component (e.g. long lasting central engine emission, reverse shock emission, or  structured jets) causing the chromatic behaviour,  see further discussion in Wang et al. (2015).

\item Overall we find 62\% GRBs that are consistent with the standard afterglow model ($\Delta\alpha=[-0.25,0.25]$), when more advanced modeling is invoked, up to 91\% of the bursts in our sample may be consistent with the external shock model ($\Delta\alpha=[-0.5,0.5]$). Of these, 61\% correspond to a ISM medium and 39\% GRBs correspond to a wind medium.

\item The fraction of the bursts that are consistent with the external shock model is independent of the observational epochs (measured in the rest frame of GRBs).

\item Approximately half of all bursts have light-curve for which the break time of the shallow-to-normal decay is similar to that of the normal-to-jet decay, making an immediate change from the plateau phase to the jet phase in both the X-ray and the optical. We suggest that the normal decay segment for these bursts is short and therefore not detected.

\item The average values of the post-break decay-slopes, in the X-ray and optical band, are consistent with the case of an ISM medium and slow cooling electrons ($\nu_{\rm m}<\nu_{\rm o}<\nu_{\rm x}<\nu_{\rm c}$).

\item We do not detect one single case in our sample with a temporal break that can be interpreted as the cooling frequency crossing the X-ray or optical bands.

\end{itemize}

The models we have tested are the simplest analytical external shock models of GRBs. More sophisticated afterglow modeling has been carried out, which gives somewhat different predictions than the simplest models. For example, the recent numerical simulations of Duffell \& MacFadyen (2014)
simulated hydrodynamical evolution of a jet starting from a collapsar engine to the late afterglow phase.
The change of decay slope in their simulated light curve from the plateau to the following power-law decay is $\Delta\alpha_{\rm x}= 9/8$.  We investigated 48 GRBs in X-rays and 33 GRBs in optical that show such a break, and find that 23 of 48 (48\%) GRBs for the X-ray and 17 of 33 (52\%) for the optical are consistent with having $\Delta\alpha\approx9/8$, which is consistent with their simulation results.
More detailed analyses are needed to fully confront this and other more sophisticated modeling with the observational data.

One important question is whether the X-ray and optical emissions come from the same emission component, as predicted by the standard afterglow model.
One interesting fact is that we have identified 9 of 87 GRBs, whose afterglow light-curves show a genuine afterglow behavior, namely a single power-law decay sometimes with an early afterglow onset behavior, for both the X-ray and the optical emission. These GRBs are fully consistent with the simplest standard fireball model. There are no flares, no shallow decay segments, no re-brightening at late time (without extra energy injection), no post-jet breaks. The nine candidates are: GRBs 050603, 061007, 080804, 090323, 090328, 090902B, 090926A, 120815A, 130427A.

Other GRBs have more complicated features. Testing whether these GRBs are consistent with the external shock models requires more detailed modeling. In this paper, we performed a series of consistency checks, including the $\alpha-\beta$ closure relations in individual segments and epochs and $\Delta \alpha$ and $\Delta\beta$ predictions. Our results show that in general most of them are still consistent with the external shock model predictions. Nonetheless, in order to address whether these GRBs are fully consistent with the afterglow models, much more detailed analysis and modeling, including the achromaticity and global consistency with the closure relations, are needed. This has been carried out in a separate work (Wang et al. 2015). That work also suggests that most GRBs are consistent with the external shock model of GRBs, which is fully consistent with the finding of this paper.

We acknowledge the use of public data from the Swift data archive. We appreciate valuable comments and suggestions by the referee, and thank Xiang-Yu Wang for helpful discussions.
This work is supported by the Swedish National Space Board, the Swedish Research Council, the National Basic Research Program of China (973 Program, Grant No. 2014CB845800) and the National Natural Science Foundation of China (Grant No. 11322328).
Liang Li is supported by the Erasmus Mundus Joint Doctorate Program by Grant Number 2013-1471 from the EACEA of the European Commission.
XFW acknowledges support by the One-Hundred-Talent Program, the Youth Innovation Promotion Association and the Strategic Priority Research Program "The Emergence of Cosmological Structures" (Grant No. XDB09000000) of Chinese Academy of Sciences. YFH was supported by the National Natural Science Foundation of China with Grant Nos. 11473012 and 11033002. BZ acknowledges NASA NNX14AF85G for support.

\clearpage
\begin{deluxetable}{lccccccccccccc}
%\rotate
\tabletypesize{\scriptsize}
\tablecaption{XRT Observations and Fitting Results}
\tablewidth{0pt}
\tabletypesize{\tiny}
\tablehead{
\colhead{GRB}&
\colhead{$T_{\rm x1},T_{\rm x2}$\tablenotemark{a}}&
\colhead{$T_{\rm x,b}$\tablenotemark{a}}&
\colhead{$\alpha_{\rm x,1}$\tablenotemark{a}}&
\colhead{$\alpha_{\rm x,2}$\tablenotemark{a}}&
\colhead{$S_{\rm x}$\tablenotemark{b}}&
\colhead{$L_{\rm X,b}$\tablenotemark{c}}&
\colhead{$E_{\rm X,iso}$\tablenotemark{c}}&
\colhead{$\Gamma_{\rm x_{1}}$\tablenotemark{c}}&
\colhead{$\Gamma_{\rm x_{2}}$\tablenotemark{c}}&
}
\startdata
\hline
&Group I\\
\hline
050319&0.44,1390&36.8$\pm$10.8&0.49$\pm$0.04&1.56$\pm$0.20&71$\pm$24&405$\pm$65&155$\pm$52&1.92$\pm$0.08&2.17$\pm$0.23\\
050408&2.60,2540&384.0$\pm$169.6&0.82$\pm$0.03&1.61$\pm$0.20&70$\pm$26&2.80$\pm$1.3&28$\pm$10&...&...\\
050730&4.01,408&9.7$\pm$0.5&0.72$\pm$0.12&2.69$\pm$0.04&205$\pm$28&(19.2$\pm$2.2)$\times10^{3}$&622$\pm$85&1.46$\pm$0.06&1.63$\pm$0.05\\
050922C&0.12,362&3.2$\pm$1.3&0.97$\pm$0.04&1.49$\pm$0.05&82$\pm$37&1390$\pm$670&93$\pm$42&2.17$\pm$0.11&2.19$\pm$0.12\\
060210&3.83,1580&28.4$\pm$10.4&0.83$\pm$0.07&1.41$\pm$0.05&130$\pm$55&1390$\pm$570&385$\pm$165&1.51$\pm$0.06&1.98$\pm$0.09\\
060526&1.00,314&97.1$\pm$20.7&0.63$\pm$0.60&2.89$\pm$0.48&16$\pm$15&41.8$\pm$16&34$\pm$3258&1.95$\pm$0.16&1.9$\pm$0.47\\
060605&0.20,127&8.7$\pm$0.6&0.46$\pm$0.03&2.19$\pm$0.08&24$\pm$3&1130$\pm$99&68$\pm$8&1.98$\pm$0.12&2.14$\pm$0.16\\
060614&4.54,2770&50.2$\pm$2.6&0.12$\pm$0.04&1.97$\pm$0.04&56$\pm$5&0.20$\pm$0.01&...&1.77$\pm$0.11&1.9$\pm$0.14\\
060708&0.07,1228&8.9$\pm$2.2&0.60$\pm$0.05&1.32$\pm$0.05&24$\pm$7&111$\pm$27&21$\pm$7&2.16$\pm$0.2&2.15$\pm$0.19\\
060714&0.39,1250&4.6$\pm$1.6&0.44$\pm$0.11&1.31$\pm$0.05&33$\pm$13&859$\pm$300&54$\pm$22&1.86$\pm$0.15&2.05$\pm$0.17\\
060729&0.58,11800&65.8$\pm$2.2&0.15$\pm$0.02&1.42$\pm$0.01&398$\pm$18&16.5$\pm$0.54&30$\pm$1&2.04$\pm$0.04&2.03$\pm$0.05\\
060912A&0.12,440&0.3$\pm$0.3&0.64$\pm$0.34&1.13$\pm$0.05&12$\pm$12&145$\pm$140&2$\pm$2&1.53$\pm$0.45&1.91$\pm$0.16\\
061021&0.08,4236&16.6$\pm$3.2&0.60$\pm$0.03&1.16$\pm$0.02&97$\pm$22&3.82$\pm$0.67&3$\pm$1&1.98$\pm$0.1&1.93$\pm$0.07\\
070411&0.50,695&43.9$\pm$41.8&0.95$\pm$0.09&1.49$\pm$0.22&30$\pm$20&73.3$\pm$82&57$\pm$37&2.36$\pm$0.16&2.31$\pm$0.36\\
070518&0.80,914&145.0$\pm$264.3&0.59$\pm$0.13&1.24$\pm$0.62&5$\pm$7&0.52$\pm$0.77&2$\pm$3&2.14$\pm$0.43&2.66$\pm$1.25\\
080310&1.35,378&10.7$\pm$0.9&0.13$\pm$0.09&1.63$\pm$0.06&21$\pm$3&330$\pm$31&28$\pm$4&2.06$\pm$0.17&1.91$\pm$0.15\\
080413B&0.14,582&100.0$\pm$16.9&0.90$\pm$0.01&1.76$\pm$0.12&73$\pm$16&6.35$\pm$1.2&23$\pm$5&1.93$\pm$0.07&1.86$\pm$0.29\\
080710&3.17,560&15.1$\pm$4.3&0.81$\pm$0.15&1.91$\pm$0.14&19$\pm$8&14.7$\pm$5.8&4$\pm$2&1.79$\pm$0.16&2.16$\pm$0.18\\
080721&0.11,1400&3.1$\pm$0.2&0.80$\pm$0.01&1.65$\pm$0.01&693$\pm$44&(24.1$\pm$1.3)$\times10^{3}$&1054$\pm$67&1.81$\pm$0.02&1.96$\pm$0.06\\
081029&2.76,288&18.2$\pm$1.2&0.43$\pm$0.06&2.70$\pm$0.15&16$\pm$2&559$\pm$55&47$\pm$6&1.92$\pm$0.1&2.04$\pm$0.21\\
090618&0.40,2840&8.4$\pm$0.5&0.72$\pm$0.01&1.51$\pm$0.01&490$\pm$37&135$\pm$8.4&37$\pm$3&1.93$\pm$0.04&1.86$\pm$0.05\\
090727&4.08,2140&372.7$\pm$141.3&0.51$\pm$0.07&1.60$\pm$0.26&34$\pm$9&...&...&2.09$\pm$0.29&1.92$\pm$0.54\\
091018&0.07,520&0.5$\pm$0.1&0.31$\pm$0.06&1.24$\pm$0.02&67$\pm$11&1070$\pm$140&16$\pm$3&1.84$\pm$0.15&2.04$\pm$0.09\\
091127&3.22,4030&29.3$\pm$6.9&1.02$\pm$0.04&1.58$\pm$0.02&481$\pm$170&29.3$\pm$9.3&30$\pm$10&1.8$\pm$0.11&1.77$\pm$0.08\\
\hline
&Group II\\
\hline
050401&0.13,548&5.1$\pm$0.6&0.55$\pm$0.01&1.47$\pm$0.06&247$\pm$35&7070$\pm$650&451$\pm$65&1.78$\pm$0.21&1.79$\pm$0.15\\
050502A&2.87,1050&18.8$\pm$2.5&0.66$\pm$0.06&1.66$\pm$0.05&....&...&0$\pm$29&...&...\\
050824&6.31,2030&74.9$\pm$58.4&0.32$\pm$0.14&0.94$\pm$0.13&30$\pm$17&2.19$\pm$1.0&5$\pm$3&1.85$\pm$0.19&2.48$\pm$0.38\\
051221A&6.03,848&26.7$\pm$5.2&0.18$\pm$0.15&1.43$\pm$0.06&12$\pm$3&1.70$\pm$0.32&1$\pm$0&1.87$\pm$0.22&1.97$\pm$0.16\\
060111B&0.20,222&7.6$\pm$4.0&0.86$\pm$0.07&1.58$\pm$0.14&20$\pm$9&...&...&2.03$\pm$0.2&2.07$\pm$0.32\\
060418&0.36,537&1.3$\pm$0.4&0.87$\pm$0.13&1.57$\pm$0.05&47$\pm$19&1390$\pm$590&26$\pm$11&1.98$\pm$0.22&1.95$\pm$0.22\\
060906&0.20,258&12.6$\pm$1884.1&0.29$\pm$0.06&1.82$\pm$0.17&12$\pm$2&343$\pm$51&33$\pm$6&2.07$\pm$0.21&1.77$\pm$0.25\\
060908&0.08,450&0.7$\pm$0.2&0.44$\pm$0.10&1.56$\pm$0.07&36$\pm$9&3360$\pm$920&31$\pm$8&2.05$\pm$0.29&1.98$\pm$0.22\\
060927&0.09,103&4.2$\pm$2212.5&0.68$\pm$0.12&1.99$\pm$0.46&7$\pm$5&1340$\pm$980&36$\pm$24&1.86$\pm$0.25&1.96$\pm$0.35\\
061121&0.30,2100&6.7$\pm$0.5&0.41$\pm$0.03&1.48$\pm$0.02&261$\pm$27&849$\pm$79&116$\pm$12&1.95$\pm$0.12&1.82$\pm$0.06\\
070318&5.59,708&341.2$\pm$62.2&0.85$\pm$0.03&2.41$\pm$0.51&40$\pm$8&0.818$\pm$0.17&7$\pm$1&2.02$\pm$0.12&1.82$\pm$0.49\\
070420&0.25,525&3.4$\pm$0.3&0.26$\pm$0.06&1.47$\pm$0.02&113$\pm$11&...&...&2.07$\pm$0.15&1.92$\pm$0.12\\
071031&6.13,344&104.2&0.97$\pm$0.08&1.75$\pm$0.48&4$\pm$1&4.68$\pm$0.69&6$\pm$1&1.73$\pm$0.3&1.26$\pm$0.98\\
080810&3.85,366&17.2$\pm$9.6&1.17$\pm$0.14&1.87$\pm$0.11&32$\pm$22&525$\pm$440&73$\pm$52&2.06$\pm$0.12&2.2$\pm$0.22\\
081008&0.52,261&18.5$\pm$4.7&0.86$\pm$0.04&1.85$\pm$0.14&60$\pm$20&169$\pm$52&56$\pm$19&2.05$\pm$0.13&1.83$\pm$0.32\\
081126&0.07,368&5.8$\pm$1.2&0.60$\pm$0.04&1.56$\pm$0.06&85$\pm$15&...&...&1.89$\pm$0.17&2.28$\pm$0.19\\
081203A&0.14,345&14.9$\pm$2.8&1.16$\pm$0.01&2.14$\pm$0.12&87$\pm$27&190$\pm$50&92$\pm$28&2.02$\pm$0.11&2.08$\pm$0.35\\
090510&0.10,67&1.4$\pm$0.2&0.62$\pm$0.04&2.15$\pm$0.09&57$\pm$9&459$\pm$70&12$\pm$2&1.67$\pm$0.1&2.05$\pm$0.2\\
100418A&0.40,2090&125.0$\pm$17089.6&-0.01$\pm$0.05&1.66$\pm$0.10&49$\pm$8&2.31$\pm$0.31&5$\pm$1&1.89$\pm$0.29&1.84$\pm$0.17\\
100621A&0.40,1770&99.3$\pm$20.6&0.86$\pm$0.03&1.66$\pm$0.08&371$\pm$51&6.37$\pm$1.5&28$\pm$4&2.34$\pm$0.1&3.06$\pm$0.31\\
100814A&3.88,4850&148.9$\pm$7.33&0.50$\pm$0.02&2.09$\pm$0.06&215$\pm$15&51.52.6&113$\pm$8&1.896$\pm$0.048&2.08$\pm$0.13\\
100906A&0.30,202&13.2$\pm$1.3&0.76$\pm$0.03&2.12$\pm$0.80&172$\pm$28&560$\pm$80&127$\pm$20&1.96$\pm$0.1&2.05$\pm$0.15\\
120119A&0.20,317&32.5$\pm$7.0&1.00$\pm$0.02&2.03$\pm$0.18&86$\pm$26&66.7$\pm$17&64$\pm$19&1.61$\pm$0.11&1.92$\pm$0.38\\
121217A&5.55,1340&24.9$\pm$4.5&0.47$\pm$0.09&1.43$\pm$0.05&134$\pm$22&...&...&2.06$\pm$0.16&2.05$\pm$0.16\\
\hline
&Group III\\
\hline
050416A&1.00,5800&...&...&0.88$\pm$0.01&47$\pm$6&...&5$\pm$1&...&2.27$\pm$0.1\\
050525A&5.92,1080&...&...&1.54$\pm$0.04&32$\pm$10&...&3$\pm$1&...&2.09$\pm$0\\
050801&0.19,442&...&...&1.07$\pm$0.06&6$\pm$3&...&4$\pm$2&...&1.86$\pm$0.21\\
051109A&3.56,1520&...&...&1.20$\pm$0.01&96$\pm$11&...&122$\pm$15&...&2.05$\pm$0.07\\
051111&5.63,48&...&...&1.61$\pm$0.06&6$\pm$3&...&4$\pm$2&...&2.23$\pm$0.17\\
061126&1.61,2460&...&...&1.34$\pm$0.01&151$\pm$14&...&53$\pm$5&...&1.93$\pm$0.07\\
080413A&0.14,13&...&...&1.21$\pm$0.03&8$\pm$2&...&11$\pm$3&...&2.15$\pm$0.24\\
090426&0.13,497&...&...&1.02$\pm$0.04&8$\pm$2&...&12$\pm$3&...&2.12$\pm$0.17\\
130702A&89.14,15100&...&...&1.16$\pm$0.01&418$\pm$71&...&2$\pm$0&...&1.94$\pm$0.08\\
\hline
&Group IV\\
\hline
050603&34.81,1050&...&...&1.71$\pm$0.06&15$\pm$9&...&26$\pm$15&...&1.97$\pm$0.13\\
050721&0.20,136&...&...&1.32$\pm$0.02&71$\pm$5&...&...&...&1.89$\pm$0.19\\
050820A&4.64,3680&...&...&1.24$\pm$0.01&236$\pm$28&...&361$\pm$44&...&1.98$\pm$0.05\\
060124&11.44,1960&...&...&1.33$\pm$0.01&202$\pm$30&...&248$\pm$37&...&1.98$\pm$0.05\\
060206&10.00,3430&...&...&1.25$\pm$0.02&50$\pm$10&...&156$\pm$32&...&1.9$\pm$0.19\\
061007&0.09,407&...&...&1.73$\pm$0.01&471$\pm$21&...&193$\pm$9&...&2.02$\pm$0.09\\
070125&46.83,1020&...&...&1.63$\pm$0.12&32$\pm$46&...&19$\pm$28&...&2.06$\pm$0.15\\
070311&7.10,54&...&...&1.09$\pm$0.06&10$\pm$4&...&...&...&...\\
071010A&34.51,472&...&...&1.89$\pm$0.14&20$\pm$2&...&5$\pm$1&...&1.85$\pm$0.21\\
071025&0.15,257&...&...&1.54$\pm$0.01&113$\pm$10&...&462$\pm$39&...&2.24$\pm$0.13\\
071112C&0.09,170&...&...&1.33$\pm$0.01&41$\pm$5&...&7$\pm$1&...&1.7$\pm$0.13\\
080319B&0.06,2730&...&...&1.61$\pm$0.00&3001$\pm$68&...&689$\pm$16&...&1.91$\pm$0.07\\
080804&0.11,426&...&...&1.11$\pm$0.01&31$\pm$3&...&35$\pm$3&...&1.82$\pm$0.1\\
080928&4.28,222&...&...&1.59$\pm$0.03&20$\pm$6&...&14$\pm$4&...&2.14$\pm$0.1\\
081109A&0.33,477&...&...&1.22$\pm$0.02&47$\pm$6&...&12$\pm$1&...&2.2$\pm$0.13\\
090102&0.40,696&...&...&1.30$\pm$0.01&110$\pm$8&...&66$\pm$5&...&1.8$\pm$0.08\\
090323&70.13,1050&...&...&1.62$\pm$0.09&11$\pm$9&...&28$\pm$23&...&2.2$\pm$0.28\\
090328&57.28,909&...&...&1.68$\pm$0.08&26$\pm$17&...&4$\pm$2&...&2.09$\pm$0.26\\
090902B&45.21,1460&...&...&1.40$\pm$0.03&49$\pm$17&...&40$\pm$14&...&1.82$\pm$0.12\\
090926A&46.68,1810&...&...&1.41$\pm$0.04&32$\pm$15&...&34$\pm$16&...&2.15$\pm$0.13\\
091029&0.70,1830&...&...&1.27$\pm$0.05&39$\pm$9&...&65$\pm$16&...&2.16$\pm$0.09\\
100728B&3.55,105&...&...&1.51$\pm$0.07&6$\pm$2&...&6$\pm$2&...&2.08$\pm$0.18\\
100901A&8.00,1560&...&...&1.48$\pm$0.03&77$\pm$6&...&39$\pm$3&...&2.29$\pm$0.09\\
101024A&4.28,137&...&...&1.37$\pm$0.04&12$\pm$5&...&...&...&1.72$\pm$0.16\\
110205A&0.73,246&...&...&1.67$\pm$0.01&54$\pm$6&...&62$\pm$7&...&1.95$\pm$0.08\\
110213A&0.15,485&...&...&2.04$\pm$0.04&165$\pm$18&...&89$\pm$10&...&2.00$\pm$0.06\\
110918A&107.45,3880&...&...&1.61$\pm$0.03&302$\pm$73&...&76$\pm$19&...&1.96$\pm$0.09\\
120404A&0.80,21&...&...&1.79$\pm$0.12&16$\pm$4&...&28$\pm$7&...&1.91$\pm$0.13\\
120815A&2.73,41&...&...&0.92$\pm$0.04&19$\pm$7&...&24$\pm$9&...&1.83$\pm$0.12\\
130427A&0.35,15700&...&...&1.28$\pm$0.00&4837$\pm$98&...&139$\pm$3&...&1.79$\pm$0.04\\
\enddata
\tablenotetext{a}{The time intervals, for the fit results of the X-ray afterglow for each GRB; and the break time for the X-ray break cases, in units of ks; the decay slopes before and after break time.}
\tablenotetext{b}{X-ray fluence, which is calculated by integrating the flux over the time intervals, in units of $10^{-8}$ erg cm$^{-2}$. Note that
 we corrected the X-ray fluence from absorbed to unabsorbed by calculating the ratios of the absorbed and unabsorbed flux, which collect from the Swift archive.}
\tablenotetext{c}{The break luminosity in the rest frame, in units of $10^{45}$ erg s$^{-1}$; and the X-ray isotropic total energy release, in units of $10^{50}$ erg; the photon spectral index of X-ray before and after break time.}
\end{deluxetable}

\begin{deluxetable}{cccccccccccccccccccccccccc}
%\rotate
\tabletypesize{\scriptsize}
\tablecaption{Optical Afterglow Observations and Fitting Results}
\tablewidth{0pt}
\tabletypesize{\tiny}
\tablehead{
\colhead{GRB(Band)}&
\colhead{$T_{\rm o1},T_{\rm o2}$\tablenotemark{a}}&
\colhead{$T_{\rm o,b}$\tablenotemark{a}}&
\colhead{$\alpha_{\rm o,1}$\tablenotemark{b}}&
\colhead{$\alpha_{\rm o,2}$\tablenotemark{b}}&
\colhead{$S_{\rm R}$\tablenotemark{c}}&
\colhead{$L_{\rm R,b}$\tablenotemark{c}}&
\colhead{$E_{\rm R,iso}$\tablenotemark{c}}&
\colhead{$\beta_{\rm o}$\tablenotemark{d}}&
\colhead{$z$\tablenotemark{d}}&
}
\startdata
\hline
&Group I\\
\hline
050319(R)&0.04,994&249.6$\pm$13.72&0.54$\pm$0.004&1.94&304$\pm$21&2.81$\pm$0.13&667$\pm$46&0.74$\pm$0.42&$3.24^{(1)}$\\
050408(R)&3.35,215&33.8$\pm$2.33&0.46&1.12&30$\pm$2&0.22$\pm$0.01&12$\pm$1&0.28$\pm$0.33&$1.2357^{(2)}$\\
050730(R)&0.07,73&8.5$\pm$0.66&0.47$\pm$0.02&1.45$\pm$0.03&276$\pm$28&117.34$\pm$7.49&839$\pm$84&0.52$\pm$0.05&$3.969^{(1)}$\\
050922C(R)&0.74,606&6.4$\pm$0.85&0.46$\pm$0.03&1.50$\pm$0.02&386$\pm$73&46.41$\pm$6.19&439$\pm$83&0.51$\pm$0.05&$2.198^{(3)}$\\
060210(R)&0.06,7&1.1$\pm$1.08&0.11$\pm$0.40&1.22$\pm$0.13&18$\pm$23&82.24$\pm$98.43&55$\pm$70&0.37$\pm$0.1&$3.91^{(3)}$\\
060526(R)&0.06,894&124.2$\pm$2.74&0.73$\pm$0.01&2.24$\pm$0.02&535$\pm$24&6.81$\pm$0.20&1156$\pm$52&0.51$\pm$0.32&$3.21^{(3)}$\\
060605(R)&1.76,112&25.0$\pm$2.03&0.94$\pm$0.04&2.59$\pm$0.13&240$\pm$49&32.66$\pm$5.44&675$\pm$139&1.06&$3.78^{(3)}$\\
060614(R)&6.16,1280&66.3$\pm$0.91&0.16$\pm$0.03&2.01$\pm$0.02&283$\pm$9&(100.00$\pm$1.00)$\times10^{-4}$&1$\pm$0&0.47$\pm$0.04&$0.125^{(3)}$\\
060708(R)&0.06,208&0.7$\pm$0.21&0.06$\pm$0.10&0.85$\pm$0.03&358$\pm$105&96.09$\pm$18.37&320$\pm$94&...&$0.249^{(4)}$\\
060714(R)&3.86,185&7.7&0.20&1.17$\pm$0.05&93$\pm$6&20.20$\pm$0.40&151$\pm$9&0.44$\pm$0.04&$2.71^{(3)}$\\
060729(U)&0.70,662&53.4$\pm$6.12&0.16$\pm$0.10&1.39$\pm$0.06&3946$\pm$646&2.43$\pm$0.24&296$\pm$48&0.78$\pm$0.03&$0.54^{(1)}$\\
060912A(R)&0.10,24&0.2(fixed)&0.78&1.32$\pm$0.02&62$\pm$2&46.61$\pm$1.15&14$\pm$0&0.62&$0.8^{(1)}$\\
061021(R)&0.09,298&88.8$\pm$13.25&0.65$\pm$0.01&2.01$\pm$0.24&330$\pm$66&0.03$\pm$0.00&10$\pm$2&...&$0.3463^{(5)}$\\
070411(R)&0.70,517&286.6$\pm$4.17&0.95$\pm$0.00&2.27&197$\pm$6&0.70$\pm$0.02&371$\pm$11&...&$2.954^{(1)}$\\
070518(R)&2.11,312&30.0$\pm$38.86&0.60$\pm$0.12&1.80$\pm$0.12&14$\pm$21&0.10$\pm$0.01&5$\pm$7&0.8&$1.16^{(5)}$\\
080310(R)&0.30,124&2.1$\pm$0.19&0$\pm$0.03&1.12$\pm$0.02&198$\pm$18&88.59$\pm$4.08&267$\pm$24&0.42$\pm$0.12&$2.43^{(5)}$\\
080413B(R)&0.08,1901&372.7$\pm$7.77&0.50$\pm$0.01&3.07&400$\pm$50&0.19$\pm$0.01&126$\pm$16&0.25$\pm$0.07&$1.1^{(7)}$\\
080710(R)&2.54,267&8.7$\pm$0.08&0.45$\pm$0.004&1.52$\pm$0.01&382$\pm$7&4.93$\pm$0.05&71$\pm$1&0.8$\pm$0.09&$0.85^{(5)}$\\
080721(R)&0.16,2640&20.0(fixed)&1.01$\pm$0.01&1.50&460$\pm$9&13.41$\pm$0.30&700$\pm$13&0.68$\pm$0.02&$2.602^{(8)}$\\
081029(R)&4.95,253&16.9$\pm$2.80&0.48$\pm$0.19&2.08$\pm$0.09&326$\pm$98&125.76$\pm$26.15&943$\pm$283&1.08$\pm$0.02&$3.847^{(9)}$\\
090618(R)&0.08,73&30.5$\pm$1.46&0.70$\pm$0.002&1.51$\pm$0.06&836$\pm$51&0.70$\pm$0.03&63$\pm$4&0.5&$0.54^{(10)}$\\
090727(R)&0.47,52&3.3$\pm$3.10&0.07$\pm$0.33&0.78$\pm$0.16&15$\pm$9&0.90$\pm$0.41&7$\pm$4&0.81$\pm$0.28&$...$\\
091018(R)&0.15,273&64.1$\pm$1.79&0.90$\pm$0.00&1.67&468$\pm$20&0.49$\pm$0.01&115$\pm$5&0.58$\pm$0.07&$0.971^{(11)}$\\
091127(I)&7.93,531&27.5$\pm$7.73&0.45$\pm$0.09&1.48$\pm$0.13&774$\pm$310&0.86$\pm$0.20&48$\pm$19&0.18&$0.49^{(12)}$\\
\hline
&Group II\\
\hline
050401(R)&0.04,1120&...&...&0.75$\pm$0.04&33$\pm$16&...&60$\pm$29&0.39$\pm$0.05&$2.9^{(8)}$\\
050502A(V)&0.05,18&...&...&1.11$\pm$0.02&115$\pm$17&...&326$\pm$49&0.76$\pm$0.16&$3.793^{(2)}$\\
050824(R)&0.63,8990&...&...&0.56&725$\pm$13&...&131$\pm$2&0.4$\pm$0.04&$0.83^{(1)}$\\
051221A(R)&11.12,445&...&...&0.95$\pm$0.03&30$\pm$11&...&2$\pm$1&0.64$\pm$0.05&$0.5465^{(1)}$\\
060111B(R)&0.07,13700&...&...&1.12$\pm$0.08&30$\pm$20&...&...&0.7&$...$\\
060418(H)&0.08,8&...&...&1.27$\pm$0.02&469$\pm$28&...&263$\pm$16&0.78$\pm$0.09&$1.489^{(3)}$\\
060906(R)&0.66,5&...&...&0.83$\pm$0.17&9$\pm$2&...&23$\pm$4&0.56$\pm$0.02&$3.685^{(3)}$\\
060908(R)&0.06,1110&...&...&1.09$\pm$0.01&112$\pm$10&...&97$\pm$9&0.3$\pm$0.06&$1.8836^{(8)}$&\\
060927(V)&0.77,1&...&...&1.17$\pm$0.04&9$\pm$4&...&49$\pm$20&0.860.03&$5.6^{(3)}$\\
061121(R)&0.05,1&...&...&1.07$\pm$0.23&165$\pm$17&...&73$\pm$7&0.95&$1.314^{(13)}$\\
070318(V)&0.06,87&...&...&1.14$\pm$0.06&263$\pm$42&...&48$\pm$8&0.78&$0.836^{(7)}$\\
070420(R)&0.12,11&...&...&0.90$\pm$0.08&167$\pm$33&...&...&...&$...$\\
071031(R)&0.29,26&...&...&0.84$\pm$0.001&48$\pm$0&...&77$\pm$0&0.64$\pm$0.01&$2.69^{(6)}$\\
080810(R)&0.04,498&...&...&1.20$\pm$0.01&898$\pm$102&...&2079$\pm$235&0.44&$3.35^{(10)}$\\
081008(R)&0.11,185&...&...&0.96$\pm$0.00&98$\pm$2&...&92$\pm$2&1.08$\pm$0.02&$1.967^{(14)}$\\
081126(R)&0.10,1&...&...&0.39$\pm$0.01&49$\pm$0&...&...&...&$...$\\
081203A(U)&0.08,6&...&...&1.47$\pm$0.00&1244$\pm$6&...&1306$\pm$7&0.596&$2.1^{(6)}$\\
090510(R)&0.11,104&...&...&0.98$\pm$0.12&4$\pm$2&...&1$\pm$0&0.76$\pm$0.14&$0.903^{(6)}$\\
100418A(R)&1.10,1370&...&...&1.37$\pm$0.13&158$\pm$80&...&16$\pm$8&0.98$\pm$0.09&$0.6235^{(15)}$\\
100621A(R)&0.26,4&...&...&0.46$\pm$0.06&5$\pm$1&...&...&0.8$\pm$0.1&$0.542^{(16)}$\\
100814A(R)&0.52,18&...&...&1.90$\pm$0.05&28$\pm$35&...&...&0.180.08&$1.44^{(17)}$\\
100906A(R)&0.05,11&...&...&1.07$\pm$0.02&487$\pm$35&...&246$\pm$17&...&$1.727^{(18)}$\\
120119A(R)&0.08,17&...&...&1.36$\pm$0.03&69$\pm$6&...&51$\pm$5&0.89$\pm$0.01&$1.728^{(19)}$\\
121217A(R)&1.34,347&...&...&0.78$\pm$0.01&147$\pm$3&...&...&0.87$\pm$0.04&$3.1^{(20)}$\\
\hline
&Group III\\
\hline
050416A(R)&4.09,33&15.3$\pm$6.84&0.39$\pm$0.15&1.32$\pm$0.32&7$\pm$4&0.05$\pm$0.02&1$\pm$0&1.3&$0.65^{(3)}$\\
050525A(V)&0.07,316&10.4$\pm$2.33&0.78$\pm$0.01&1.52$\pm$0.07&202$\pm$61&0.49$\pm$0.10&19$\pm$6&0.97$\pm$0.1&$0.606^{(3)}$\\
050801(R)&0.02,22&0.2$\pm$0.01&0.04$\pm$0.02&1.20$\pm$0.01&131$\pm$5&196.77$\pm$4.77&80$\pm$3&1$\pm$0.16&$1.56^{(1)}$\\
051109A(R)&0.04,1040&0.4$\pm$0.32&0.40$\pm$0.08&1.04$\pm$0.04&313$\pm$193&210.95$\pm$109.42&398$\pm$246&0.7&$2.346^{(3)}$\\
051111(R)&0.03,8&3.1$\pm$0.67&0.81$\pm$0.01&2.10$\pm$0.70&299$\pm$86&31.38$\pm$5.75&181$\pm$52&0.76$\pm$0.07&$1.55^{(8)}$\\
061126(R)&1.06,156&30.4$\pm$8.94&0.72$\pm$0.05&1.39$\pm$0.07&114$\pm$46&0.63$\pm$0.20&40$\pm$16&0.95&$1.1588^{(3)}$\\
080413A(R)&0.01,1&0.4$\pm$0.07&0.64$\pm$0.03&1.82$\pm$0.29&136$\pm$33&477.21$\pm$88.04&185$\pm$44&0.67&$2.433^{(7)}$\\
090426(R)&0.09,2&0.2$\pm$0.05&0.32$\pm$0.14&1.24$\pm$0.06&14$\pm$3&113.06$\pm$17.79&21$\pm$5&0.76$\pm$0.14&$2.609^{(6)}$\\
130702A(R)&15.13,280&91.3$\pm$81.29&0.44$\pm$0.22&1.22$\pm$0.29&1259$\pm$1067&0.03$\pm$0.02&6$\pm$5&0.7$\pm$0.1&$0.145^{(21)}$\\
\hline
&Group IV\\
\hline
050603(V)&34.09,220&...&...&1.80&318$\pm$24&...&556$\pm$42&0.2$\pm$0.1&$2.821^{(3)}$\\
050721(R)&1.48,8&...&...&1.32$\pm$0.03&26$\pm$7&...&...&1.16$\pm$0.35&$...$\\
050820A(R)&0.23,663&...&...&0.91$\pm$0.00&981$\pm$14&...&1503$\pm$21&0.72$\pm$0.03&$2.612^{(3)}$\\
060124(R)&3.34,1980&...&...&0.88$\pm$0.02&1015$\pm$247&...&1246$\pm$303&0.73$\pm$0.08&$2.296^{(3)}$\\
060206(R)&18.41,202&...&...&1.32$\pm$0.01&339$\pm$43&...&1061$\pm$135&0.73$\pm$0.05&$4.408^{(3)}$\\
061007(R)&0.03,15&...&...&1.67&2657$\pm$53&...&1090$\pm$22&0.78$\pm$0.02&$1.26^{(8)}$\\
070125(R)&105.86,349&...&...&1.64$\pm$0.15&349$\pm$1201&...&210$\pm$724&0.55$\pm$0.04&$0.82^{(22)}$\\
070311(R)&0.07,92&...&...&0.73$\pm$0.02&341$\pm$86&...&...&1$\pm$0.2&$0.98^{(6)}$&\\
071010A(R)&12.24,523&...&...&1.91$\pm$0.02&256$\pm$22&...&64$\pm$5&0.68&$0.98^{(6)}$\\
071025(J)&0.17,15&...&...&1.03$\pm$0.04&131$\pm$11&...&532$\pm$46&0.42$\pm$0.08&$4.2^{(6)}$\\
071112C(R)&0.13,70&...&...&0.88$\pm$0.02&52$\pm$6&...&9$\pm$1&0.63$\pm$0.29&$0.82^{(23)}$\\
080319B(R)&0.18,4590&...&...&1.31$\pm$0.00&1564$\pm$50&...&359$\pm$11&0.66&$0.94^{(6)}$\\
080804(R)&1.16,26&...&...&0.93$\pm$0.02&29$\pm$7&...&33$\pm$8&0.43&$2.2^{(6)}$\\
080928(R)&7.77,233&...&...&2.13$\pm$0.08&187$\pm$138&...&133$\pm$98&1.08$\pm$0.02&$1.69^{(10)}$\\
081109A(H)&0.17,67&...&...&0.94$\pm$0.03&86$\pm$18&...&21$\pm$5&...&$...$\\
090102(R)&0.94,265&...&...&1.02$\pm$0.01&49$\pm$6&...&30$\pm$4&0.74&$1.547^{(6)}$\\
090323(R)&162.43,769&...&...&1.77$\pm$0.19&75$\pm$1603&...&192$\pm$4114&0.65$\pm$0.13&$3.57^{(10)}$\\
090328(U)&57.89,263&...&...&1.54$\pm$0.12&228$\pm$475&...&32$\pm$67&1.19$\pm$0.2&$0.736^{(10)}$\\
090902B(R)&62.29,564&...&...&1.00$\pm$0.06&98$\pm$74&...&80$\pm$60&0.68$\pm$0.11&$1.822^{(24)}$\\
090926A(R)&84.92,102&...&...&1.47$\pm$0.01&74$\pm$8&...&78$\pm$9&0.72$\pm$0.17&$2.1062^{(25)}$\\
091029(I)&0.31,5&...&...&0.59$\pm$0.01&14$\pm$0&...&24$\pm$0&0.57$\pm0.12$&$2.752^{(26)}$\\
100728B(R)&0.16,6&...&...&1.01$\pm$0.02&13$\pm$2&...&14$\pm$2&...&$2.106^{(27)}$\\
100901A(R)&10.19,543&...&...&1.38$\pm$0.001&949$\pm$30&...&480$\pm$15&0.82&$1.408^{(28)}$\\
101024A(R)&1.42,160&...&...&1.02$\pm$0.08&22$\pm$15&...&...&...&$...$\\
110205A(R)&0.54,384&...&...&1.47$\pm$0.00&355$\pm$3&...&410$\pm$3&0.33&$2.22^{(29)}$\\
110213A(R)&0.10,183&...&...&0.87$\pm$0.08&1515$\pm$421&...&820$\pm$228&1.12$\pm$0.24&$1.46^{(29)}$\\
110918A(R)&122.43,805&...&...&1.31$\pm$0.04&434$\pm$252&...&110$\pm$64&0.7$\pm$0.02&$0.984^{(30)}$\\
120404A(R)&0.73,57&...&...&1.47$\pm$0.02&177$\pm$6&...&228$\pm$7&1.05$\pm$0.09&$2.87^{(31)}$\\
120815A(R)&0.17,11&...&...&0.64$\pm$0.001&40$\pm$1&...&51$\pm$1&0.78$\pm$0.01&$2.358^{(32)}$\\
130427A(R)&0.31,456&...&...&1.04$\pm$0.001&5855$\pm$33&...&168$\pm$1&0.69$\pm$0.01&$0.34^{(33)}$\\
\enddata
\tablerefs{
(1) Butler et al.(2007);
(2) Nysewander et al.(2009);
(3) Nava et al.(2008);
(4) Jakobsson et al(2006GCN,5319)
(5) Fynbo et al. (2009);
(6) Butler et al.(2010)
(7) Krimm et al.(2009)
(8) Ukwatta et al. (2010);
(9) Nardini et al. (2011)
(10) Guetta et al.(2011);
(11) Wiersema et al.(2012)
(12) Vergani  et al. (2011)
(13) Page et al.(2007)
(14) Yuan et al.(2010);
(15) Marshall et al.(2011)
(16) Golenetskii et al. (2010)
(17) Nardini et al.(2014)
(18) Tanvir  et al.(2010)
(19) Morgan et al.(2014)
(20) Elliott et al.(2014)
(21) Kelly et al.(2013)
(22) Bellm et al.(2008)
(23) Krim et al.(2007)
(24) Pandey et al.(2010)
(25) Rau et al.(2010)
(26) Filgas et al.(2012)
(27) Barthelmy et al.(2010GCN11023)
(28) Hartoog et al.(2013)
(29) Cucchiara et al.(2011)
(30) Elliott  et al.(2013)
(31) Guidorzi et al.(2014)
(32) Kruehler et al.(2013)
(33) Vestrand et al.(2014)								
}
\tablenotetext{a}{The time intervals, for the fit results of the optical afterglow for each GRB; and the break time for the optical break samples, in units of ks.}
\tablenotetext{b}{The decay slopes before and after break time, note that although we also fit the rise slope for the GRBs which have the afterglow onset behaviors (such as GRB 071010A, 100901A), we did not summarize the information of the rise slope and peak time in our tables, since, in this paper, we focus on the decay behaviors.}
\tablenotetext{c}{Optical fluence, which is calculated by integrating to flux over the time interval, in units of $10^{-10}$ erg cm$^{-2}$; the luminosity at the break time, in units of $10^{45}$ erg s$^{-1}$; and the isotropic energy release in the R-band, in units of $10^{48}$ erg.}
\tablenotetext{d}{The spectral index of the optical afterglow and the redshift of gamma-ray bursts.}
\end{deluxetable}

\clearpage
\begin{deluxetable}{cccccccccccccccccccccccccc}
%\rotate
\tabletypesize{\scriptsize}
\tablecaption{The Relation Parameters of X-ray and Optical Afterglows.}
\tablewidth{0pt}
\tabletypesize{\tiny}
\tablehead{
\colhead{GRB}&
\colhead{$T_{\rm x,b}/T_{\rm o,b}$\tablenotemark{a}}&
\colhead{$\Delta\alpha_{\rm 1}$\tablenotemark{b}}&
\colhead{$\Delta\alpha_{\rm 2}$\tablenotemark{b}}&
\colhead{$\Delta\alpha_{\rm x}$\tablenotemark{b}}&
\colhead{$\Delta\alpha_{\rm o}$\tablenotemark{b}}&
\colhead{$\Delta\beta_{\rm x}$\tablenotemark{b}}&
\colhead{$L_{\rm X,b}/L_{\rm R}$\tablenotemark{c}}&
\colhead{$E_{\rm X,b}/E_{\rm R}$\tablenotemark{c}}&
\colhead{$F_{\rm X}/F_{\rm o}$\tablenotemark{c}}&
}
\startdata
\hline
&Group I\\
\hline
050319&0.15$\pm$0.05&-0.06$\pm$0.04&-0.39$\pm$0.20&1.07$\pm$0.20&1.40$\pm$0.00&0.25$\pm$0.24&144$\pm$30&23$\pm$9&42$\pm$7\\
050408&11.35$\pm$5.79&0.37$\pm$0.03&0.49$\pm$0.20&0.79$\pm$0.21&0.67$\pm$0.14&...&13$\pm$6&237$\pm$102&146$\pm$68\\
050730&1.15$\pm$0.15&0.25$\pm$0.12&1.24$\pm$0.05&1.97$\pm$0.12&0.98$\pm$0.03&0.17$\pm$0.08&164$\pm$29&74$\pm$18&217$\pm$28\\
050922C&0.51$\pm$0.26&0.51$\pm$0.05&-0.01$\pm$0.05&0.53$\pm$0.06&1.04$\pm$0.04&0.02$\pm$0.16&30$\pm$18&21$\pm$14&13$\pm$28\\
060210&27.00$\pm$37.64&0.72$\pm$0.41&0.19$\pm$0.14&0.58$\pm$0.09&1.12$\pm$0.42&0.47$\pm$0.11&17$\pm$27&704$\pm$1195&324$\pm$318\\
060526&0.78$\pm$0.18&-0.10$\pm$0.60&0.64$\pm$0.48&2.26$\pm$0.77&1.51$\pm$0.03&-0.05$\pm$0.50&6$\pm$2&3$\pm$282&3$\pm$1\\
060605&0.35$\pm$0.05&-0.48$\pm$0.06&-0.40$\pm$0.15&1.73$\pm$0.09&1.65$\pm$0.13&0.16$\pm$0.20&35$\pm$9&10$\pm$3&7$\pm$1\\
060614&0.76$\pm$0.05&-0.04$\pm$0.05&-0.03$\pm$0.05&1.85$\pm$0.06&1.84$\pm$0.03&0.13$\pm$0.18&28$\pm$2&...&22$\pm$3\\
060708&12.00$\pm$6.31&0.53$\pm$0.11&0.46$\pm$0.05&0.72$\pm$0.07&0.79$\pm$0.10&-0.01$\pm$0.28&1$\pm$1&7$\pm$4&7$\pm$11\\
060714&0.59$\pm$0.21&0.23$\pm$0.11&0.14$\pm$0.06&0.88$\pm$0.12&0.97$\pm$0.05&0.19$\pm$0.23&43$\pm$16&36$\pm$16&30$\pm$10\\
060729&1.23$\pm$0.18&-0.01$\pm$0.10&0.03$\pm$0.06&1.27$\pm$0.02&1.23$\pm$0.12&-0.01$\pm$0.06&7$\pm$1&10$\pm$2&6$\pm$1\\
060912A&2.18$\pm$2.15&-0.14$\pm$0.34&-0.19$\pm$0.05&0.49$\pm$0.34&0.54$\pm$0.02&0.38$\pm$0.48&3$\pm$3&14$\pm$15&18$\pm$16\\
061021&0.19$\pm$0.06&-0.05$\pm$0.03&-0.85$\pm$0.24&0.55$\pm$0.04&1.35$\pm$0.24&-0.05$\pm$0.12&113$\pm$36&29$\pm$12&33$\pm$16\\
070411&0.15$\pm$0.15&0.00$\pm$0.09&-0.78$\pm$0.22&0.54$\pm$0.24&1.32$\pm$0.00&-0.05$\pm$0.39&105$\pm$119&15$\pm$10&11$\pm$12\\
070518&4.83$\pm$15.07&-0.01$\pm$0.18&-0.56$\pm$0.64&0.65$\pm$0.64&1.20$\pm$0.17&0.52$\pm$1.32&5$\pm$8&36$\pm$104&11$\pm$17\\
080310&4.99$\pm$0.88&0.13$\pm$0.09&0.51$\pm$0.06&1.50$\pm$0.11&1.12$\pm$0.03&-0.15$\pm$0.23&4$\pm$1&11$\pm$2&7$\pm$3\\
080413B&0.27$\pm$0.05&0.39$\pm$0.01&-1.31$\pm$0.12&0.86$\pm$0.12&2.56$\pm$0.01&-0.07$\pm$0.30&34$\pm$8&18$\pm$6&21$\pm$4\\
080710&1.74$\pm$0.51&0.36$\pm$0.15&0.39$\pm$0.14&1.11$\pm$0.21&1.08$\pm$0.01&0.37$\pm$0.24&3$\pm$1&5$\pm$2&5$\pm$2\\
080721&0.16$\pm$0.01&-0.20$\pm$0.01&0.17$\pm$0.01&0.85$\pm$0.02&0.47$\pm$0.01&0.15$\pm$0.06&1794$\pm$134&151$\pm$12&180$\pm$10\\
081029&1.08$\pm$0.25&-0.05$\pm$0.20&0.62$\pm$0.17&2.27$\pm$0.16&1.60$\pm$0.21&0.12$\pm$0.23&4$\pm$1&5$\pm$2&4$\pm$1\\
090618&0.28$\pm$0.03&0.02$\pm$0.01&-0.00$\pm$0.06&0.79$\pm$0.02&0.81$\pm$0.06&-0.07$\pm$0.06&194$\pm$20&59$\pm$8&61$\pm$5\\
090727&112.67$\pm$148.33&0.44$\pm$0.34&0.82$\pm$0.30&1.09$\pm$0.27&0.71$\pm$0.37&-0.17$\pm$0.61&...&...&22$\pm$11\\
091018&0.01$\pm$0.00&-0.59$\pm$0.06&-0.43$\pm$0.02&0.93$\pm$0.06&0.77$\pm$0.00&0.20$\pm$0.17&2188$\pm$359&14$\pm$3&13$\pm$5\\
091127&1.07$\pm$0.55&0.57$\pm$0.10&0.10$\pm$0.13&0.56$\pm$0.05&1.03$\pm$0.16&-0.03$\pm$0.14&34$\pm$19&62$\pm$47&103$\pm$40\\
\hline
&Group II\\
\hline
050401&...&...&0.72$\pm$0.07&0.92$\pm$0.07&...&0.01$\pm$0.26&...&749$\pm$472&2184$\pm$888\\
050502A&...&...&0.56$\pm$0.05&1.00$\pm$0.08&...&...&...&0$\pm$9&60$\pm$12\\
050824&...&...&0.38$\pm$0.13&0.63$\pm$0.19&...&0.63$\pm$0.42&...&4$\pm$2&8$\pm$4\\
051221A&...&...&0.48$\pm$0.07&1.25$\pm$0.17&...&0.10$\pm$0.27&...&40$\pm$25&11$\pm$4\\
060111B&...&...&0.46$\pm$0.16&0.72$\pm$0.16&...&0.04$\pm$0.38&...&...&85$\pm$63\\
060418&...&...&0.30$\pm$0.06&0.70$\pm$0.14&...&-0.03$\pm$0.31&...&10$\pm$5&8$\pm$4\\
060906&...&...&0.98$\pm$0.23&1.53$\pm$0.18&...&-0.30$\pm$0.33&...&143$\pm$51&24$\pm$40\\
060908&...&...&0.47$\pm$0.07&1.12$\pm$0.12&...&-0.07$\pm$0.36&...&32$\pm$11&26$\pm$8\\
060927&...&...&0.82$\pm$0.46&1.31$\pm$0.47&...&0.10$\pm$0.43&...&74$\pm$79&8$\pm$7\\
061121&...&...&0.41$\pm$0.23&1.07$\pm$0.03&...&-0.13$\pm$0.13&...&158$\pm$33&37$\pm$25\\
070318&...&...&1.27$\pm$0.51&1.56$\pm$0.51&...&-0.20$\pm$0.50&...&15$\pm$5&6$\pm$7\\
070420&...&...&0.57$\pm$0.09&1.21$\pm$0.06&...&-0.15$\pm$0.19&...&...&43$\pm$27\\
071031&...&...&0.91$\pm$0.48&0.78$\pm$0.49&...&-0.47$\pm$1.02&...&8$\pm$1&7$\pm$1\\
080810&...&...&0.67$\pm$0.11&0.70$\pm$0.18&...&0.14$\pm$0.25&...&4$\pm$3&10$\pm$13\\
081008&...&...&0.89$\pm$0.14&0.98$\pm$0.15&...&-0.22$\pm$0.35&...&61$\pm$22&64$\pm$94\\
081126&...&...&1.17$\pm$0.06&0.96$\pm$0.07&...&0.39$\pm$0.25&...&...&6$\pm$6\\
081203A&...&...&0.68$\pm$0.12&0.98$\pm$0.12&...&0.06$\pm$0.37&...&7$\pm$2&5$\pm$2\\
090510&...&...&1.17$\pm$0.15&1.53$\pm$0.10&...&0.38$\pm$0.22&...&1540$\pm$894&736$\pm$276\\
100418A&...&...&0.29$\pm$0.17&1.68$\pm$0.11&...&-0.05$\pm$0.34&...&31$\pm$20&17$\pm$67\\
100621A&...&...&1.20$\pm$0.10&0.80$\pm$0.09&...&0.72$\pm$0.33&...&...&496$\pm$466\\
100814A&...&...&0.19$\pm$0.08&1.59$\pm$0.06&...&0.18$\pm$0.14&...&...&261$\pm$230\\
100906A&...&...&1.05$\pm$0.80&1.36$\pm$0.80&...&0.09$\pm$0.18&...&52$\pm$12&18$\pm$14\\
120119A&...&...&0.67$\pm$0.18&1.03$\pm$0.18&...&0.31$\pm$0.40&...&126$\pm$49&61$\pm$19\\
121217A&...&...&0.66$\pm$0.05&0.96$\pm$0.10&...&-0.01$\pm$0.23&...&...&55$\pm$96\\
\hline
&Group III\\
\hline
050416A&...&...&-0.43$\pm$0.32&...&0.93$\pm$0.35&...&...&631$\pm$388&95$\pm$30\\
050525A&...&...&0.01$\pm$0.08&...&0.75$\pm$0.07&...&...&16$\pm$9&54$\pm$22\\
050801&...&...&-0.13$\pm$0.06&...&1.16$\pm$0.02&...&...&5$\pm$2&3$\pm$4\\
051109A&...&...&0.15$\pm$0.04&...&0.65$\pm$0.09&...&...&31$\pm$23&56$\pm$26\\
051111&...&...&-0.49$\pm$0.70&...&1.29$\pm$0.70&...&...&2$\pm$2&9$\pm$7\\
061126&...&...&-0.05$\pm$0.07&...&0.67$\pm$0.09&...&...&133$\pm$66&166$\pm$54\\
080413A&...&...&-0.62$\pm$0.30&...&1.18$\pm$0.30&...&...&6$\pm$3&15$\pm$11\\
090426&...&...&-0.21$\pm$0.07&...&0.91$\pm$0.15&...&...&55$\pm$29&25$\pm$45\\
130702A&...&...&-0.06$\pm$0.29&...&0.78$\pm$0.37&...&...&33$\pm$34&163$\pm$128\\
\hline
&Group IV\\
\hline
050603&...&...&-0.09$\pm$0.06&...&...&...&...&5$\pm$3&3$\pm$2\\
050721&...&...&0.00$\pm$0.03&...&...&...&...&...&32$\pm$7\\
050820A&...&...&0.33$\pm$0.01&...&...&...&...&24$\pm$3&56$\pm$41\\
060124&...&...&0.45$\pm$0.02&...&...&...&...&20$\pm$8&91$\pm$18\\
060206&...&...&-0.07$\pm$0.02&...&...&...&...&15$\pm$5&4$\pm$1\\
061007&...&...&0.06$\pm$0.01&...&...&...&...&18$\pm$1&8$\pm$3\\
070125&...&...&-0.01$\pm$0.19&...&...&...&...&9$\pm$45&3$\pm$3\\
070311&...&...&0.36$\pm$0.07&...&...&...&...&...&9$\pm$6\\
071010A&...&...&-0.03$\pm$0.14&...&...&...&...&8$\pm$1&8$\pm$6\\
071025&...&...&0.51$\pm$0.04&...&...&...&...&87$\pm$15&22$\pm$32\\
071112C&...&...&0.45$\pm$0.02&...&...&...&...&80$\pm$18&42$\pm$56\\
080319B&...&...&0.31$\pm$0.00&...&...&...&...&192$\pm$10&76$\pm$2\\
080804&...&...&0.18$\pm$0.02&...&...&...&...&107$\pm$35&39$\pm$7\\
080928&...&...&-0.53$\pm$0.08&...&...&...&...&11$\pm$11&2$\pm$6\\
081109A&...&...&0.29$\pm$0.03&...&...&...&...&55$\pm$18&38$\pm$28\\
090102&...&...&0.28$\pm$0.01&...&...&...&...&223$\pm$46&175$\pm$19\\
090323&...&...&-0.15$\pm$0.21&...&...&...&...&15$\pm$328&3$\pm$8\\
090328&...&...&0.14$\pm$0.14&...&...&...&...&11$\pm$31&11$\pm$18\\
090902B&...&...&0.40$\pm$0.07&...&...&...&...&50$\pm$55&124$\pm$103\\
090926A&...&...&-0.06$\pm$0.04&...&...&...&...&44$\pm$25&2$\pm$1\\
091029&...&...&0.68$\pm$0.05&...&...&...&...&272$\pm$72&4$\pm$15\\
100728B&...&...&0.50$\pm$0.07&...&...&...&...&44$\pm$25&74$\pm$26\\
100901A&...&...&0.10$\pm$0.03&...&...&...&...&8$\pm$1&34$\pm$94\\
101024A&...&...&0.35$\pm$0.09&...&...&...&...&...&120$\pm$105\\
110205A&...&...&0.20$\pm$0.01&...&...&...&...&15$\pm$2&7$\pm$7\\
110213A&...&...&1.17$\pm$0.09&...&...&...&...&11$\pm$4&23$\pm$89\\
110918A&...&...&0.29$\pm$0.05&...&...&...&...&70$\pm$57&112$\pm$63\\
120404A&...&...&0.32$\pm$0.12&...&...&...&...&12$\pm$4&7$\pm$66\\
120815A&...&...&0.28$\pm$0.04&...&...&...&...&48$\pm$19&40$\pm$54\\
130427A&...&...&0.25$\pm$0.00&...&...&...&...&83$\pm$2&83$\pm$1\\

\enddata
\tablenotetext{a}{The temporal ratio of break times (Group I samples).}
\tablenotetext{b}{The various values of pre-break decay ($\alpha_{\rm x,1}-\alpha_{\rm o,1}$)and post-break decay($\alpha_{\rm x,2}-\alpha_{\rm o,2}$); the value of decay slope of pre- and post break of X-ray (Group I and Group II) and optical break samples (Group I and Group III); and the various values of the spectral index for X-ray break samples (Group I and Group II).}
\tablenotetext{c}{The ratio of break luminosity, isotropic energy release and flux at 1 hour between X-ray and optical band.}
\end{deluxetable}

\clearpage
\begin{deluxetable}{ccccccccccccccccccccccccc}
%\rotate
\tabletypesize{\scriptsize}
\tablecaption{Test the Temporal Indices for Various External Shock Models}
\tablewidth{520pt}
\tabletypesize{\tiny}
\tablehead{
\colhead{}&
\colhead{Total\tablenotemark{a}}&
\colhead{Chromatic?\tablenotemark{b}}&
\colhead{Injection\tablenotemark{c}}&
\colhead{Standard\tablenotemark{d}}&
\colhead{Standard\tablenotemark{d}}&
\colhead{Injection\tablenotemark{c}}&
\colhead{Chromatic?\tablenotemark{b}}&
\colhead{External\tablenotemark{e}}&
\colhead{Standard\tablenotemark{e}}&\\
&
&
\colhead{wind}&
\colhead{wind}&
\colhead{wind}&
\colhead{ISM}&
\colhead{ISM}&
\colhead{ISM}&
\colhead{ISM and wind}&
\colhead{ISM and wind}&
&
&\\
&
&
\colhead{$<-0.5$}&
\colhead{$(-0.5, -0.25)$}&
\colhead{$(-0.25, 0.0)$}&
\colhead{$(0.0, 0.25)$}&
\colhead{$(0.25,0.5)$}&
\colhead{$>0.5$}&
\colhead{$(-0.5,0.5)$}&
\colhead{$(-0.25,0.25)$}&
\\&
}
\startdata
\hline
Group Samples\\
\hline
\hline
Group I\\
\hline
$\Delta\alpha(\alpha_{\rm x,1}-\alpha_{\rm o,1})$&24&1&1&8&5&4&5&18(75\%)/23(96\%)\tablenotemark{f}&13(54\%)/16(67\%)\\
$\Delta\alpha(\alpha_{\rm x,2}-\alpha_{\rm o,2})$&24&3&3&5&5&3&5&16(67\%)/21(88\%)&10(42\%)/16(67\%)\\
\hline
Group II\\
\hline
$\Delta\alpha(\alpha_{\rm x,1}-\alpha_{\rm o,2})$&24&7&11&3&2&1&0&17(71\%)/20(83\%)&5(21\%)/10(42\%)\\
$\Delta\alpha(\alpha_{\rm x,2}-\alpha_{\rm o,2})$&24&0&0&0&1&7&16&8(33\%)/14(58\%)&1(4\%)/7(29\%)\\
\hline
Group III\\
\hline
$\Delta\alpha(\alpha_{\rm x,2}-\alpha_{\rm o,1})$&9&0&0&0&0&3&6&3(33\%)/6(67\%)&0(0)/1(11\%)\\
$\Delta\alpha(\alpha_{\rm x,2}-\alpha_{\rm o,2})$&9&1&2&4&2&0&0&8(89\%)/9(100\%)&6(67\%)/8(89\%)\\
\hline
Group IV\\
\hline
$\Delta\alpha(\alpha_{\rm x,2}-\alpha_{\rm o,2})$&30&1&0&6&7&14&2&27(90\%)/29(97\%)&13(43\%)/20(67\%)\\
\hline
\hline
Epoch Samples\\
\hline
\hline
epoch 1\\
\hline
$\Delta\alpha(\alpha_{\rm x}-\alpha_{\rm o})$&50&6&7&9&7&10&11&33(66\%)/42(84\%)&16(32\%)/23(46\%)\\
\hline
epoch 2\\
\hline
$\Delta\alpha(\alpha_{\rm x}-\alpha_{\rm o})$&97&8&13&15&15&22&24&65(67\%)/76(78\%)&30(31\%)/45(46\%)\\
\hline
epoch 3\\
\hline
$\Delta\alpha(\alpha_{\rm x}-\alpha_{\rm o})$&80&12&6&13&13&15&21&47(59\%)/60(75\%)&26(33\%)/37(46\%)\\
\hline
epoch 4\\
\hline
$\Delta\alpha(\alpha_{\rm x}-\alpha_{\rm o})$&35&4&2&5&7&12&5&26(74\%)/28(80\%)&12(34\%)/19(54\%)\\
\hline
\enddata
\tablenotetext{a}{Total numbers of data points for each Group or epoch samples.}
\tablenotetext{b}{Numbers of the data points for which are $\Delta\alpha<-0.5$ or $\Delta\alpha>-0.5$, which do not fall into our external shock model ranges.}
\tablenotetext{c}{Numbers of the data points for energy injection model, ISM and wind-like medium.}
\tablenotetext{d}{Numbers of the data points for standard afterglow model, ISM and wind-like medium.}
\tablenotetext{e}{Numbers of the data points (brackets correspond with the fraction) for satisfied with the external shock and standard afterglow model ranges for each Groups or epochs samples.}
\tablenotetext{f}{Not consider error bar/consider error bar.}
\end{deluxetable}

\begin{deluxetable}{cccccccccccccccccc}
\tabletypesize{\scriptsize}
\tablecaption{The Probability of the K-S Test for Our Epochs Samples}
\tablewidth{0pt}
\tabletypesize{\tiny}
\tablehead{
\colhead{epochs}&
\colhead{$P_{\rm K-S}(\alpha_{\rm x,obs},\alpha_{\rm x,Sim})\tablenotemark{a}$}&
\colhead{$P_{\rm K-S}(\alpha_{\rm o,obs},\alpha_{\rm o,Sim})\tablenotemark{b}$}&
}
\startdata
epoch 1	&	0.02	&	1.32$\times10^{-10}$\\
epoch 2	&	8.36$\times10^{-3}$	&	2.28$\times10^{-9}$\\
epoch 3	&	0.24	&	7.32$\times10^{-7}$\\
epoch 4	&	0.03	&	0.49	\\
\enddata
\tablenotetext{a}{The probability of the K-S test for observed decay slope and simulated decay slope for the X-ray band.}
\tablenotetext{b}{The probability of the K-S test for observed decay slope and simulated decay slope for the optical band.}
\end{deluxetable}

\clearpage
\begin{deluxetable}{ccccccccccccccccccccccccc}
\rotate
\tabletypesize{\scriptsize}
\tablecaption{The Closure Relation of Afterglows in Gamma-ray Bursts}
\tablewidth{0pt}
\tabletypesize{\tiny}
\tablehead{
\colhead{$$}&
\colhead{$\alpha(q=1)$}&
\colhead{$\alpha(0\leq q<1)$}&
\colhead{$\alpha$}&
\colhead{$\beta$}&
\colhead{$\alpha(\beta)(q=1)$}&
\colhead{$\alpha(\beta)(0\leq q<1)$}&
\colhead{$\alpha(\beta)$}\\&
\colhead{Non-Injection}&
\colhead{Injection}&
\colhead{Jet}&
&
\colhead{Non-Injection}&
\colhead{Injection}&
\colhead{Jet}&
}
\startdata
\hline
& ISM, Slow cooling\\
\hline
$\nu<\upsilon_{m}$&$\frac{1}{2}$&$\frac{8-5q}{6}$&$\frac{1}{2}$&$\frac{1}{3}$&$\alpha=\frac{3\beta}{2}$&$\alpha=(q-1)+\frac{(2+q)\beta}{2}$&$\alpha=\frac{3\beta}{2}$\\
$\upsilon_{m}<\nu<\upsilon_{c}$&$\frac{3(1-p)}{4}$&$\frac{(6-2p)-(p+3)q}{4}$&$\sim-p$&$\frac{1-p}{2}$&$\alpha=\frac{3\beta}{2}$&$\alpha=(q-1)+\frac{(2+q)\beta}{2}$&$\alpha=2\beta+1$\\
$\nu>\upsilon_{c}$&$\frac{2-3p}{4}$&$\frac{(4-2p)-(p+2)q}{4}$&$\sim-p$&$-\frac{p}{2}$&$\alpha=\frac{3\beta-1}{2}$&$\alpha=\frac{q-2}{2}+\frac{(2+q)\beta}{2}$&$\alpha=2\beta$\\
\hline
& ISM, Fast cooling\\
\hline
$\nu<\upsilon_{c}$&$\frac{1}{6}$&$\frac{8-7q}{6}$&...&$\frac{1}{3}$&$\alpha=\frac{\beta}{2}$&$\alpha=(q-1)+\frac{(2-q)\beta}{2}$&...\\
$\upsilon_{c}<\nu<\upsilon_{m}$&$-\frac{1}{4}$&$\frac{2-3q}{4}$&...&$-\frac{1}{2}$&$\alpha=\frac{\beta}{2}$&$\alpha=(q-1)+\frac{(2-q)\beta}{2}$&...\\
$\nu>\upsilon_{m}$&$\frac{2-3p}{4}$&$\frac{(4-2p)-(p+2)q}{4}$&...&$-\frac{p}{2}$&$\alpha=\frac{3\beta-1}{2}$&$\alpha=\frac{q-2}{2}+\frac{(2+q)\beta}{2}$&...\\
\hline
& Wind, Slow cooling\\
\hline
$\nu<\upsilon_{m}$&$0$&$\frac{(1-q)}{3}$&$\frac{1}{2}$&$\frac{(q-1)}{3}$&$\frac{1}{2}$&$\alpha=\frac{q}{2}+\frac{(2+q)\beta}{2}$&$\alpha=\frac{3\beta}{2}$\\
$\upsilon_{m}<\nu<\upsilon_{c}$&$\frac{1-3p}{4}$&$\frac{(2-2p)-(p+1)q}{4}$&$\sim-p$&$\frac{1-p}{2}$&$\alpha=\frac{3\beta+1}{2}$&$\alpha=\frac{q}{2}+\frac{(2+q)\beta}{2}$&$\alpha=2\beta+1$\\
$\nu>\upsilon_{c}$&$\frac{2-3p}{4}$&$\frac{(4-2p)-(p+2)q}{4}$&$\sim-p$&$-\frac{p}{2}$&$\alpha=\frac{3\beta-1}{2}$&$\alpha=\frac{q-2}{2}+\frac{(2+q)\beta}{2}$&$\alpha=2\beta$\\
\hline
& Wind, Fast cooling\\
\hline
$\nu<\upsilon_{c}$&$-\frac{2}{3}$&$-\frac{(1+q)}{3}$&$...$&$\frac{1}{3}$&$\alpha=\frac{1-\beta}{2}$&$\alpha=\frac{q}{2}-\frac{(2-q)\beta}{2}$&...\\
$\upsilon_{c}<\nu<\upsilon_{m}$&$-\frac{1}{4}$&$\frac{2-3q}{4}$&$...$&$-\frac{1}{2}$&$\alpha=\frac{1-\beta}{2}$&$\alpha=\frac{q}{2}-\frac{(2-q)\beta}{2}$&...\\
$\nu>\upsilon_{m}$&$\frac{2-3p}{4}$&$\frac{(4-2p)-(p+2)q}{4}$&$...$&$-\frac{p}{2}$&$\alpha=\frac{3\beta-1}{2}$&$\alpha=\frac{q-2}{2}+\frac{(2+q)\beta}{2}$&...\\
\hline
& $p=2.5(p=2,p=3),q=0$\\
\hline
& ISM, Slow cooling\\
\hline
$\nu<\upsilon_{m}$&$0.5(0.5,0.5)$&$1.3(1.3,1.3)$&$0.5(0.5,0.5)$&$0.3(0.3,0.3)$\\
$\upsilon_{m}<\nu<\upsilon_{c}$&$-1.1(-0.75,-1.5)$&$0.25(0.5,0.0)$&$-2.5(-2.0,-3.0)$&$-0.75(-0.75,-0.75)$\\
$\nu>\upsilon_{c}$&$-1.4(-1.0,-1.75)$&$-0.25(0.0,-0.5)$&$-2.5(-2.0,-3.0)$&$-1.25(-1.25,-1.25)$\\
\hline
& ISM, Fast cooling\\
\hline
$\nu<\upsilon_{c}$&$0.2(0.2,0.2)$&$1.3(1.3,1.3)$&...&$0.3(0.3,0.3)$\\
$\upsilon_{c}<\nu<\upsilon_{m}$&$-0.25(-0.25,-0.25)$&$0.5(0.5,0.5)$&...&$-0.5(-0.5,-0.5)$\\
$\nu>\upsilon_{m}$&$-1.4(-1.0,-1.75)$&$-0.25(0.0,-0.5)$&...&$-1.25(-1.25,-1.25)$\\
\hline
& Wind, Slow cooling\\
\hline
$\nu<\upsilon_{m}$&$0.0(0.0,0.0)$&$0.3(0.3,0.3)$&$0.5(0.5,0.5)$&$0.3(0.3,0.3)$\\
$\upsilon_{m}<\nu<\upsilon_{c}$&$-1.6(-1.25,-2.0)$&$-0.75(-0.5,-1.0)$&$-2.5(-2.0,-3.0)$&$-0.75(-0.75,-0.75)$\\
$\nu>\upsilon_{c}$&$-1.4(-1.0,-1.75)$&$-0.25(0.0,-0.5)$&$-2.5(-2.0,-3.0)$&$-1.25(-1.25,-1.25)$\\
\hline
& Wind, Fast cooling\\
\hline
$\nu<\upsilon_{c}$&$-0.7(-0.7,-0.7)$&$-0.3(-0.3,-0.3)$&$...$&$0.3(0.3,0.3)$\\
$\upsilon_{c}<\nu<\upsilon_{m}$&$-0.25(-0.25,-0.25)$&$-0.5(-0.2,-0.5)$&$...$&$-0.5(-0.5,-0.5)$\\
$\nu>\upsilon_{m}$&$-1.4(-1.0,-1.75)$&$-0.25(0.0,-0.5)$&...&$-1.25(-1.25,-1.25)$\\
\hline
\enddata

Notes: This table provides the ranges of relations in the temporal index and the spectral index that are expected from synchrotron emission with or without energy injection, and the post jet-break case (Zhang $\&$ M\'{e}szaros 2004; Zhang et al. 2006; Sari et al 1999; Panaitescu et al. 2006). Here $p$ is the electron index and $q$ is the luminosity index. When $q=1$ non-injection is the case and when $q=0$ energy injection is the case; and calculated numerical values for each cases with $p=2.5$ $p=2.0$, $p=3.0$, $q=1$ and $q=0$.
\end{deluxetable}

\clearpage
\begin{deluxetable}{cccccccccccccccccccccccccc}
%\rotate
\tabletypesize{\scriptsize}
\tablecaption{Diagnosis of the Properties of Afterglow Break Behaviors.}
\tablewidth{450pt}
\tabletypesize{\tiny}
\tablehead{
\colhead{GRB}&
\colhead{$\alpha_{1}<\alpha_{(\beta)}$\tablenotemark{a}}&
\colhead{$\alpha_{2}>1.5$\tablenotemark{a}}&
\colhead{Closure Relation?\tablenotemark{b}}&
\colhead{Spectral Evolution\tablenotemark{c}}&
\colhead{Spectral Break\tablenotemark{c}}&
\colhead{Injection Break?\tablenotemark{c}}&
\colhead{Jet break?\tablenotemark{c}}&
}
\startdata
\hline
&Group I\\
\hline
050319&Yes/Yes&Yes/Yes&Yes/Yes&No&No&Yes/Yes&Yes/Yes\\
050408&No/Yes&Yes/No&Yes/Yes&No&No&No/Yes&Yes/No\\
050730&Yes/Yes&Yes/No&No/Yes&Yes&No&Yes/Yes&Yes/No\\
050922C&Yes/Yes&No/No&Yes/Yes&No&No&Yes/Yes&No/No\\
060210&No/Yes&No/No&Yes/Yes&No&No&Yes/Yes&No/No\\
060526&Yes/Yes&Yes/Yes&Yes/Yes&No&No&Yes/Yes&Yes/Yes\\
060605&Yes/Yes&Yes/Yes&Yes/Yes&No&No&Yes/Yes&Yes/Yes\\
060614&Yes/Yes&Yes/Yes&Yes/Yes&No&No&Yes/Yes&Yes/Yes\\
060708&Yes/...&No/No&Yes/Yes&No&No&Yes/...&No/No\\
060714&Yes/Yes&No/No&Yes/Yes&No&No&Yes/Yes&No/No\\
060729&Yes/Yes&No/No&Yes/Yes&No&No&Yes/Yes&No/No\\
060912A&Yes/No&No/No&Yes/Yes&No&No&Yes/Yes&No/No\\
061021&Yes/...&No/Yes&Yes/Yes&No&No&Yes/...&No/Yes\\
070411&Yes/...&No/Yes&Yes/Yes&No&No&Yes/...&No/Yes\\
070518&Yes/Yes&No/No&No/Yes&No&No&Yes/Yes&No/No\\
080310&Yes/Yes&Yes/No&Yes/Yes&No&No&Yes/Yes&Yes/No\\
080413B&Yes/Yes&Yes/Yes&Yes/Yes&No&No&Yes/Yes&Yes/Yes\\
080710&Yes/Yes&Yes/Yes&Yes/Yes&No&No&Yes/Yes&Yes/Yes\\
080721&Yes/No&Yes/No&Yes/Yes&Yes&No&Yes/No&Yes/Yes\\
081029&Yes/Yes&Yes/Yes&Yes/Yes&No&No&Yes/Yes&Yes/Yes\\
090618&Yes/Yes&Yes/Yes&Yes/Yes&No&No&Yes/Yes&Yes/Yes\\
090727&Yes/Yes&Yes/No&Yes/Yes&No&No&Yes/Yes&Yes/No\\
091018&Yes/No&No/Yes&Yes/Yes&No&No&Yes/No&No/Yes\\
091127&No/Yes&Yes/No&Yes/Yes&No&No&No/Yes&Yes/No\\
\hline
&Group II\\
\hline
050401&Yes/...&No/No&Yes/Yes&No&No&Yes/...&No/No\\
050502A&Yes/...&Yes/No&Yes/Yes&...&No&Yes/...&Yes/No\\
050824&Yes/...&No/No&No/Yes&Yes&No&Yes/...&No/No\\
051221A&Yes/...&No/No&Yes/Yes&No&No&Yes/...&No/No\\
060111B&Yes/...&Yes/No&Yes/Yes&No&No&Yes/...&Yes/No\\
060418&Yes/...&Yes/No&Yes/Yes&No&No&Yes/...&Yes/No\\
060906&Yes/...&Yes/No&Yes/Yes&No&No&Yes/...&Yes/No\\
060908&Yes/...&Yes/No&Yes/Yes&No&No&Yes/...&Yes/No\\
060927&Yes/...&Yes/No&Yes/Yes&No&No&Yes/...&Yes/No\\
061121&Yes/...&No/No&Yes/Yes&No&No&Yes/...&No/No\\
070318&Yes/...&Yes/No&Yes/Yes&No&No&Yes/...&Yes/No\\
070420&Yes/...&No/No&Yes/Yes&No&No&Yes/...&No/No\\
071031&Yes/...&Yes/No&Yes/Yes&No&No&Yes/...&Yes/No\\
080810&Yes/...&Yes/No&Yes/Yes&No&No&Yes/...&Yes/No\\
081008&Yes/...&Yes/No&Yes/Yes&No&No&Yes/...&Yes/No\\
081126&Yes/...&Yes/No&Yes/Yes&No&No&Yes/...&Yes/No\\
081203A&No/...&Yes/Yes&Yes/Yes&No&No&No/...&Yes/Yes\\
090510&Yes/...&Yes/No&Yes/Yes&Yes&No&Yes/...&Yes/No\\
100418A&Yes/...&Yes/No&Yes/Yes&No&No&Yes/...&Yes/No\\
100621A&No/...&Yes/No&No/Yes&Yes&No&No/...&Yes/No\\
100814A&Yes/...&Yes/Yes&Yes/Yes&No&No&Yes/...&Yes/Yes\\
100906A&Yes/...&Yes/No&Yes/Yes&No&No&Yes/...&Yes/No\\
120119A&No/...&Yes/No&Yes/Yes&No&No&No/...&Yes/No\\
121217A&Yes/...&No/No&Yes/Yes&No&No&Yes/...&No/No\\
\hline
&Group III\\
\hline
050416A&.../Yes&No/No&Yes/Yes&...&...&.../Yes&No/No\\
050525A&.../Yes&Yes/Yes&Yes/Yes&...&...&.../Yes&Yes/Yes\\
050801&.../Yes&No/No&Yes/Yes&...&...&.../Yes&No/No\\
051109A&.../Yes&No/No&Yes/Yes&...&...&.../Yes&No/No\\
051111&.../Yes&Yes/Yes&Yes/Yes&...&...&.../Yes&Yes/Yes\\
061126&.../Yes&No/No&Yes/Yes&...&...&.../Yes&No/No\\
080413A&.../Yes&No/Yes&Yes/Yes&...&...&.../Yes&No/Yes\\
090426&.../Yes&No/No&Yes/Yes&...&...&.../Yes&No/No\\
130702A&.../Yes&No/No&Yes/Yes&...&...&.../Yes&No/No\\
\enddata

\tablenotetext{a}{X-ray/optical. We defined the shallow decay by the 'closure relation' with different spectral regimes, 3 GRBs lack the optical spectral information; we choose the jet break with the criterion of $\alpha_{2}>1.5$.}
\tablenotetext{b}{Test the data with the 'closure relation' ($\alpha_{2}-\beta_{2}$).}
\tablenotetext{c}{The spectral breaks required $\Delta\alpha=0.25$ and $\Delta\beta=0.5$; only for X-ray break samples, since no time-resolved spectral analysis for optical band available; and statistic break types.}
\end{deluxetable}

%\newpage
\begin{figure*}
\includegraphics[angle=0,scale=0.350,width=0.3\textwidth,height=0.25\textheight]{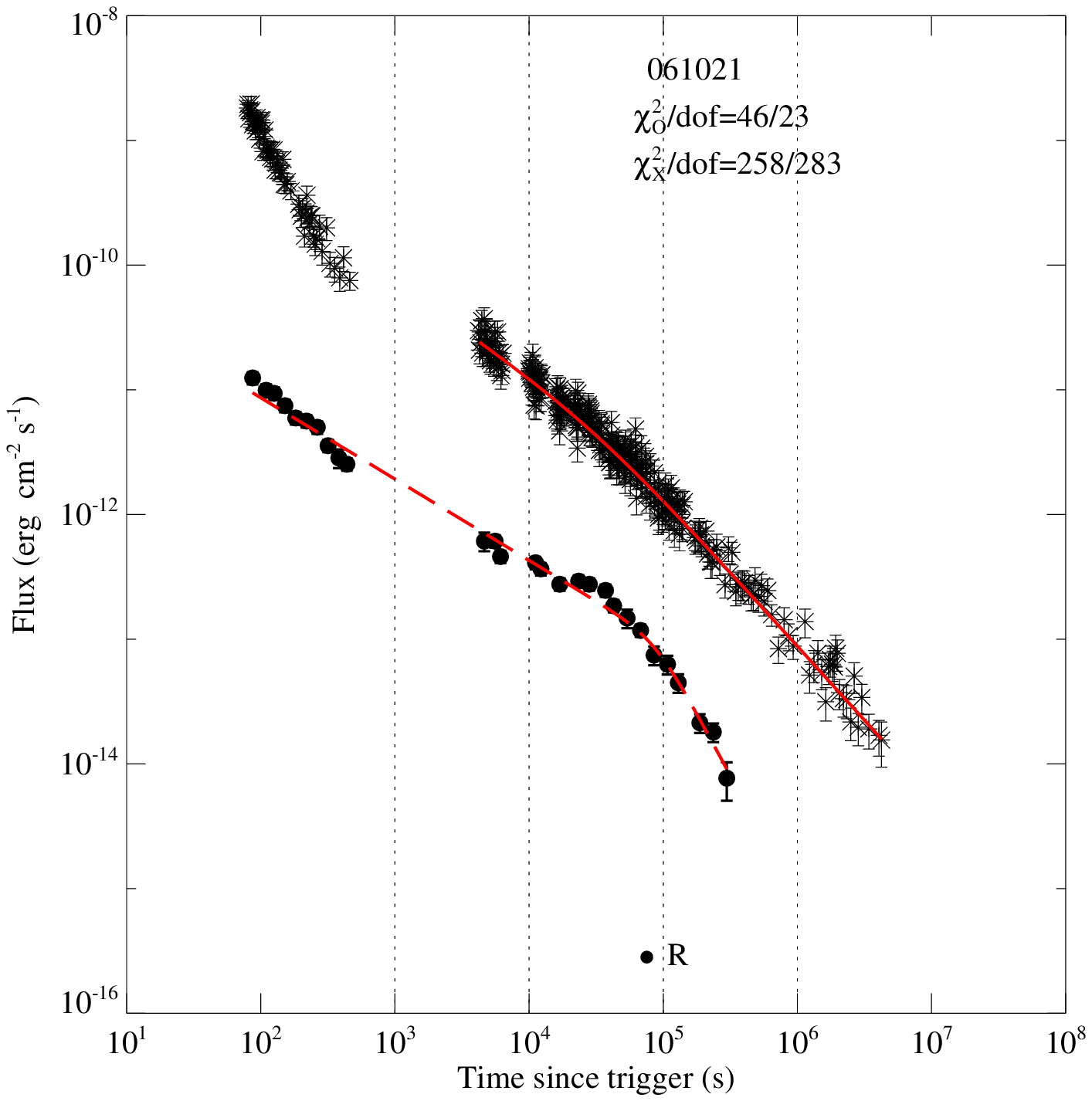}
\includegraphics[angle=0,scale=0.350,width=0.3\textwidth,height=0.25\textheight]{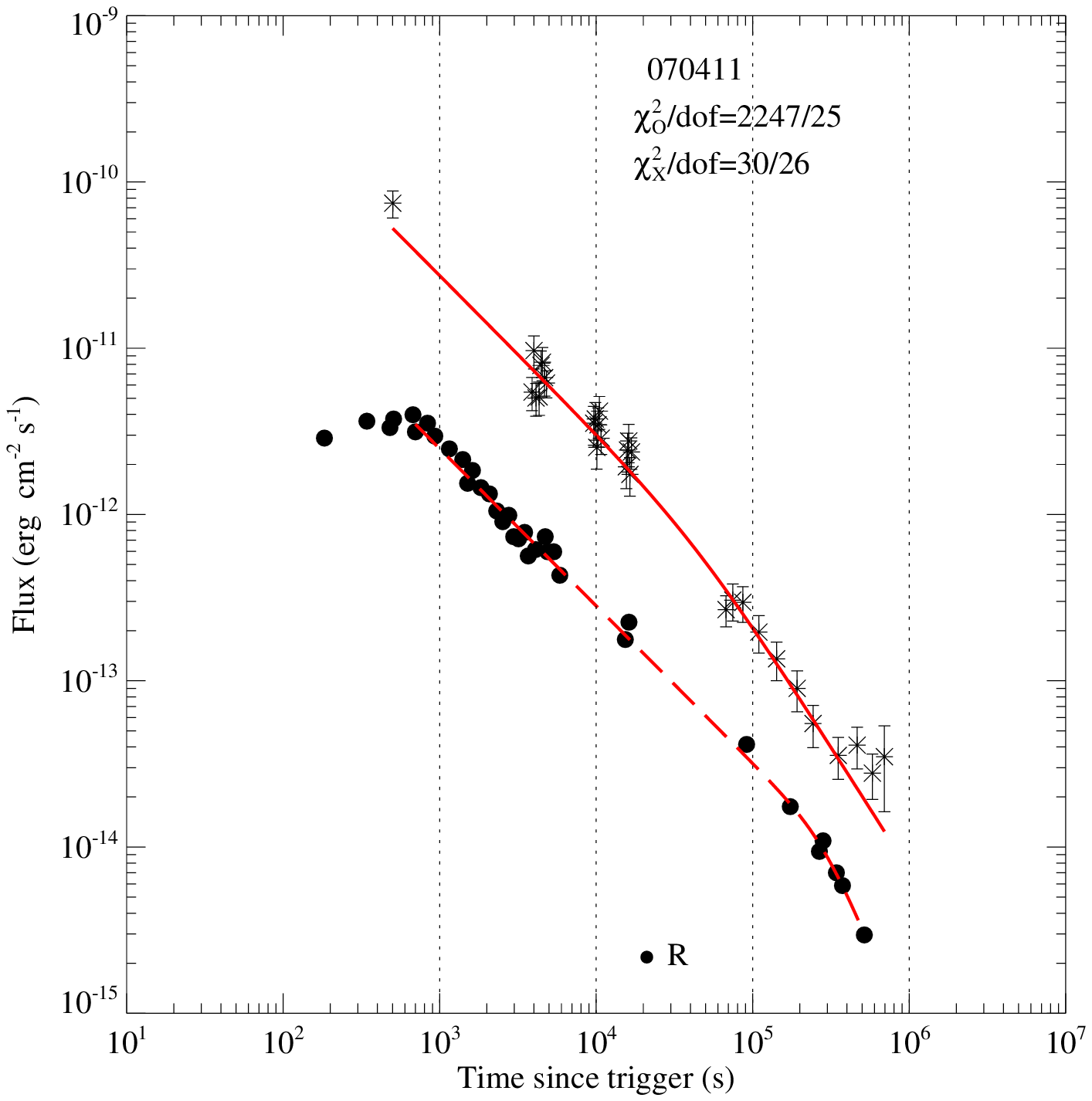}
\includegraphics[angle=0,scale=0.350,width=0.3\textwidth,height=0.25\textheight]{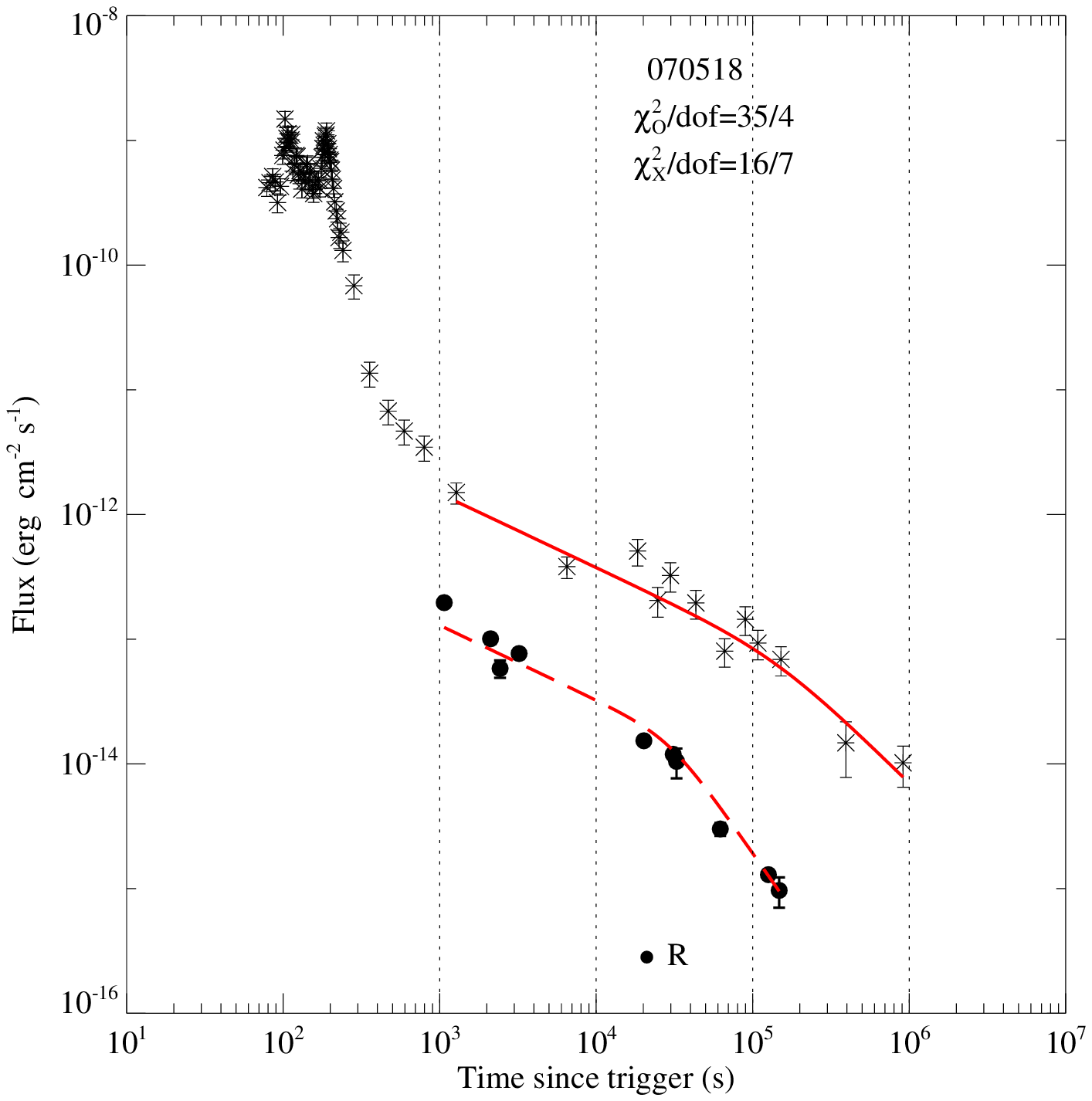}
\includegraphics[angle=0,scale=0.350,width=0.3\textwidth,height=0.25\textheight]{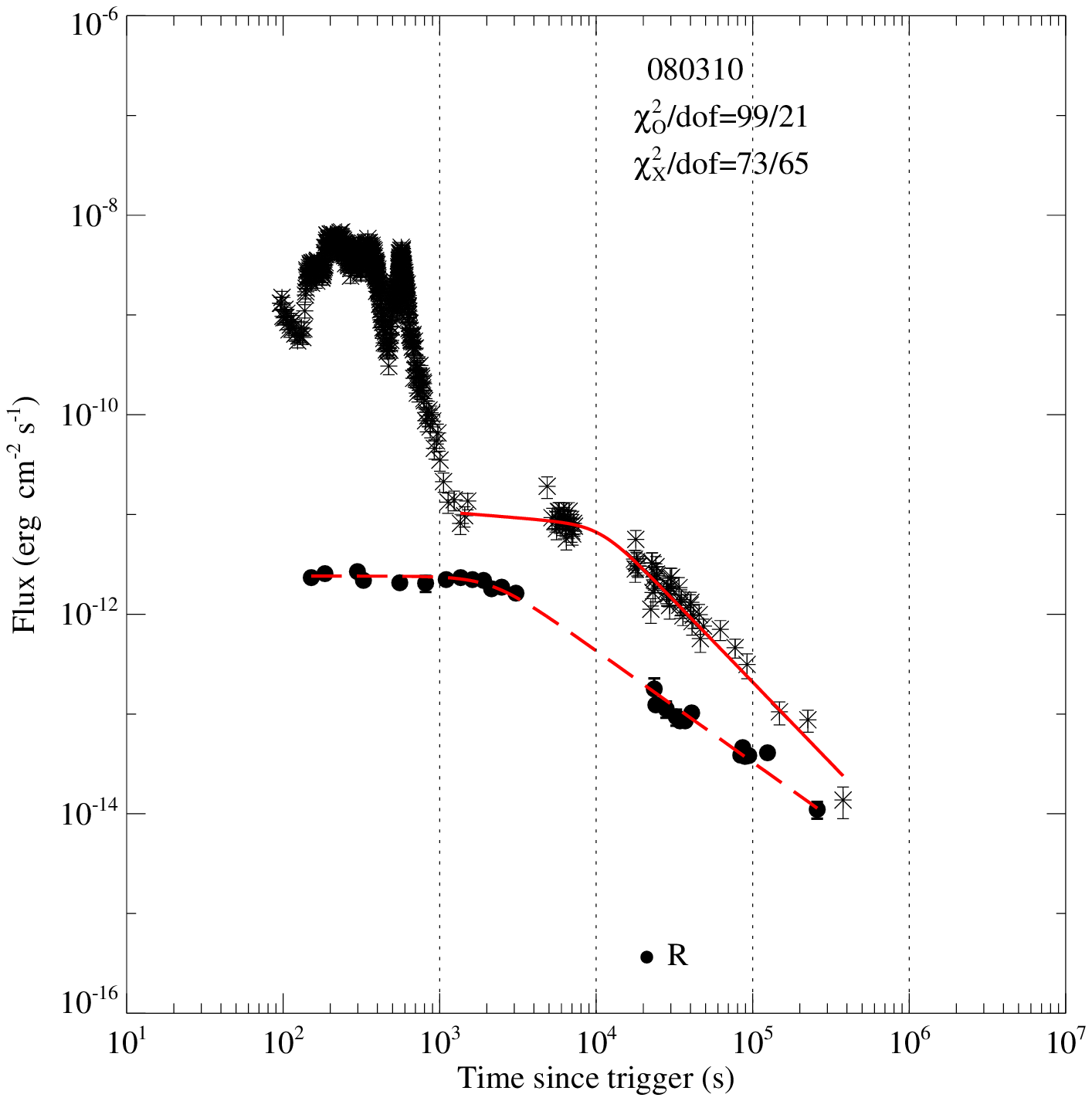}
\includegraphics[angle=0,scale=0.350,width=0.3\textwidth,height=0.25\textheight]{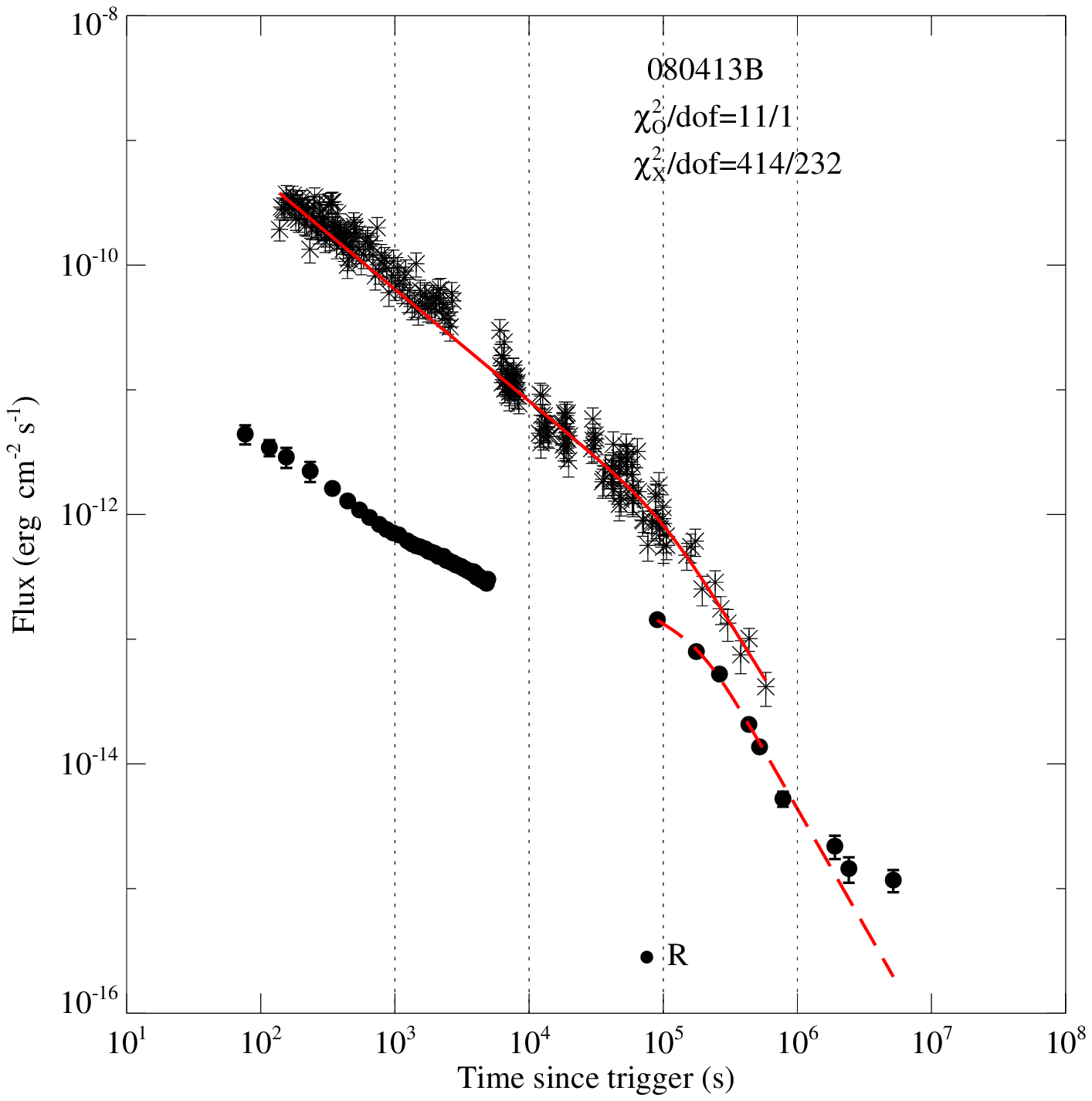}
\includegraphics[angle=0,scale=0.350,width=0.3\textwidth,height=0.25\textheight]{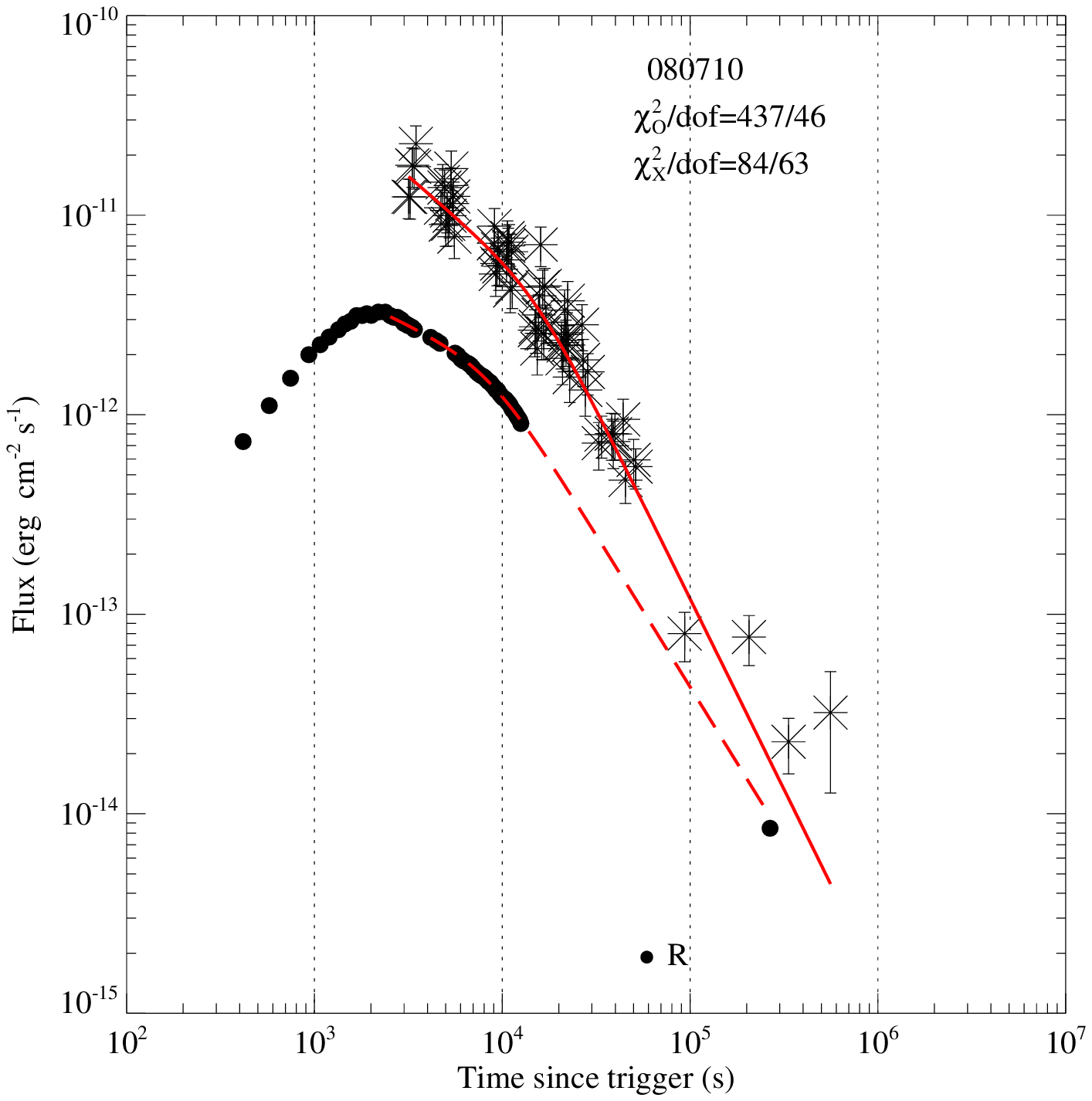}
\includegraphics[angle=0,scale=0.350,width=0.3\textwidth,height=0.25\textheight]{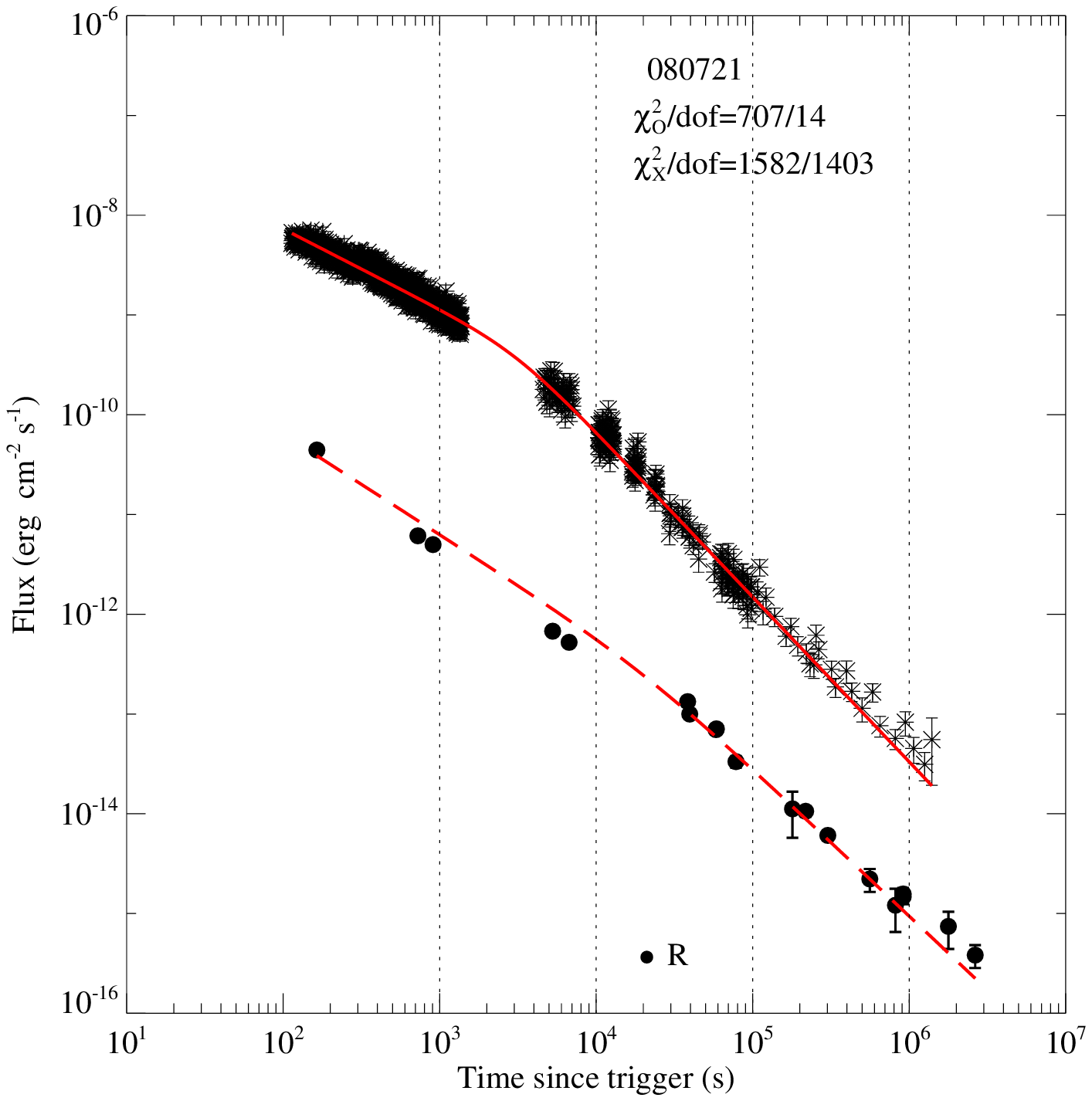}
\includegraphics[angle=0,scale=0.350,width=0.3\textwidth,height=0.25\textheight]{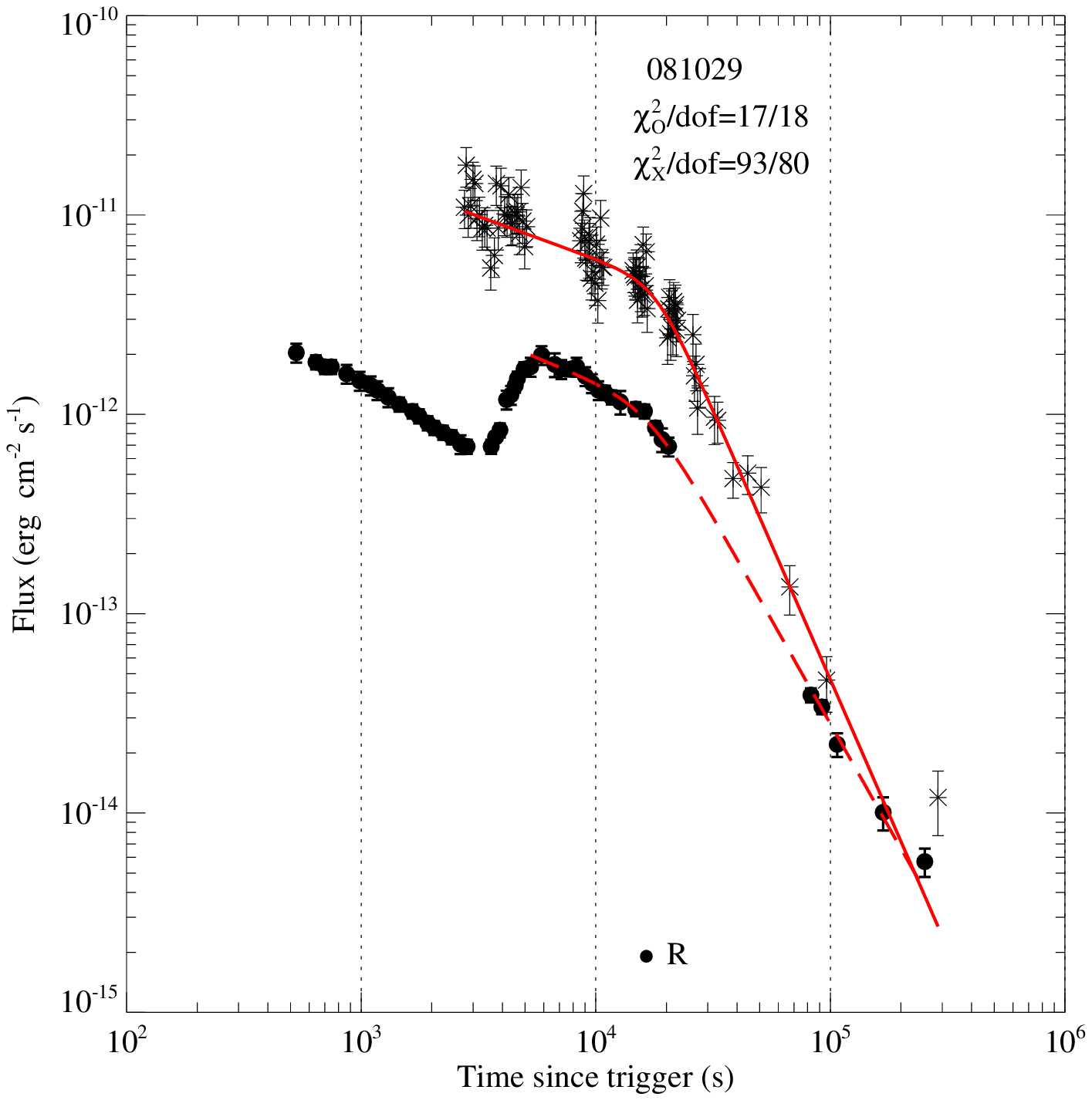}
\includegraphics[angle=0,scale=0.350,width=0.3\textwidth,height=0.25\textheight]{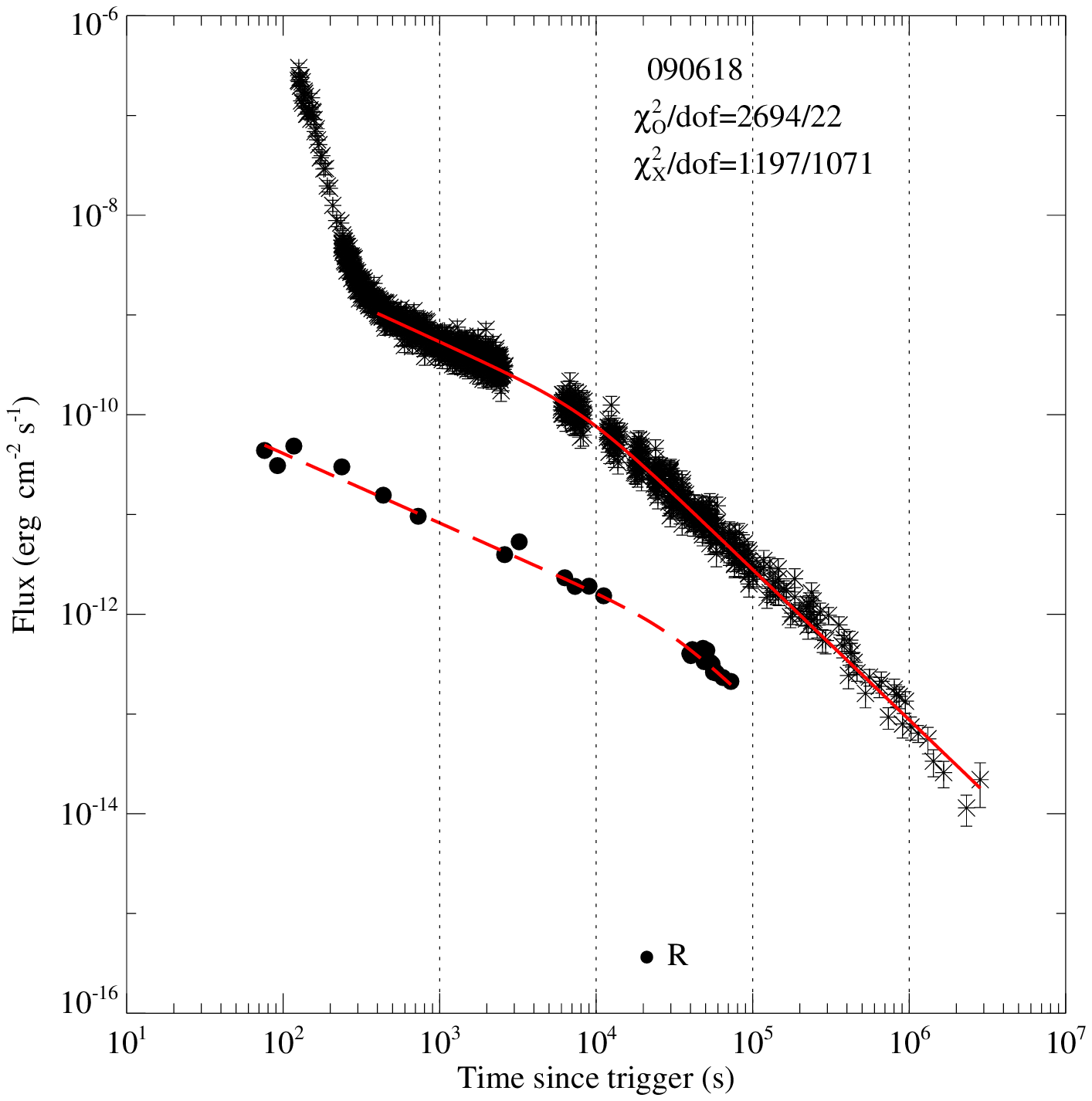}
\includegraphics[angle=0,scale=0.350,width=0.3\textwidth,height=0.25\textheight]{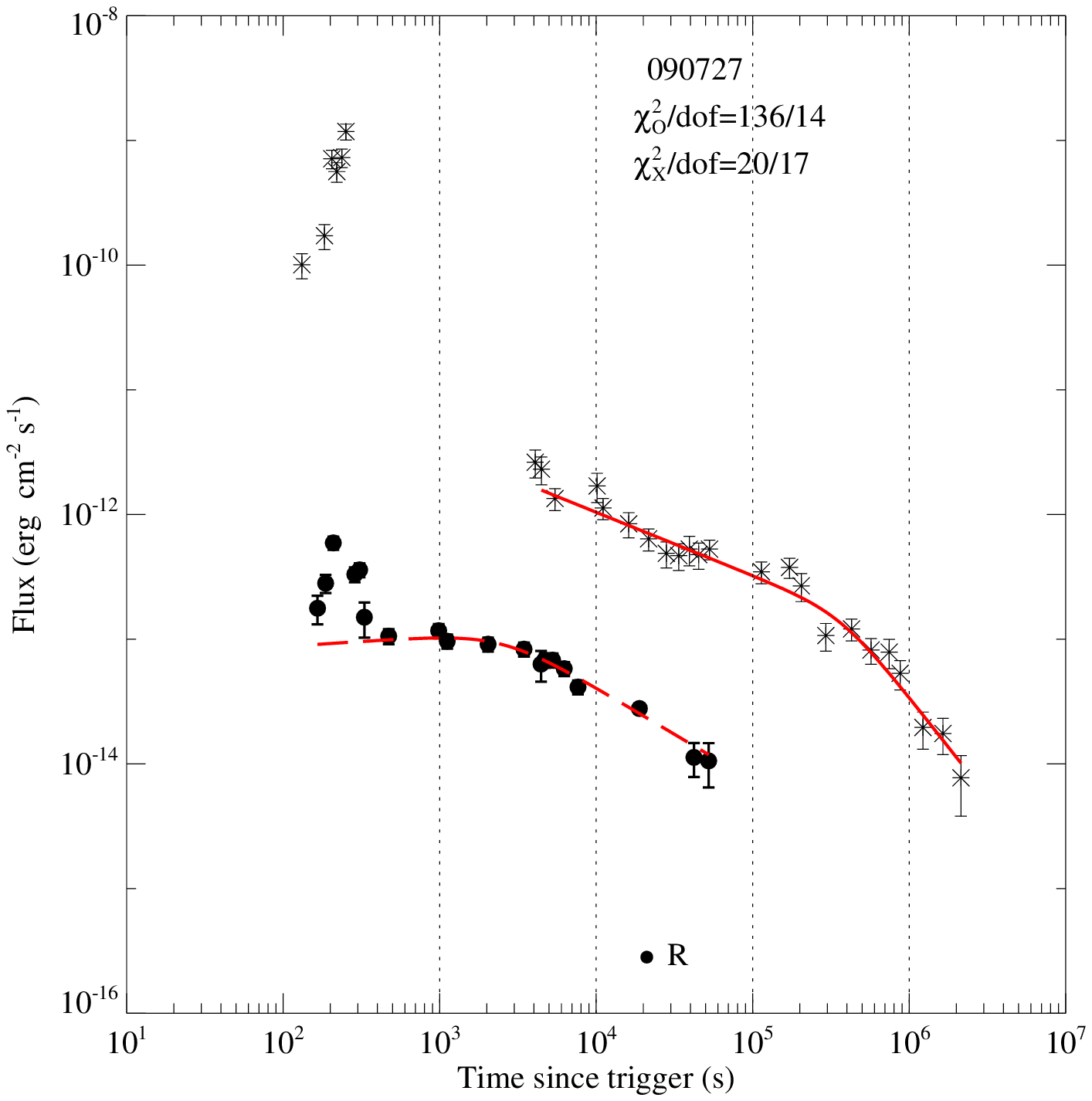}
\includegraphics[angle=0,scale=0.350,width=0.3\textwidth,height=0.25\textheight]{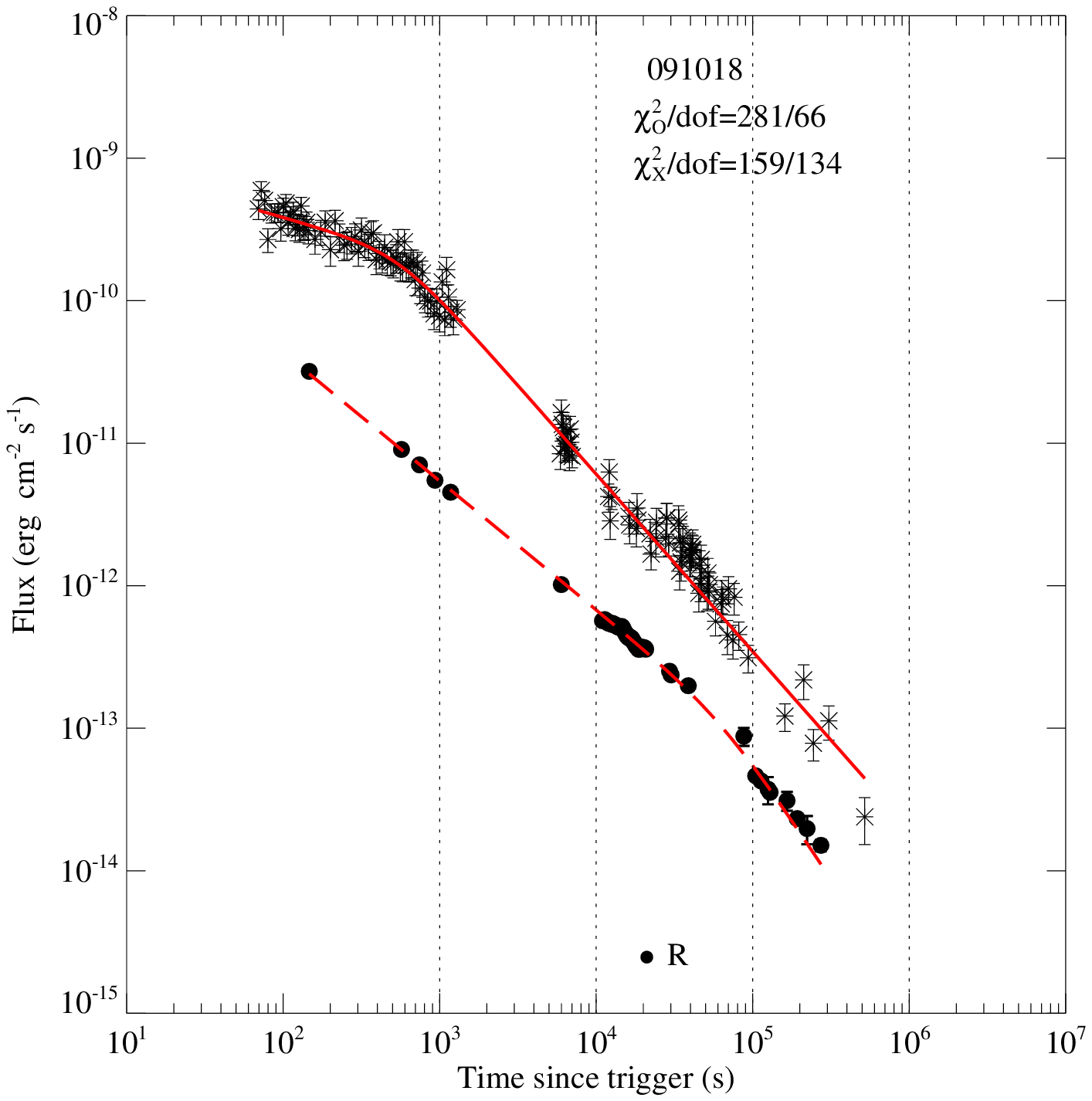}
\includegraphics[angle=0,scale=0.350,width=0.3\textwidth,height=0.25\textheight]{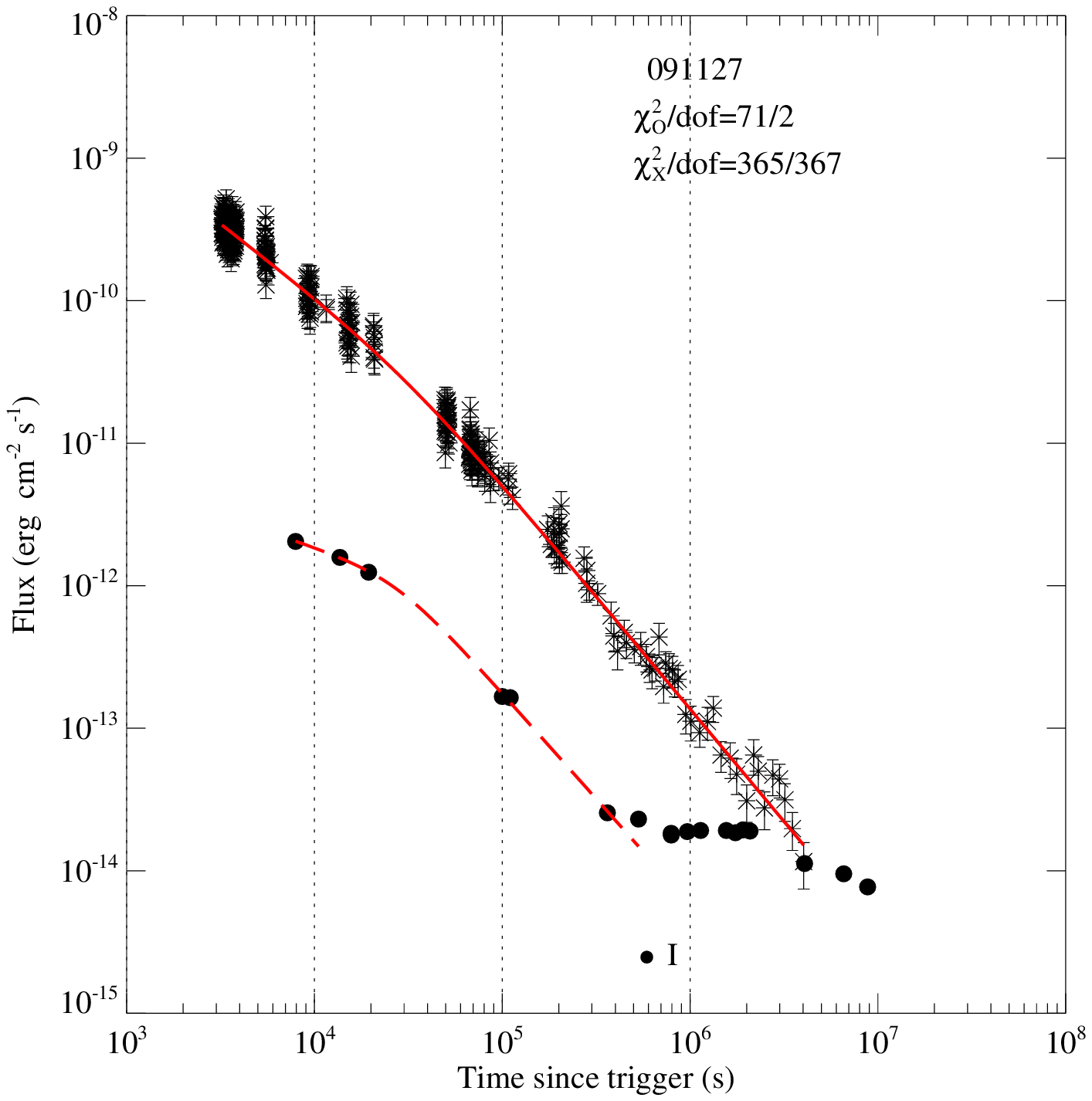}
\center{Fig. \ref{Both Break}--- Continued}
\end{figure*}

\begin{figure*}
\includegraphics[angle=0,scale=0.350,width=0.3\textwidth,height=0.25\textheight]{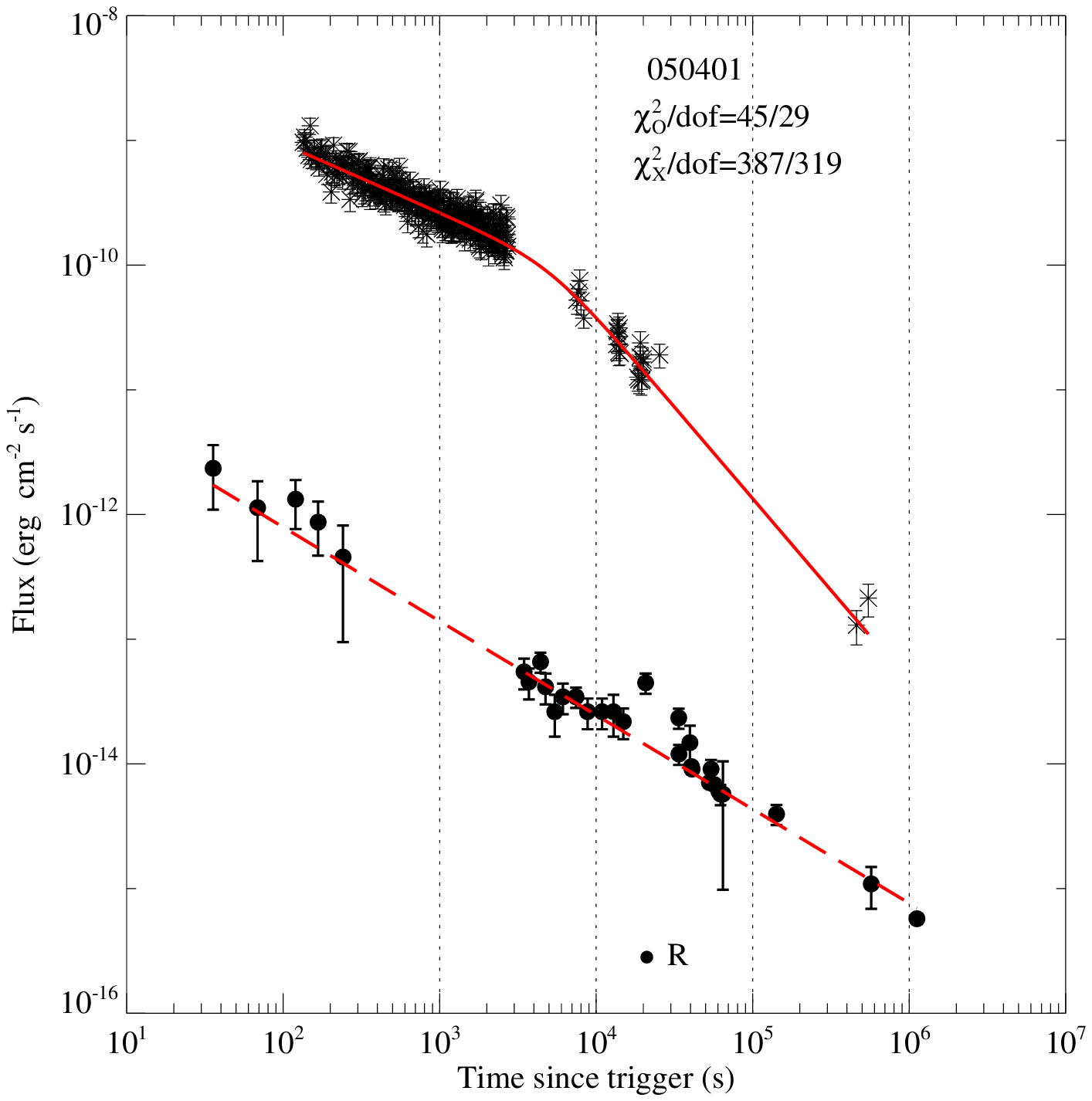}
\includegraphics[angle=0,scale=0.350,width=0.3\textwidth,height=0.25\textheight]{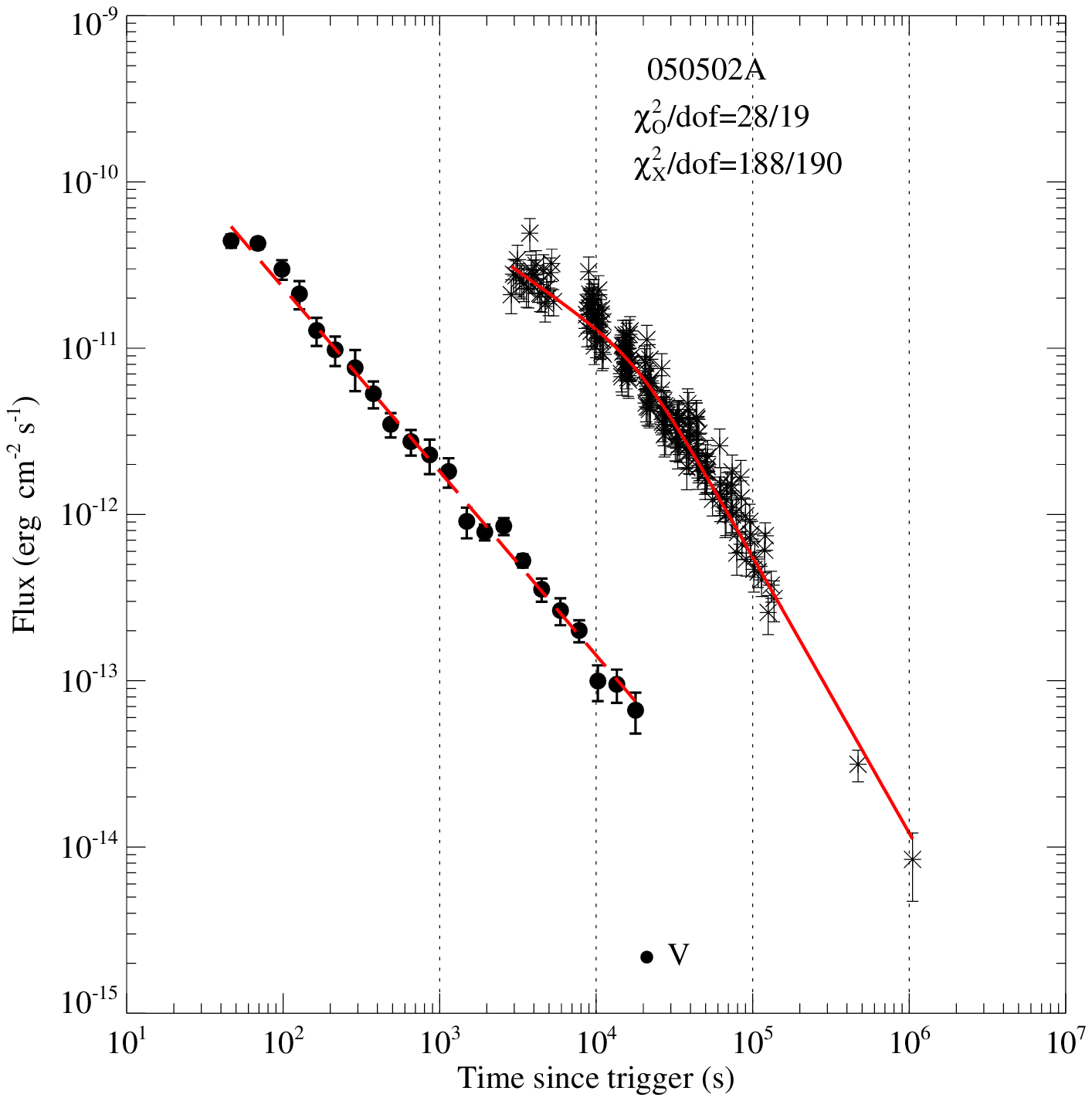}
\includegraphics[angle=0,scale=0.350,width=0.3\textwidth,height=0.25\textheight]{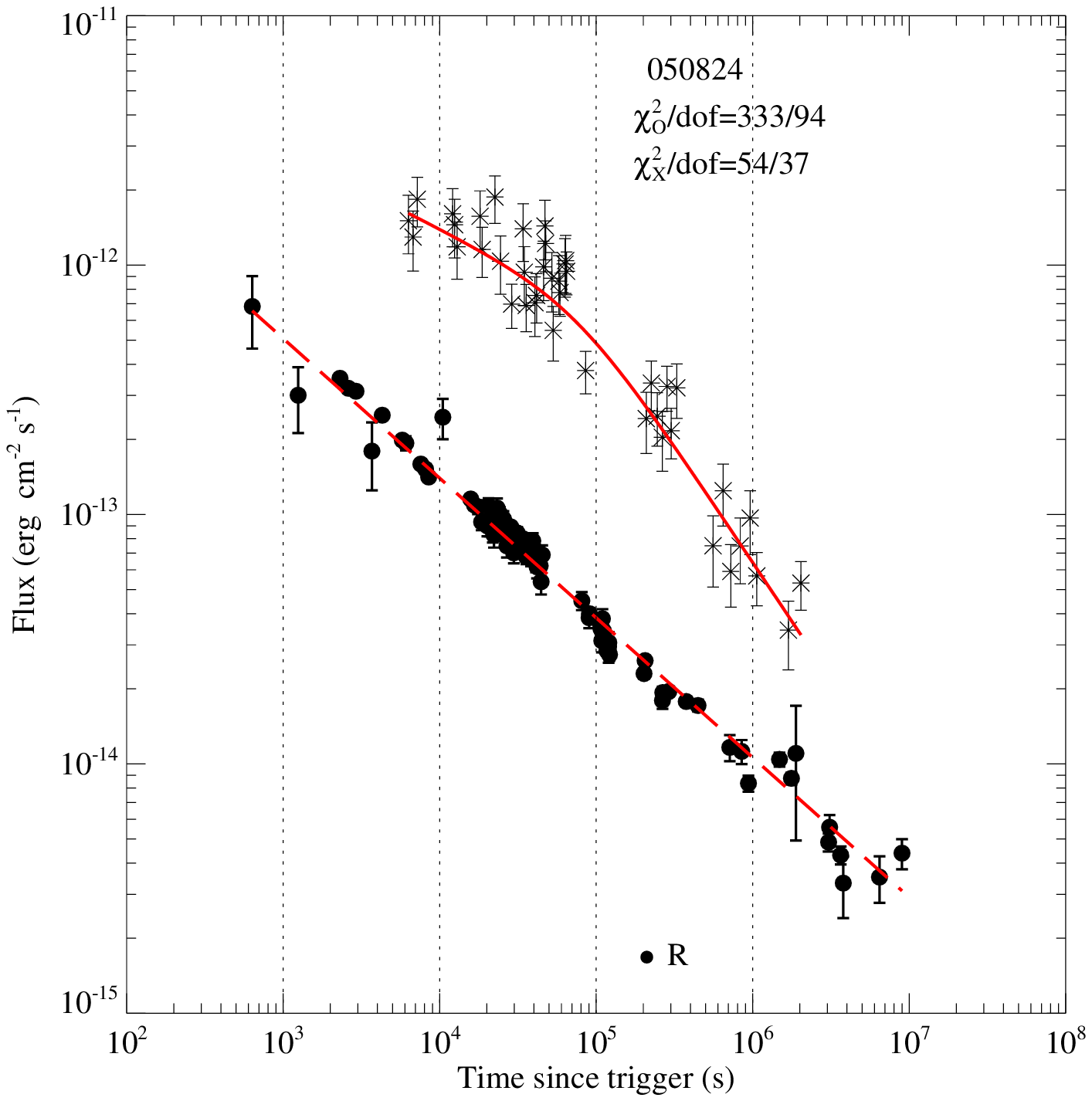}
\includegraphics[angle=0,scale=0.350,width=0.3\textwidth,height=0.25\textheight]{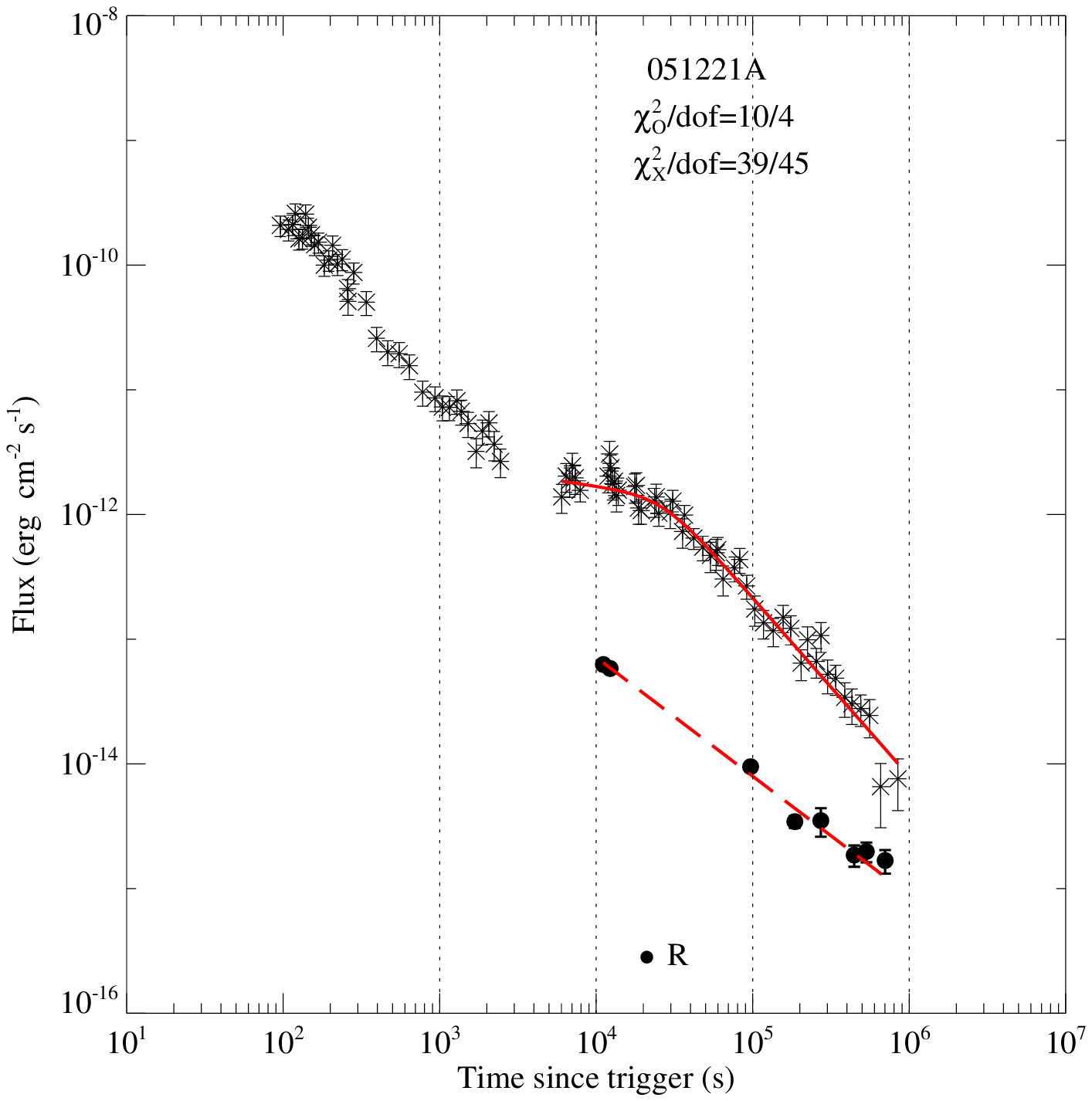}
\includegraphics[angle=0,scale=0.350,width=0.3\textwidth,height=0.25\textheight]{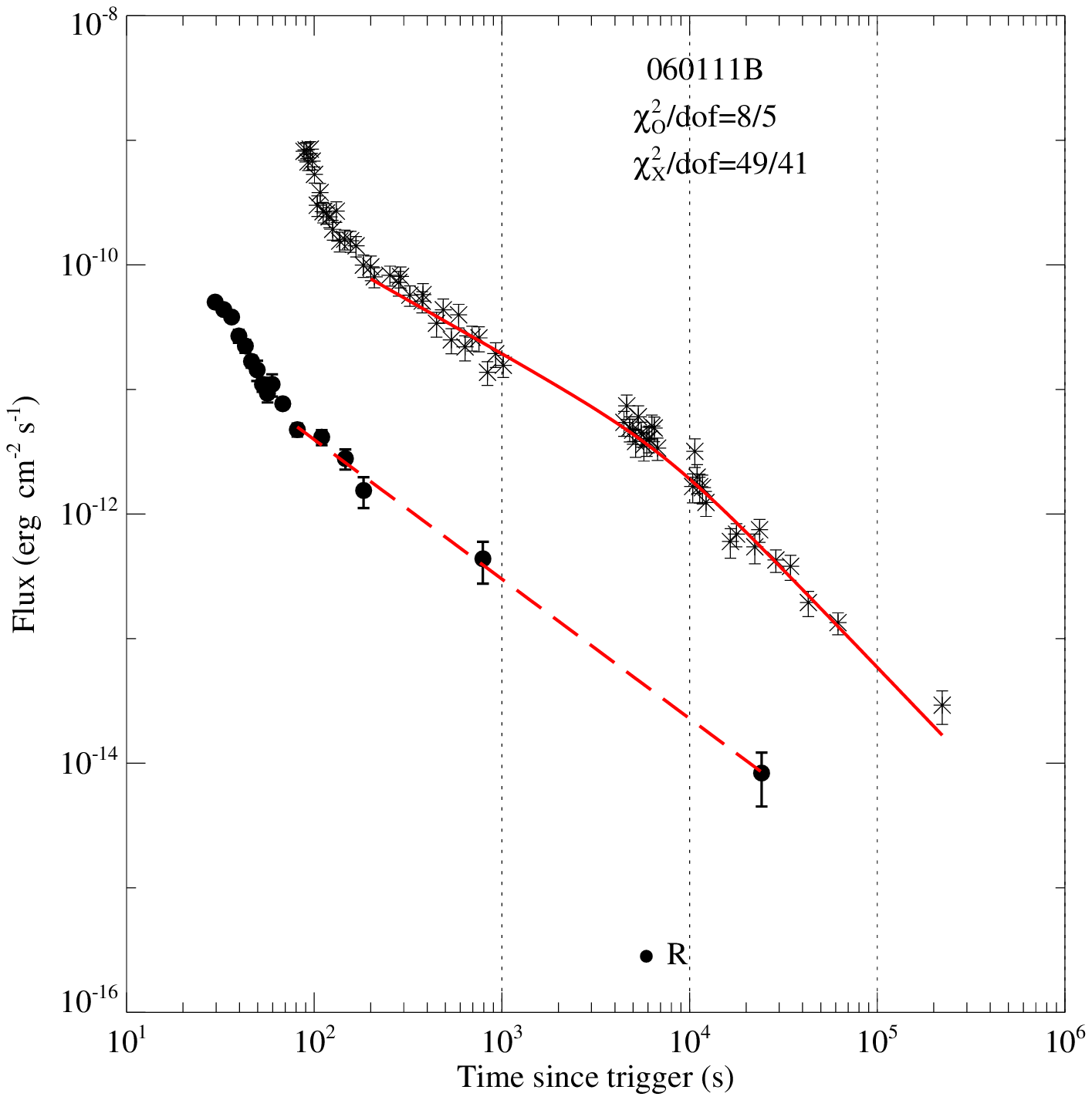}
\includegraphics[angle=0,scale=0.350,width=0.3\textwidth,height=0.25\textheight]{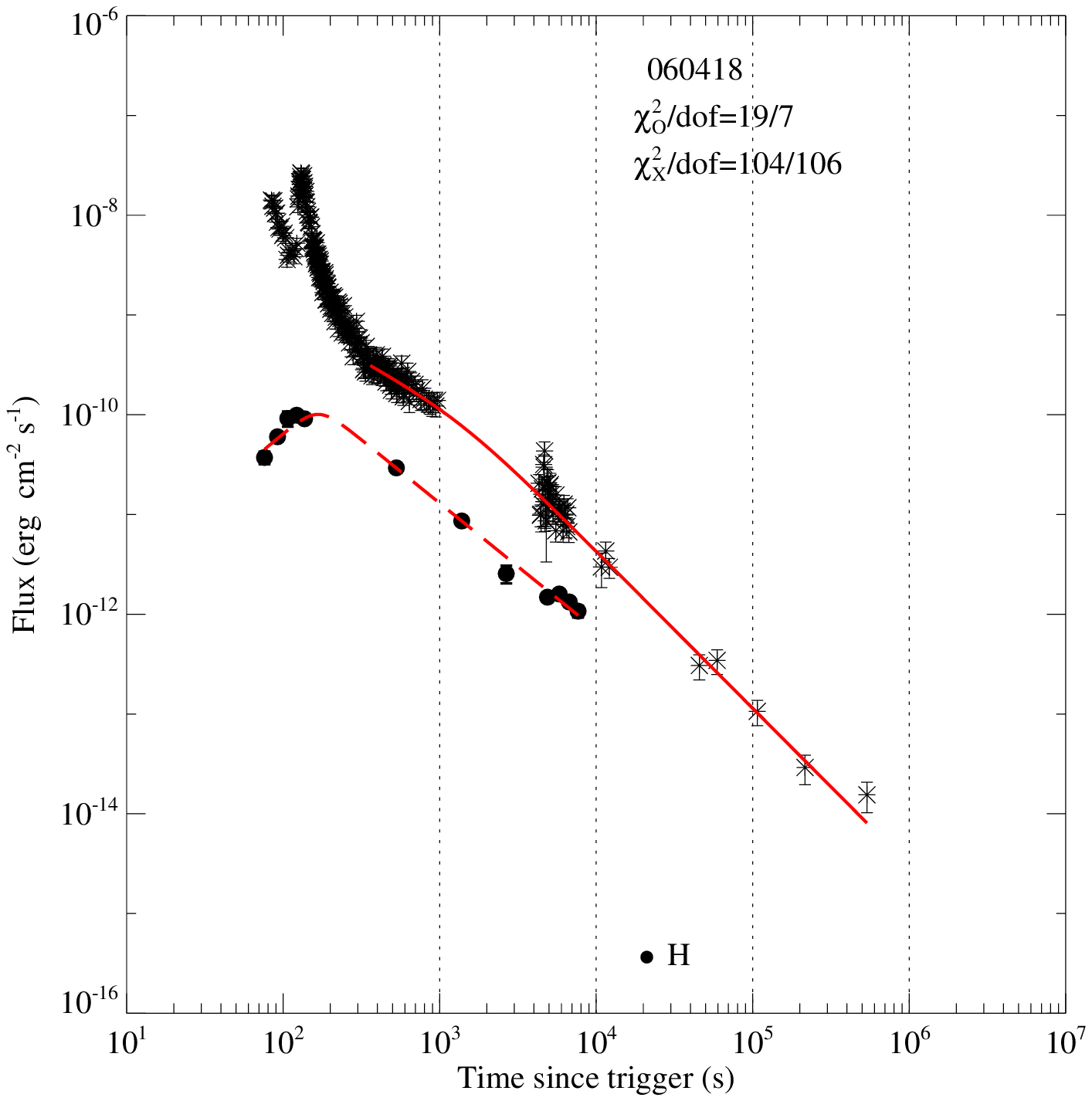}
\includegraphics[angle=0,scale=0.350,width=0.3\textwidth,height=0.25\textheight]{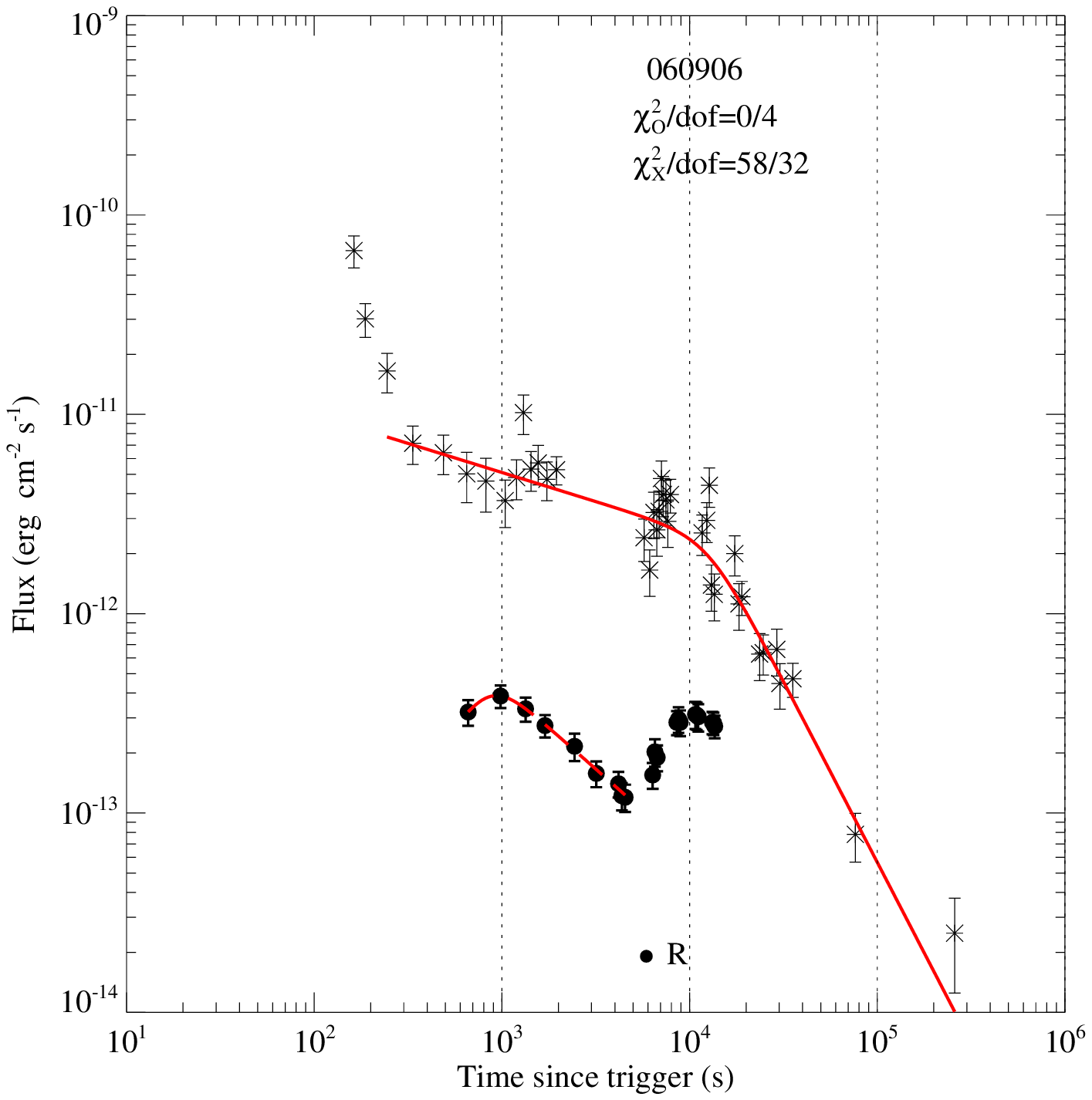}
\includegraphics[angle=0,scale=0.350,width=0.3\textwidth,height=0.25\textheight]{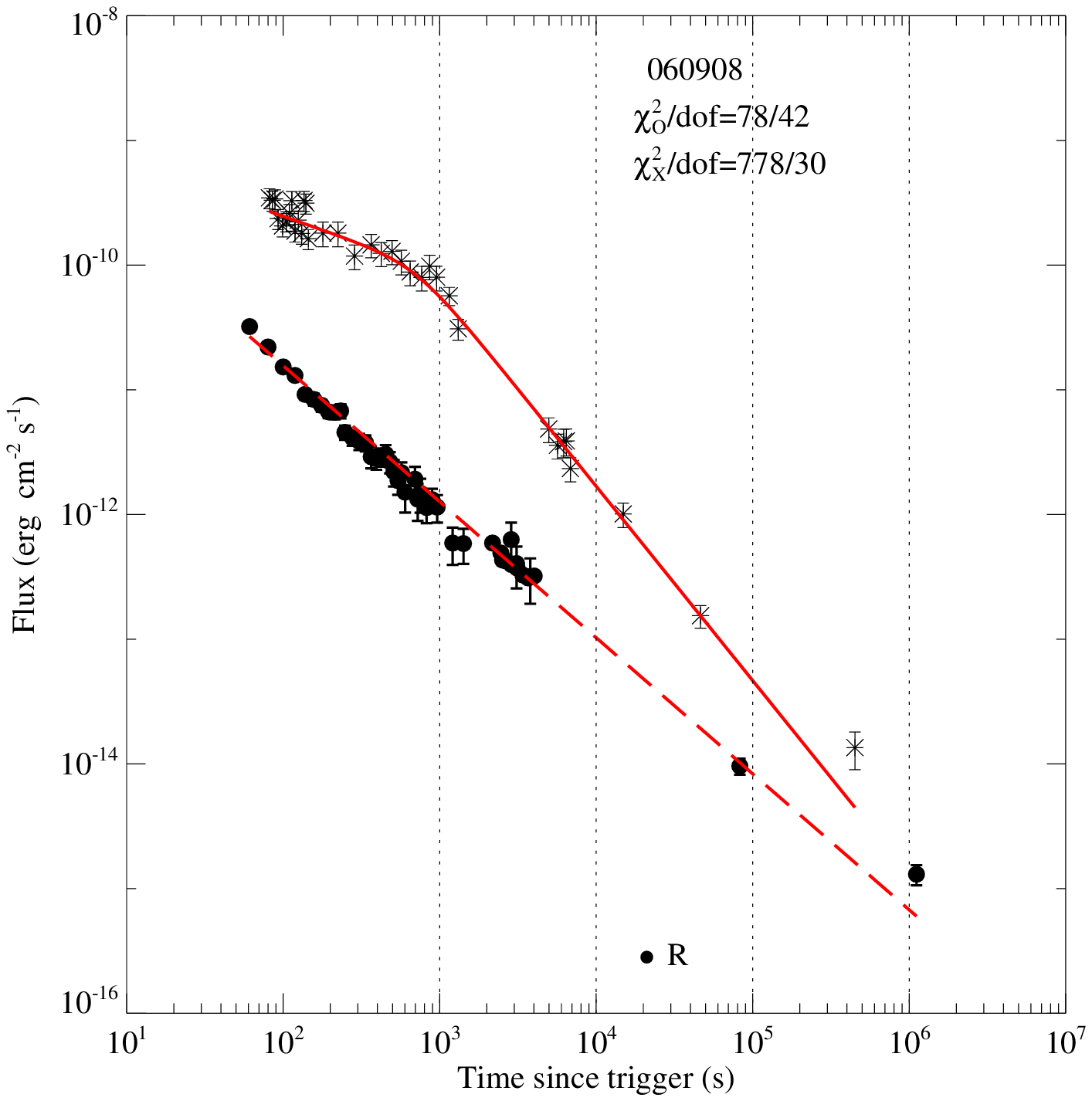}
\includegraphics[angle=0,scale=0.350,width=0.3\textwidth,height=0.25\textheight]{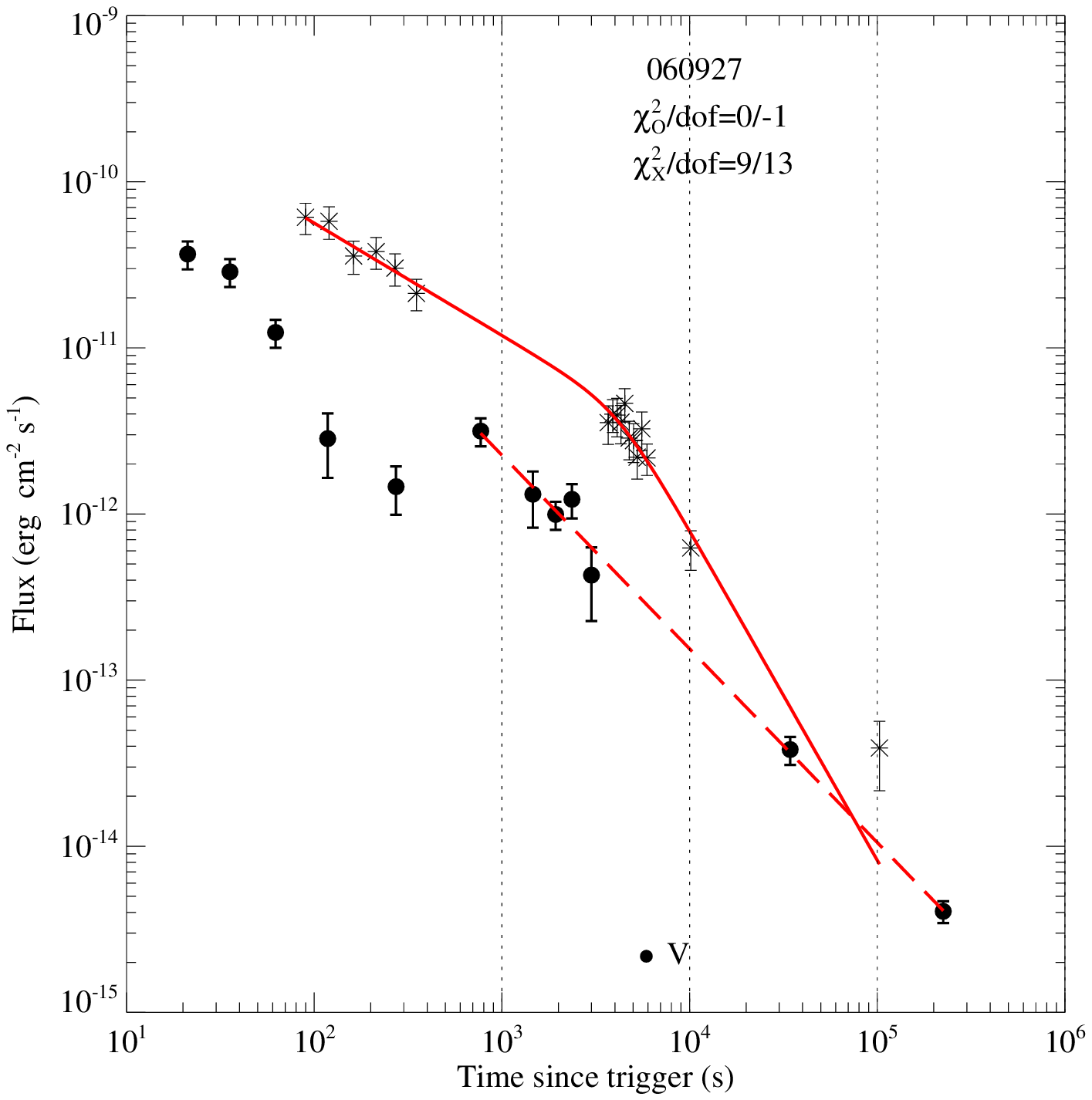}
\includegraphics[angle=0,scale=0.350,width=0.3\textwidth,height=0.25\textheight]{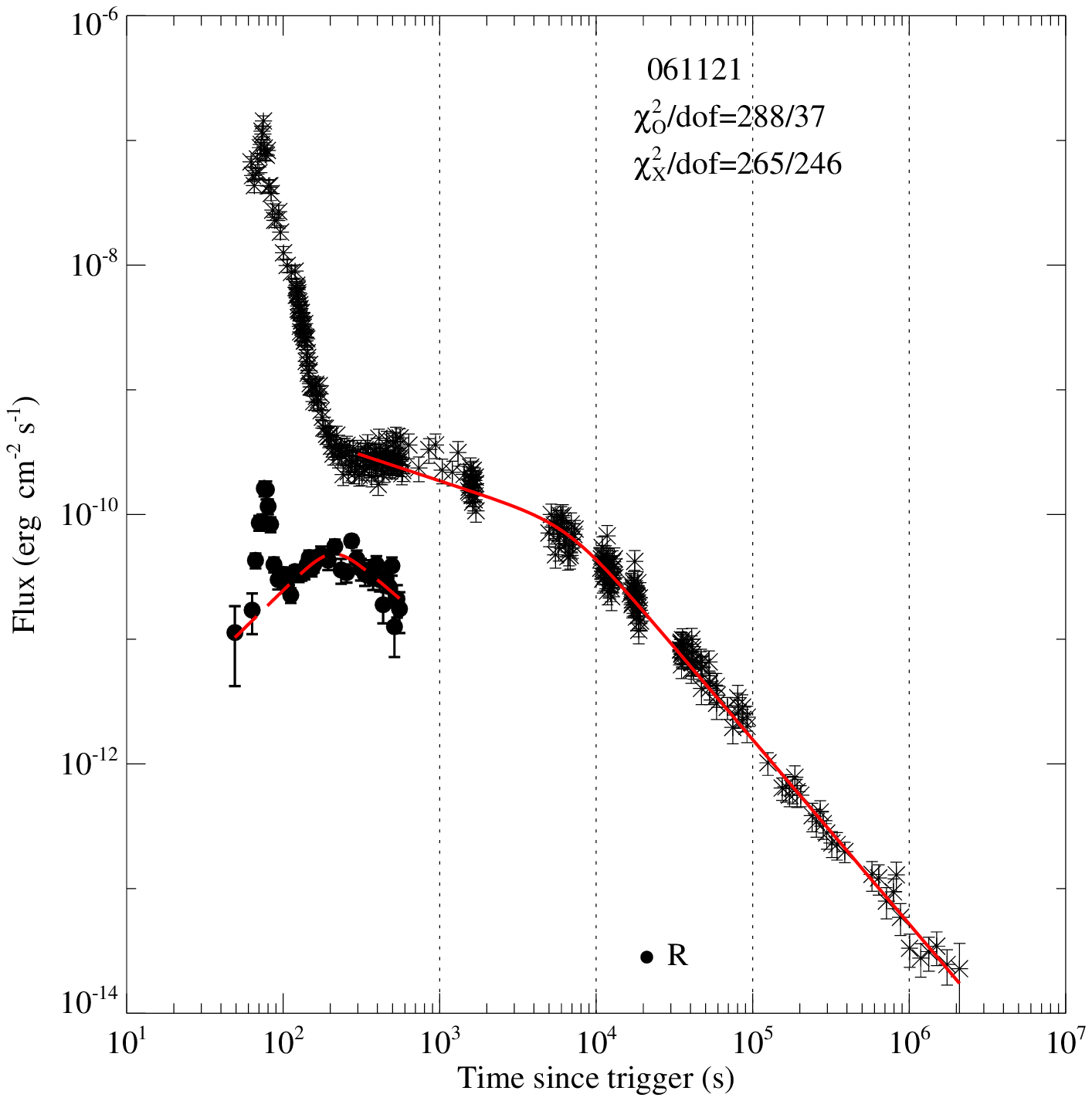}
\includegraphics[angle=0,scale=0.350,width=0.3\textwidth,height=0.25\textheight]{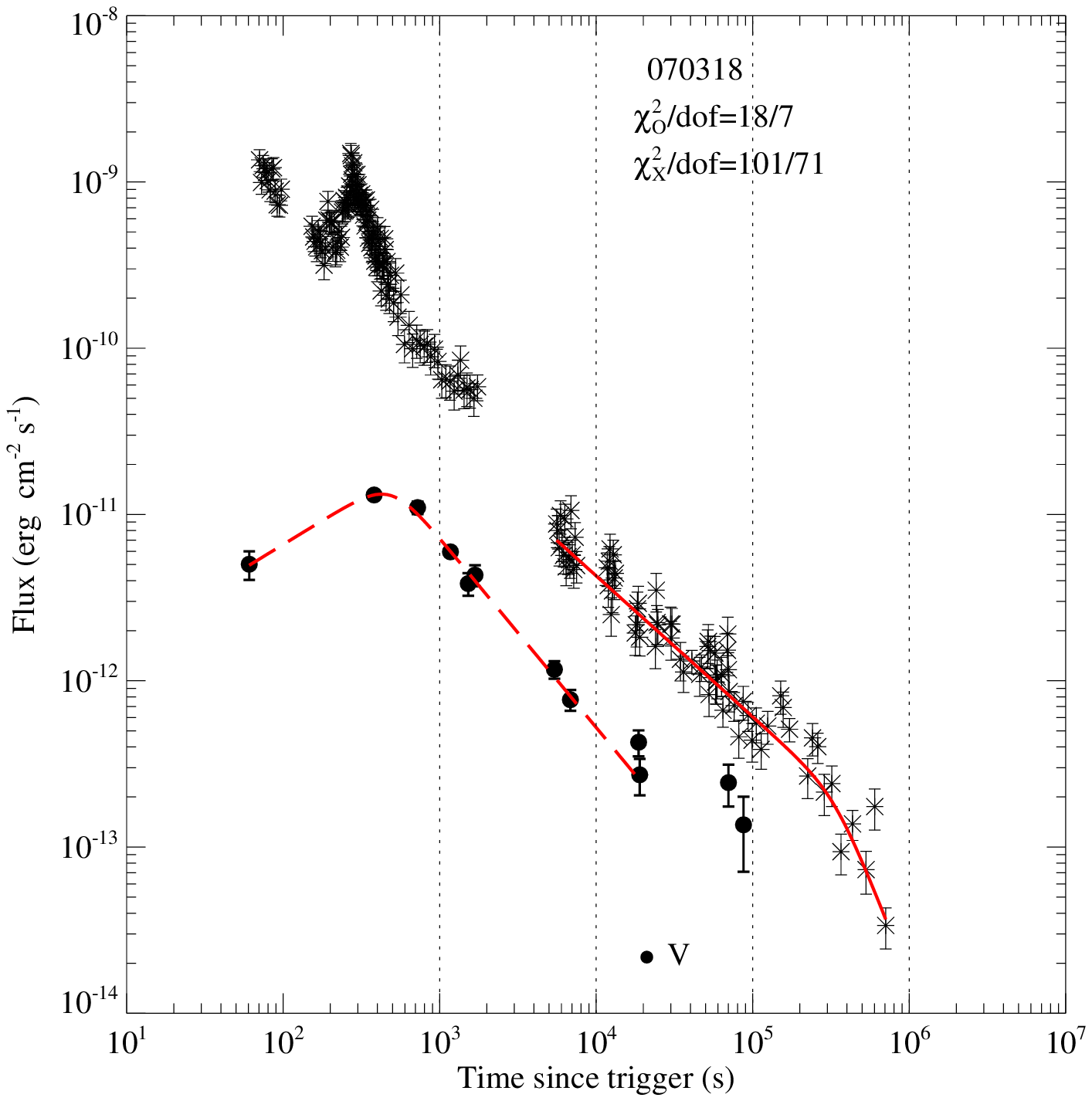}
\includegraphics[angle=0,scale=0.350,width=0.3\textwidth,height=0.25\textheight]{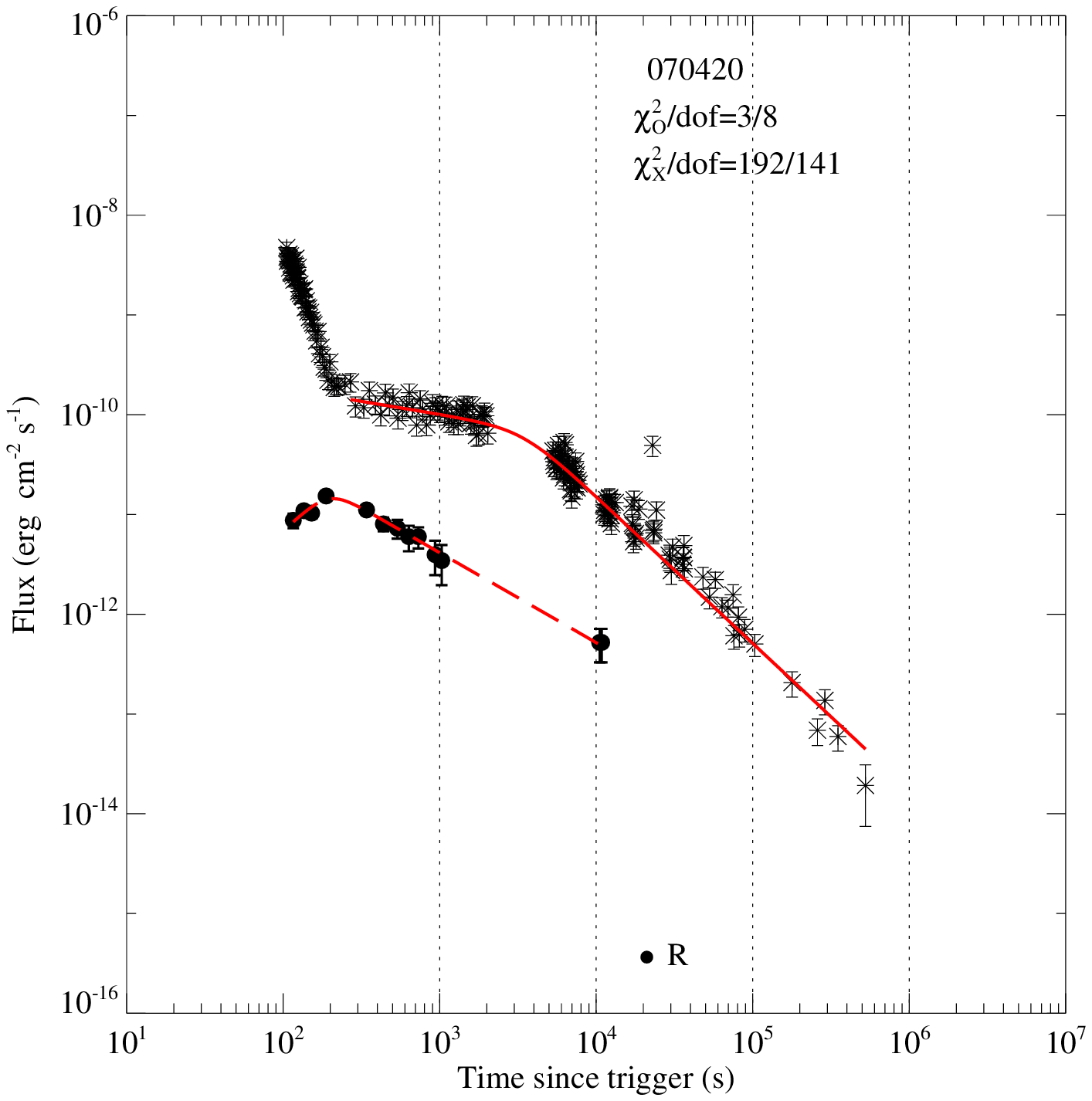}
\caption{The cases in which a breaks was detected only in the X-ray band but not in the optical band in their afterglow light-curve.}\label{Xray Break}
\end{figure*}
\begin{figure*}
\includegraphics[angle=0,scale=0.350,width=0.3\textwidth,height=0.25\textheight]{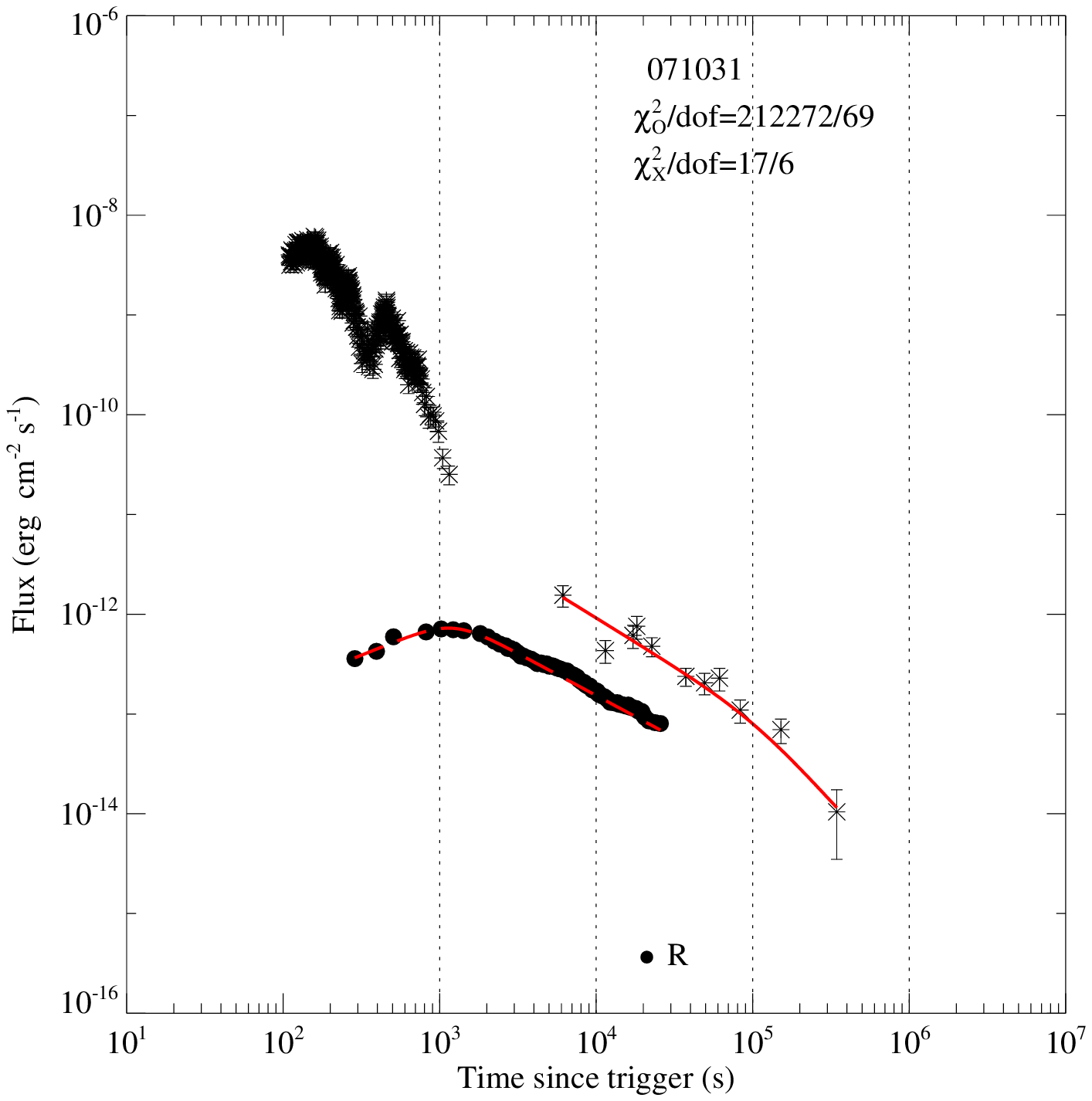}
\includegraphics[angle=0,scale=0.350,width=0.3\textwidth,height=0.25\textheight]{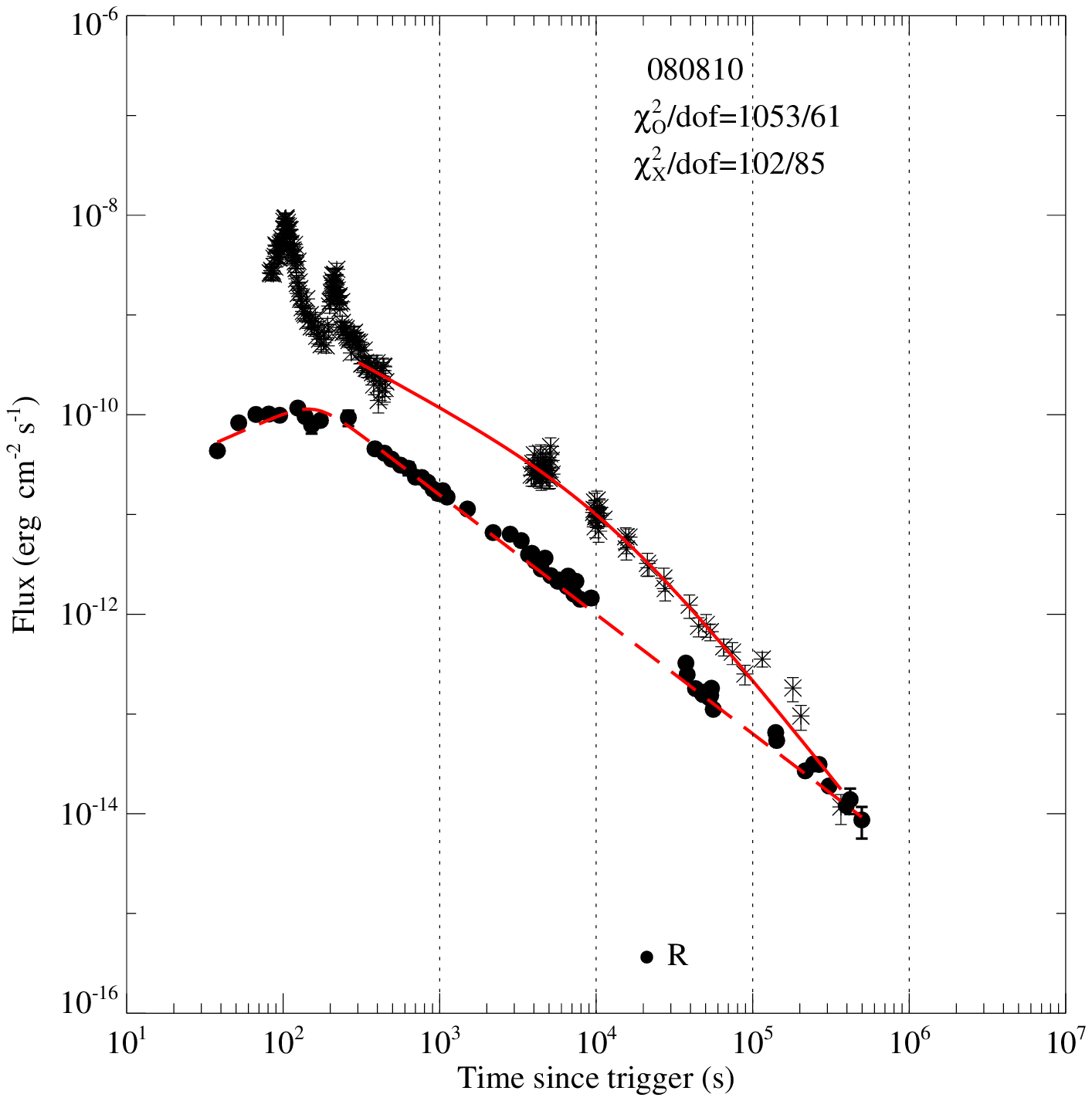}
\includegraphics[angle=0,scale=0.350,width=0.3\textwidth,height=0.25\textheight]{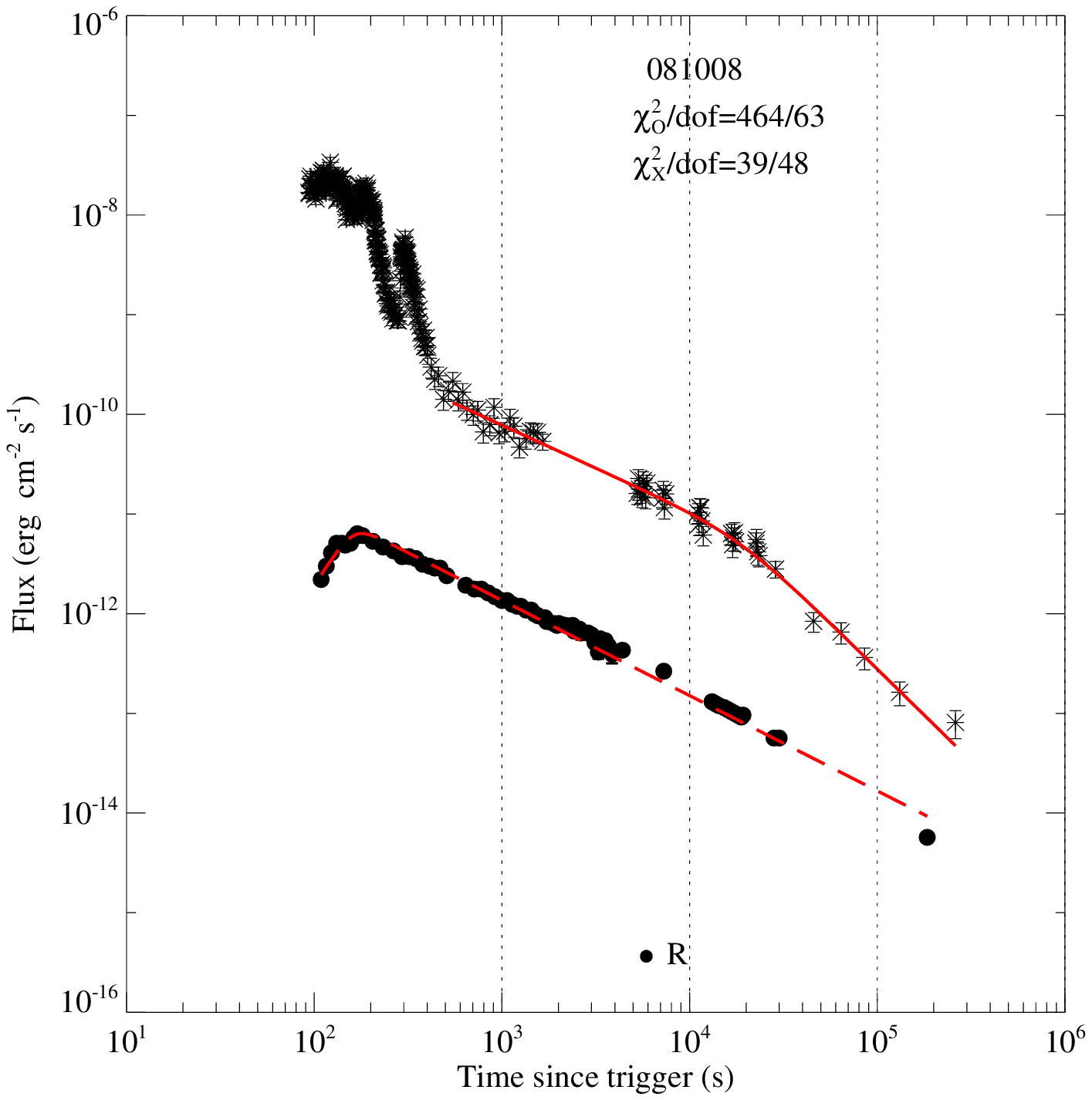}
\includegraphics[angle=0,scale=0.350,width=0.3\textwidth,height=0.25\textheight]{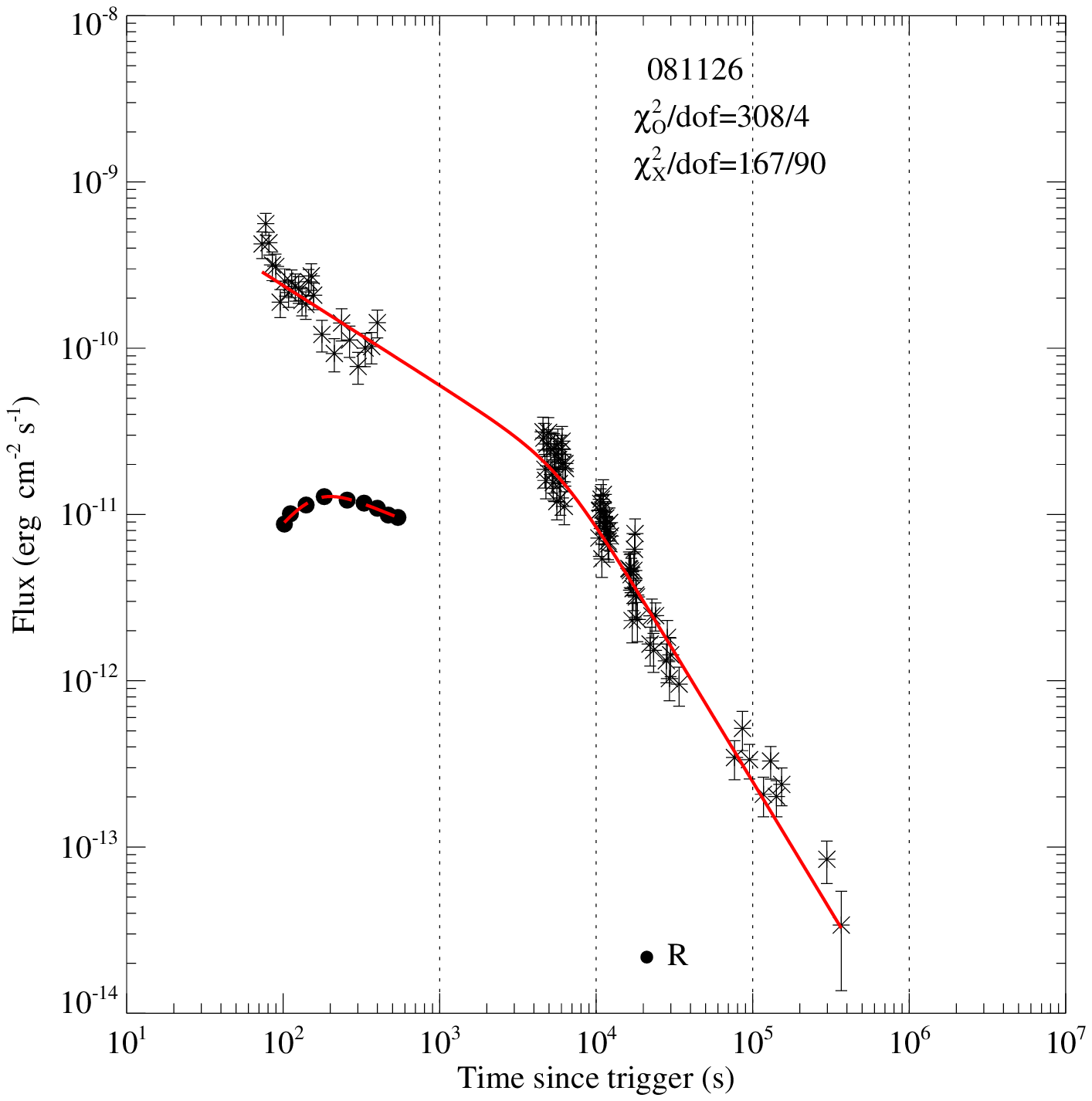}
\includegraphics[angle=0,scale=0.350,width=0.3\textwidth,height=0.25\textheight]{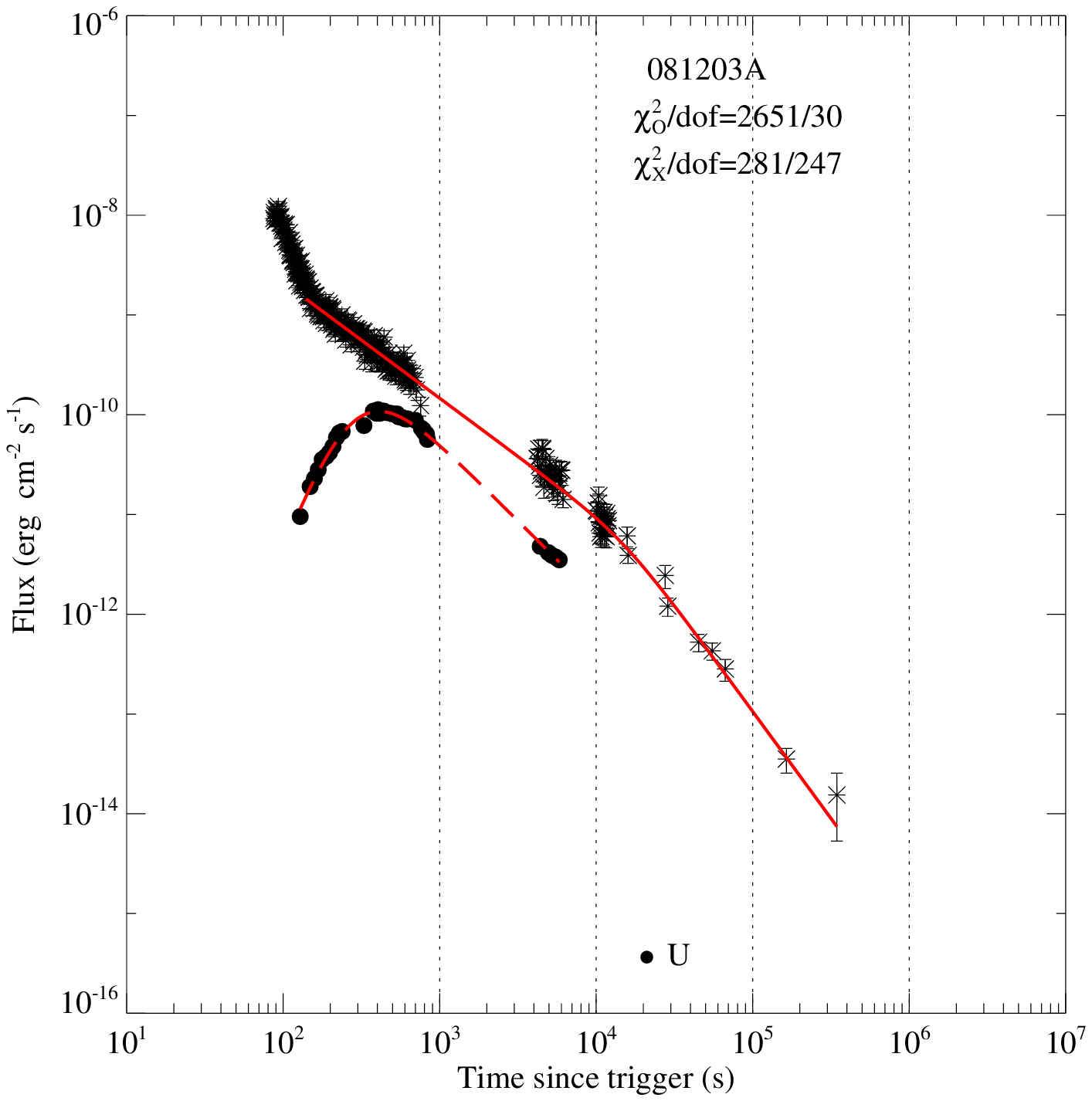}
\includegraphics[angle=0,scale=0.350,width=0.3\textwidth,height=0.25\textheight]{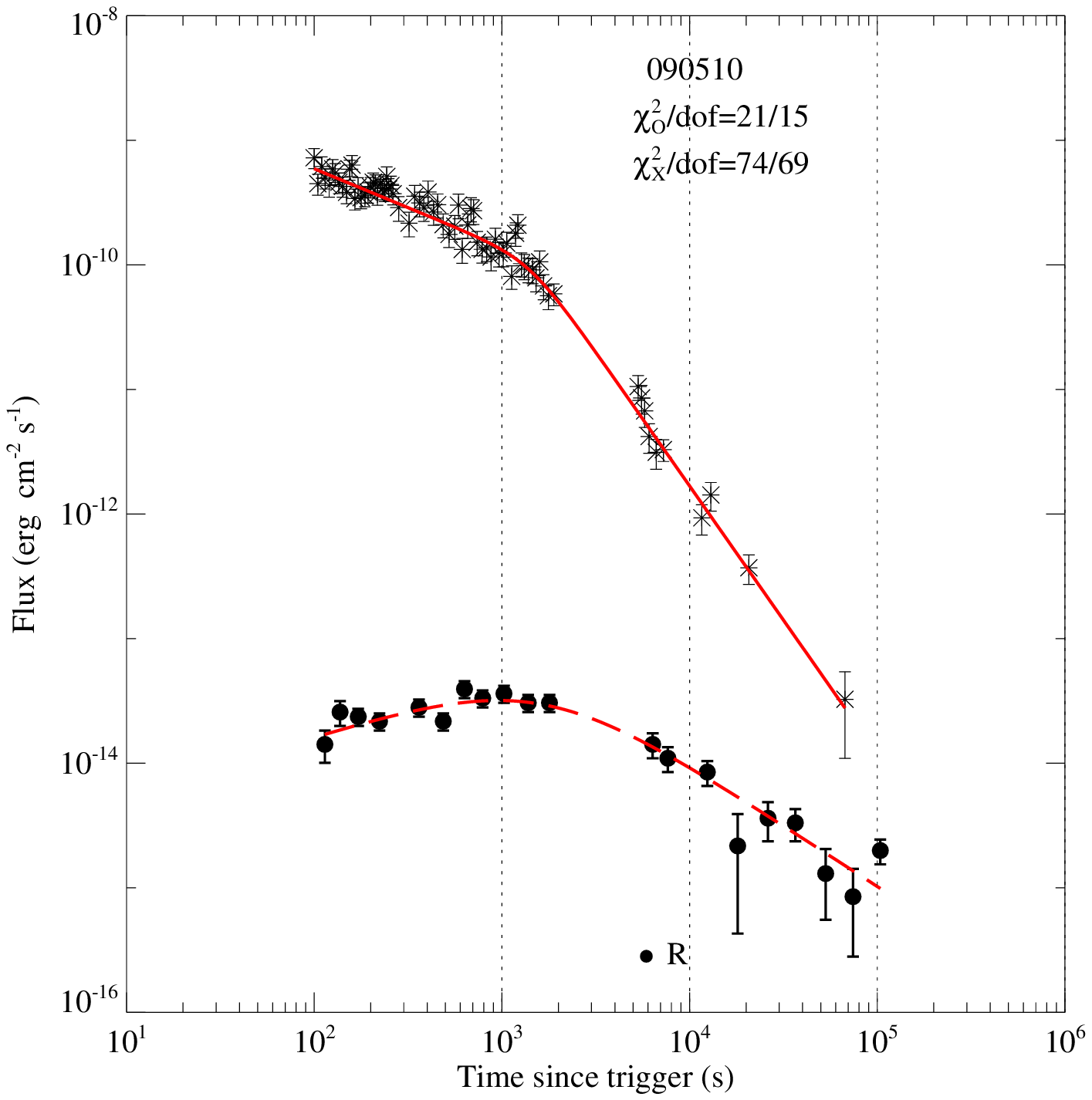}
\includegraphics[angle=0,scale=0.350,width=0.3\textwidth,height=0.25\textheight]{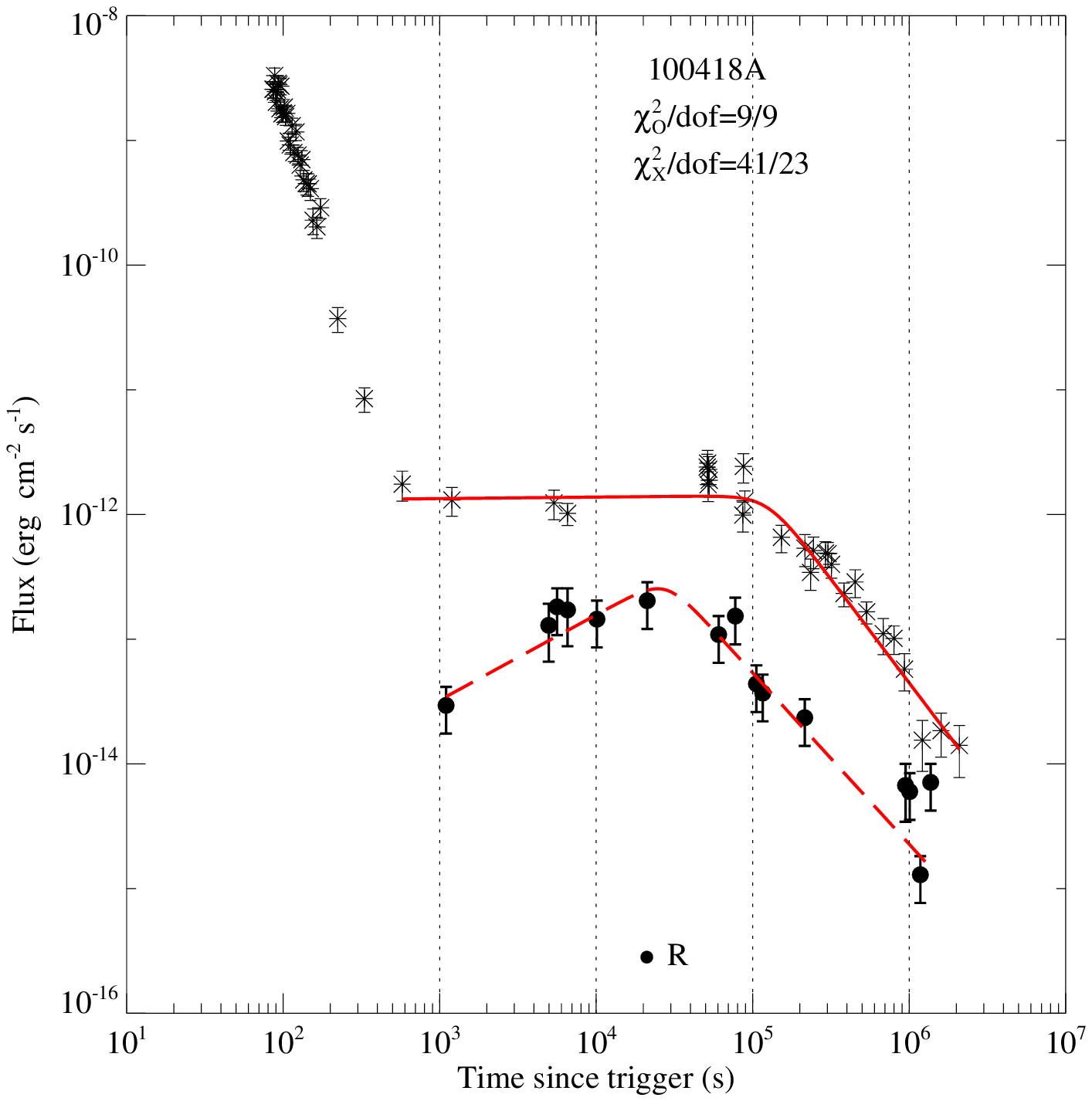}
\includegraphics[angle=0,scale=0.350,width=0.3\textwidth,height=0.25\textheight]{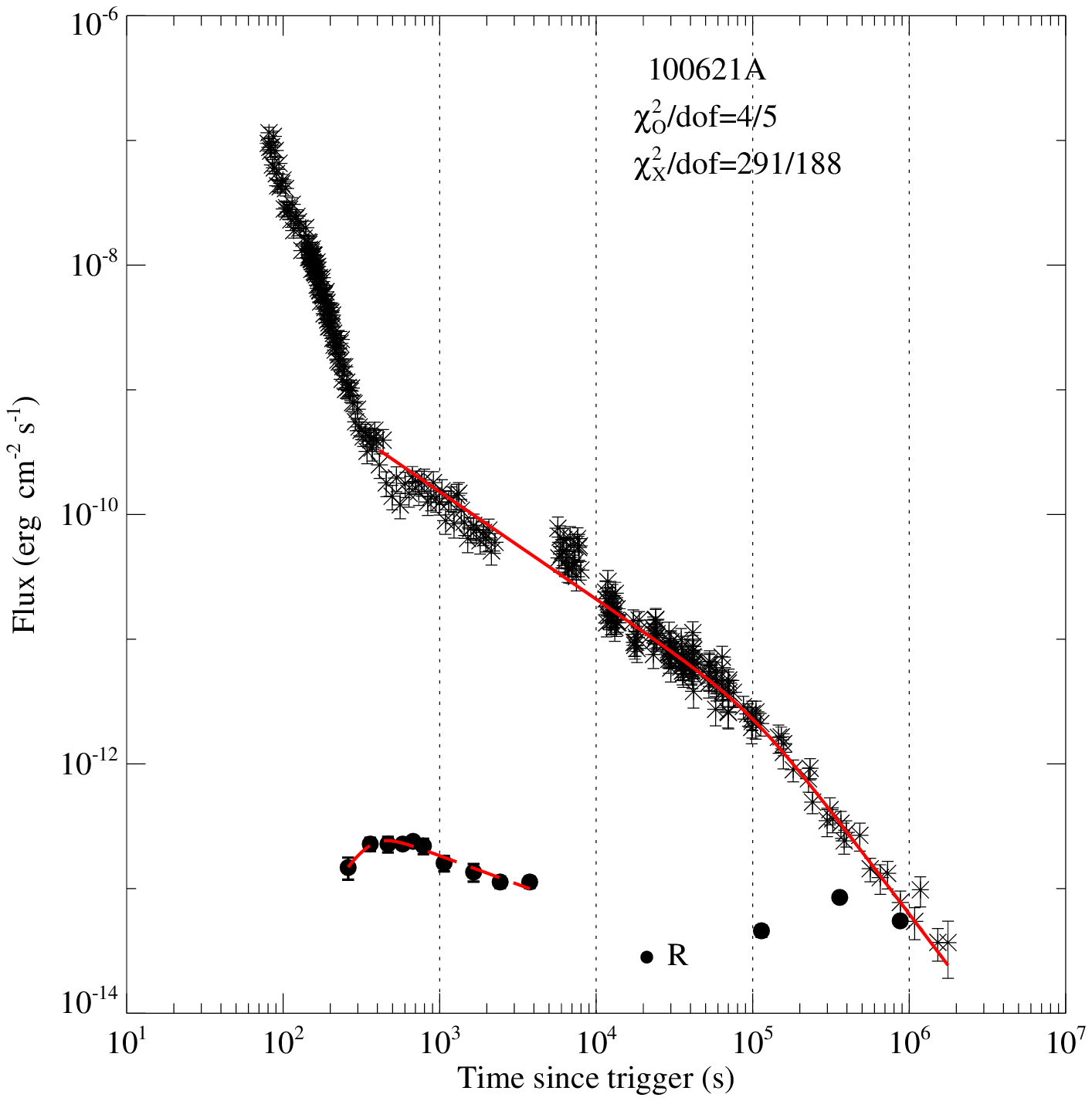}
\includegraphics[angle=0,scale=0.350,width=0.3\textwidth,height=0.25\textheight]{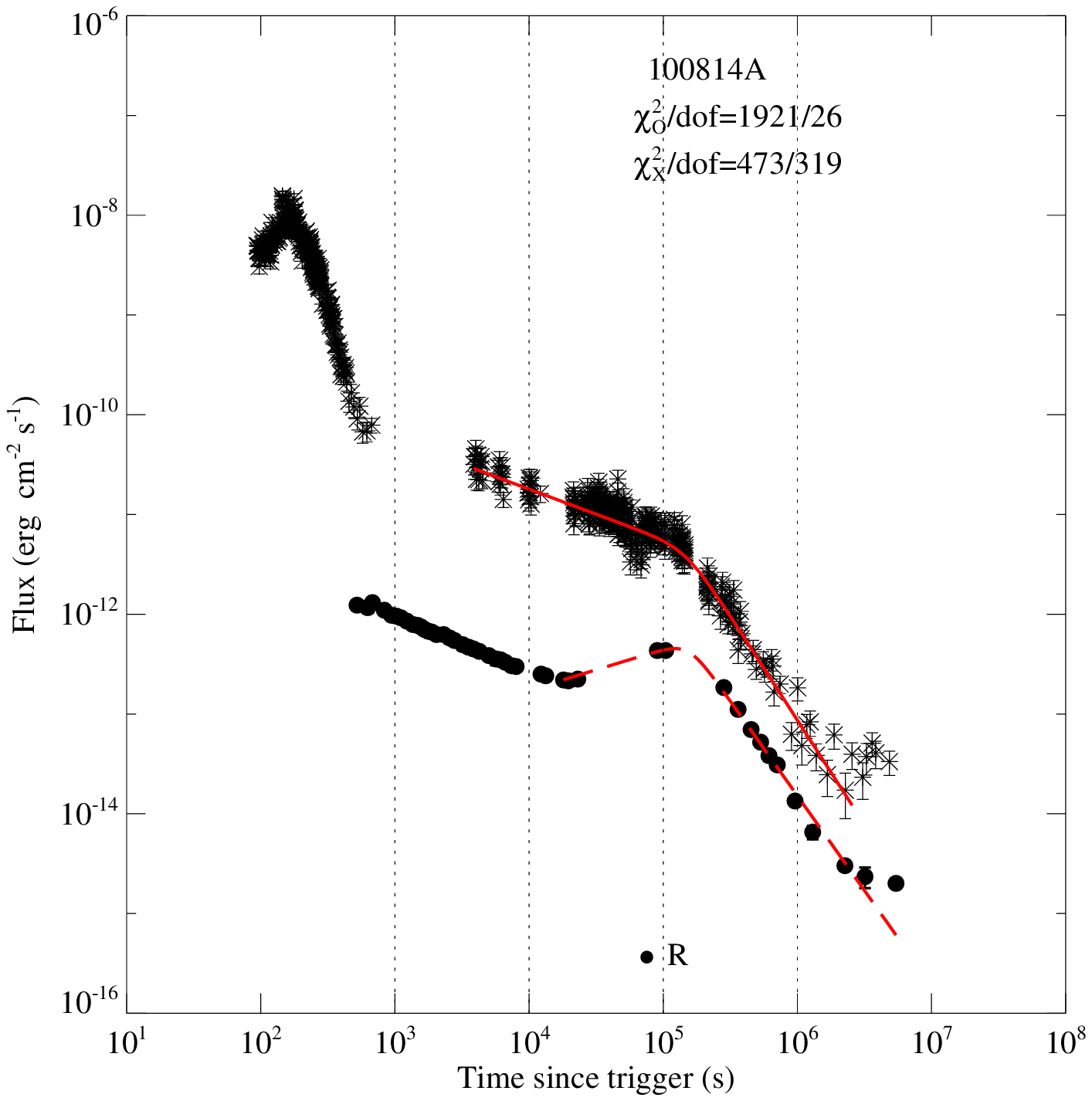}
\includegraphics[angle=0,scale=0.350,width=0.3\textwidth,height=0.25\textheight]{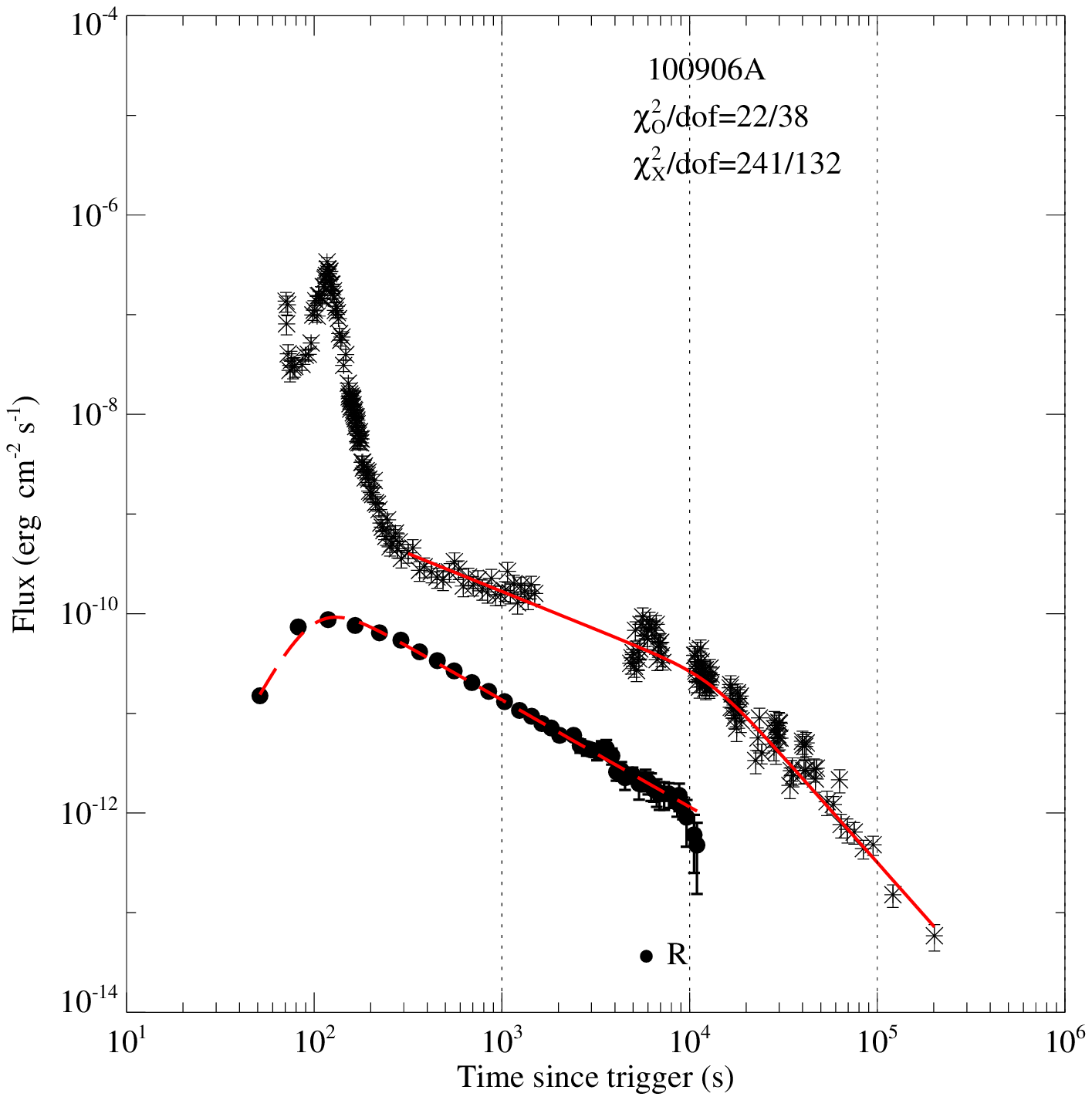}
\includegraphics[angle=0,scale=0.350,width=0.3\textwidth,height=0.25\textheight]{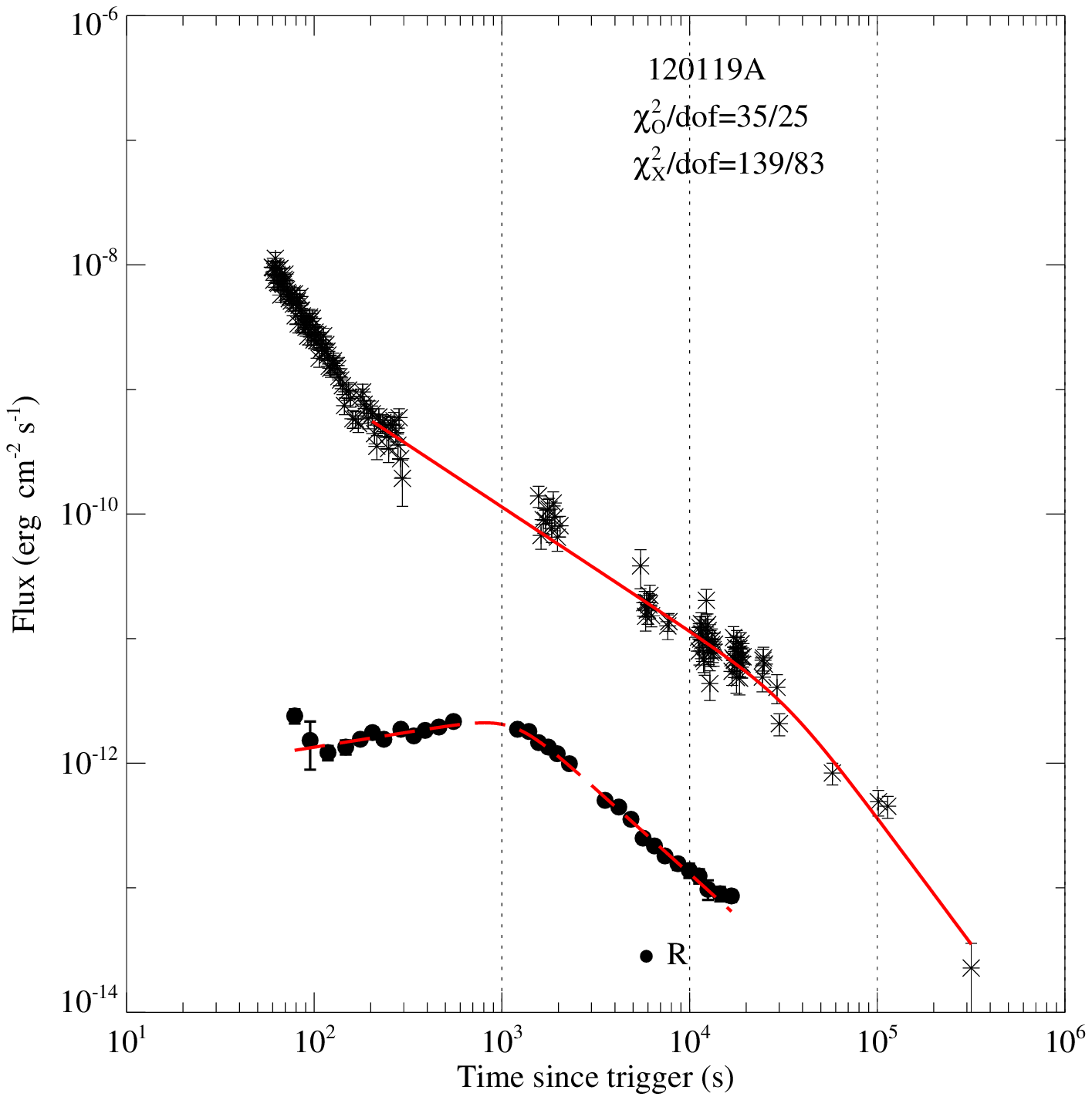}
\includegraphics[angle=0,scale=0.350,width=0.3\textwidth,height=0.25\textheight]{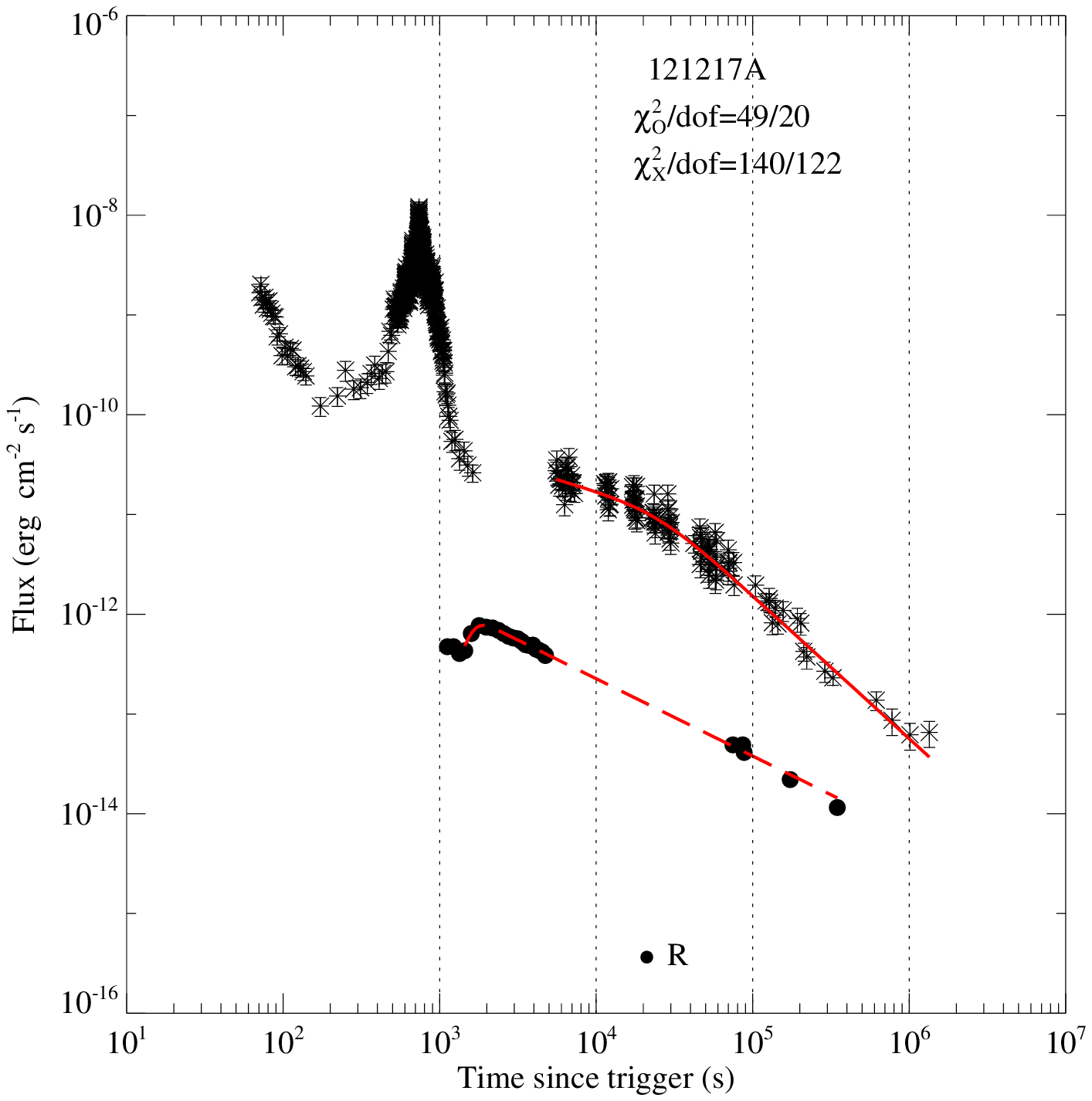}
\center{Fig. \ref{Xray Break}--- Continued}
\end{figure*}

\begin{figure*}
\includegraphics[angle=0,scale=0.350,width=0.3\textwidth,height=0.25\textheight]{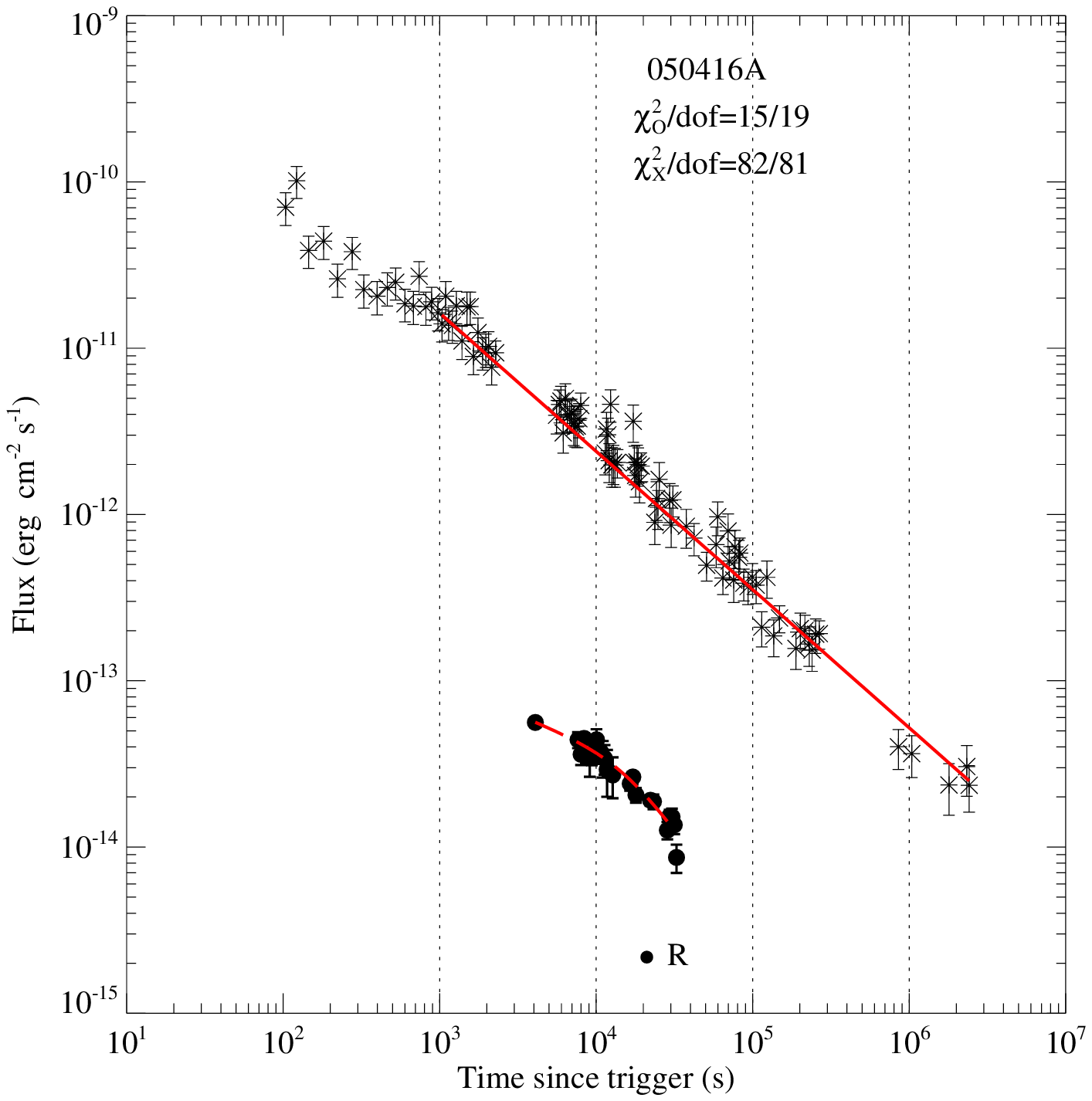}
\includegraphics[angle=0,scale=0.350,width=0.3\textwidth,height=0.25\textheight]{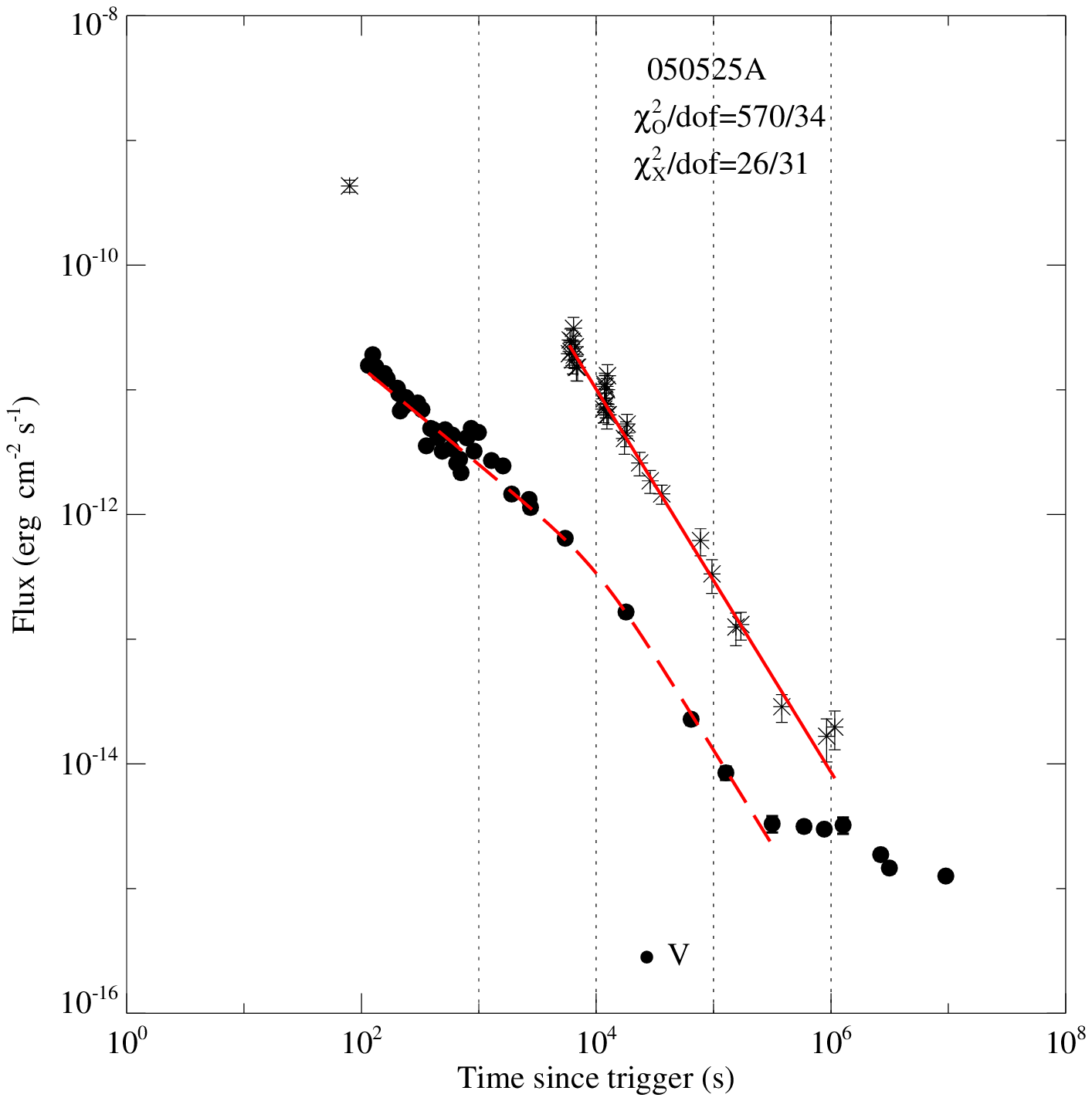}
\includegraphics[angle=0,scale=0.350,width=0.3\textwidth,height=0.25\textheight]{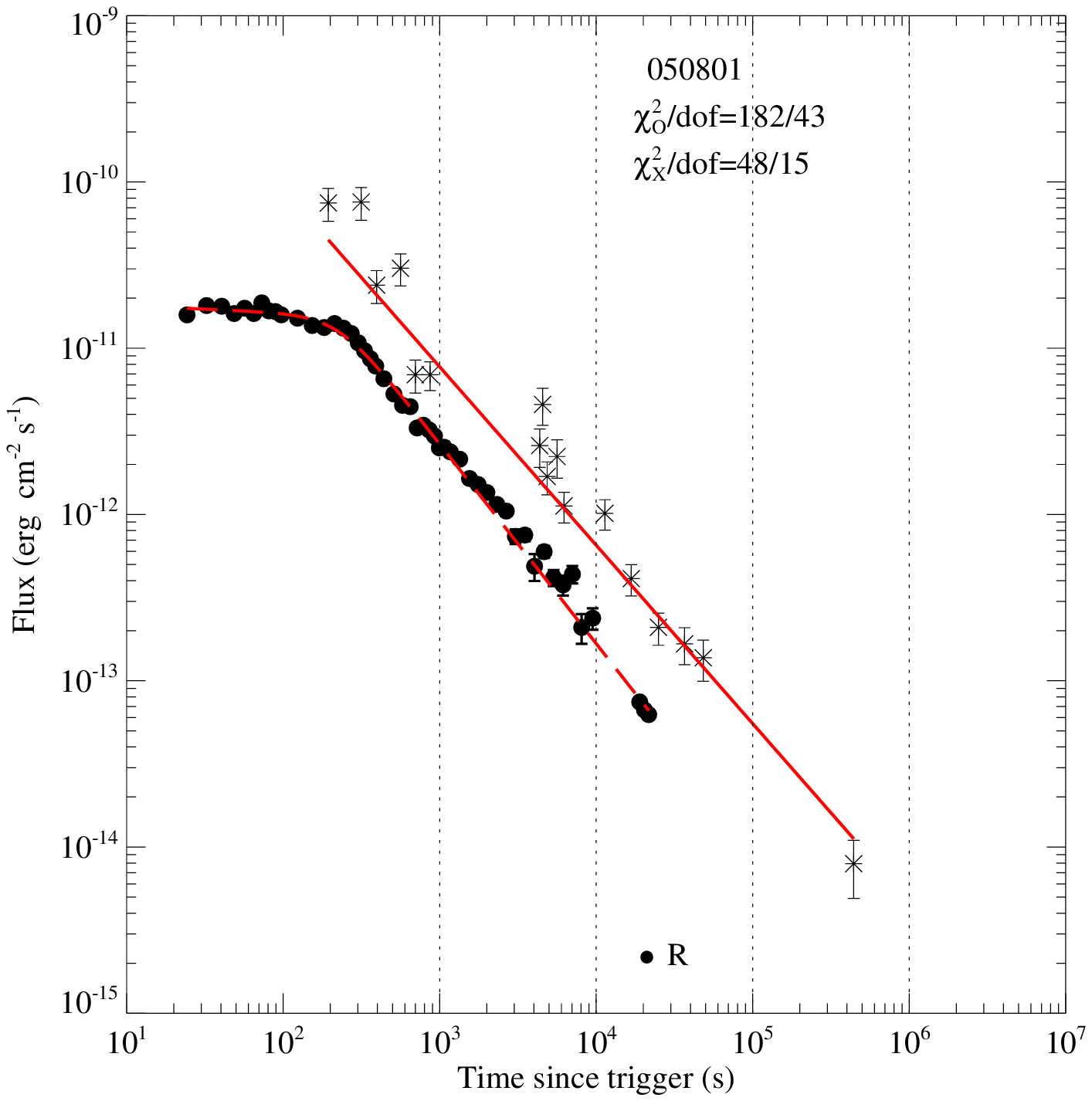}
\includegraphics[angle=0,scale=0.350,width=0.3\textwidth,height=0.25\textheight]{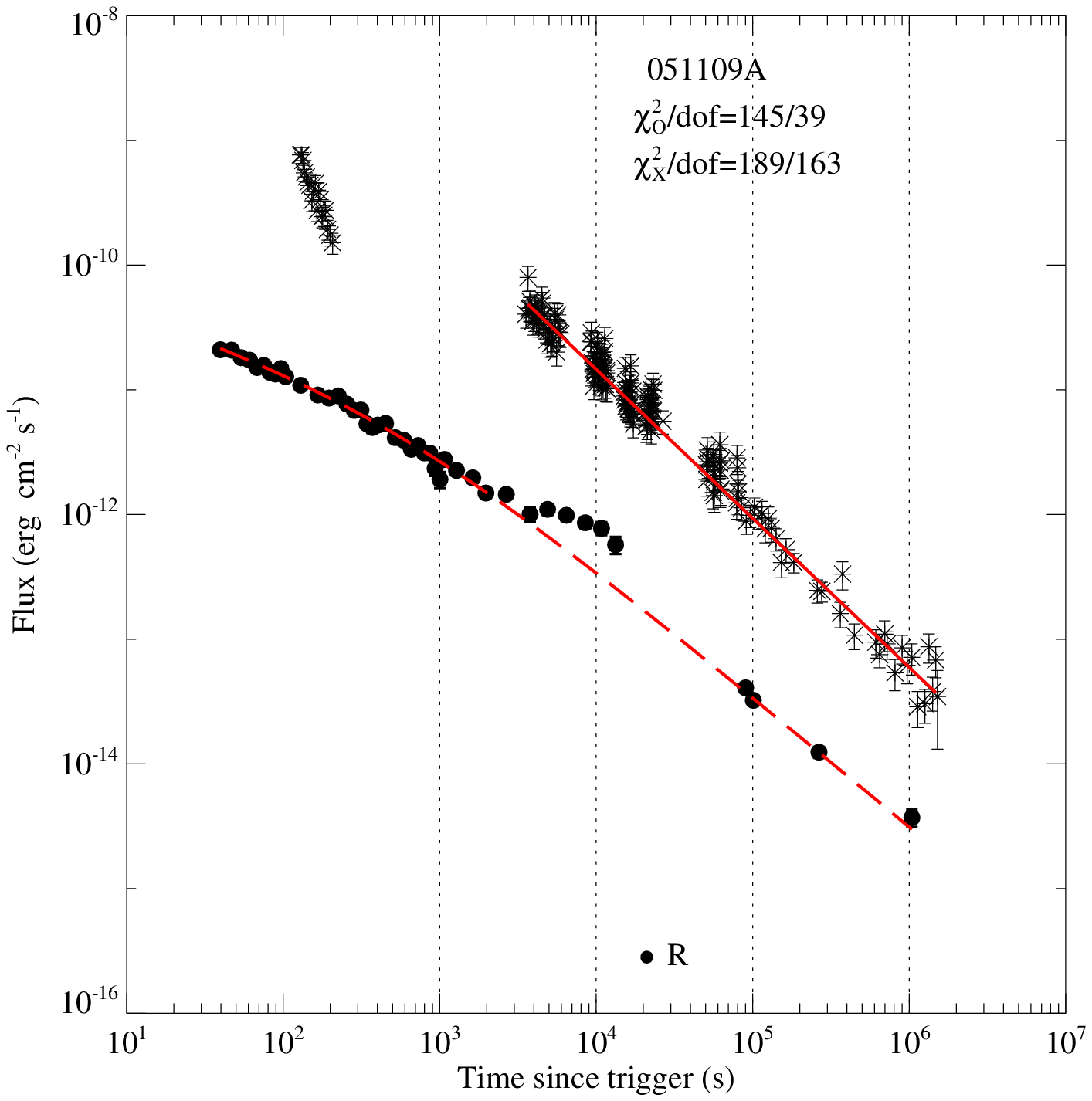}
\includegraphics[angle=0,scale=0.350,width=0.3\textwidth,height=0.25\textheight]{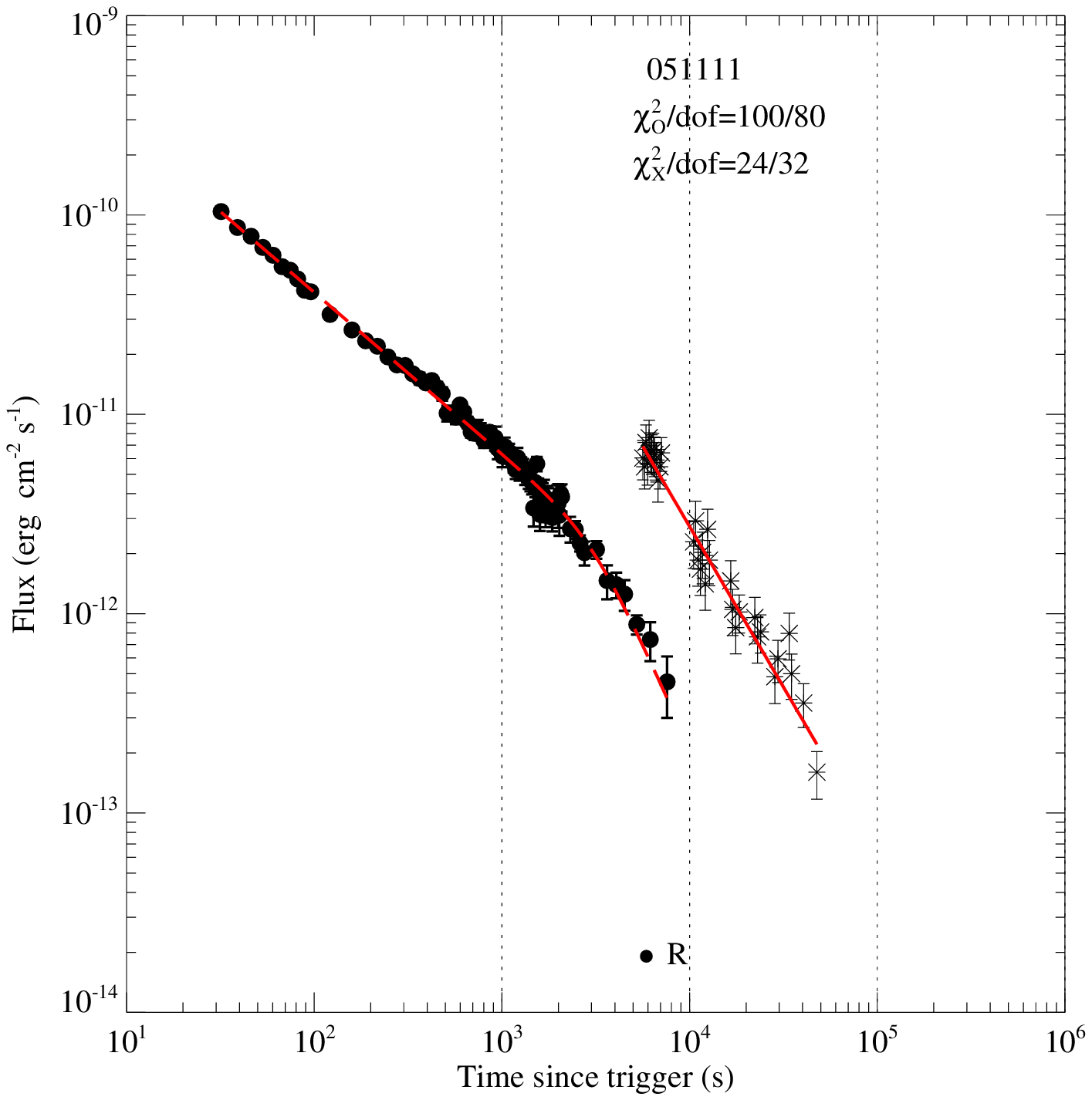}
\includegraphics[angle=0,scale=0.350,width=0.3\textwidth,height=0.25\textheight]{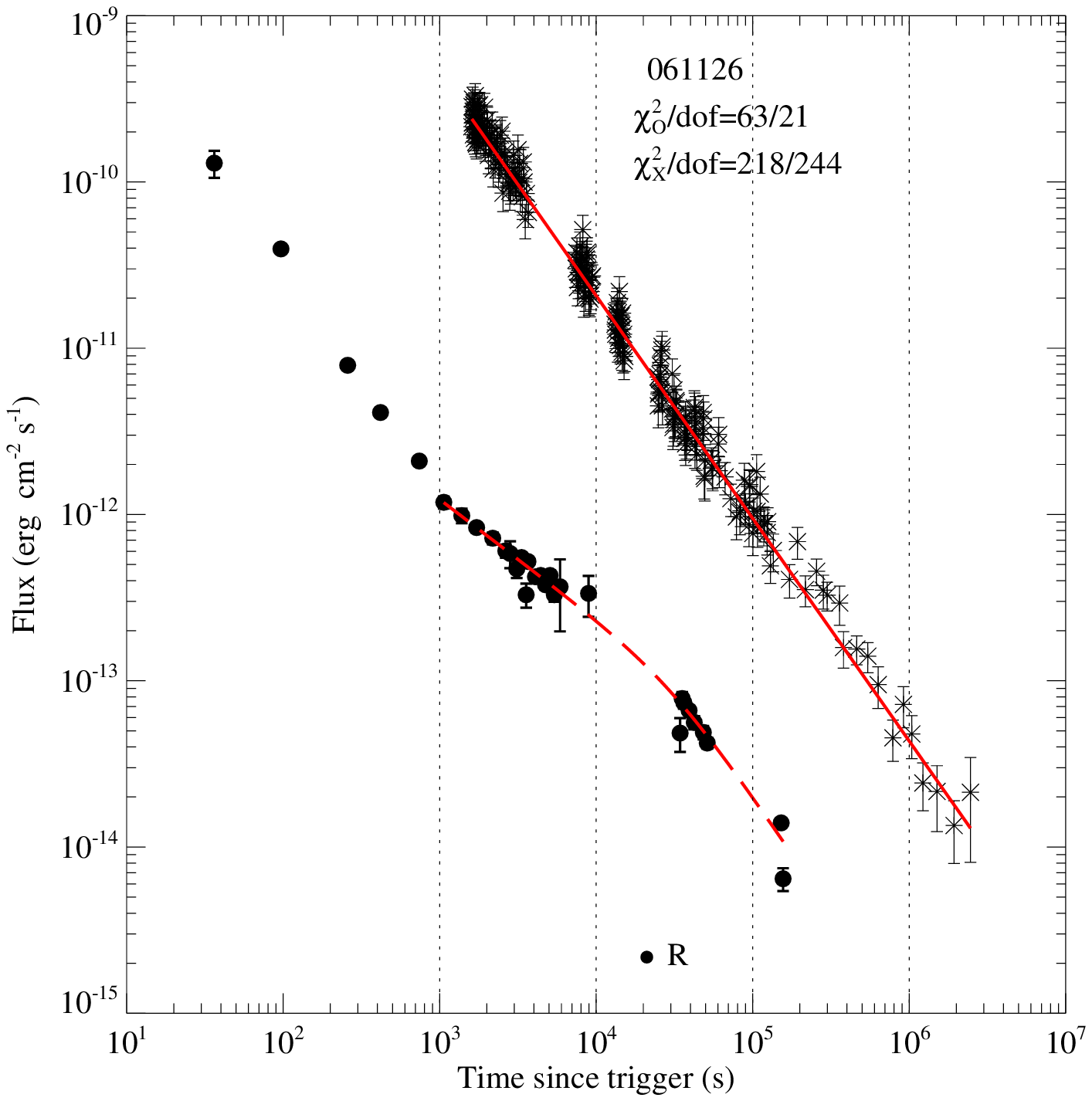}
\includegraphics[angle=0,scale=0.350,width=0.3\textwidth,height=0.25\textheight]{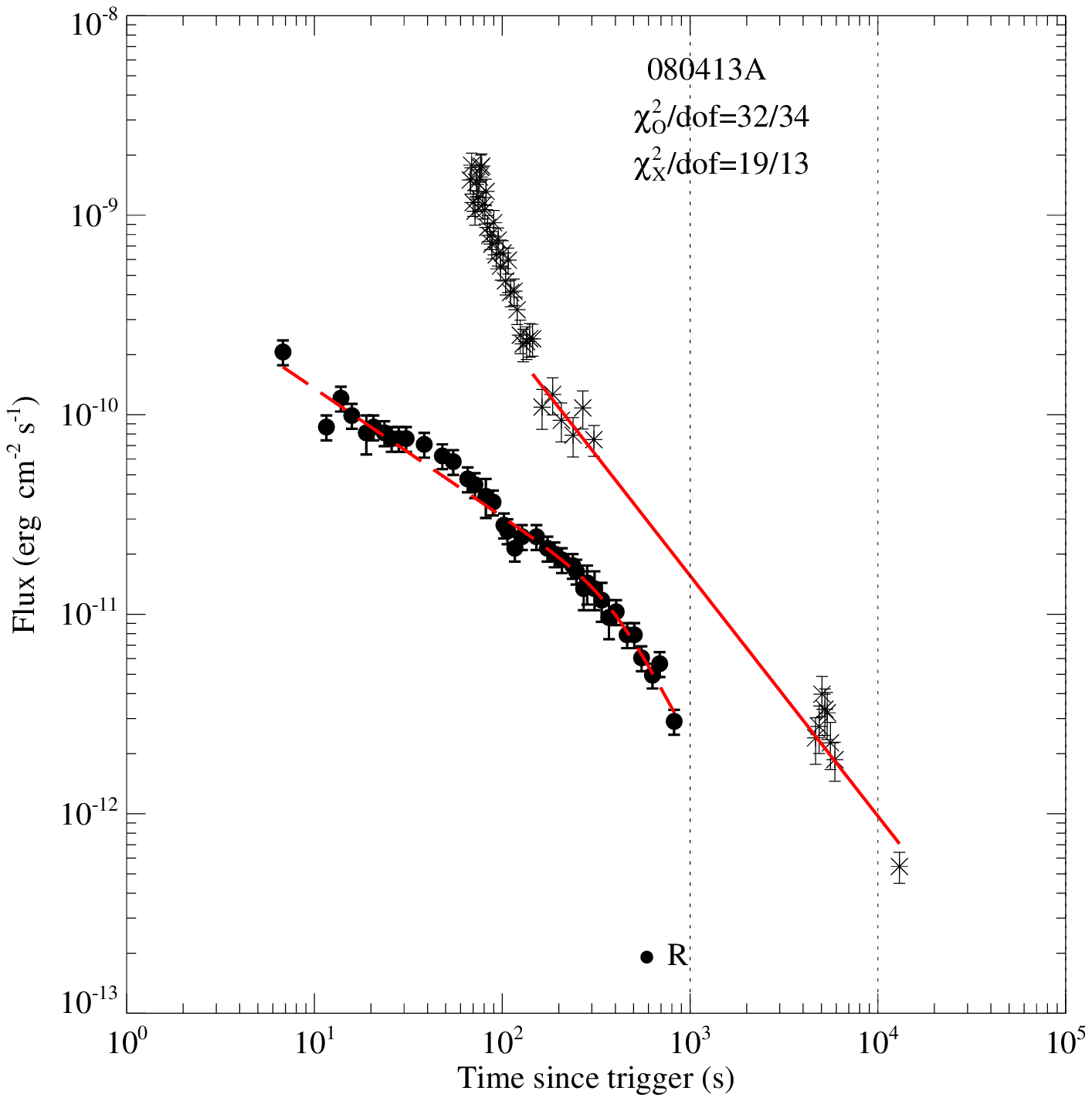}
\includegraphics[angle=0,scale=0.350,width=0.3\textwidth,height=0.25\textheight]{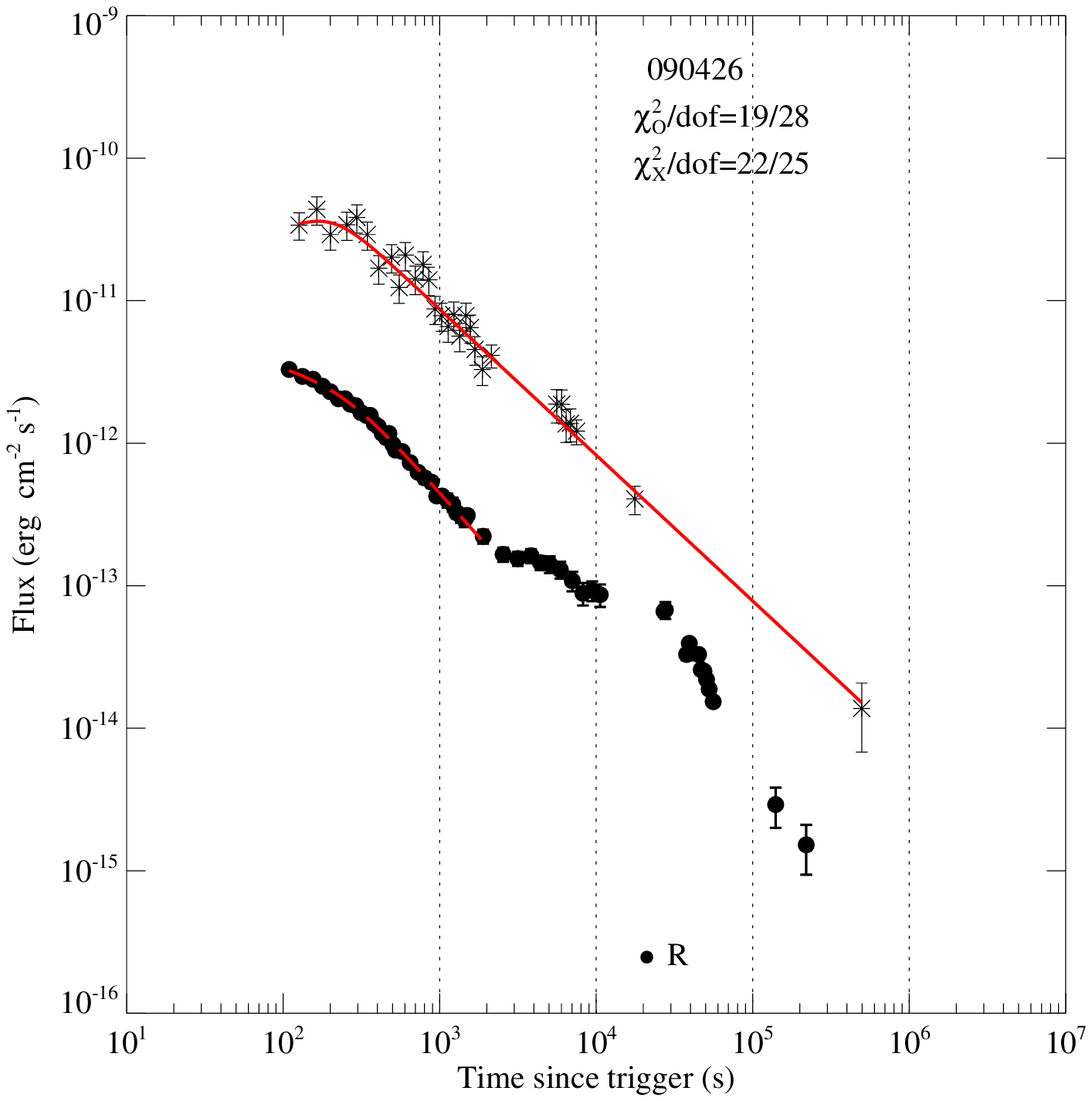}
\includegraphics[angle=0,scale=0.350,width=0.3\textwidth,height=0.25\textheight]{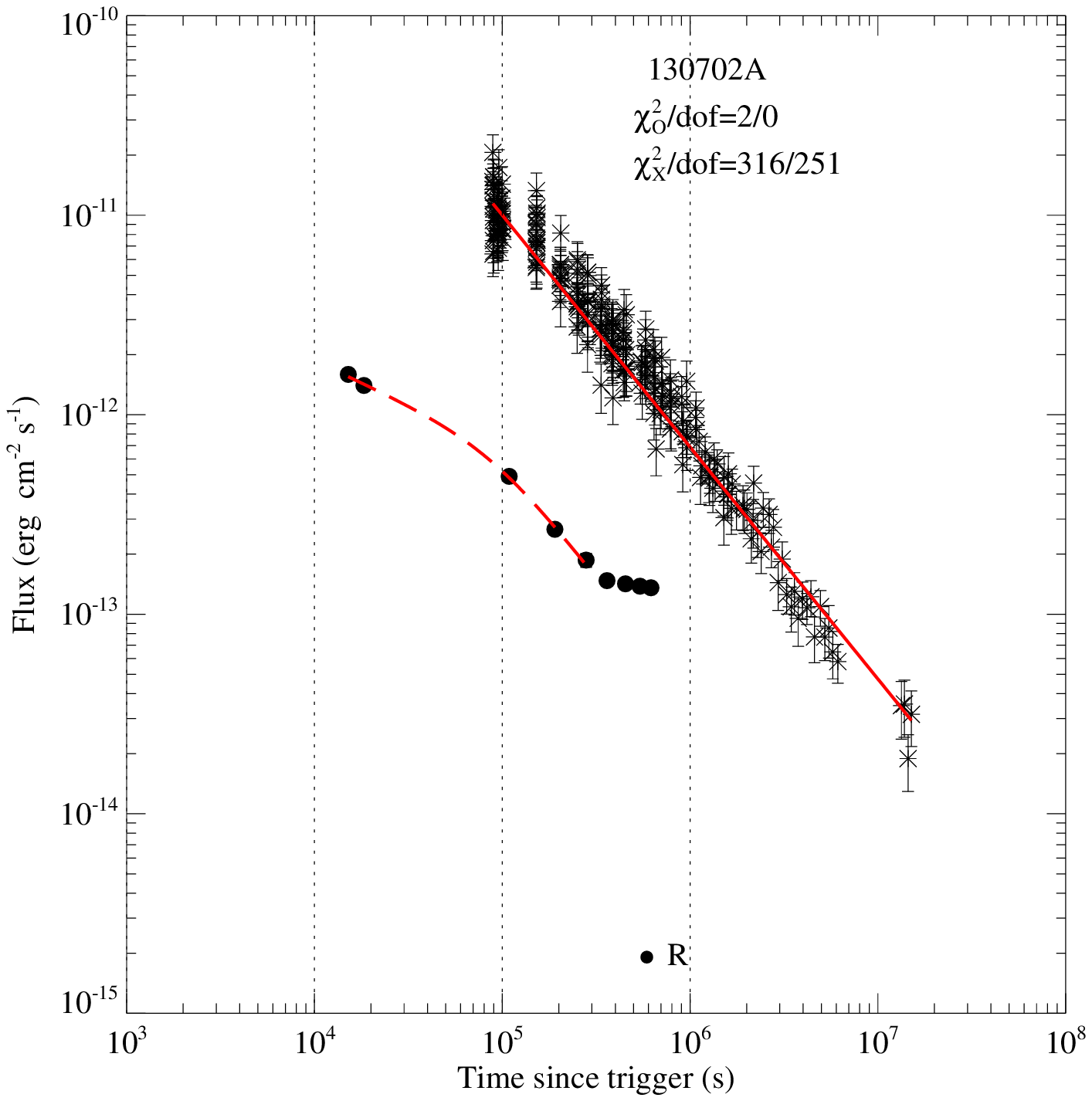}
\caption{The cases in which a breaks was detected only in the optical band but not in the X-ray band in their afterglow light-curves.}\label{Optical Break}
\end{figure*}

\begin{figure*}
\includegraphics[angle=0,scale=0.350,width=0.3\textwidth,height=0.25\textheight]{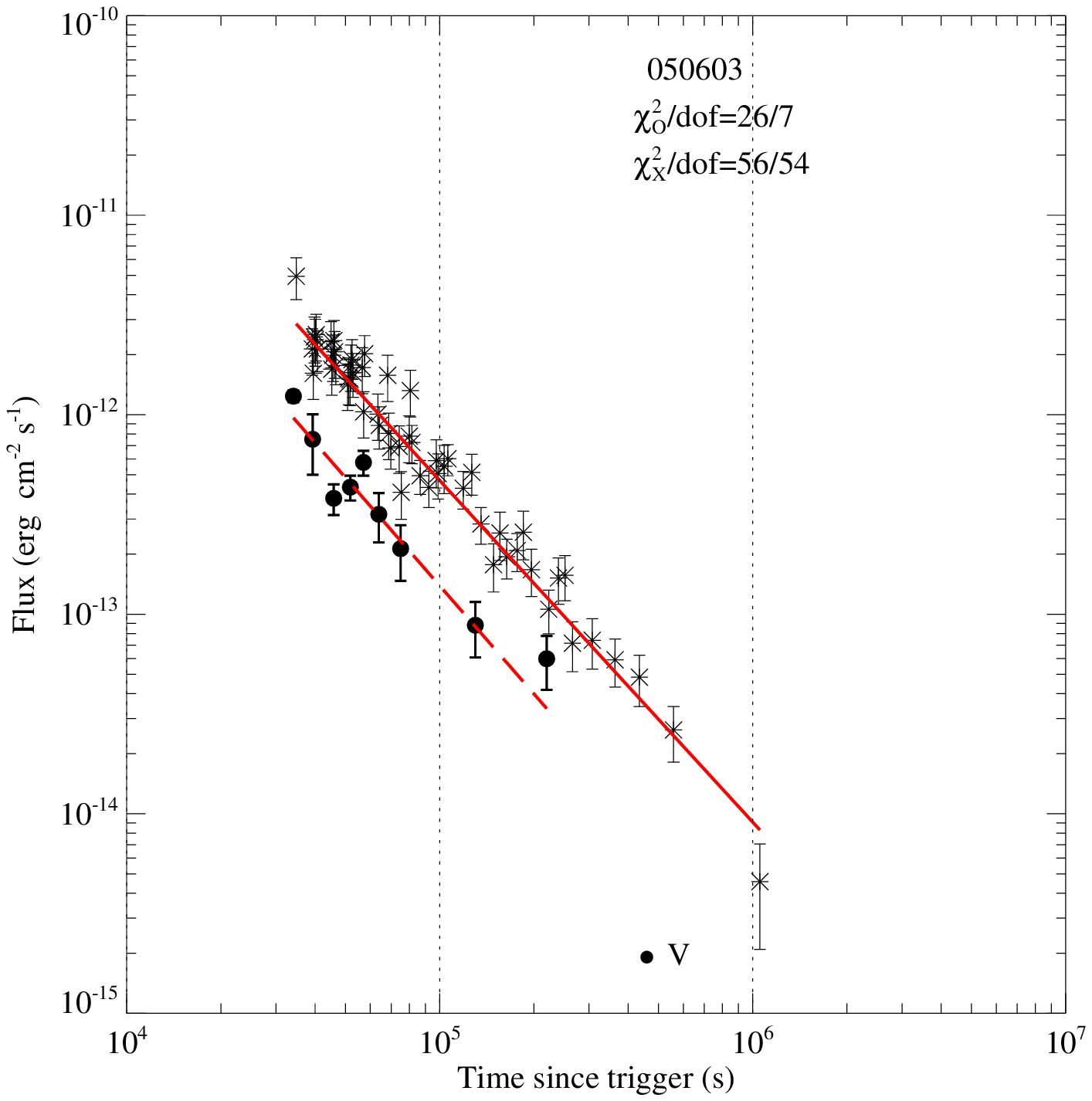}
\includegraphics[angle=0,scale=0.350,width=0.3\textwidth,height=0.25\textheight]{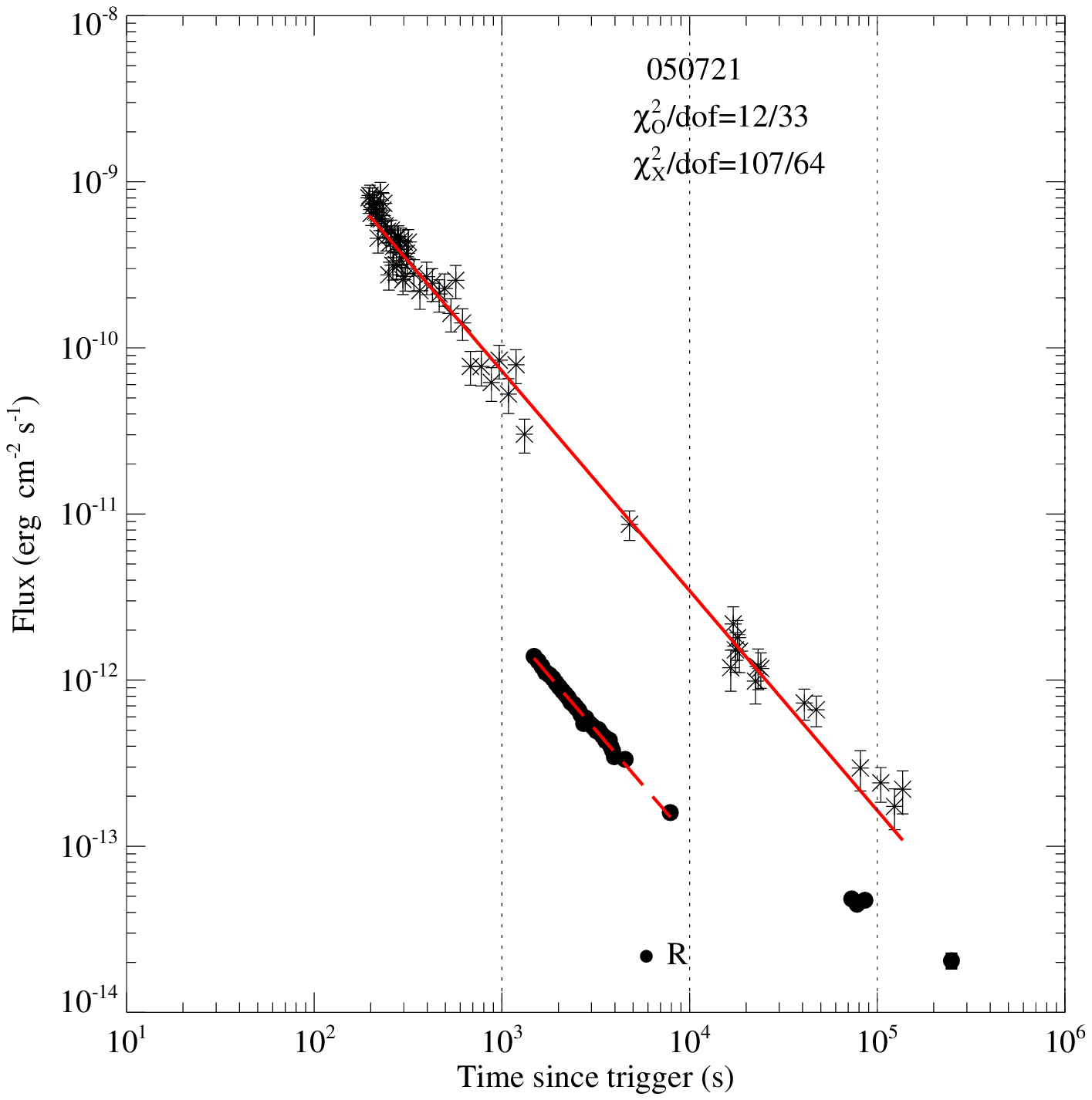}
\includegraphics[angle=0,scale=0.350,width=0.3\textwidth,height=0.25\textheight]{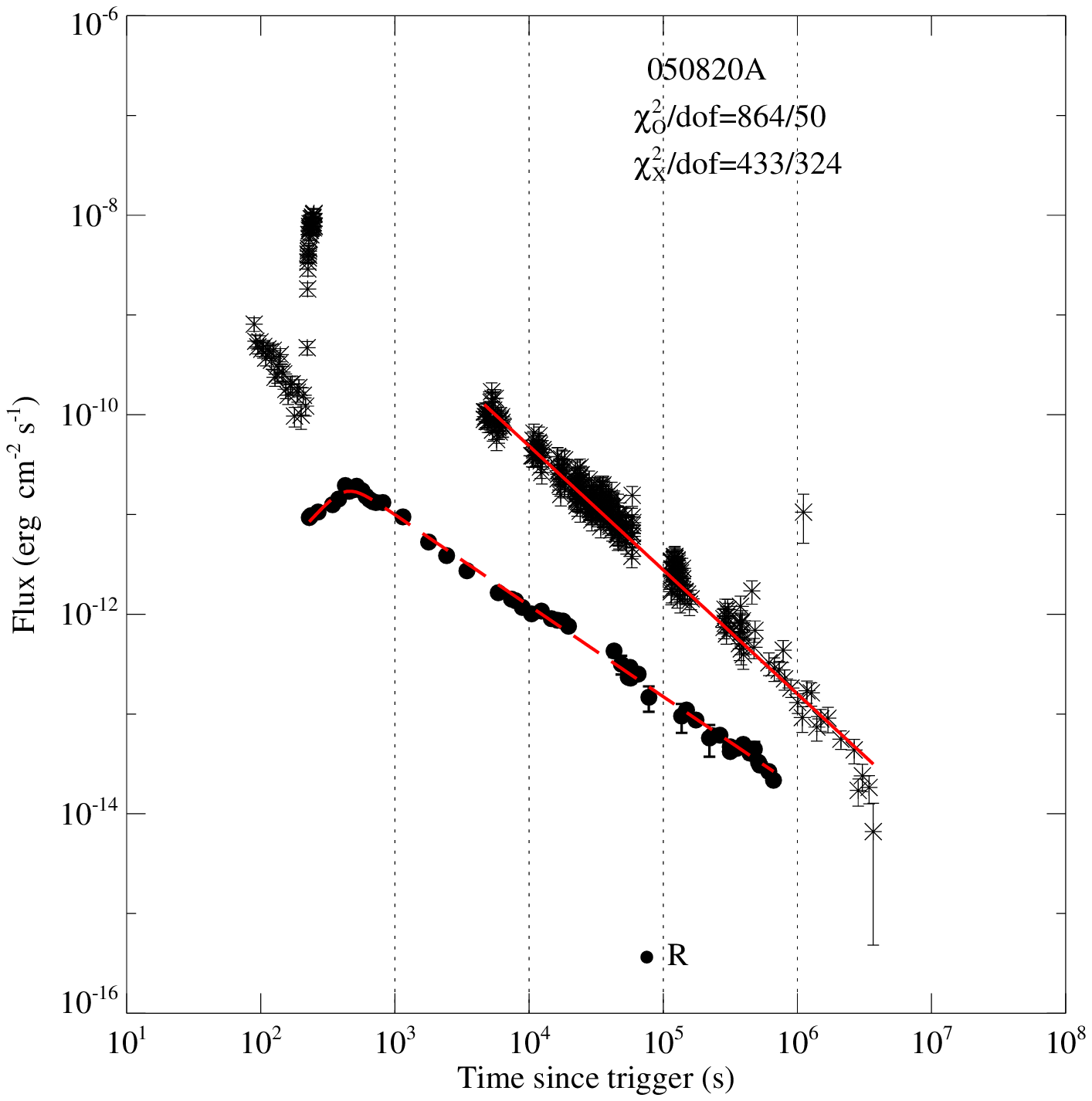}
\includegraphics[angle=0,scale=0.350,width=0.3\textwidth,height=0.25\textheight]{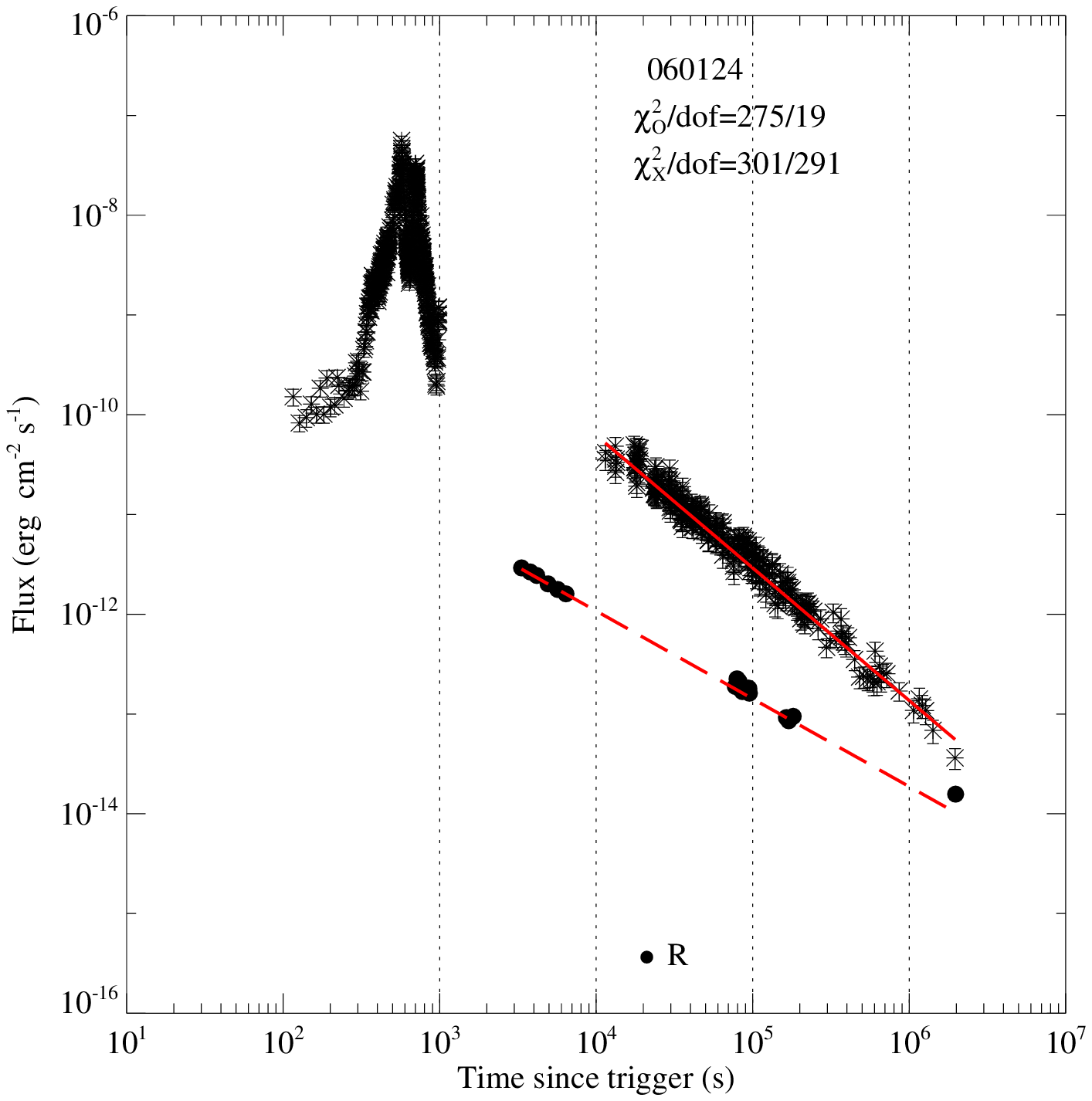}
\includegraphics[angle=0,scale=0.350,width=0.3\textwidth,height=0.25\textheight]{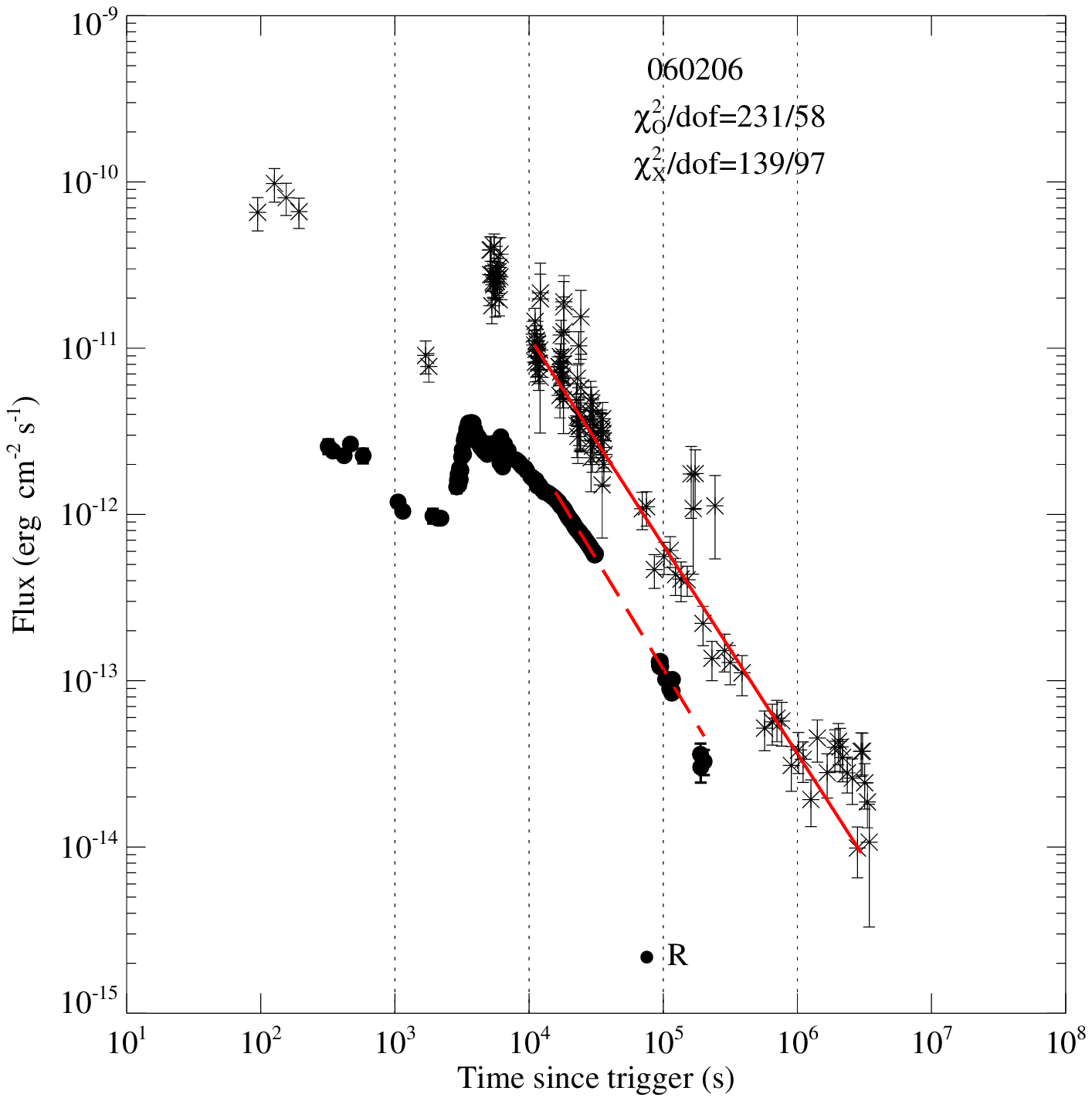}
\includegraphics[angle=0,scale=0.350,width=0.3\textwidth,height=0.25\textheight]{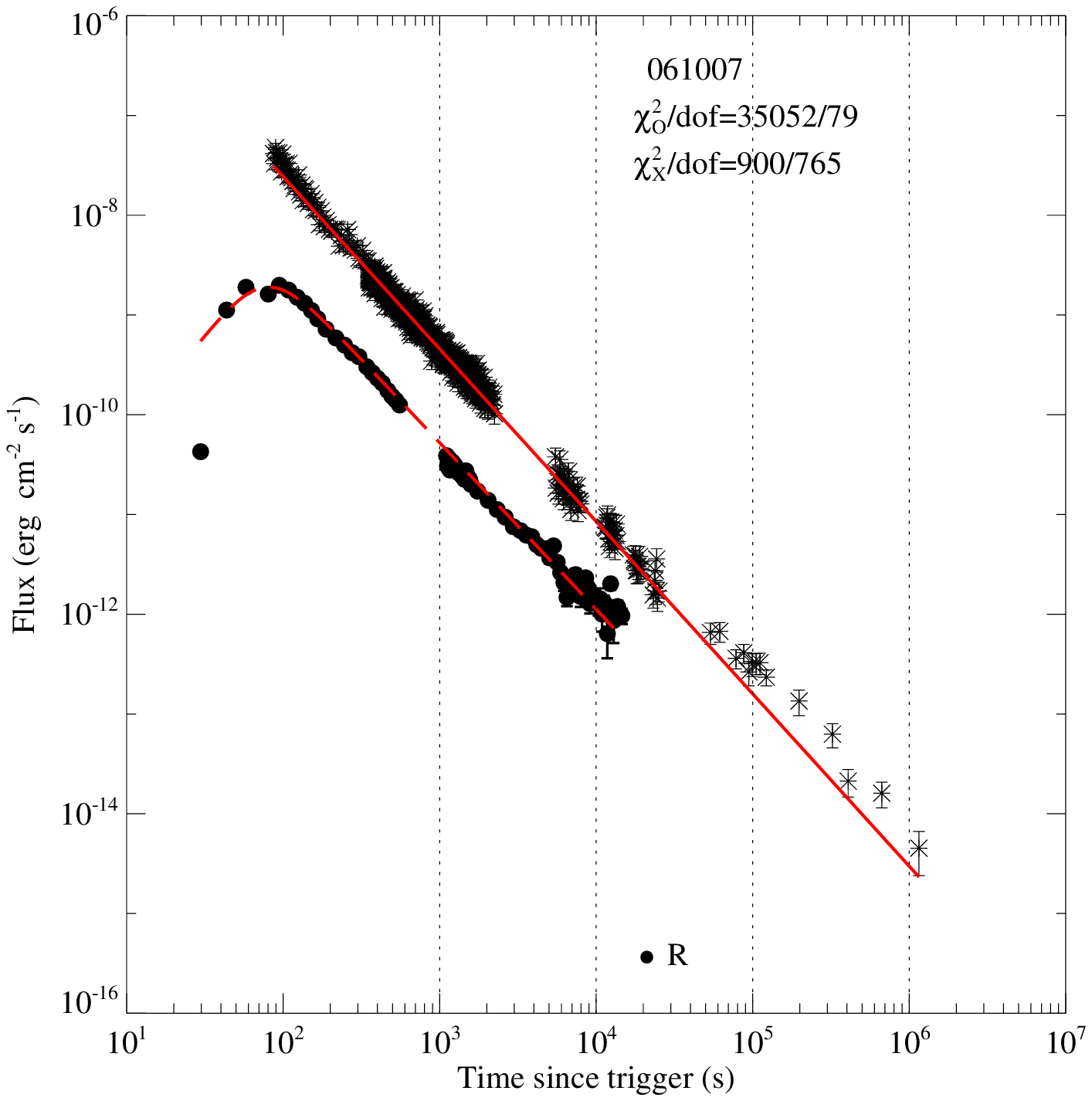}
\includegraphics[angle=0,scale=0.350,width=0.3\textwidth,height=0.25\textheight]{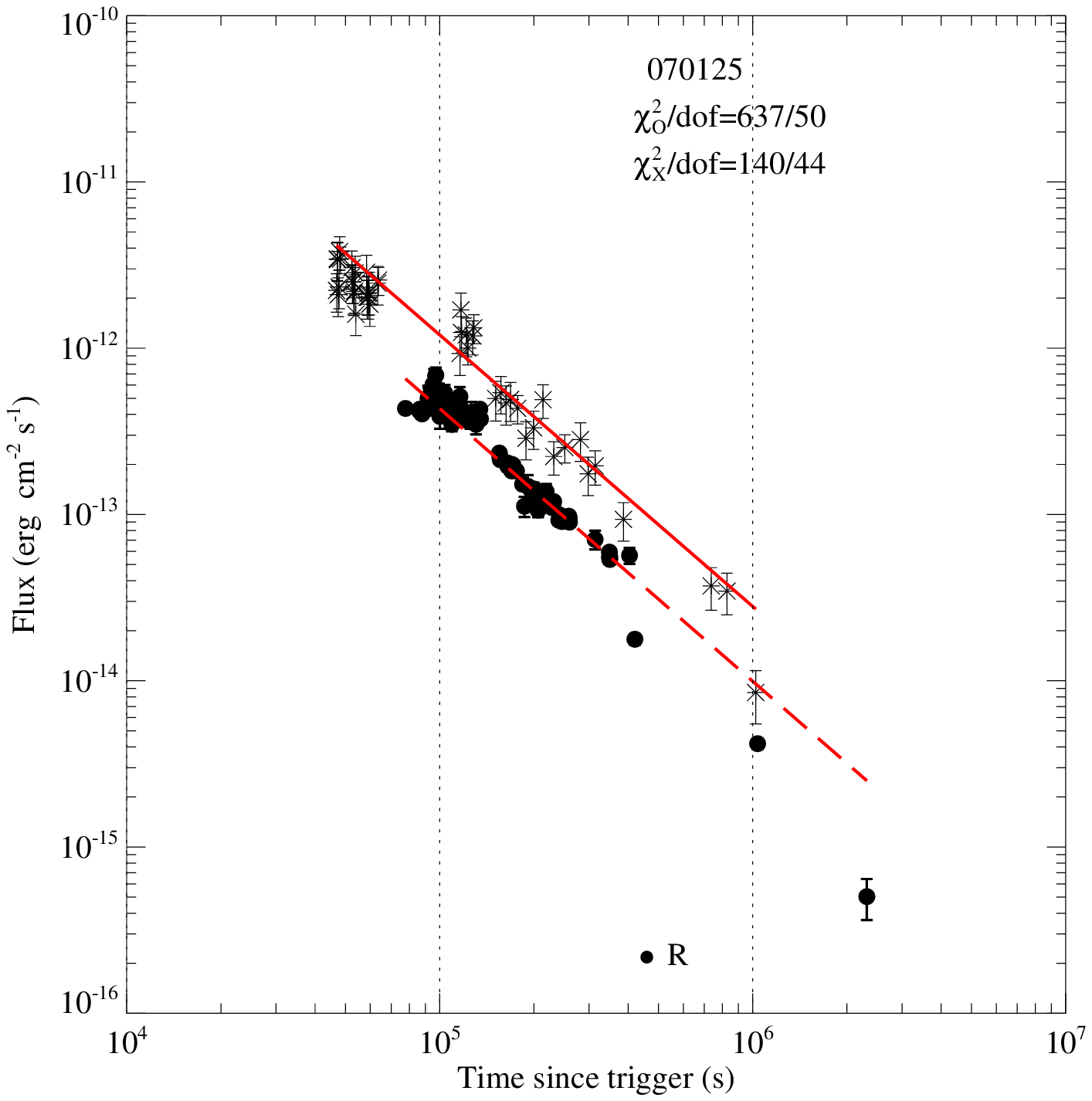}
\includegraphics[angle=0,scale=0.350,width=0.3\textwidth,height=0.25\textheight]{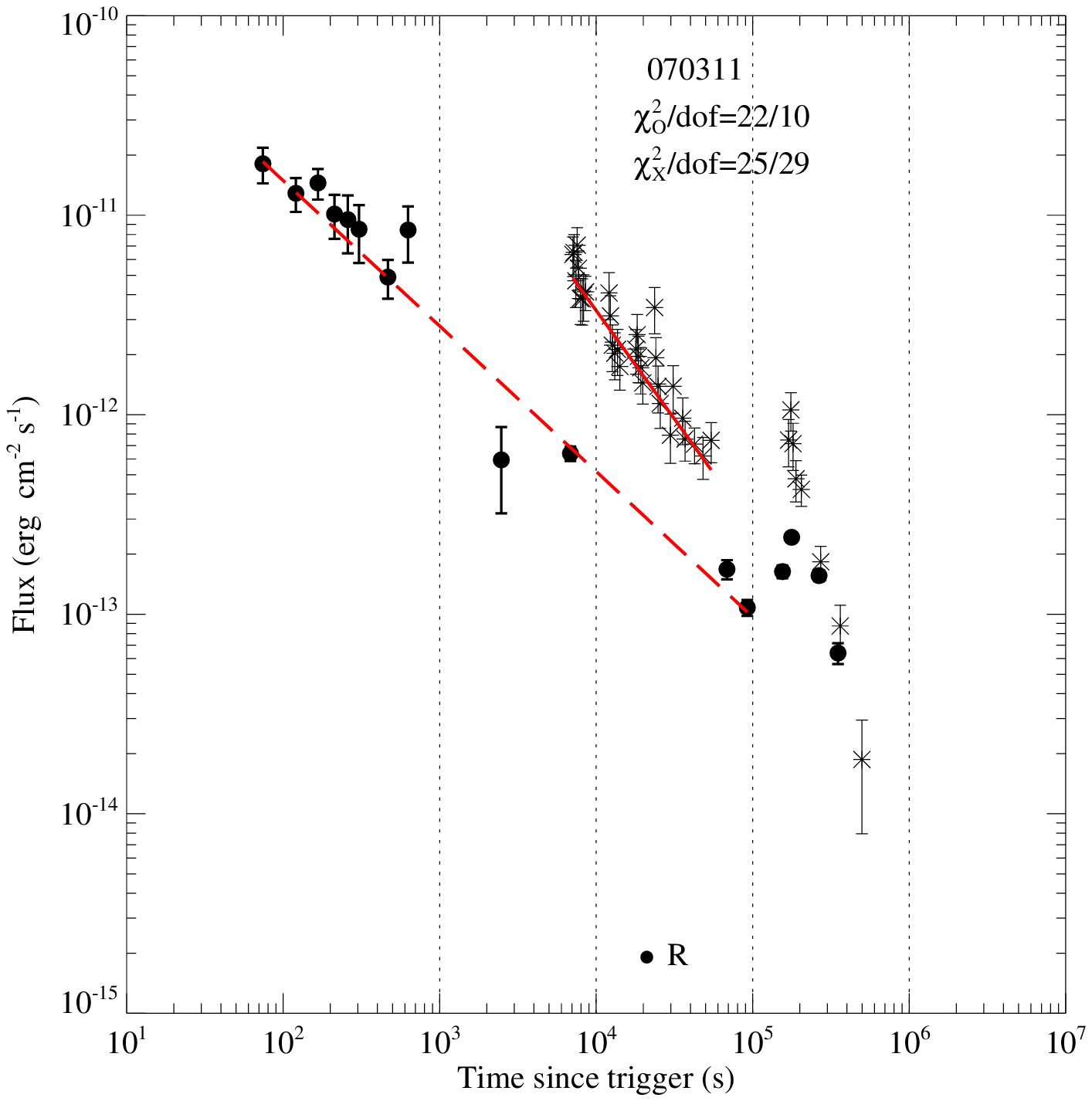}
\includegraphics[angle=0,scale=0.350,width=0.3\textwidth,height=0.25\textheight]{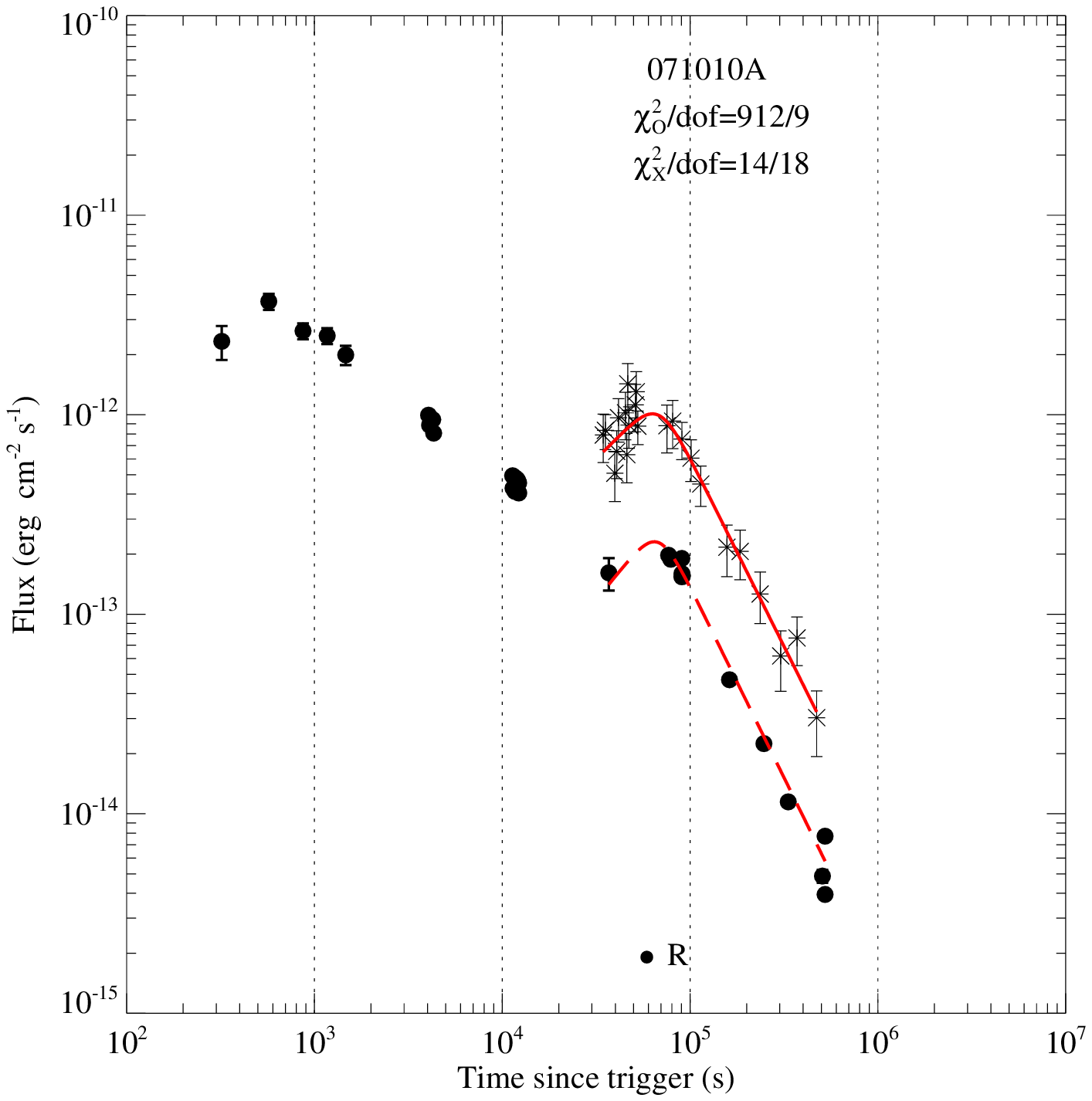}
\includegraphics[angle=0,scale=0.350,width=0.3\textwidth,height=0.25\textheight]{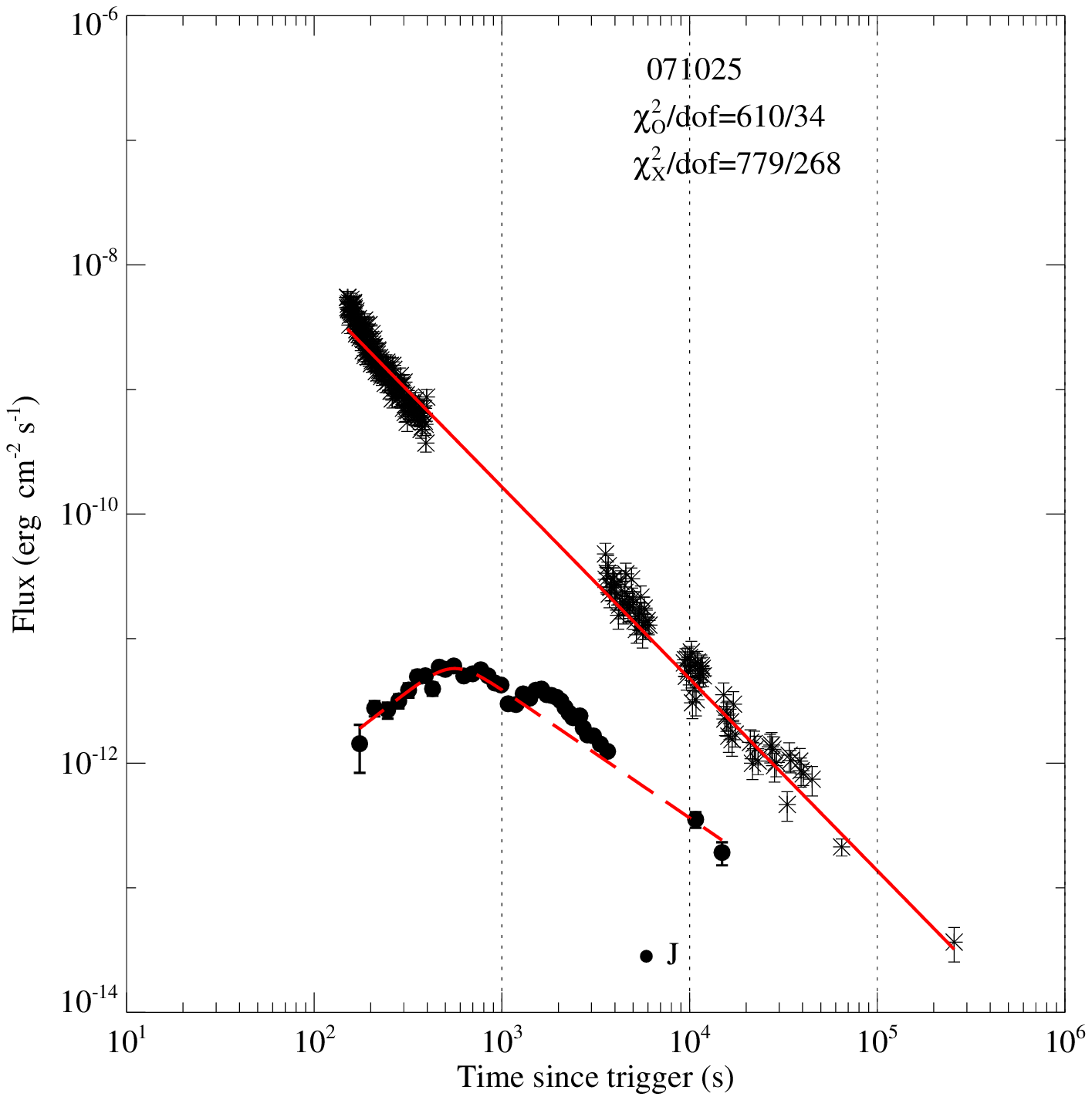}
\includegraphics[angle=0,scale=0.350,width=0.3\textwidth,height=0.25\textheight]{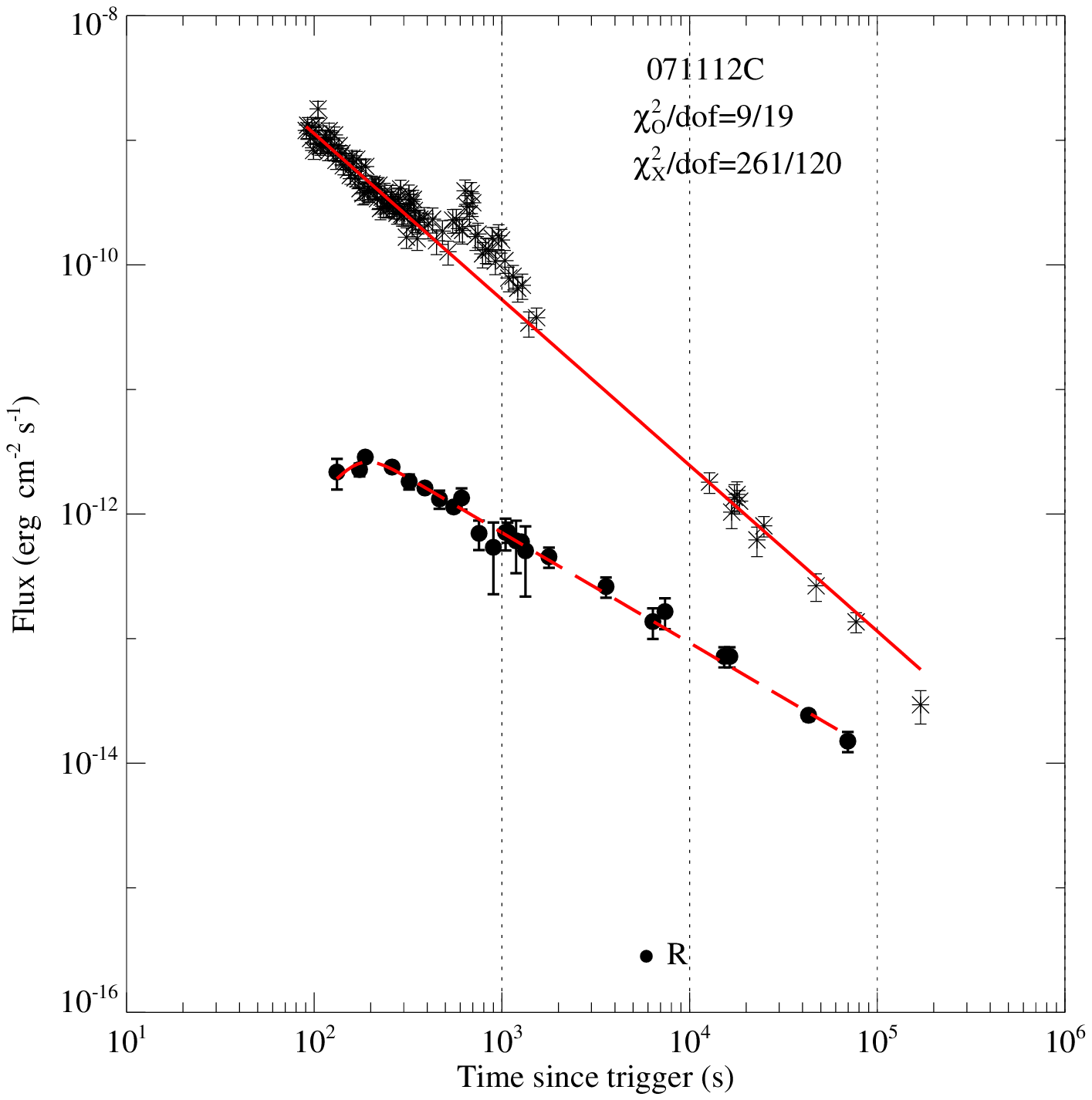}
\includegraphics[angle=0,scale=0.350,width=0.3\textwidth,height=0.25\textheight]{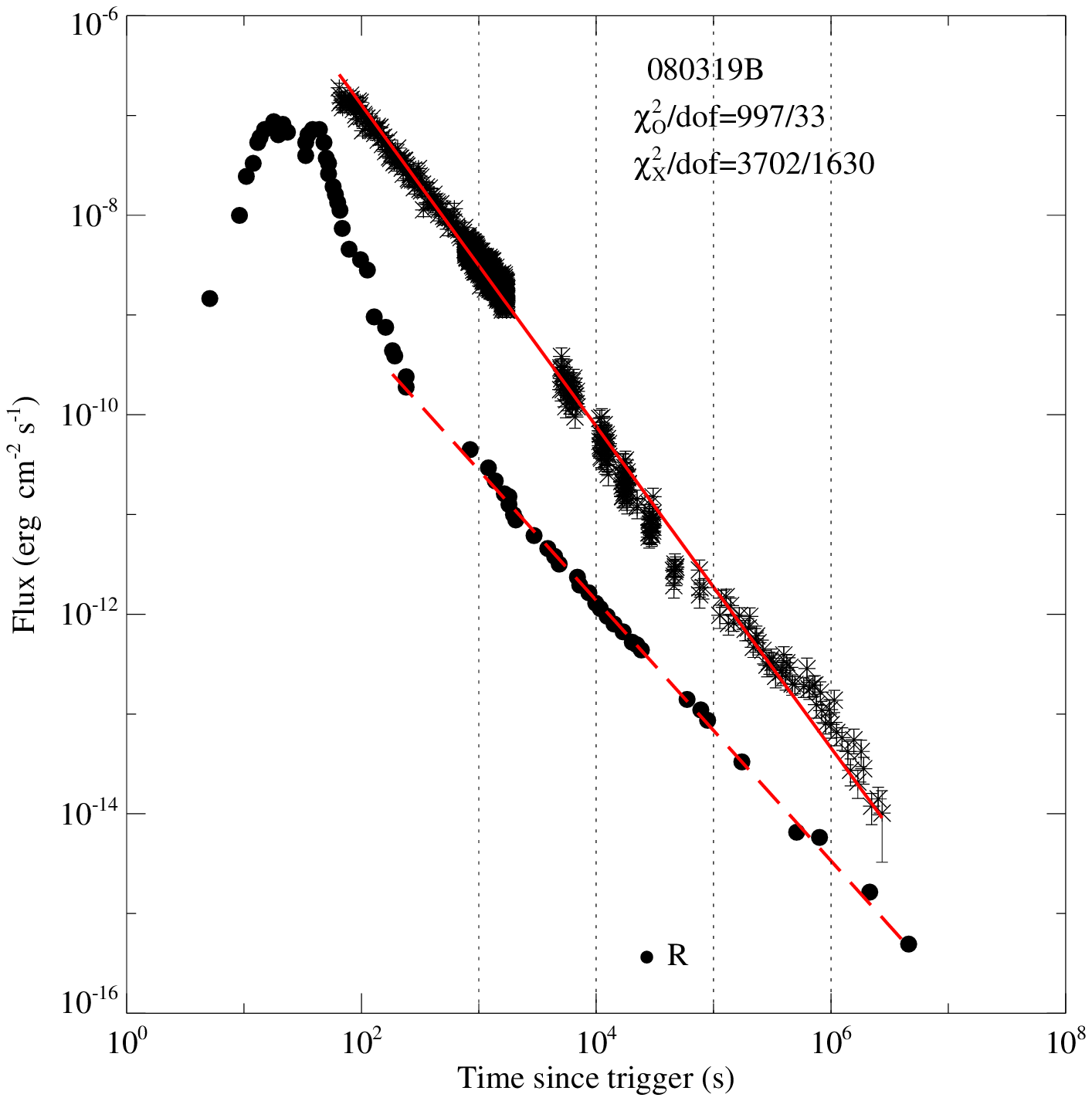}
\caption{The cases that did not have detected break behavior in neither the X-ray band nor optical band in their afterglow light-curves.} \label{Non Break}
\end{figure*}
\begin{figure*}
\includegraphics[angle=0,scale=0.350,width=0.3\textwidth,height=0.25\textheight]{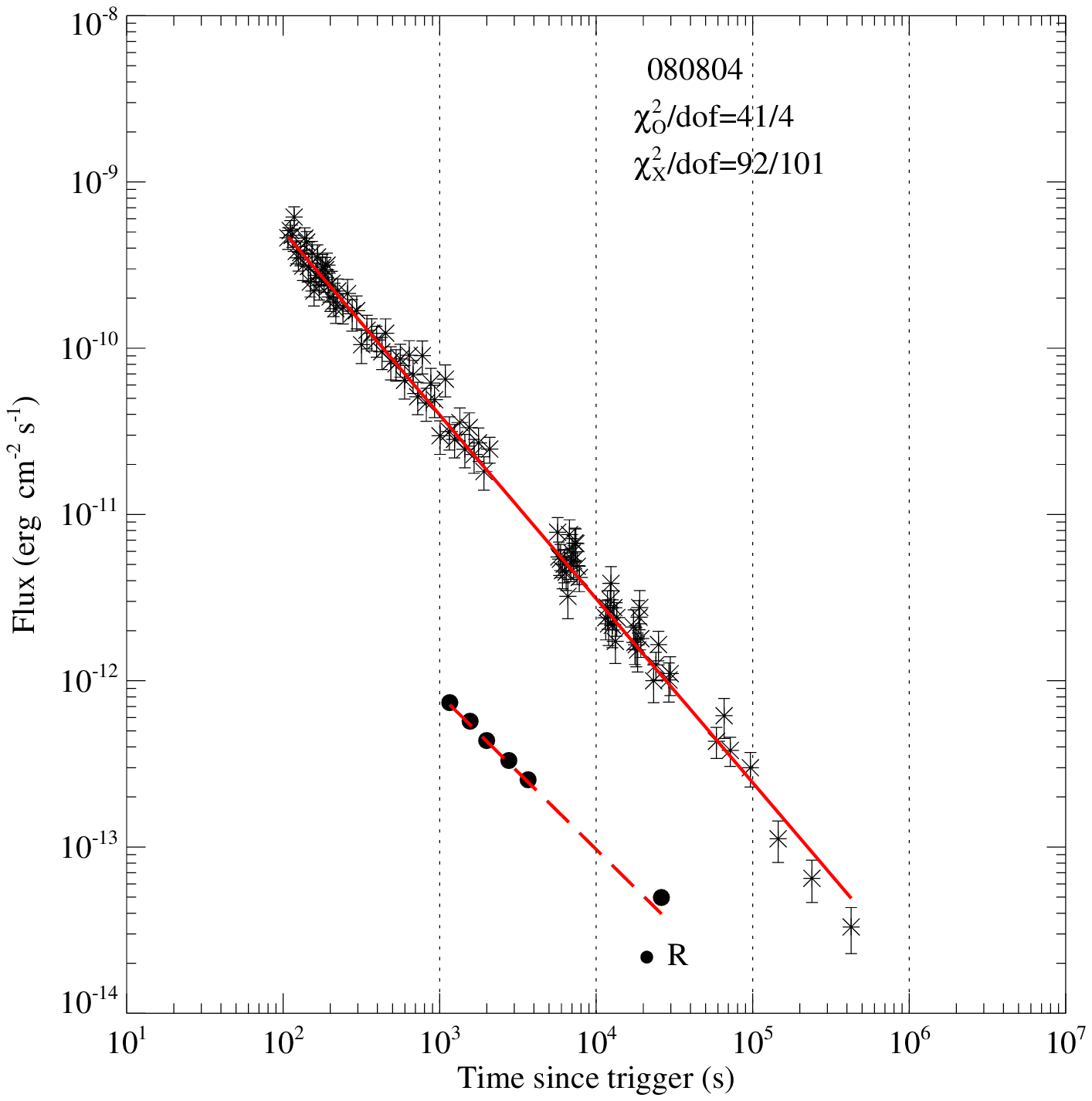}
\includegraphics[angle=0,scale=0.350,width=0.3\textwidth,height=0.25\textheight]{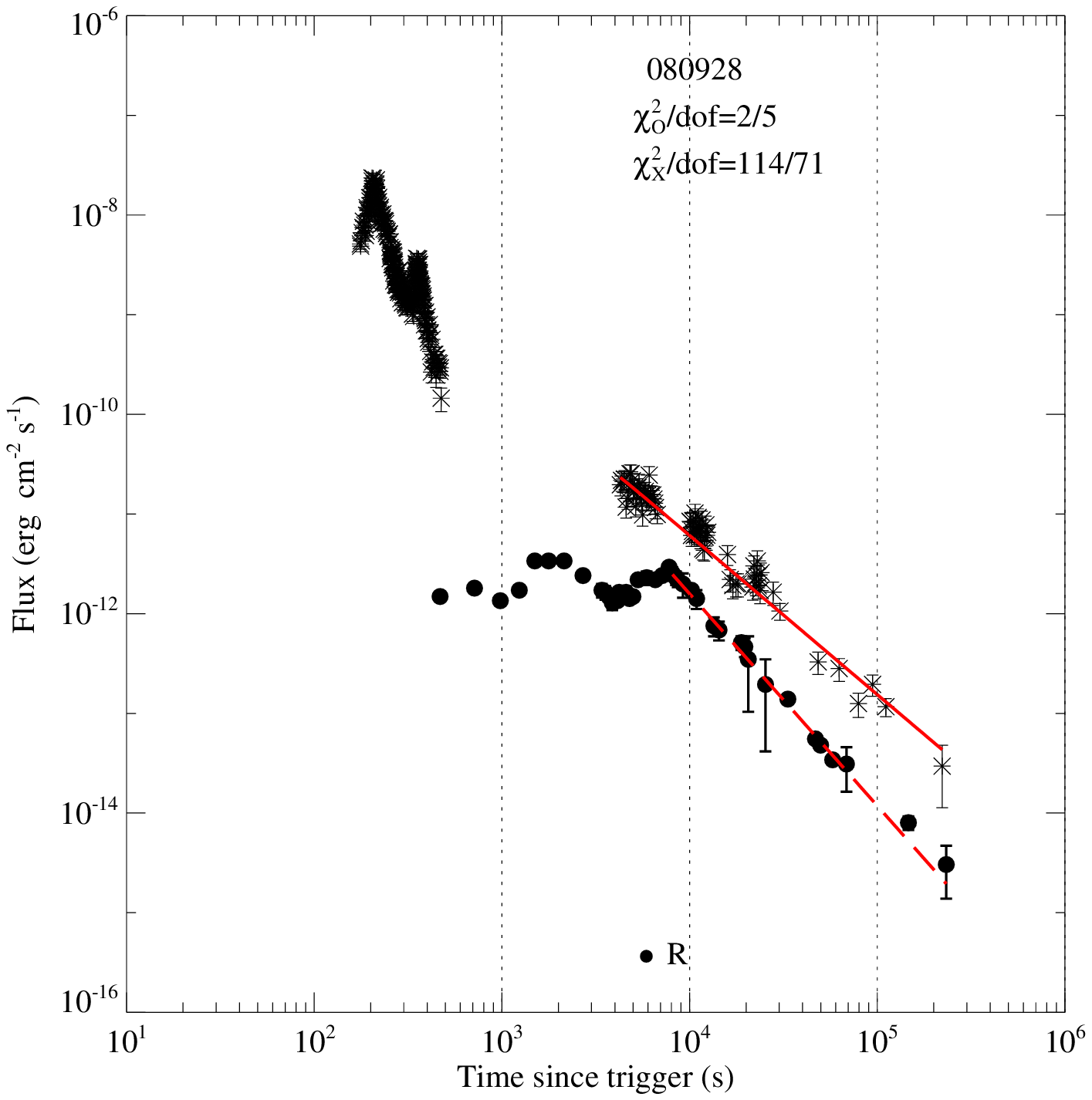}
\includegraphics[angle=0,scale=0.350,width=0.3\textwidth,height=0.25\textheight]{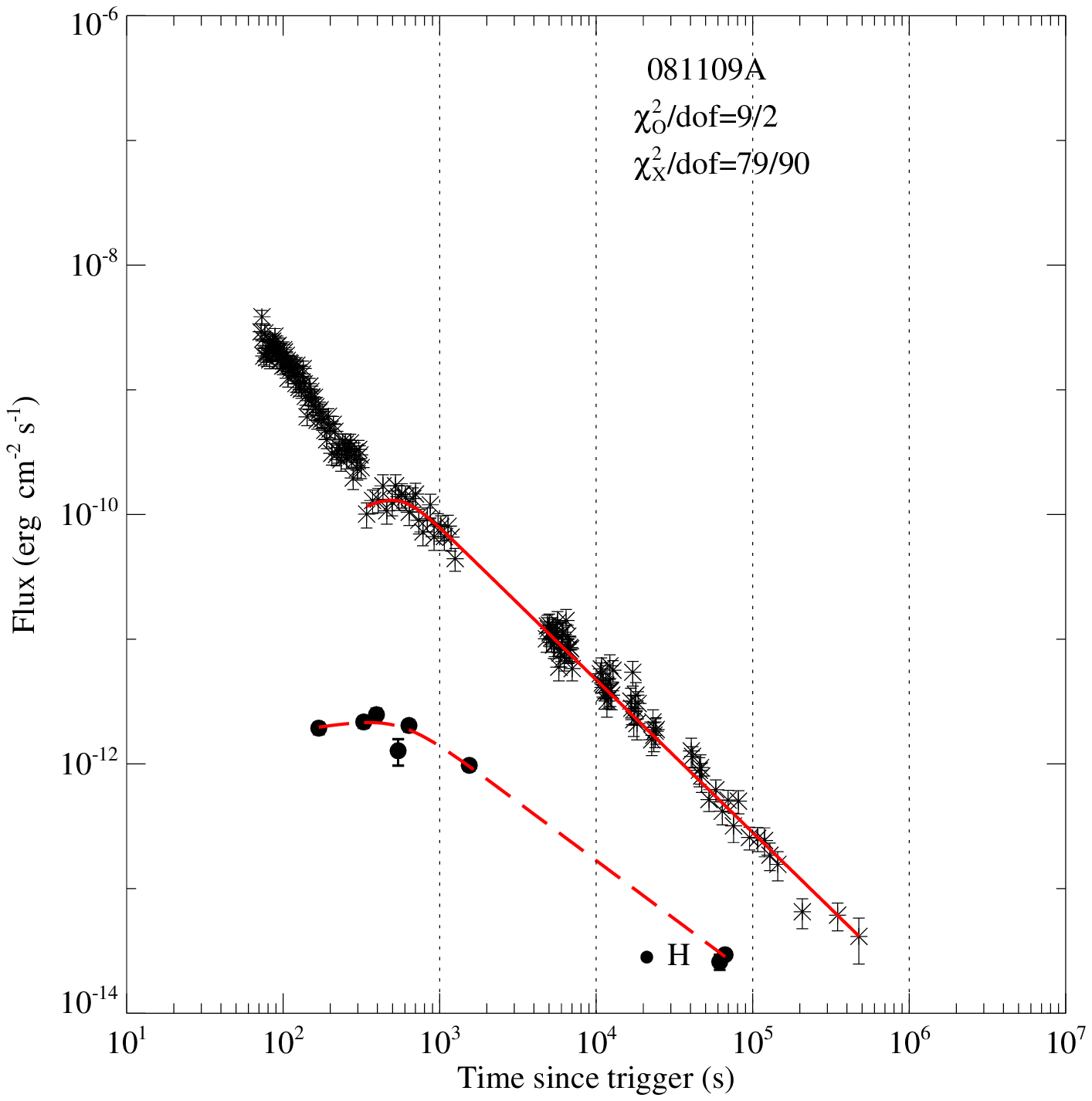}
\includegraphics[angle=0,scale=0.350,width=0.3\textwidth,height=0.25\textheight]{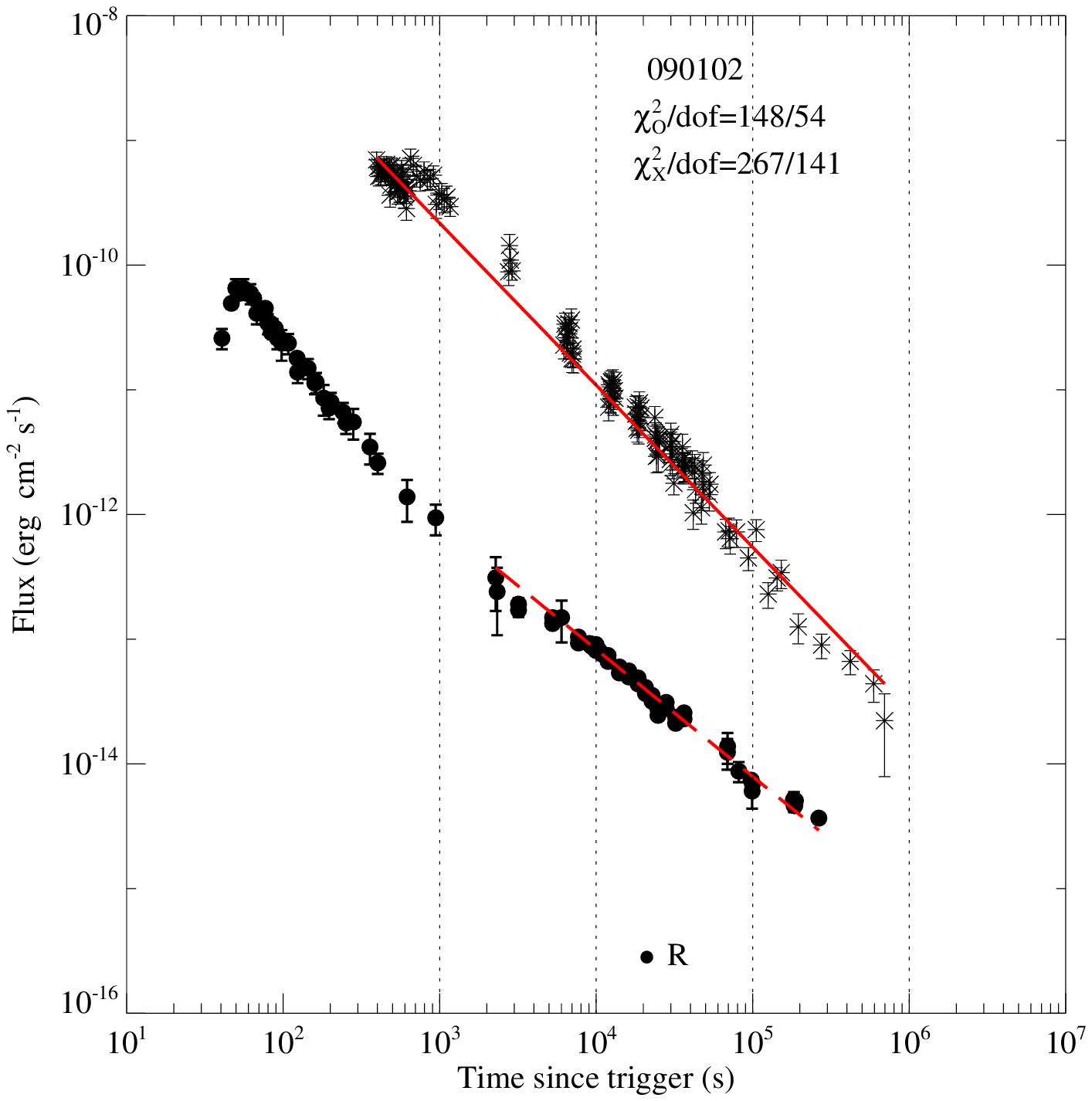}
\includegraphics[angle=0,scale=0.350,width=0.3\textwidth,height=0.25\textheight]{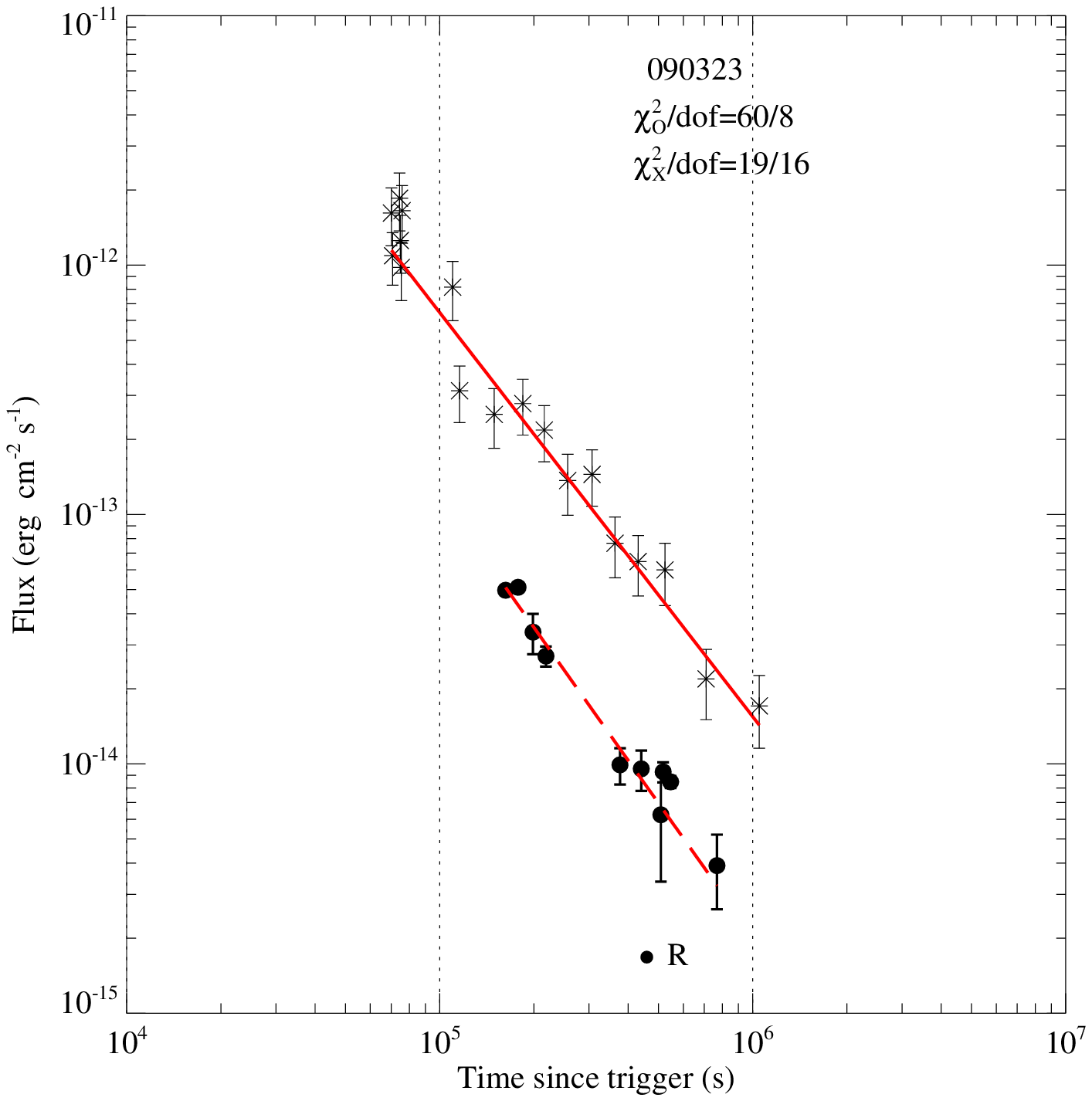}
\includegraphics[angle=0,scale=0.350,width=0.3\textwidth,height=0.25\textheight]{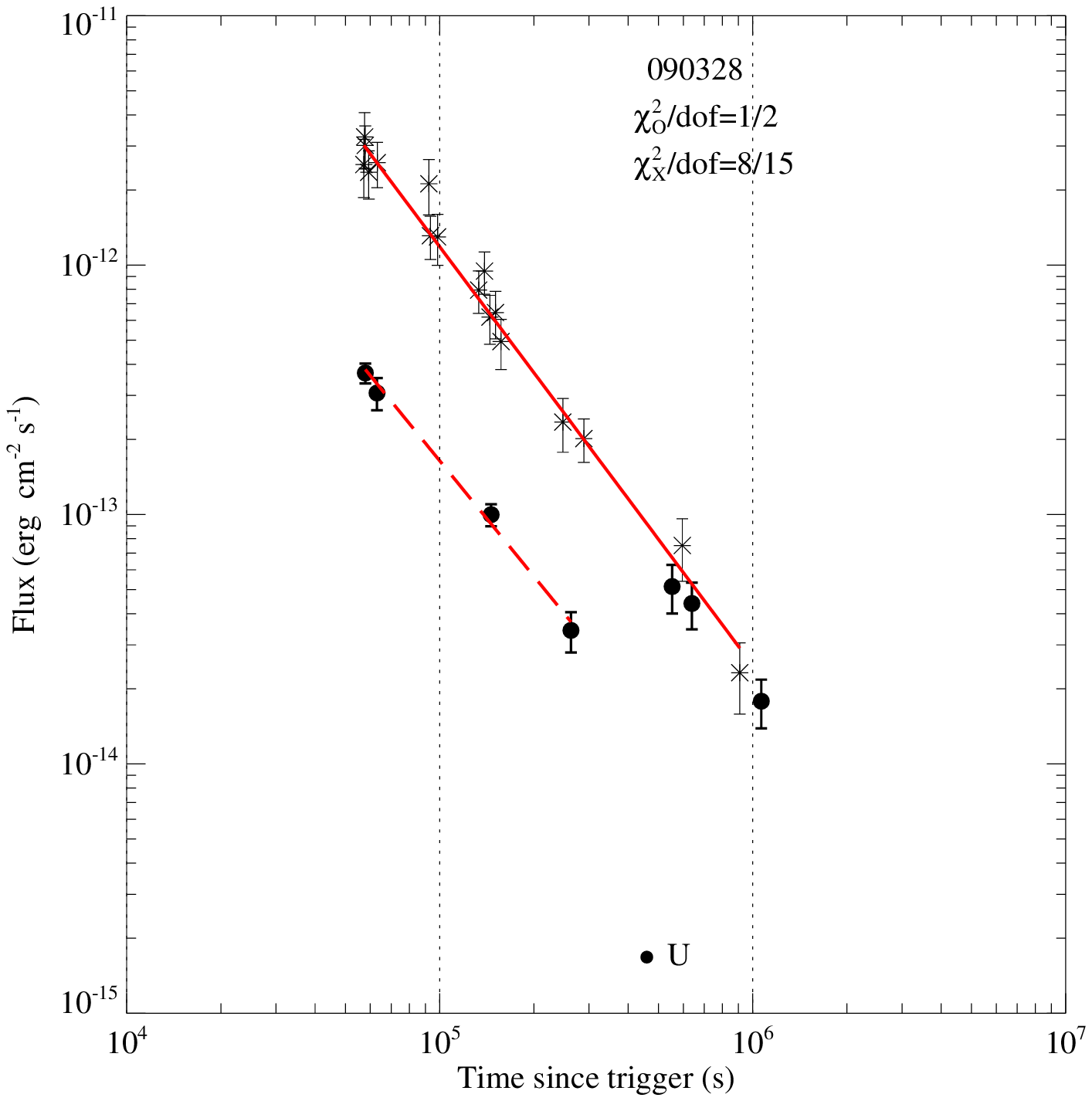}
\includegraphics[angle=0,scale=0.350,width=0.3\textwidth,height=0.25\textheight]{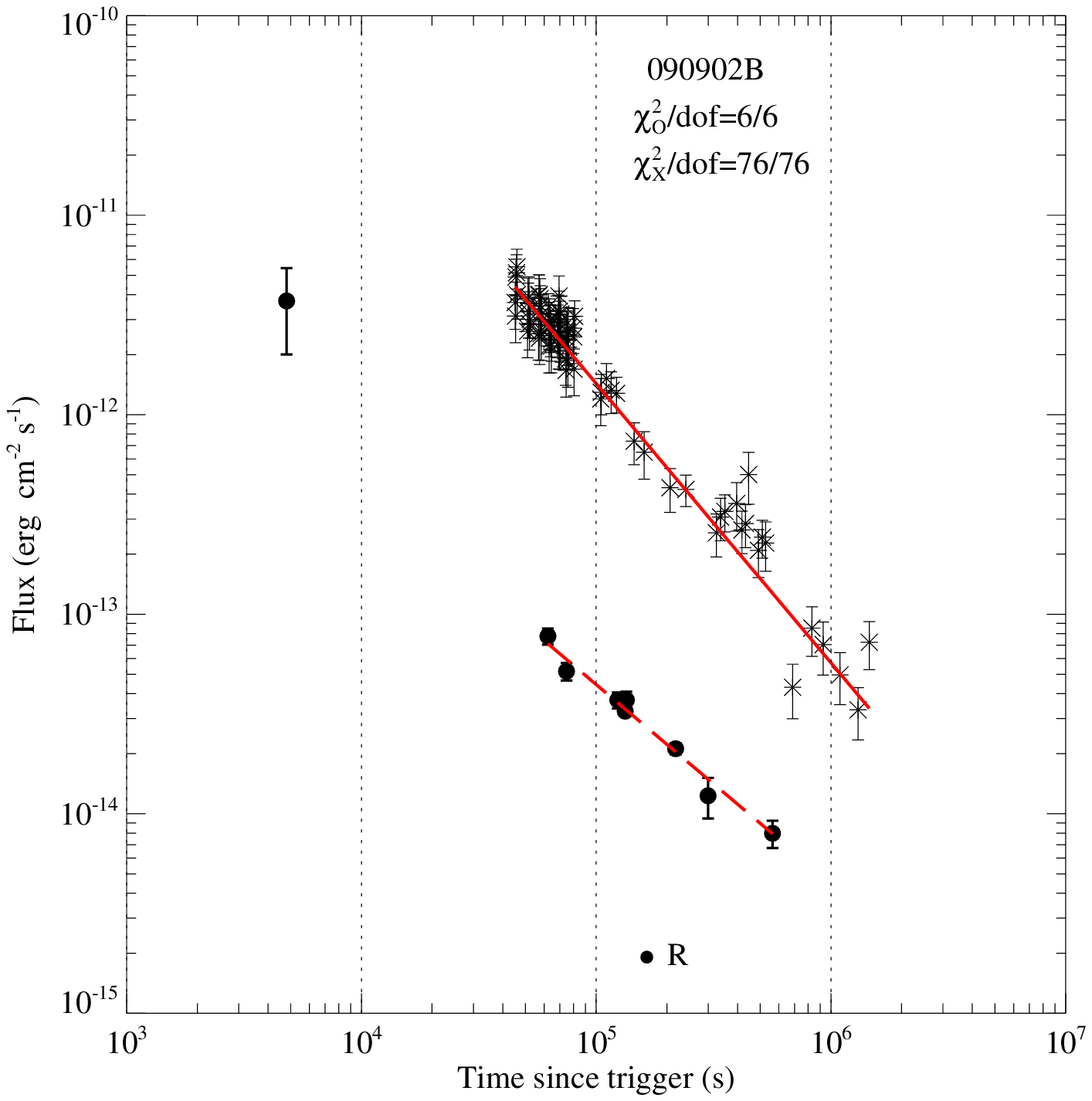}
\includegraphics[angle=0,scale=0.350,width=0.3\textwidth,height=0.25\textheight]{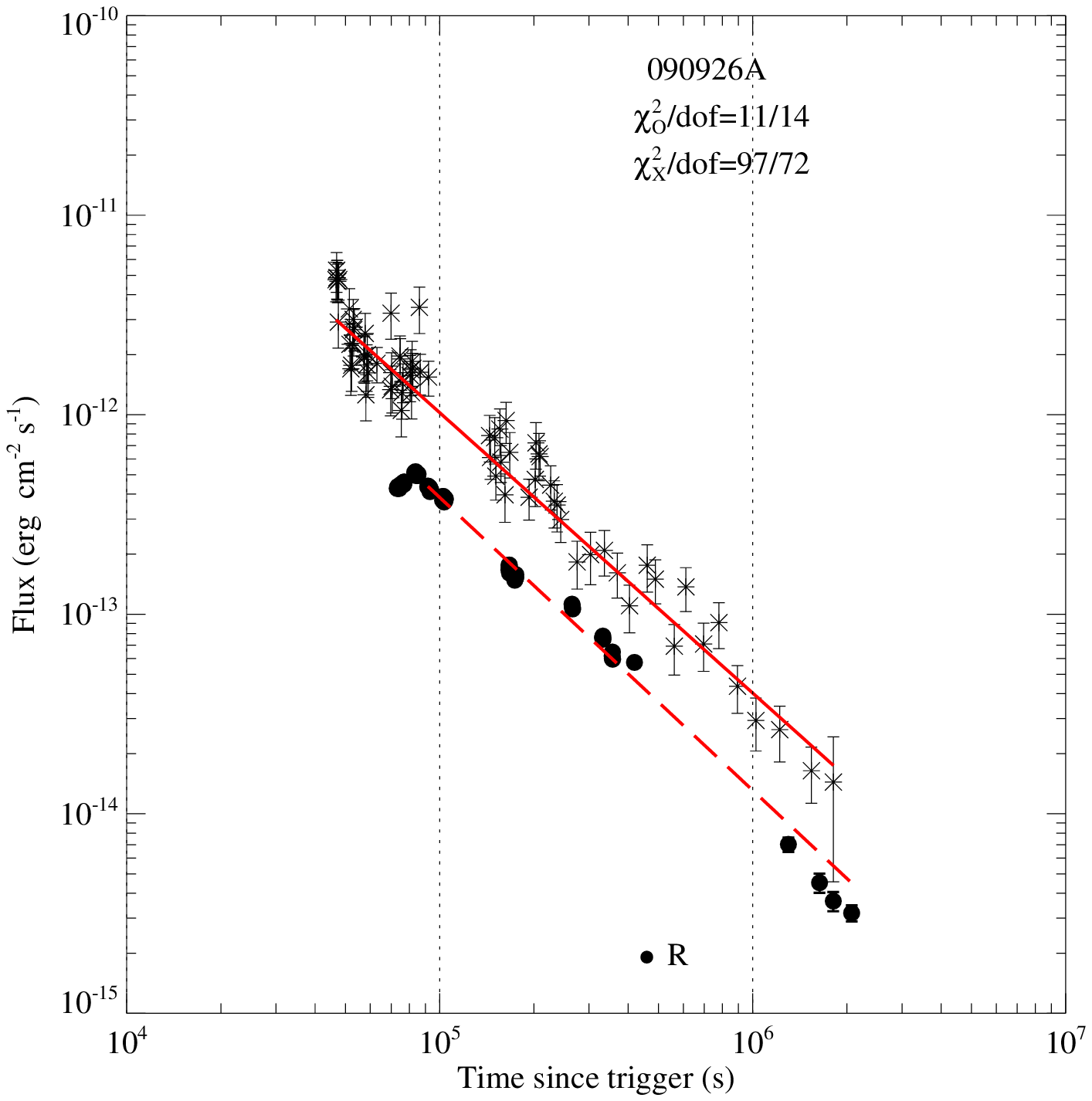}
\includegraphics[angle=0,scale=0.350,width=0.3\textwidth,height=0.25\textheight]{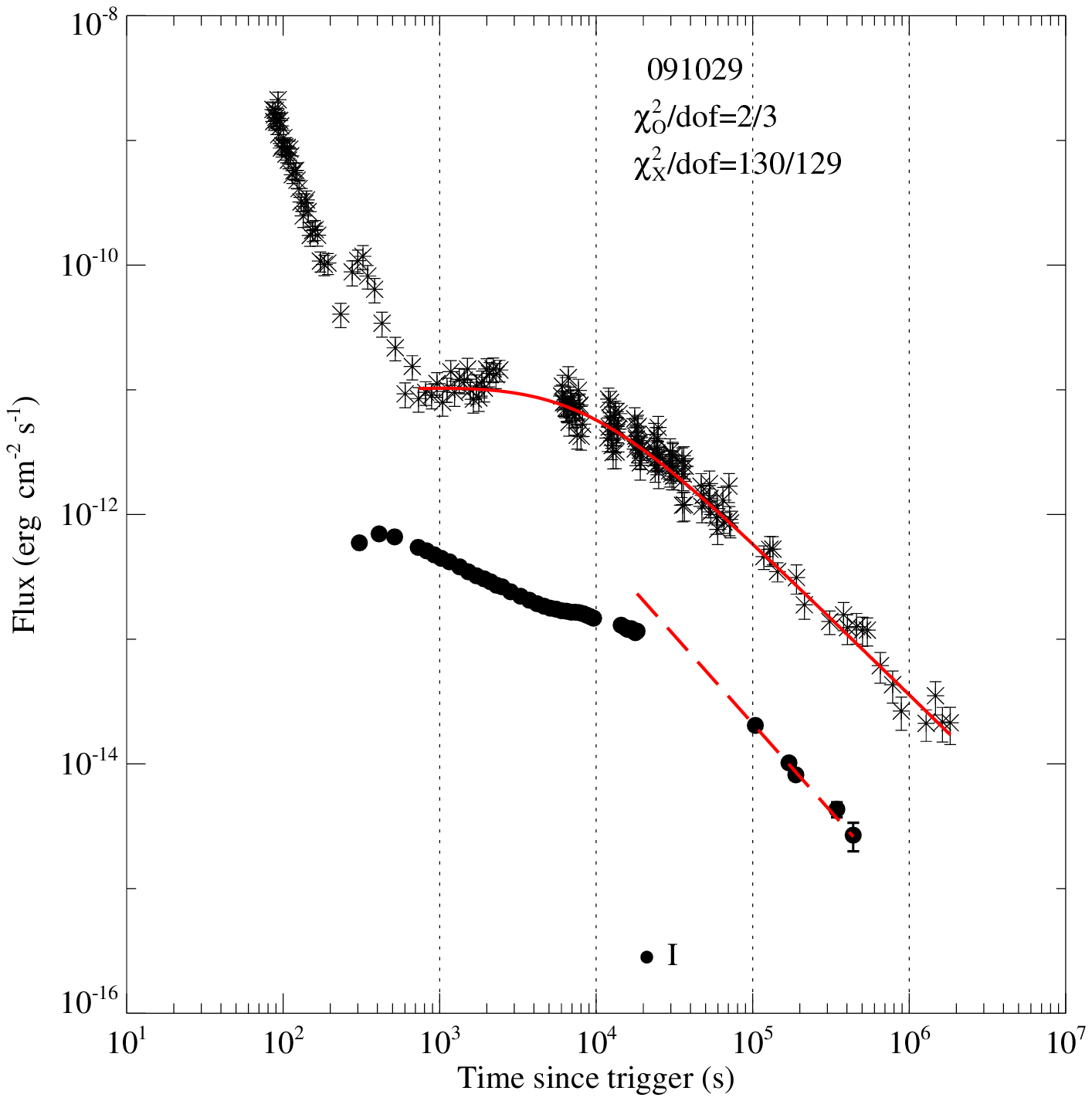}
\includegraphics[angle=0,scale=0.350,width=0.3\textwidth,height=0.25\textheight]{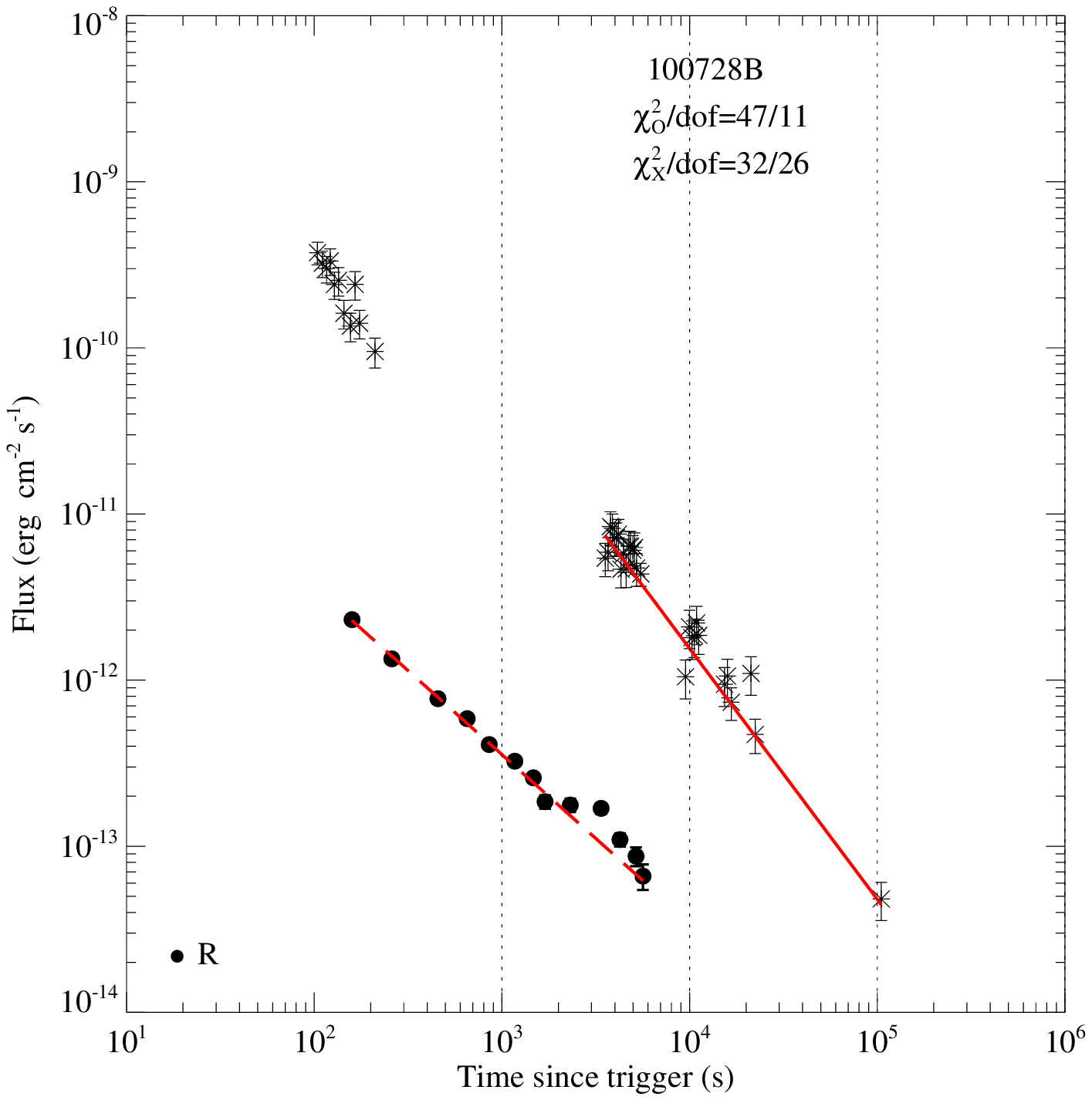}
\includegraphics[angle=0,scale=0.350,width=0.3\textwidth,height=0.25\textheight]{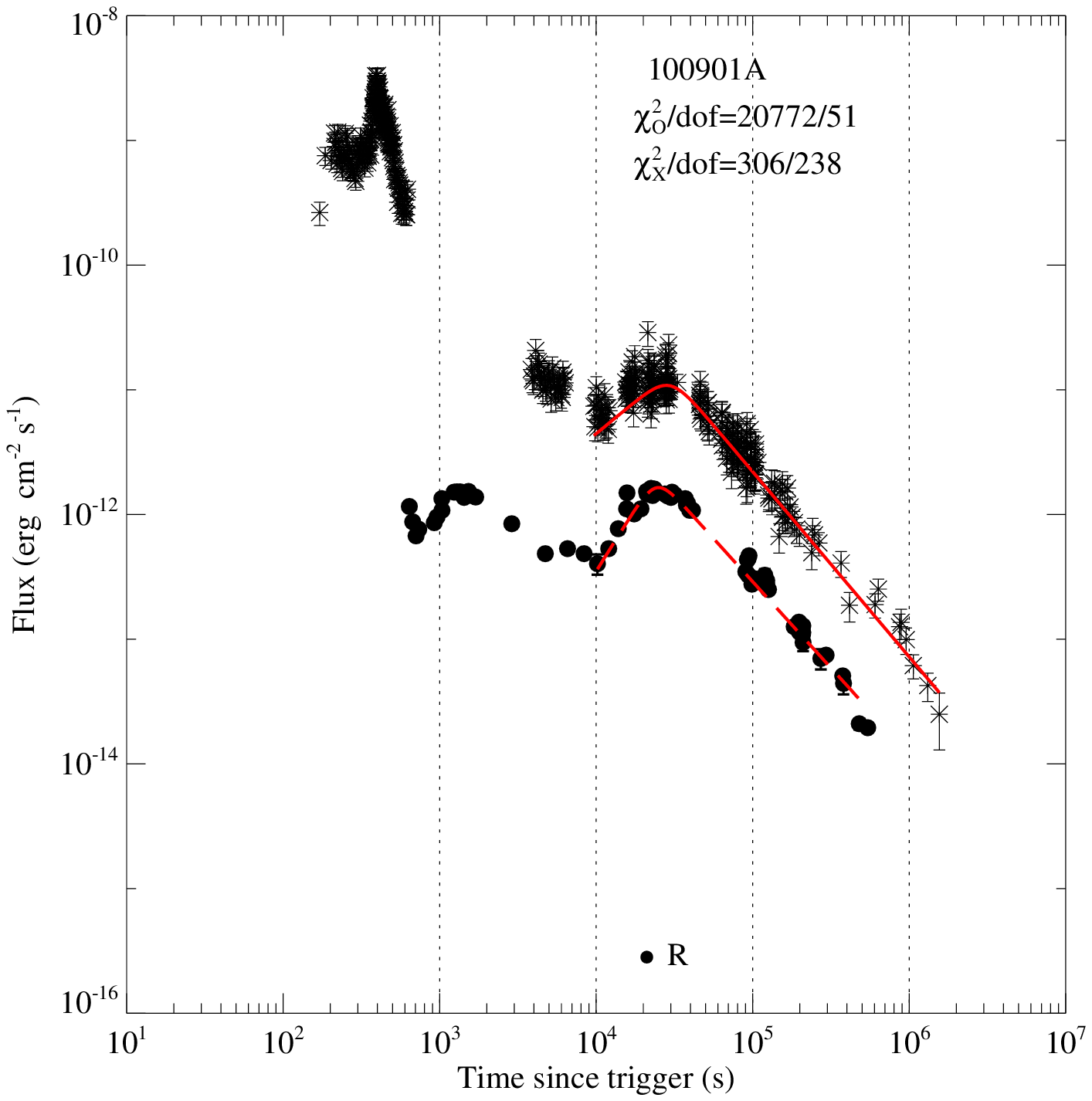}
\includegraphics[angle=0,scale=0.350,width=0.3\textwidth,height=0.25\textheight]{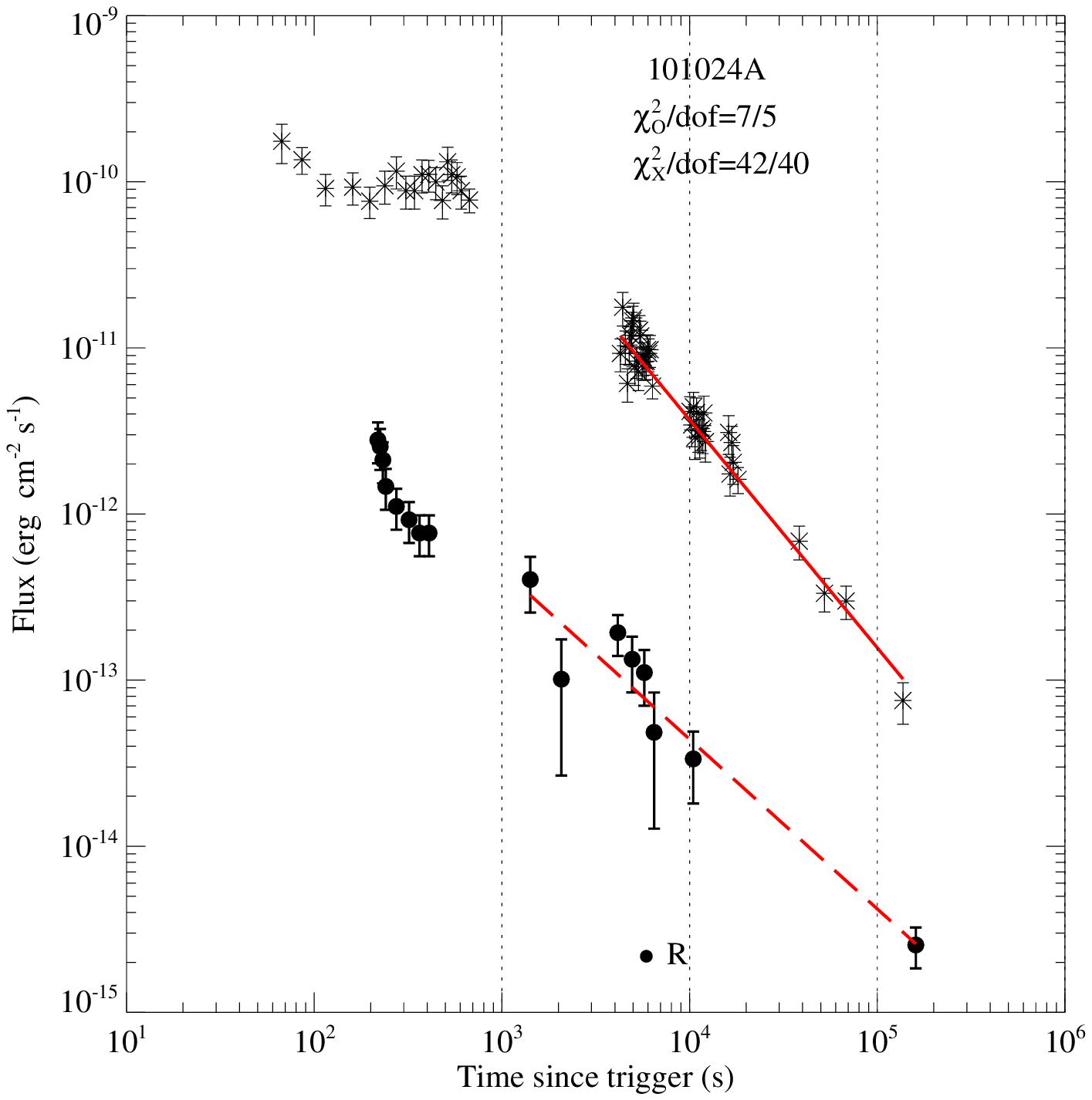}
\center{Fig. \ref{Non Break}--- Continued}
\end{figure*}
\begin{figure*}
\includegraphics[angle=0,scale=0.350,width=0.3\textwidth,height=0.25\textheight]{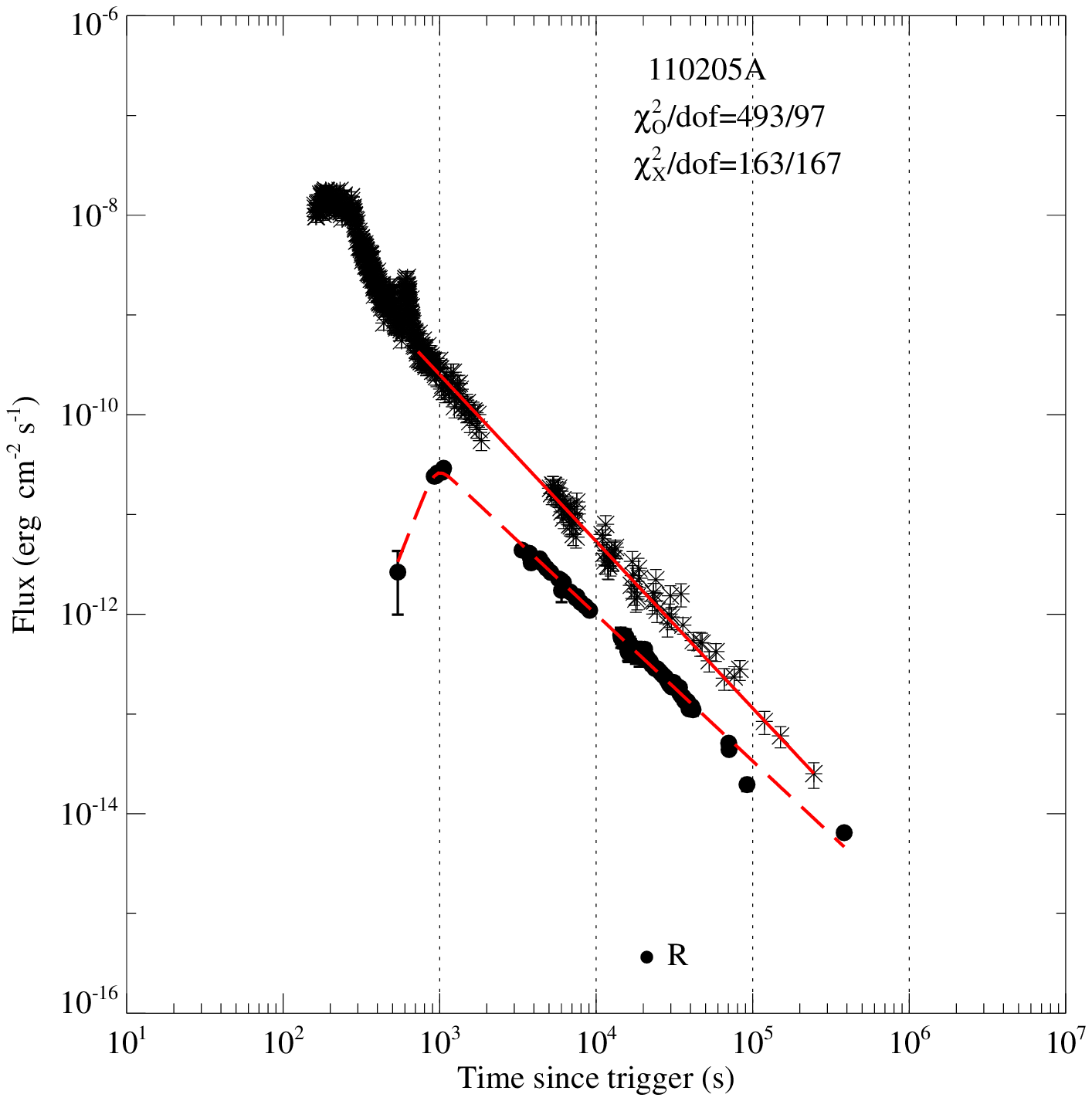}
\includegraphics[angle=0,scale=0.350,width=0.3\textwidth,height=0.25\textheight]{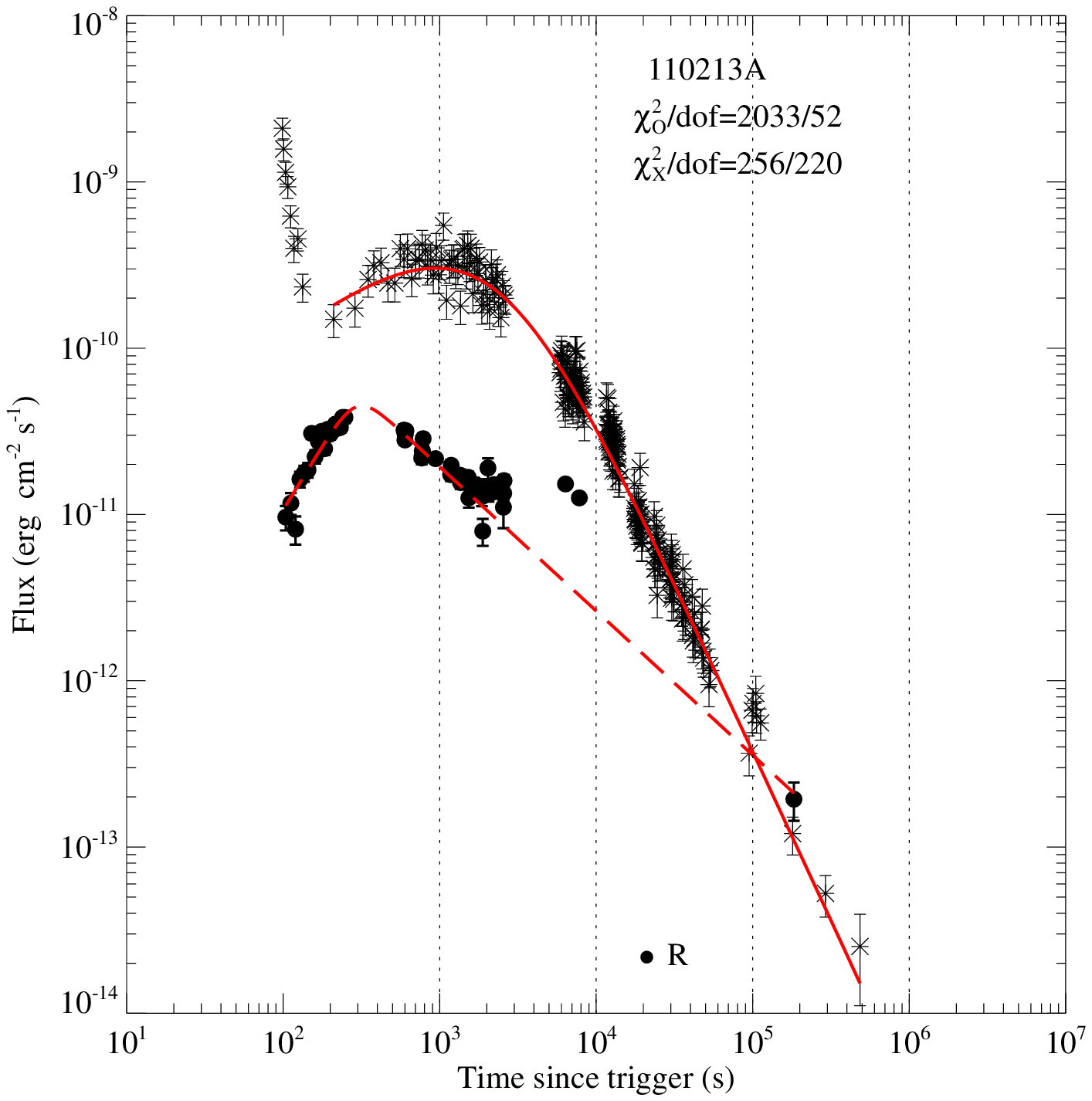}
\includegraphics[angle=0,scale=0.350,width=0.3\textwidth,height=0.25\textheight]{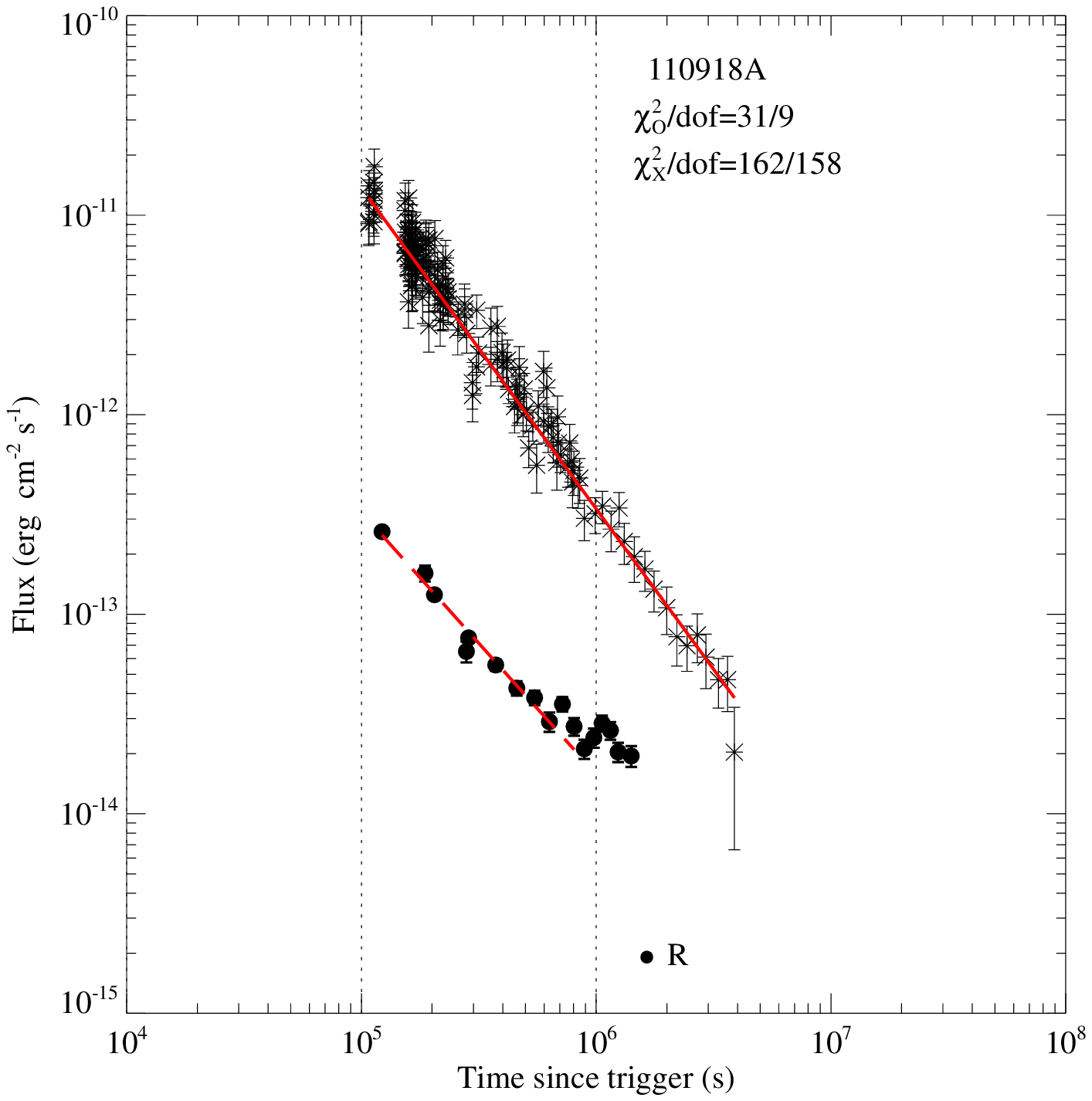}
\includegraphics[angle=0,scale=0.350,width=0.3\textwidth,height=0.25\textheight]{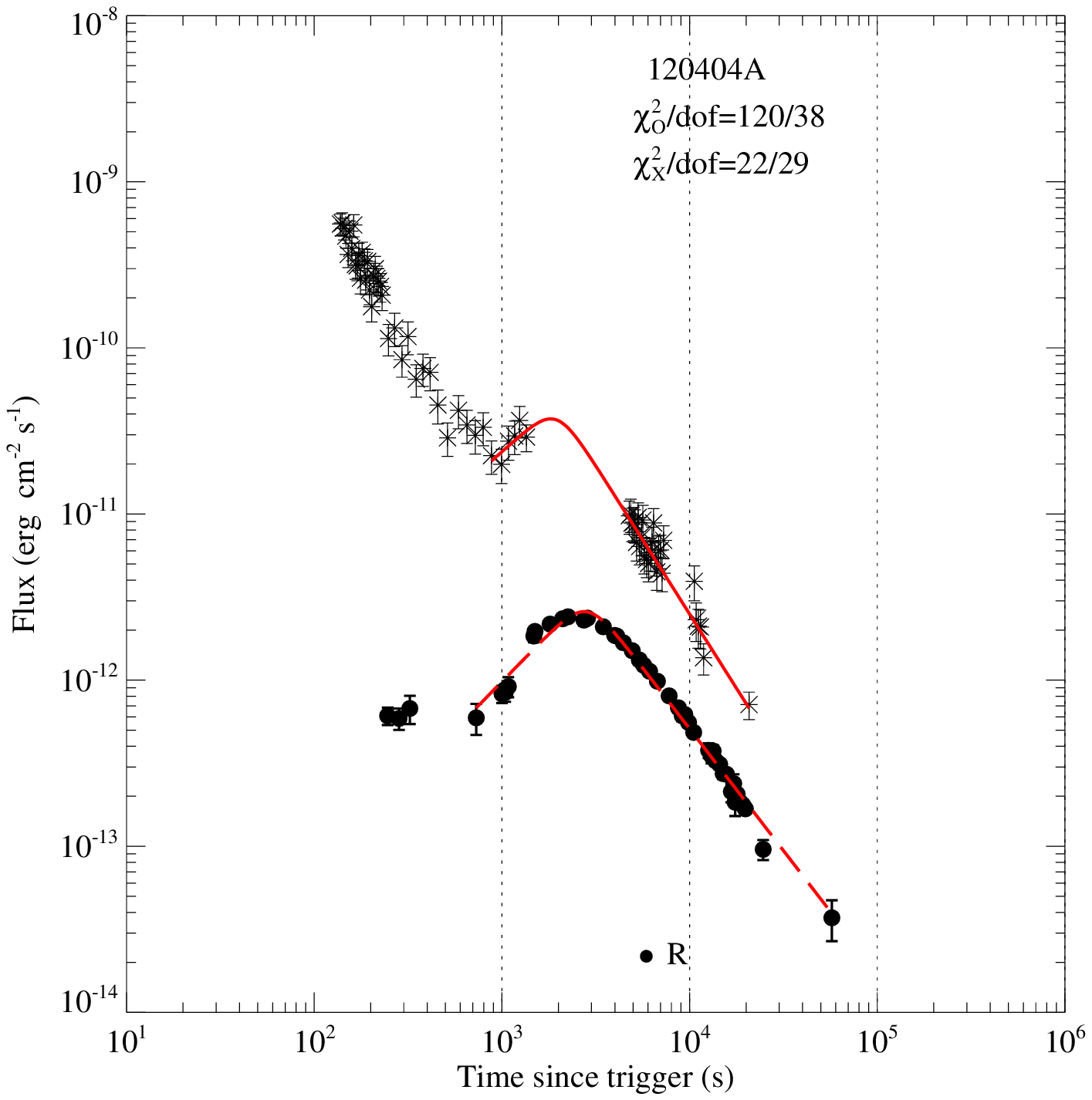}
\includegraphics[angle=0,scale=0.350,width=0.3\textwidth,height=0.25\textheight]{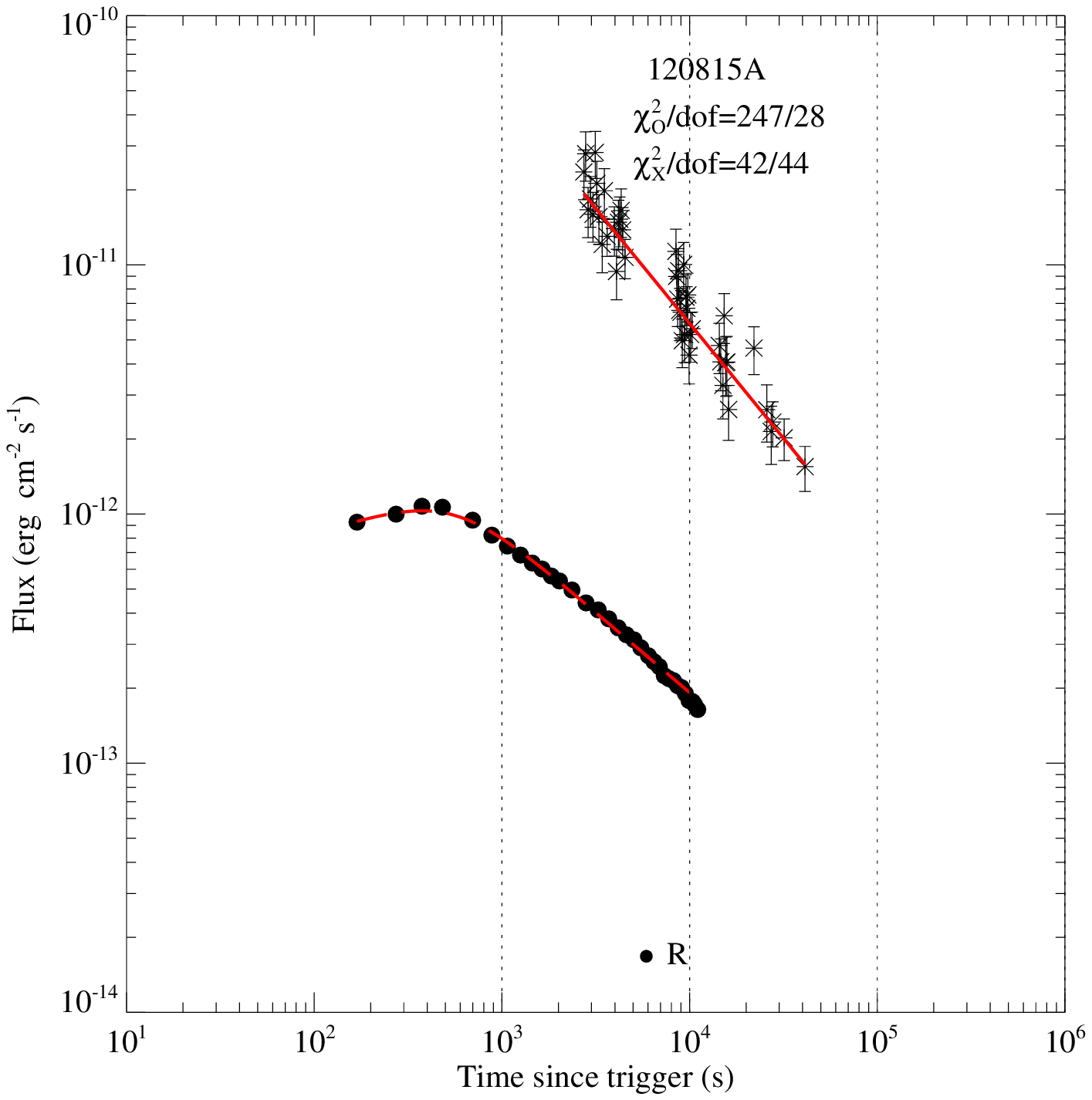}
\includegraphics[angle=0,scale=0.350,width=0.3\textwidth,height=0.25\textheight]{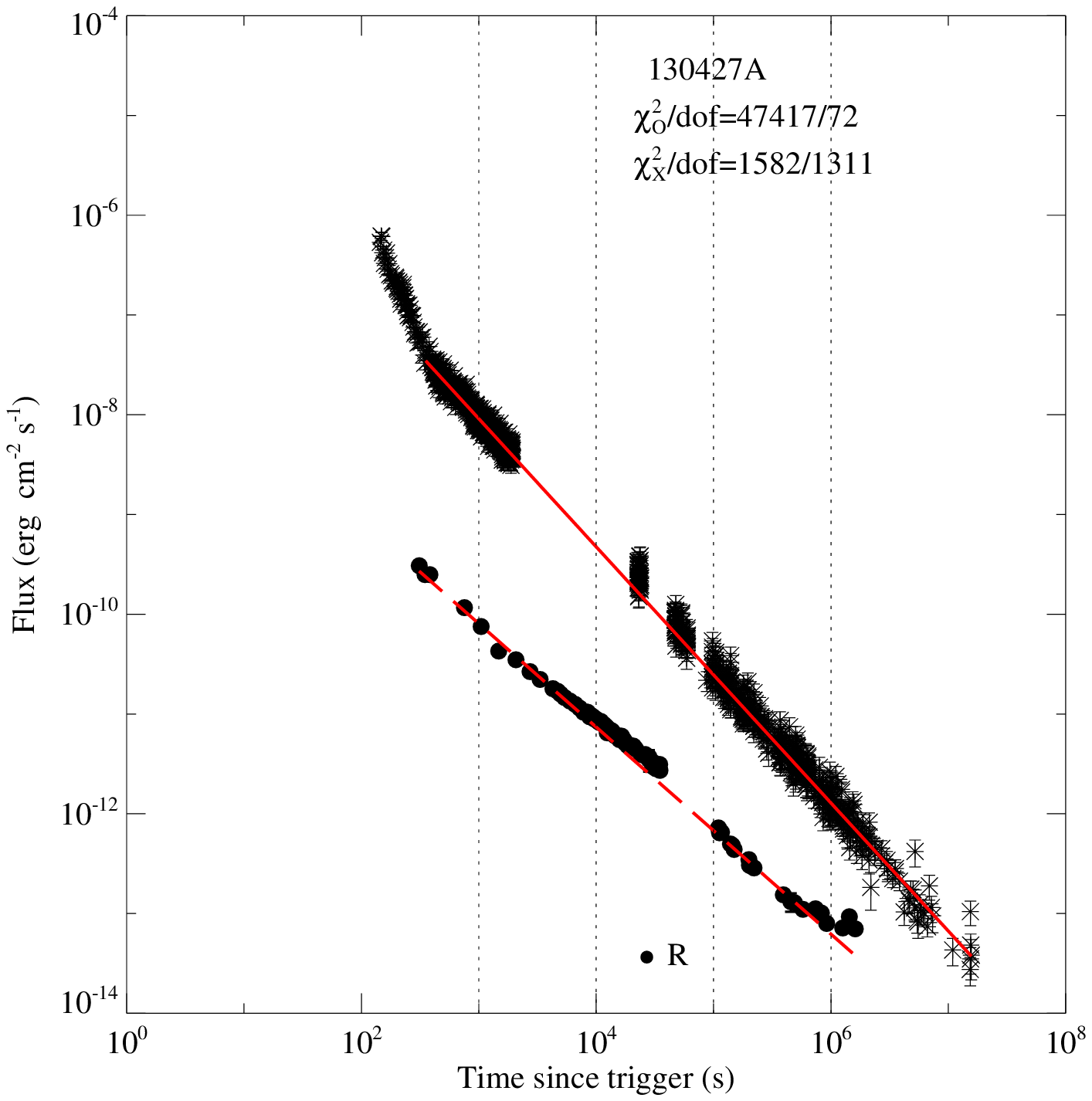}
\center{Fig. \ref{Non Break}--- Continued}
\end{figure*}

\begin{figure*}
\includegraphics[angle=0,scale=0.350,width=0.3\textwidth,height=0.25\textheight]{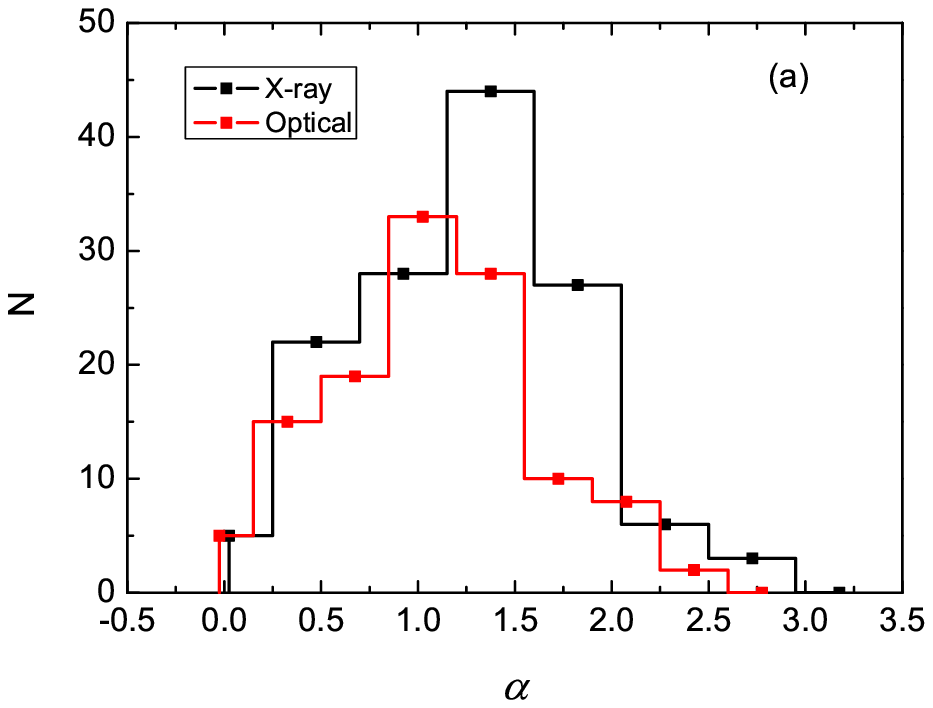}
\includegraphics[angle=0,scale=0.350,width=0.3\textwidth,height=0.25\textheight]{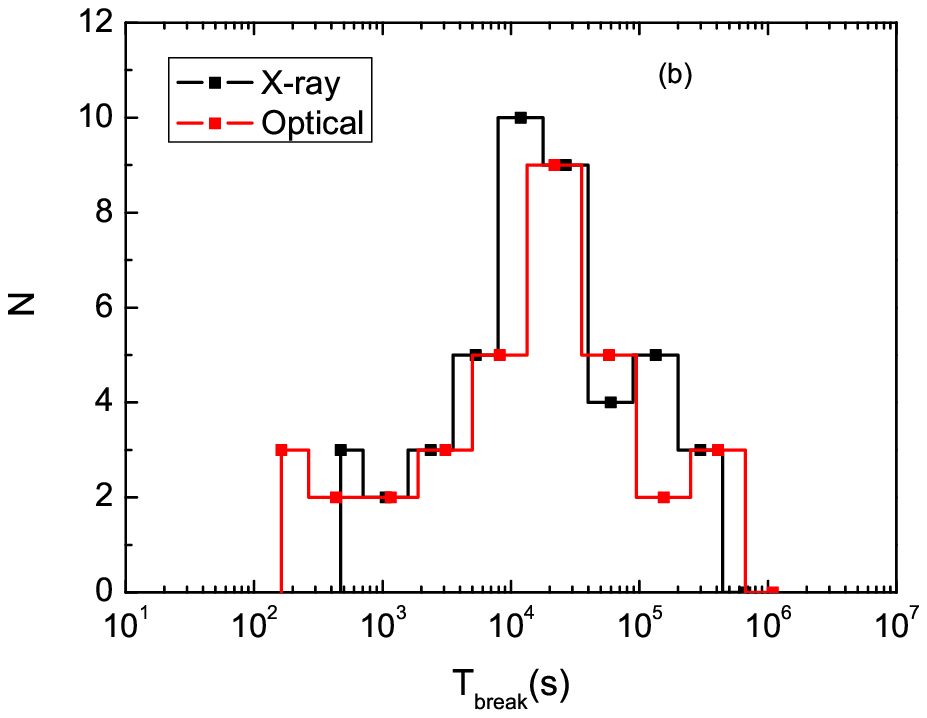}\\
\includegraphics[angle=0,scale=0.350,width=0.3\textwidth,height=0.25\textheight]{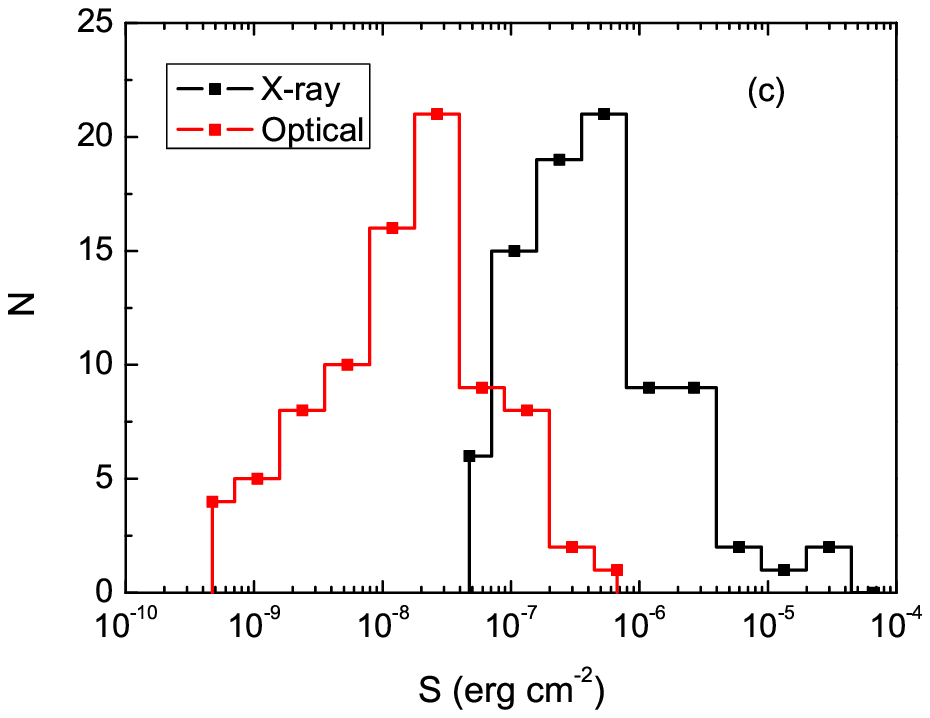}
\includegraphics[angle=0,scale=0.350,width=0.3\textwidth,height=0.25\textheight]{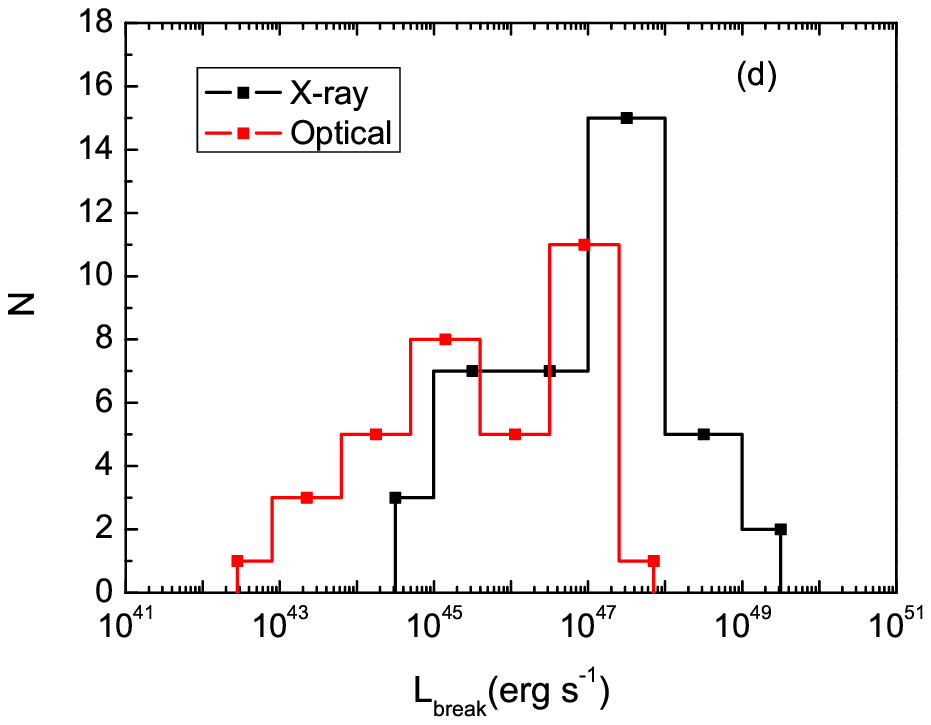}
\includegraphics[angle=0,scale=0.350,width=0.3\textwidth,height=0.25\textheight]{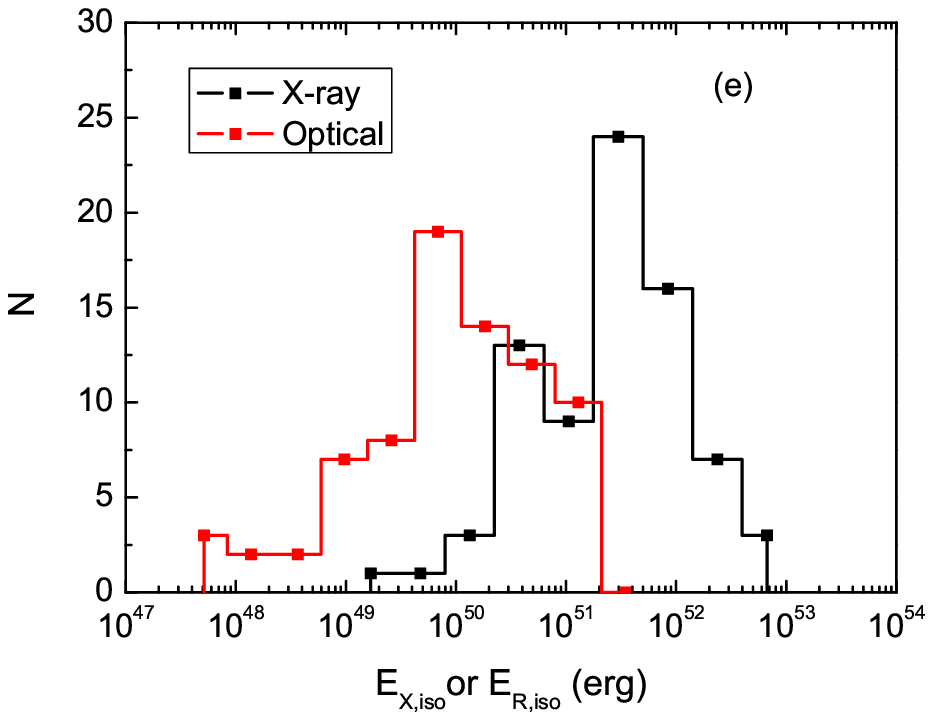}
\caption{The distributions of properties for our cases comparing the X-ray with optical emission, including the temporal index, the break time and break luminosity (for the break cases), and fluence and isotropic energy release.} \label{Distributions}
\end{figure*}

\begin{figure*}
\includegraphics[angle=0,scale=0.350,width=0.3\textwidth,height=0.25\textheight]{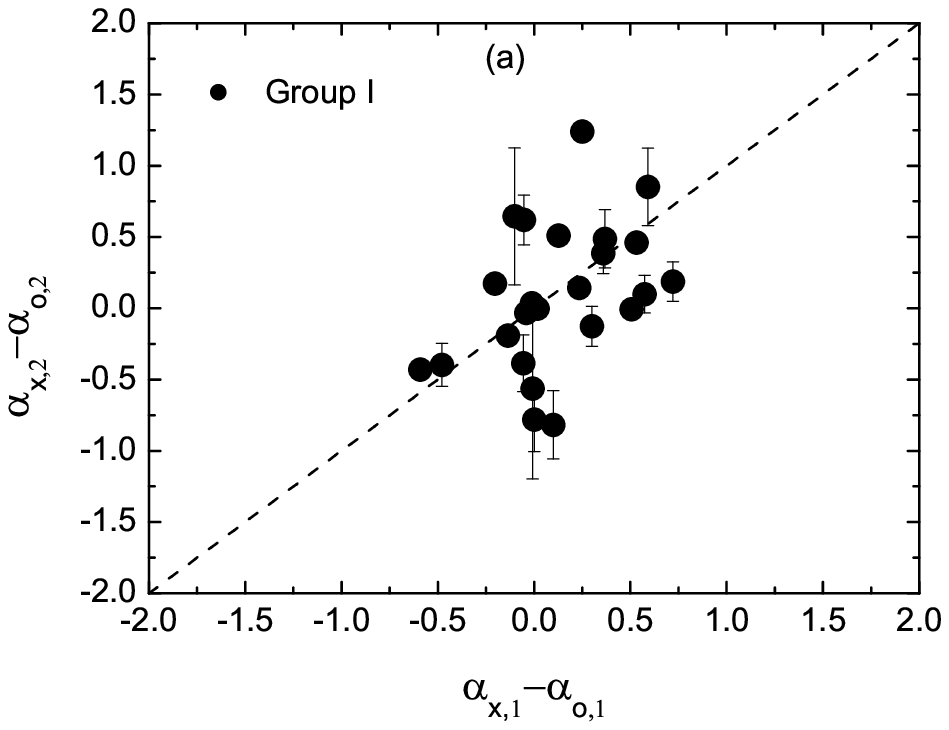}
\includegraphics[angle=0,scale=0.350,width=0.3\textwidth,height=0.25\textheight]{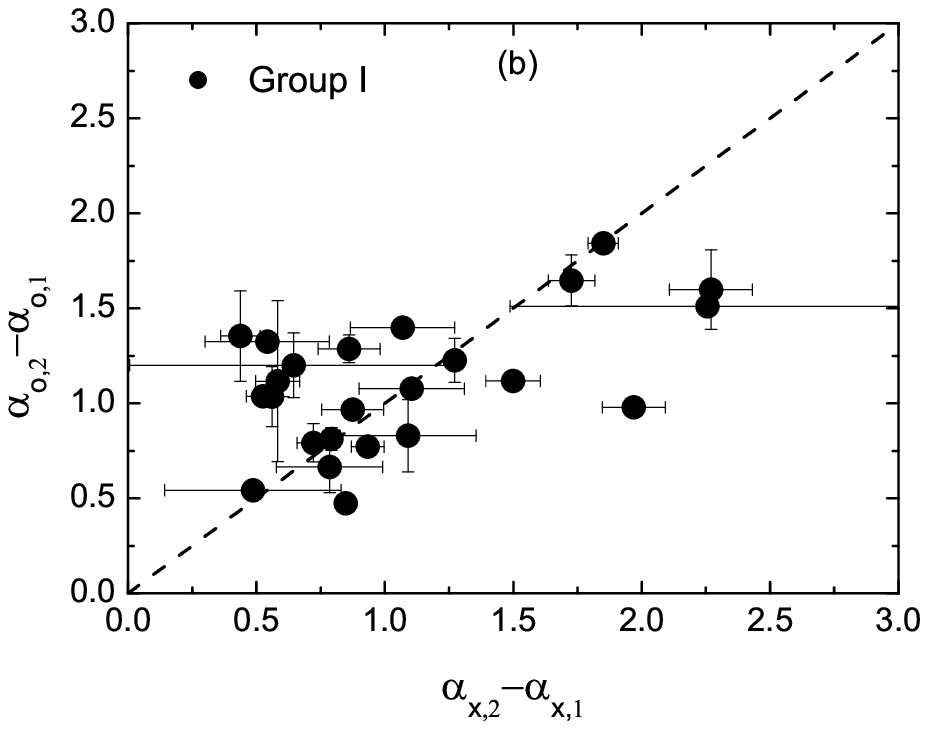}
\includegraphics[angle=0,scale=0.350,width=0.3\textwidth,height=0.25\textheight]{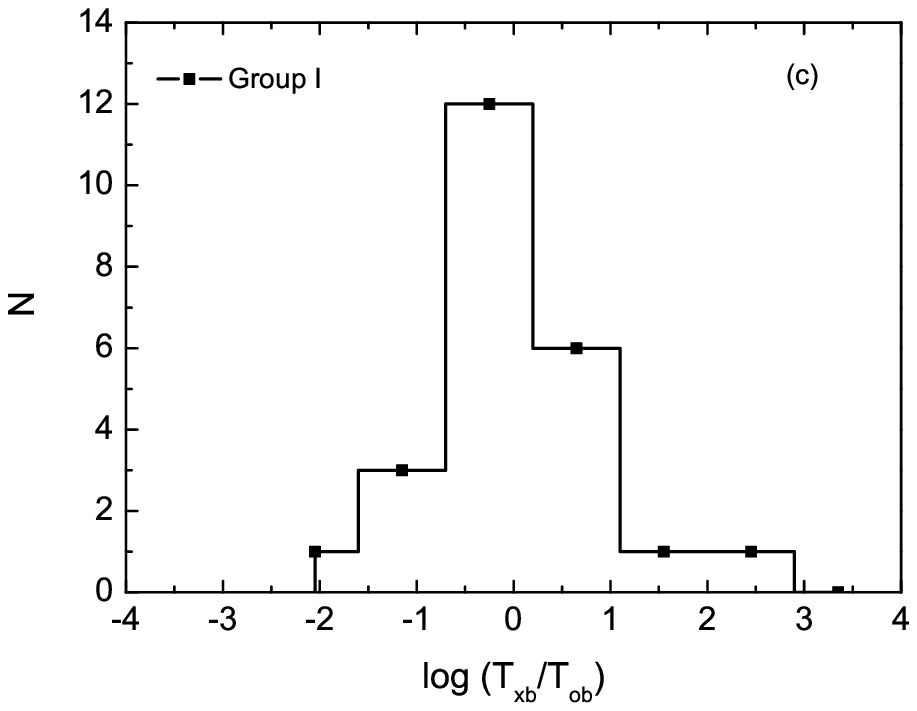}
\caption{Pair correlations comparing the X-ray with optical emission for Group I cases, including the change of X-ray and
optical pre- and post-break decay slopes and break time.} \label{Correlations}
\end{figure*}

\begin{figure*}
\includegraphics[angle=0,scale=0.350,width=0.3\textwidth,height=0.25\textheight]{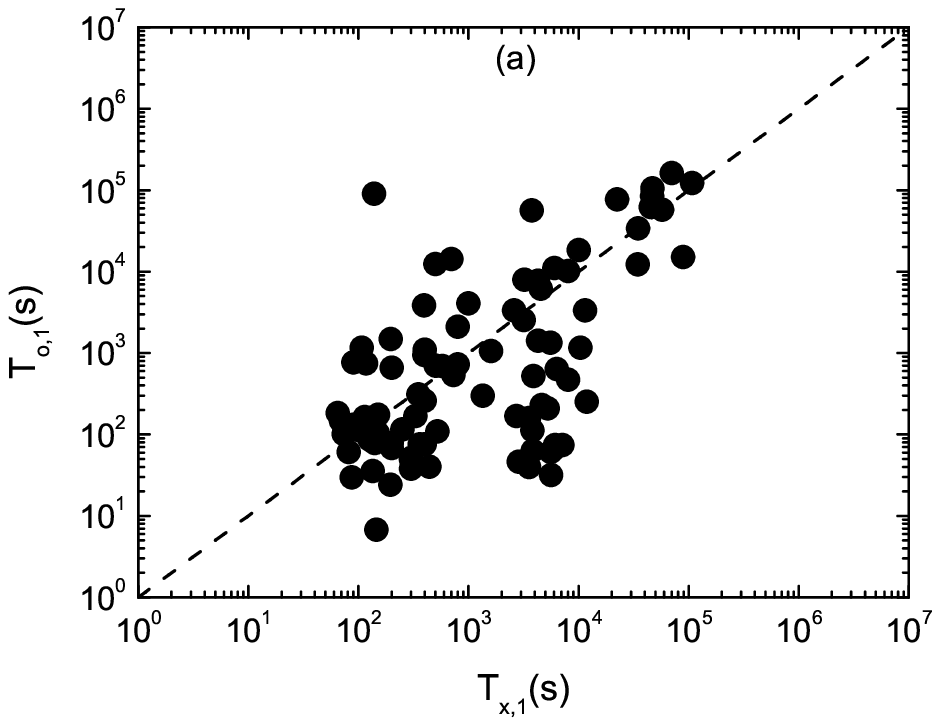}
\includegraphics[angle=0,scale=0.350,width=0.3\textwidth,height=0.25\textheight]{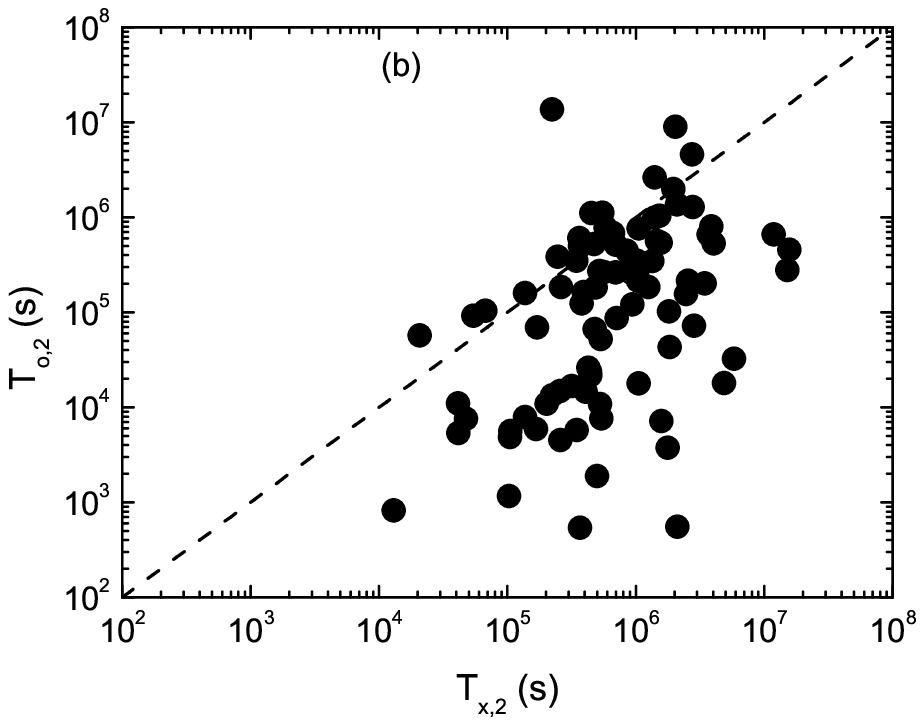}
\caption{The time interval of the optical detection as a function of the time interval of the X-ray detection for all afterglow cases. The dashed lines are the equal lines.}\label{Times}
\end{figure*}

\begin{figure*}
\includegraphics[angle=0,scale=0.350,width=0.45\textwidth,height=0.35\textheight]{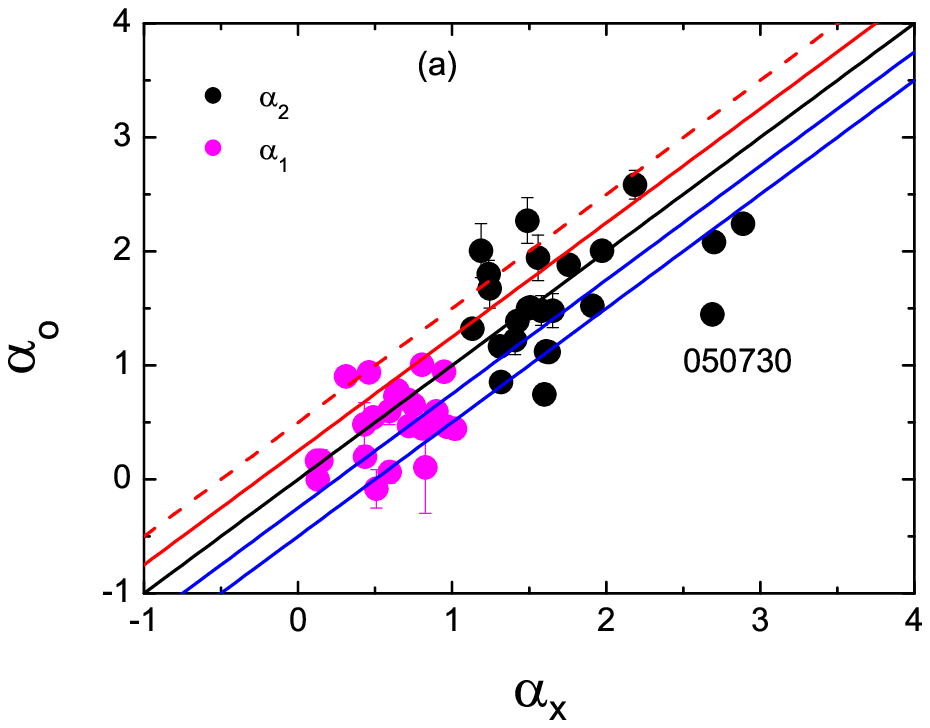}
\includegraphics[angle=0,scale=0.350,width=0.45\textwidth,height=0.35\textheight]{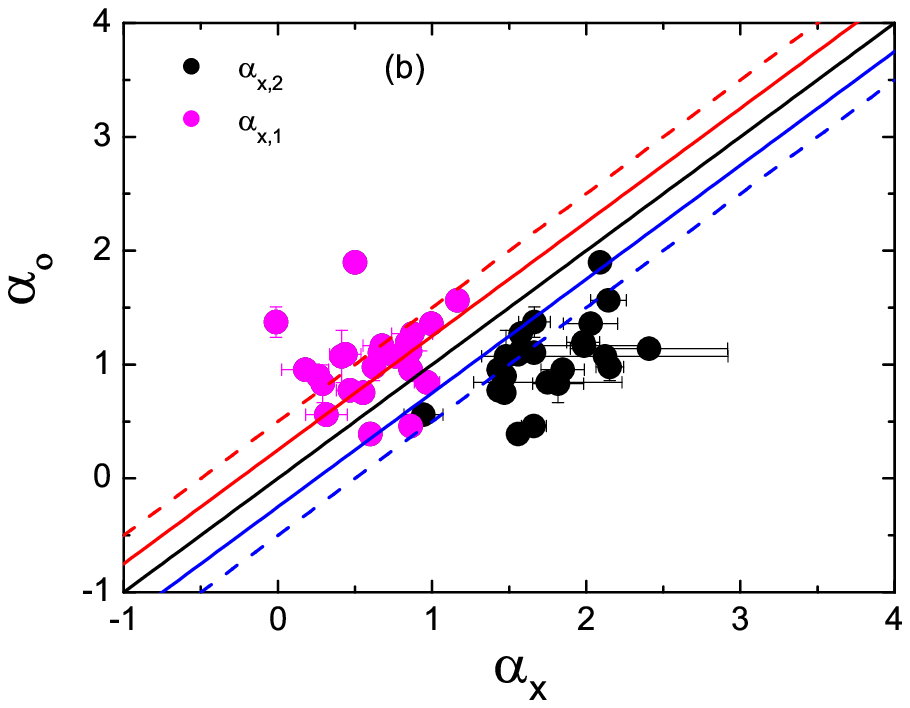}\\
\includegraphics[angle=0,scale=0.350,width=0.45\textwidth,height=0.35\textheight]{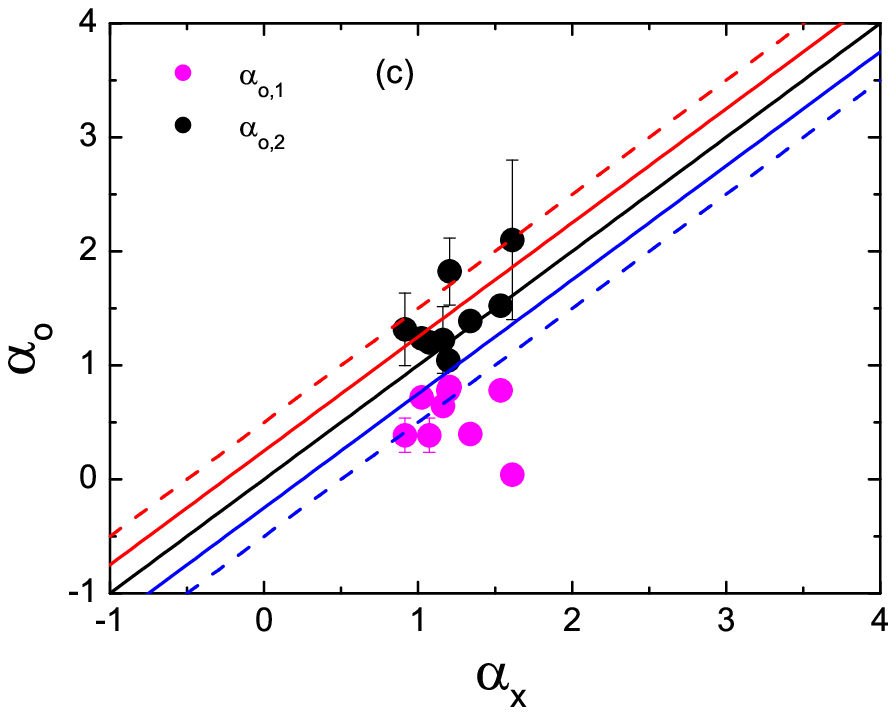}
\includegraphics[angle=0,scale=0.350,width=0.45\textwidth,height=0.35\textheight]{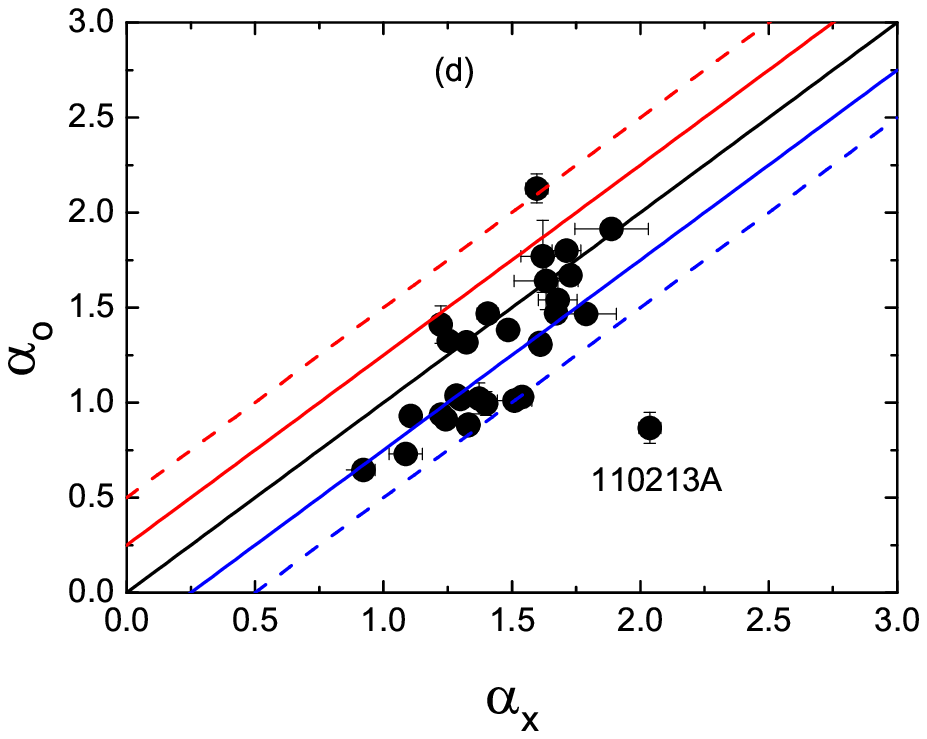}
\caption{The power-law indices for the light-curve decay. The panel from the top to bottom show Group I, II, III and IV,respectively. The black solid line indicates where are the X-ray and optical temporal indices are equal and the red and green solid line indicates where $\alpha_{\rm x}=\alpha_{\rm o}\pm 0.25$, which is expected by standard forward shock model with ISM and wind-like environment. The red and green dashed lines are $\alpha_{\rm x}=\alpha_{\rm o}\pm 0.5$, which is expected by the energy injection case for a ISM and a wind-like environment.} \label{Temporal Group}
\end{figure*}

\begin{figure*}
\includegraphics[angle=0,scale=0.350,width=0.45\textwidth,height=0.35\textheight]{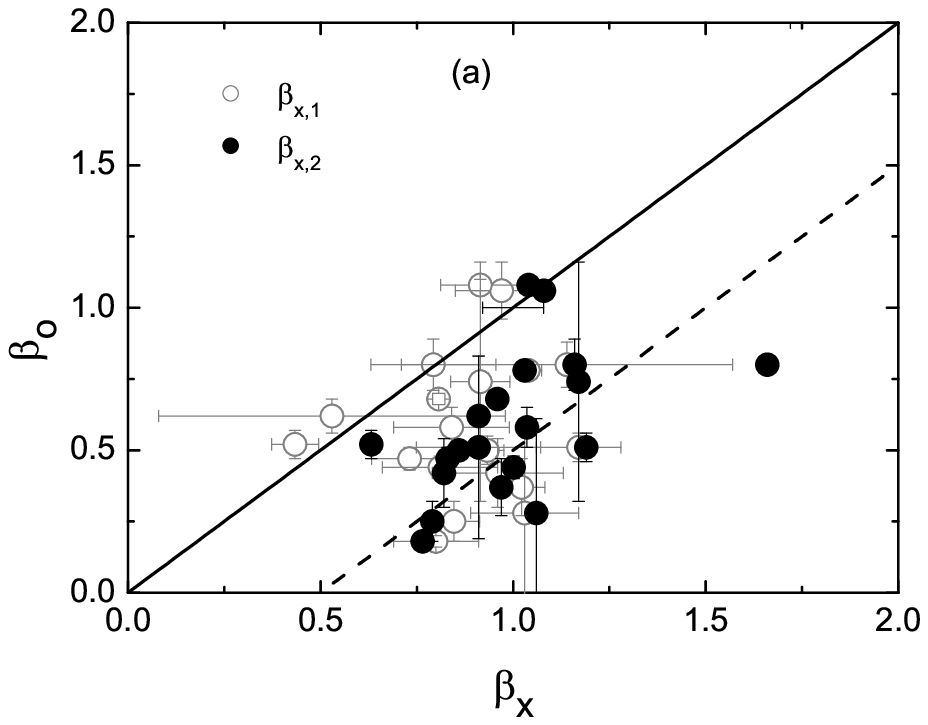}
\includegraphics[angle=0,scale=0.350,width=0.45\textwidth,height=0.35\textheight]{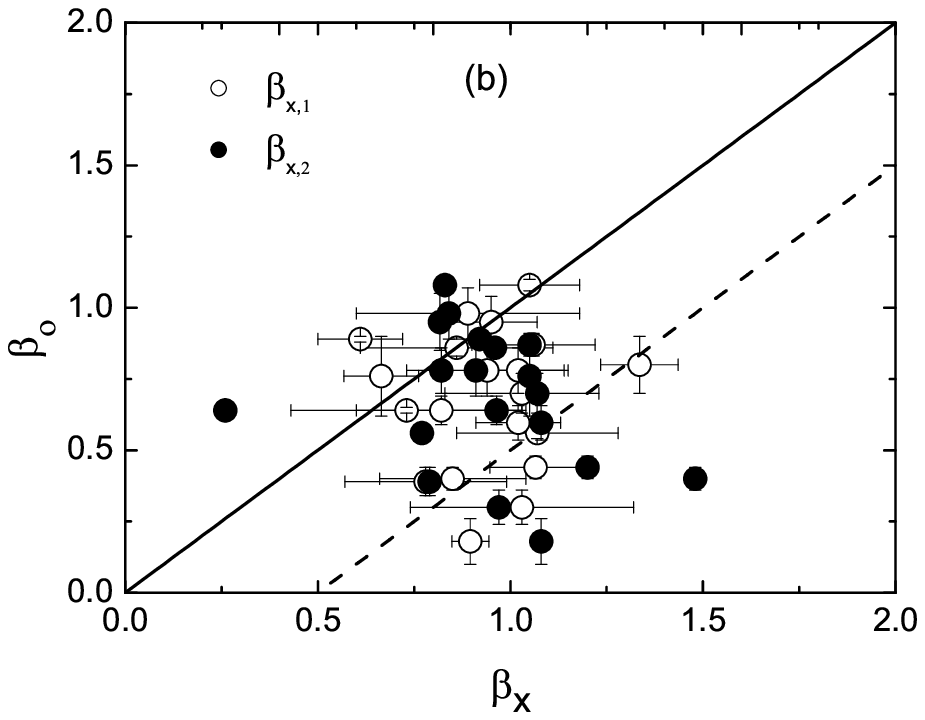}\\
\includegraphics[angle=0,scale=0.350,width=0.45\textwidth,height=0.35\textheight]{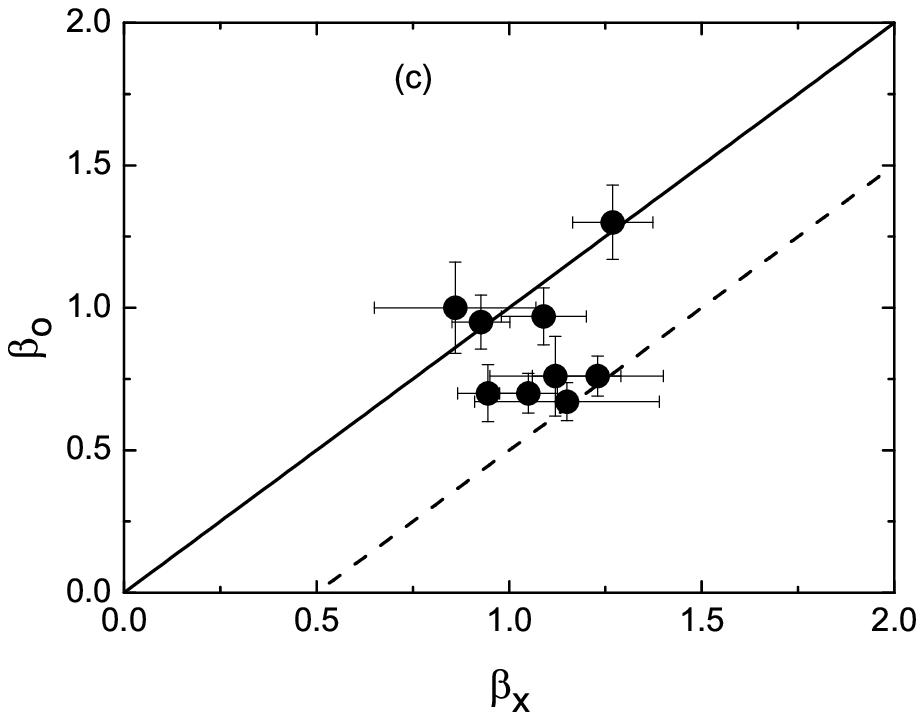}
\includegraphics[angle=0,scale=0.350,width=0.45\textwidth,height=0.35\textheight]{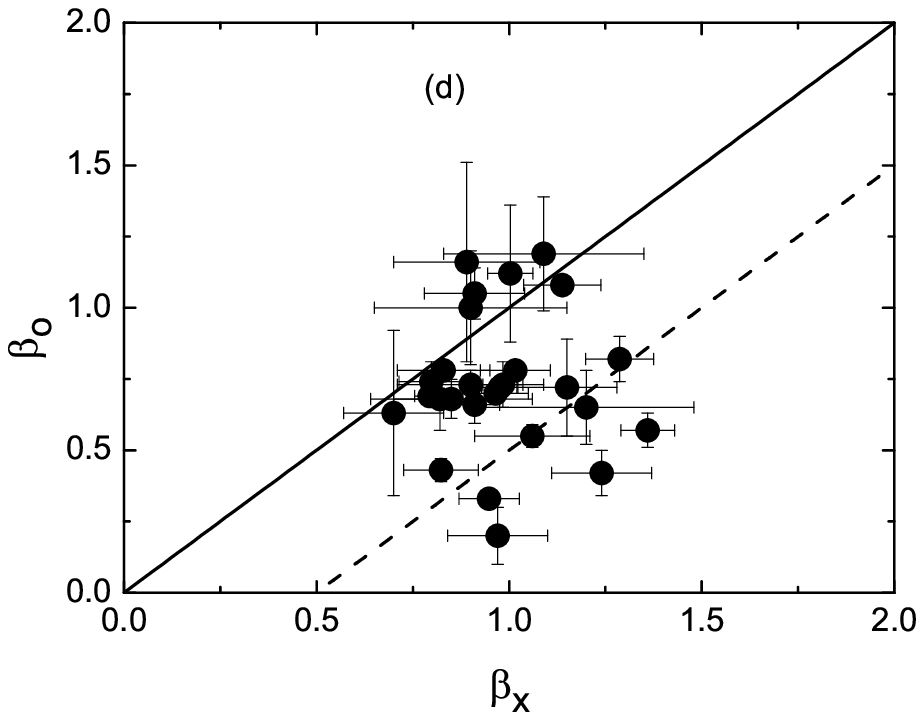}
\caption{The power-law indices of spectra. The panel from the top to bottom show Group I, II, III and IV, respectively. The black solid line indicates where are the X-ray and optical spectral indices are equal (Spectral regime I: $\nu_{\rm m}<\nu_{\rm o}<\nu_{\rm x}<\nu_{\rm c}$) and the black dashed line indicate where $\beta_{\rm x}=\beta_{\rm o}+ 0.5$ (Spectral regime II: $\nu_{\rm m}<\nu_{\rm o}<\nu_{\rm c}<\nu_{\rm x}$).} \label{Spectral Group}
\end{figure*}

\begin{figure*}
\includegraphics[angle=0,scale=0.350,width=0.3\textwidth,height=0.25\textheight]{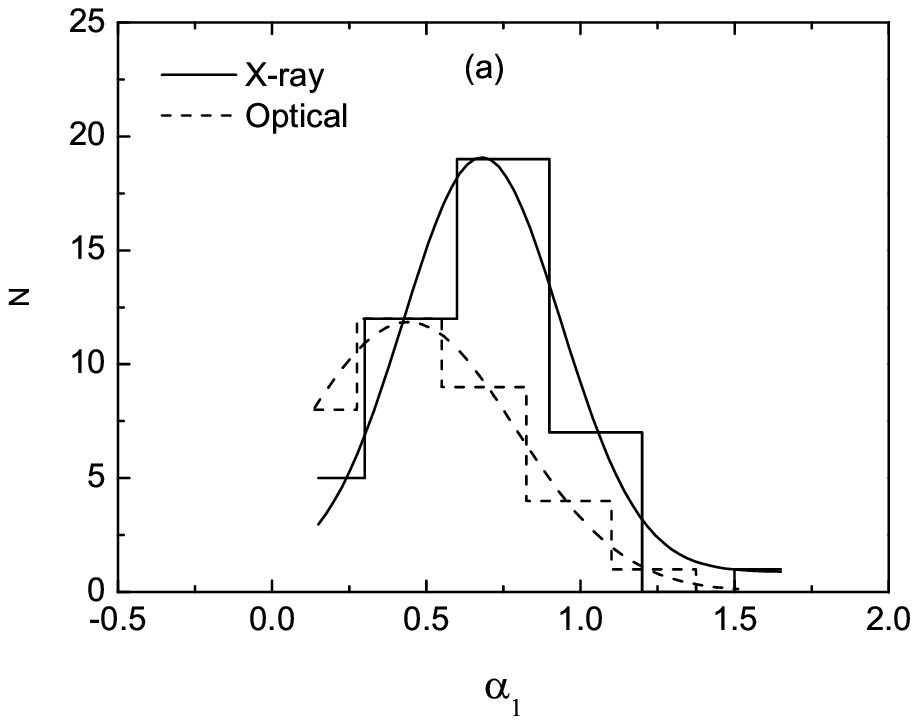}
\includegraphics[angle=0,scale=0.350,width=0.3\textwidth,height=0.25\textheight]{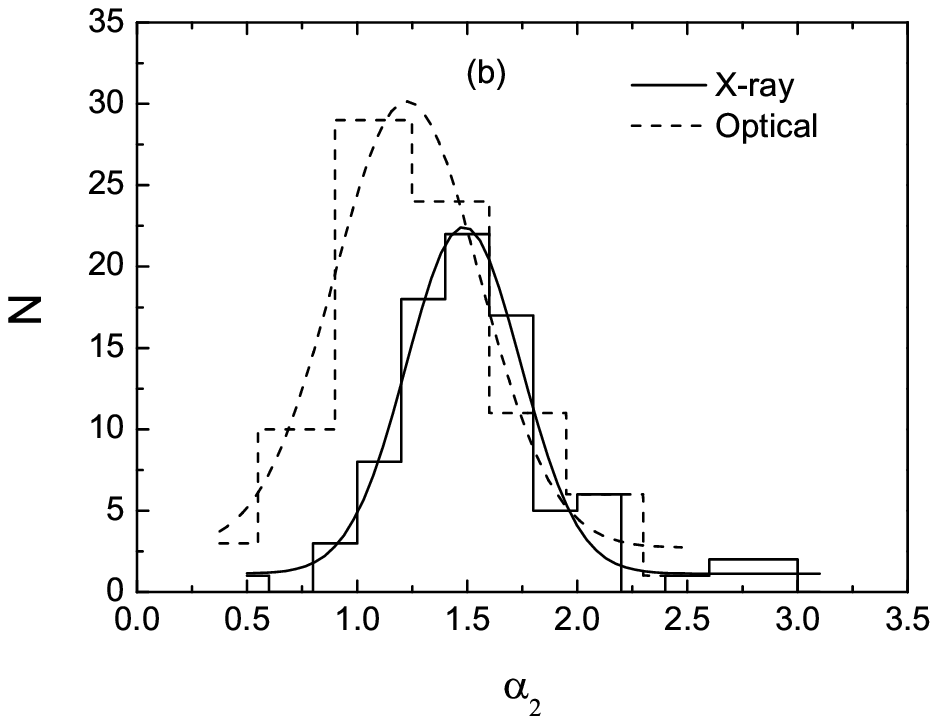}
\includegraphics[angle=0,scale=0.350,width=0.3\textwidth,height=0.25\textheight]{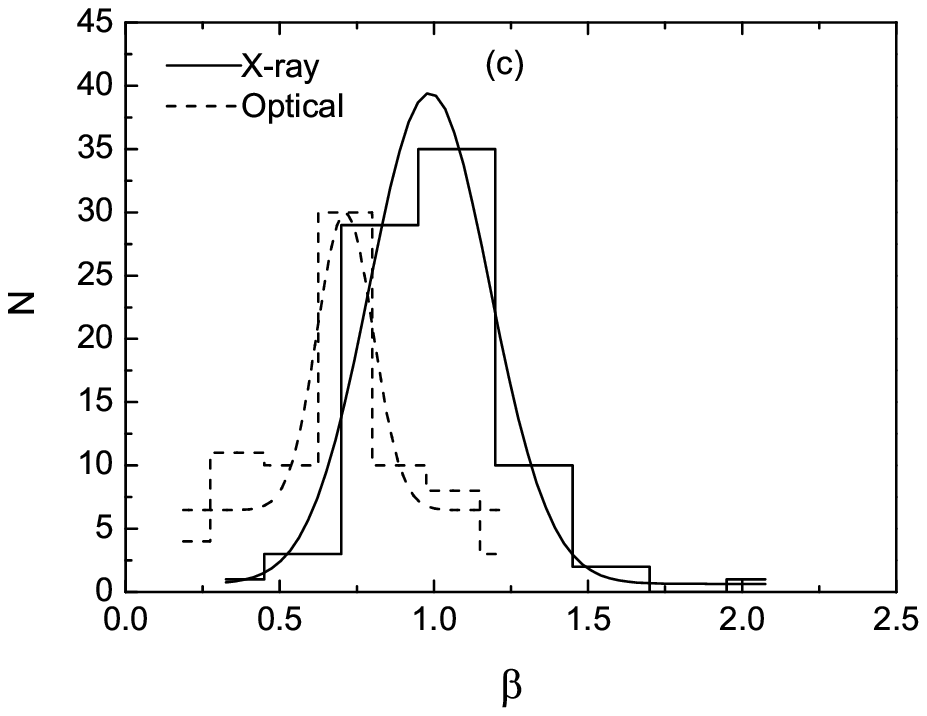}
\caption{The distribution of the temporal and spectral indices for both the X-ray and optical bands for our afterglow cases. (a): The lines are the best Gaussian fitting with the center value $\alpha_{\rm x,1(c)}$=0.68$\pm$ 0.03, $\alpha_{\rm o,1(c)}$=0.44$\pm$ 0.01 for the pre-break decay slopes and (b): $\alpha_{\rm x,2(c)}$=1.48$\pm$ 0.02, $\alpha_{\rm o,2(c)}$=1.23$\pm$ 0.03 for the post-decay slopes, respectively. The center value of all cases are consistent with case of ISM medium and slow cooling, $\nu_{\rm m}<\nu_{\rm o}<\nu_{\rm c}<\nu_{\rm x}$. (c): The best Gaussian fitting for spectral are $\beta_{\rm x,2(c)}$=0.99$\pm$ 0.01, $\beta_{\rm o(c)}$=0.69$\pm$ 0.01, respectively. This is generally consistent with the model with ISM medium and slow cooling, $\nu_{\rm m}<\nu_{\rm o}<\nu_{\rm c}<\nu_{\rm x}$.} \label{Mean}
\end{figure*}

\begin{figure*}
\includegraphics[angle=0,scale=0.350,width=0.45\textwidth,height=0.35\textheight]{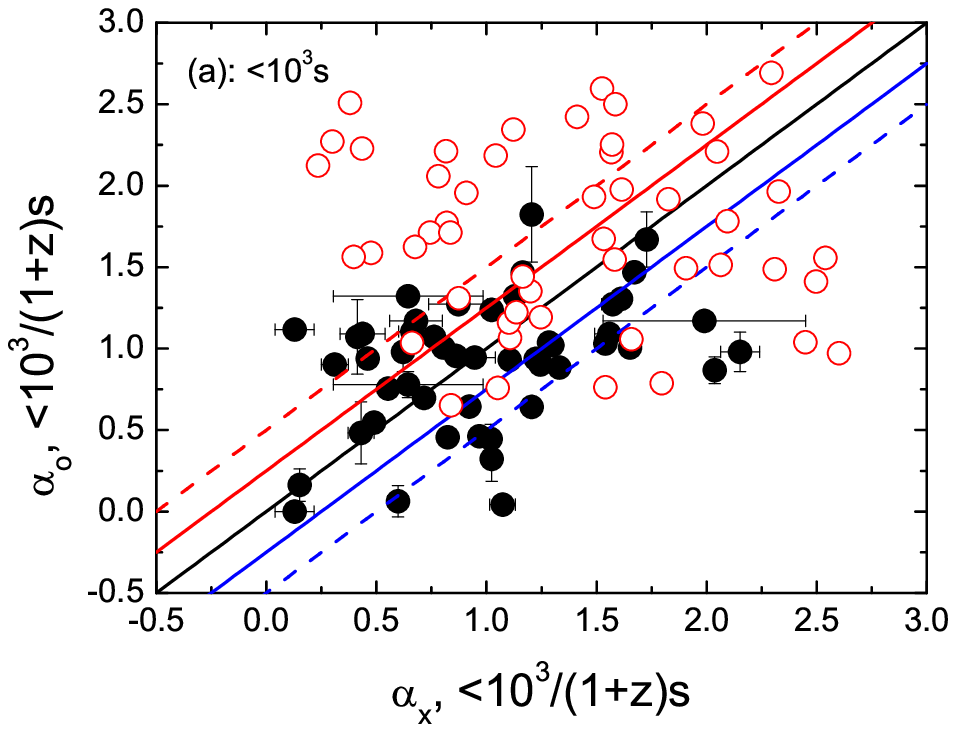}
\includegraphics[angle=0,scale=0.350,width=0.45\textwidth,height=0.35\textheight]{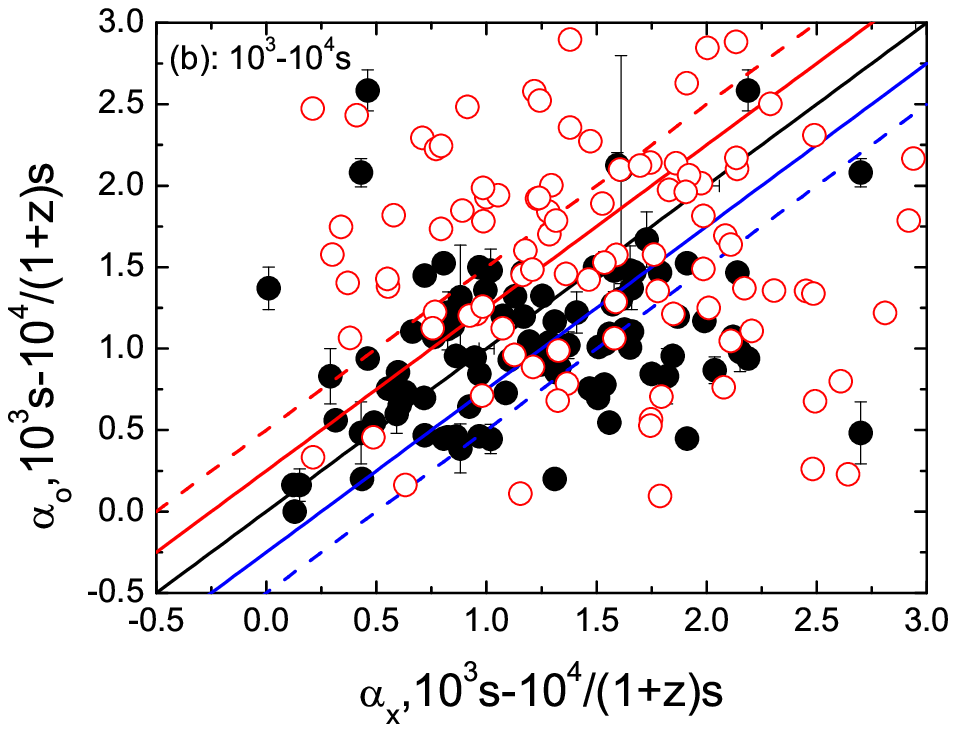}\\
\includegraphics[angle=0,scale=0.350,width=0.45\textwidth,height=0.35\textheight]{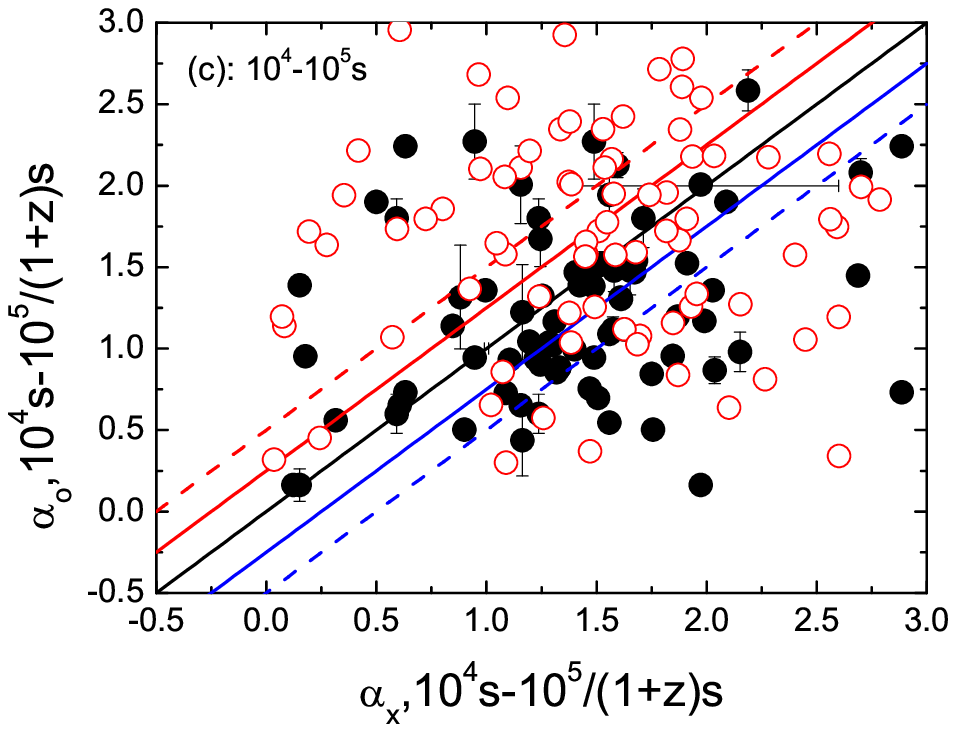}
\includegraphics[angle=0,scale=0.350,width=0.45\textwidth,height=0.35\textheight]{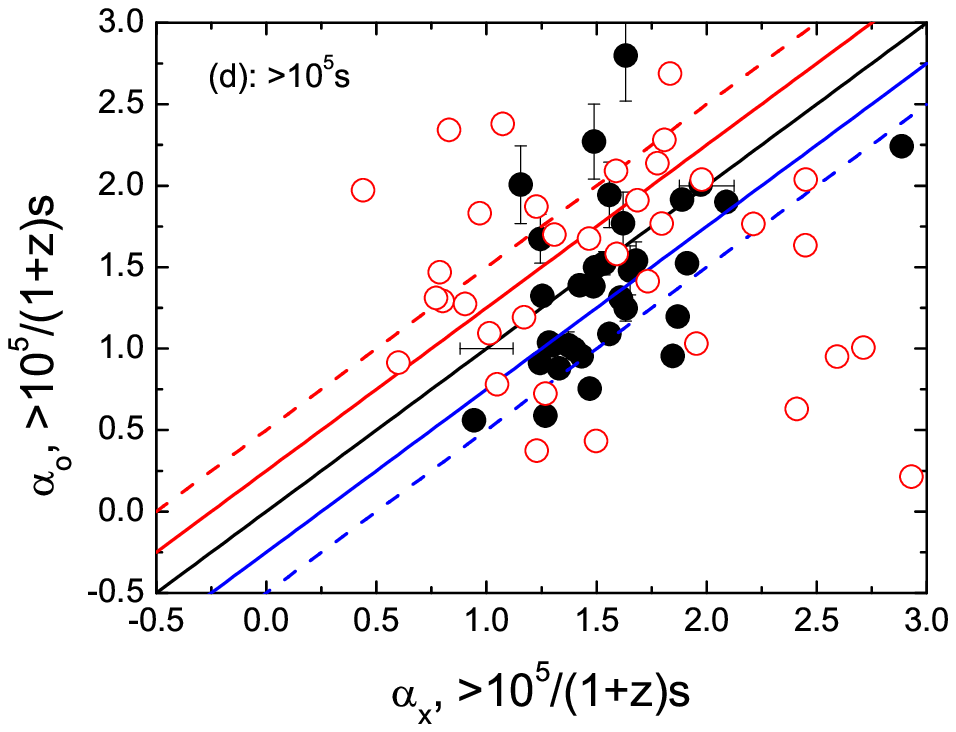}
\caption{As Figure \ref{Temporal Group}, the external shock model is compared to the power-law indices measured over different time intervals for the 4 groups. The panel from the left to right show $<10^{3}$ s, $10^{3}-10^{4}$ s, $10^{4}-10^{5}$ s, $>10^{5}$ s for each bin, respectively. Note that we have corrected them to the rest frame. The black solid points are observe data and red hollow points are simulate data.} \label{Temporal Bins}
\end{figure*}

\begin{figure*}
\includegraphics[angle=0,scale=0.350,width=0.45\textwidth,height=0.4\textheight]{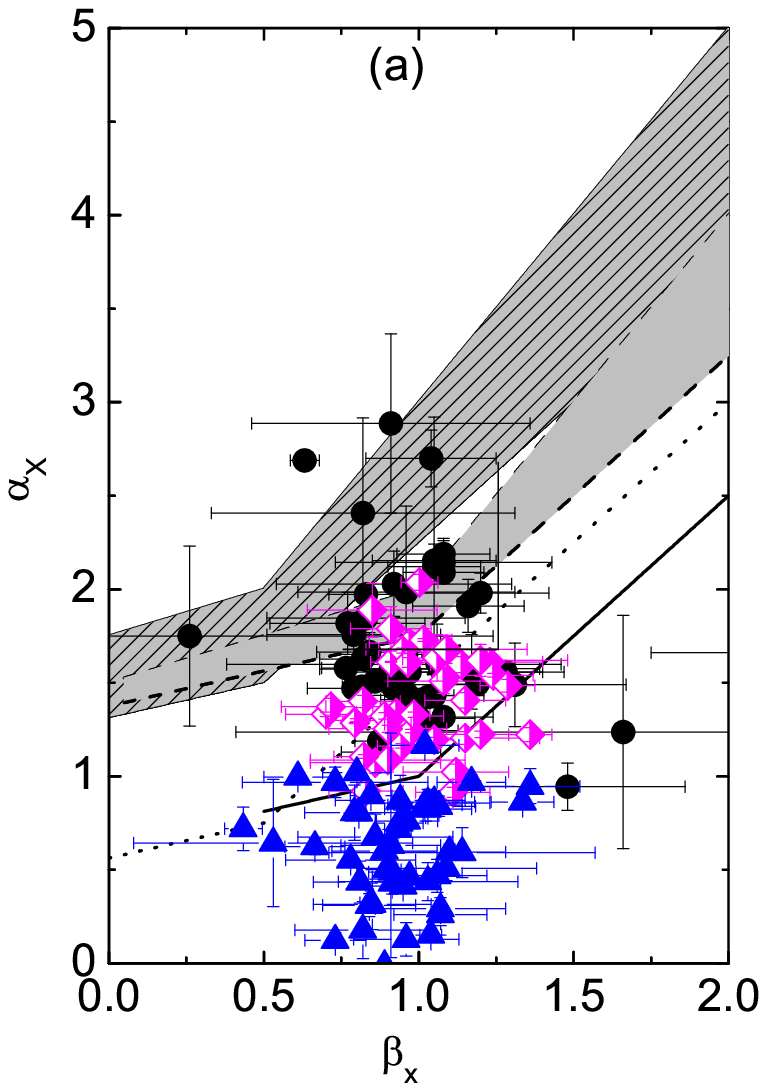}
\includegraphics[angle=0,scale=0.350,width=0.45\textwidth,height=0.4\textheight]{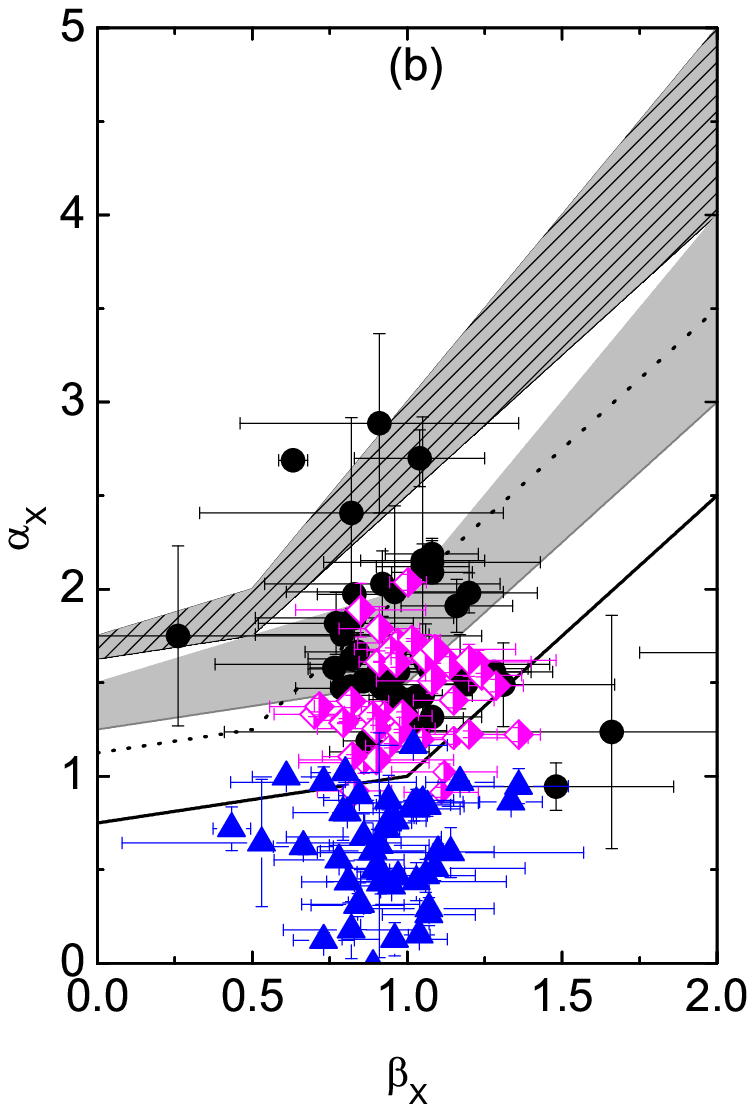}\\
\includegraphics[angle=0,scale=0.350,width=0.45\textwidth,height=0.4\textheight]{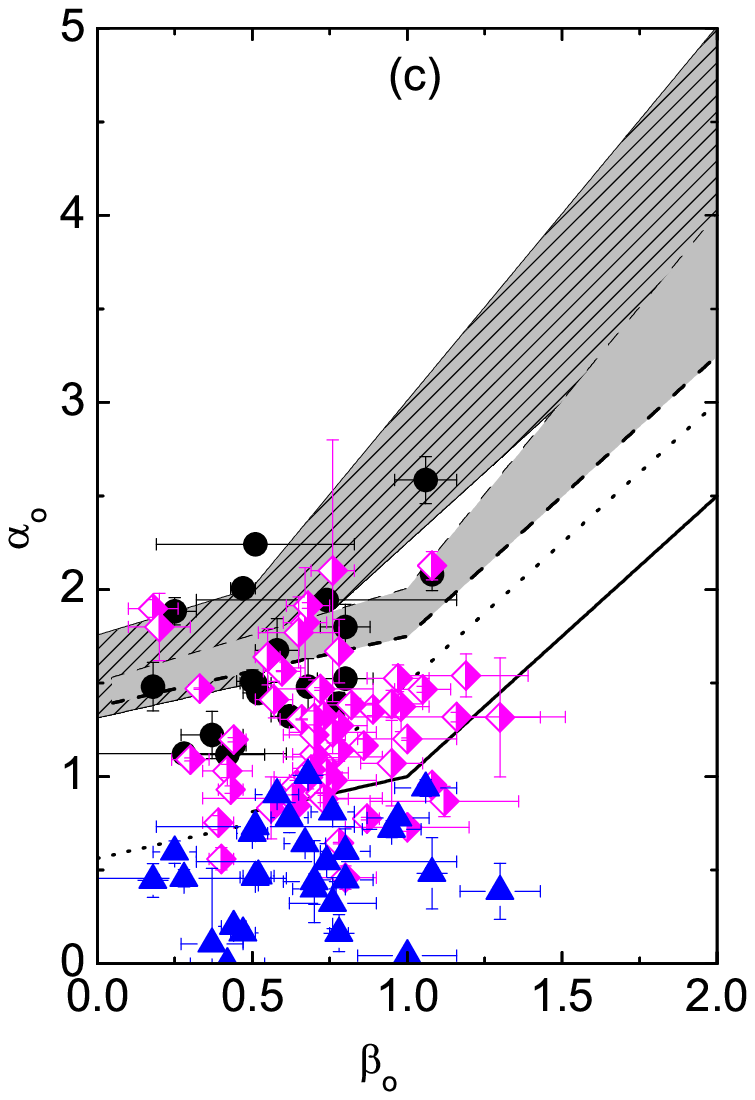}
\includegraphics[angle=0,scale=0.350,width=0.45\textwidth,height=0.4\textheight]{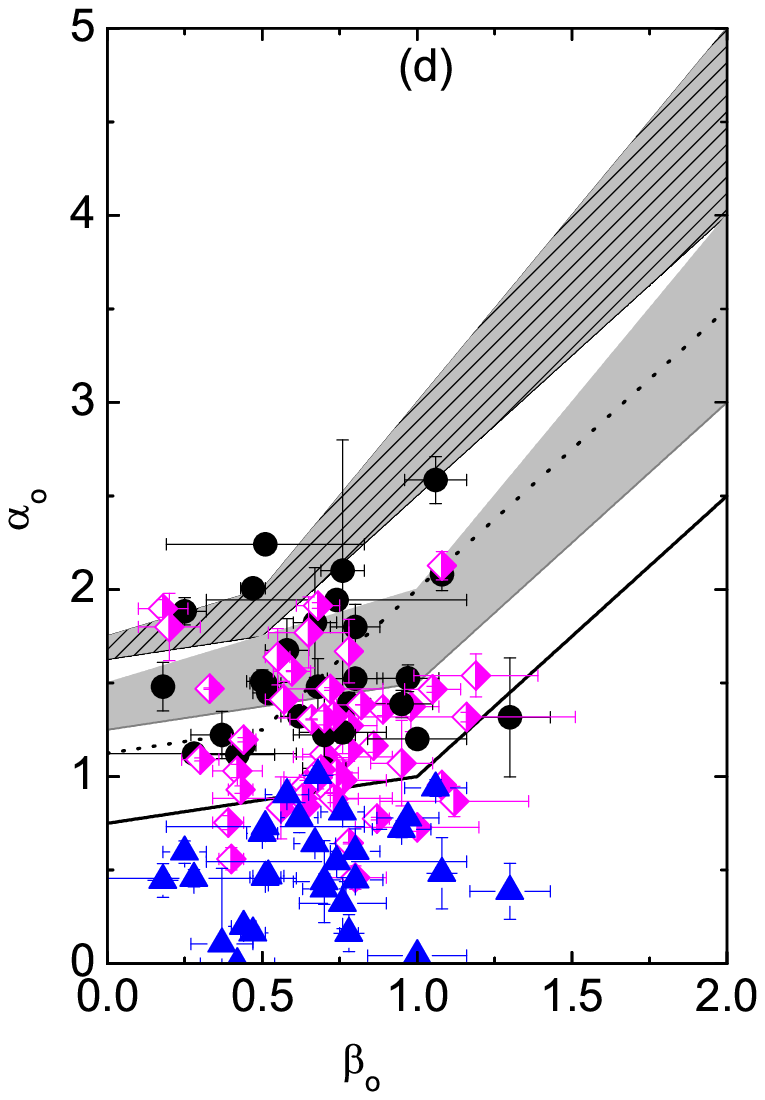}
\caption{The temporal decay index $\alpha$ versus spectral index $\beta$ along with the 'closure relations' of the external forward shock models for all our cases. Regime I ($\nu> max(\nu_{\rm m},\nu_{\rm c})$), Regime II ($\nu_{\rm m}<\nu<\nu_{\rm c}$). The thick solid lines and solid shaded regions are the closure relation of the pre- and post-break in spectral regime I. The lower and upper boundaries of the regions are defined by the 'closure relation', respectively. Similarly, the thick dashed lines and hatched regions are the spectral regime II.
The solid round and triangle dots of black and blue are the temporal decay index of pre- and post-break for our total break samples (X-ray: Group I+Group II; Optical: Group I+Group III.); the half-solid diamond of pink is the temporal index for our non-break samples (X-ray: Group III+Group IV; Optical: Group II+Group IV.). (a) The case of ISM model for XRT data. (b) The case of wind model for XRT data. (c) The case of ISM model for optical data. (d) The case of wind model for optical data.}\label{Closure Relations}
\end{figure*}

\clearpage
\begin{figure*}
\includegraphics[angle=0,scale=0.350,width=0.45\textwidth,height=0.35\textheight]{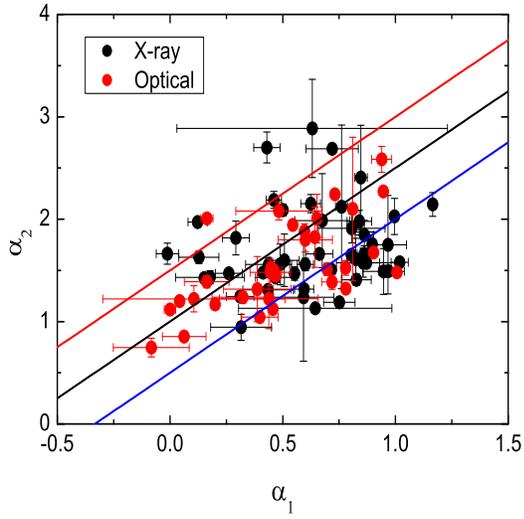}
\caption{The temporal decay indices of pre-break $\alpha_{1}$ versus post-break $\alpha_{2}$ for our break cases, obtained the optical break samples (Group I+Group III) and the X-ray break samples (Group I+Group II). The three solid lines are the various external shock models which invoking energy injection with index $\textit{q}$=0.} \label{Alpha1Alpha2}
\end{figure*}

\begin{figure*}
\includegraphics[angle=0,scale=0.350,width=0.3\textwidth,height=0.25\textheight]{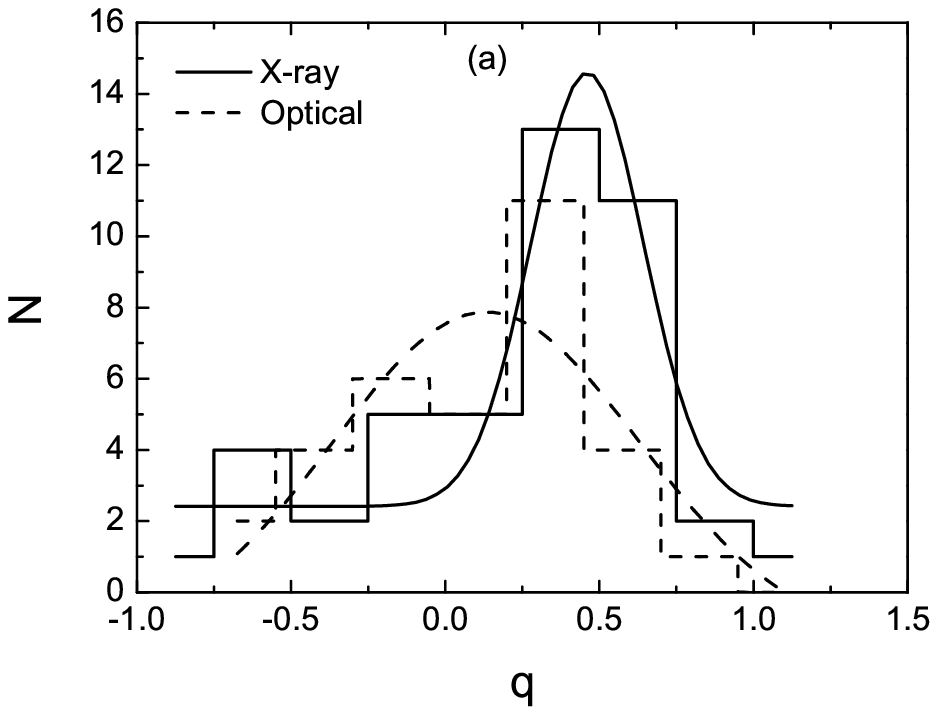}
\includegraphics[angle=0,scale=0.350,width=0.3\textwidth,height=0.25\textheight]{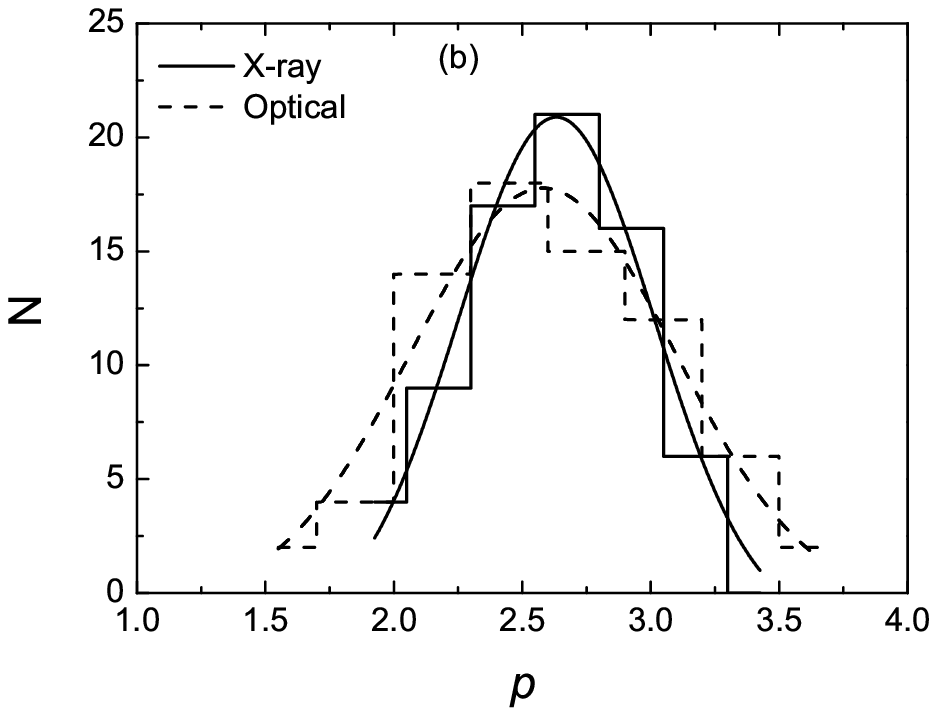}
\includegraphics[angle=0,scale=0.350,width=0.3\textwidth,height=0.25\textheight]{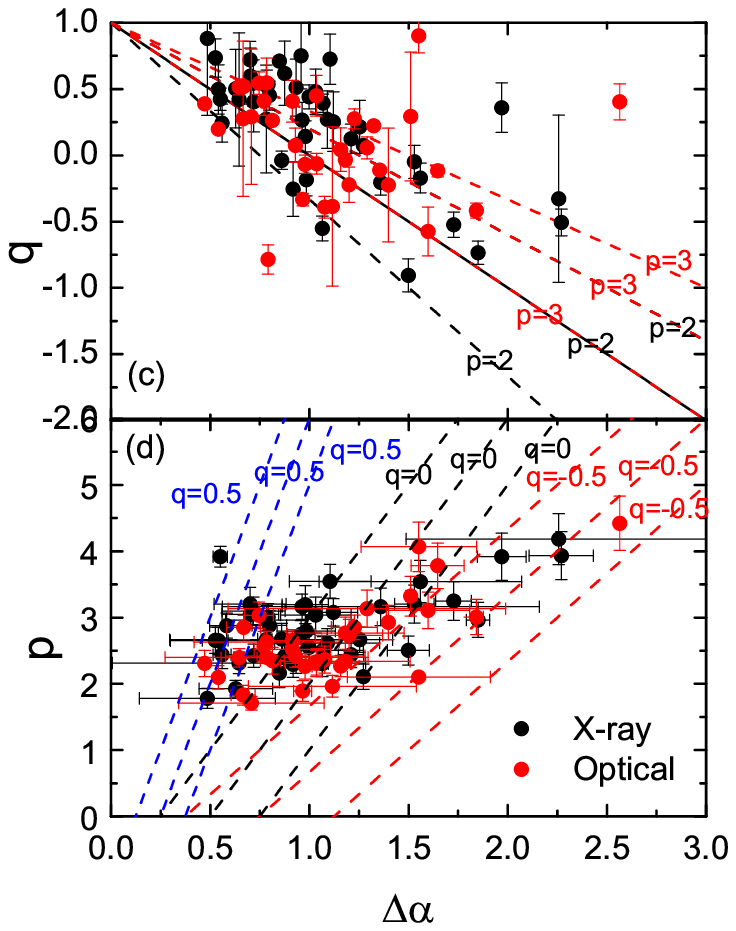}
\caption{The relation of $\Delta\alpha$, \emph{q} and \emph{p} indices.
The distributions of $\Delta\alpha=\alpha_{2}-\alpha_{1}$, luminosity injection index \textit{q} and electron spectral index \emph{p} ($\Delta\alpha$ calculated based on 44 X-ray break cases (Group I+ II) and 33 optical break cases (Group I+III), and the \textit{q} index is derived from the shallow decay segments, and the best Gaussian function fitting with the center value $q_{\rm x(c)}$=0.46$\pm$ 0.03 and $q_{\rm o(c)}$=0.13$\pm$ 0.11, respectively. The \emph{p} index is derived from the post-break decay segments for all the cases, and the best Gaussian fitting with a center value $p_{\rm x(c)}$=2.63$\pm$0.02 and $p_{\rm o(c)}$=2.58$\pm$0.04 for X-ray and optical. The solid and dashed lines are the best Gaussian fitting for X-ray and optical, respectively. The right panel is $\Delta\alpha-\textit{p}, \textit{q}$ planes, in panel (c), the black, red and blue line are the \textit{q}=0, \textit{q}=-0.5 and \textit{q}=0.5, respectively. The black and red points are the X-ray and optical data, respectively. In panel (d), the black solid line and red dash line are \textit{p}=2 and \textit{p}=3 for different external shock models.} \label{deltaalphaqp}
\end{figure*}

\newpage
\begin{figure*}
\includegraphics[angle=0,scale=0.350,width=0.3\textwidth,height=0.25\textheight]{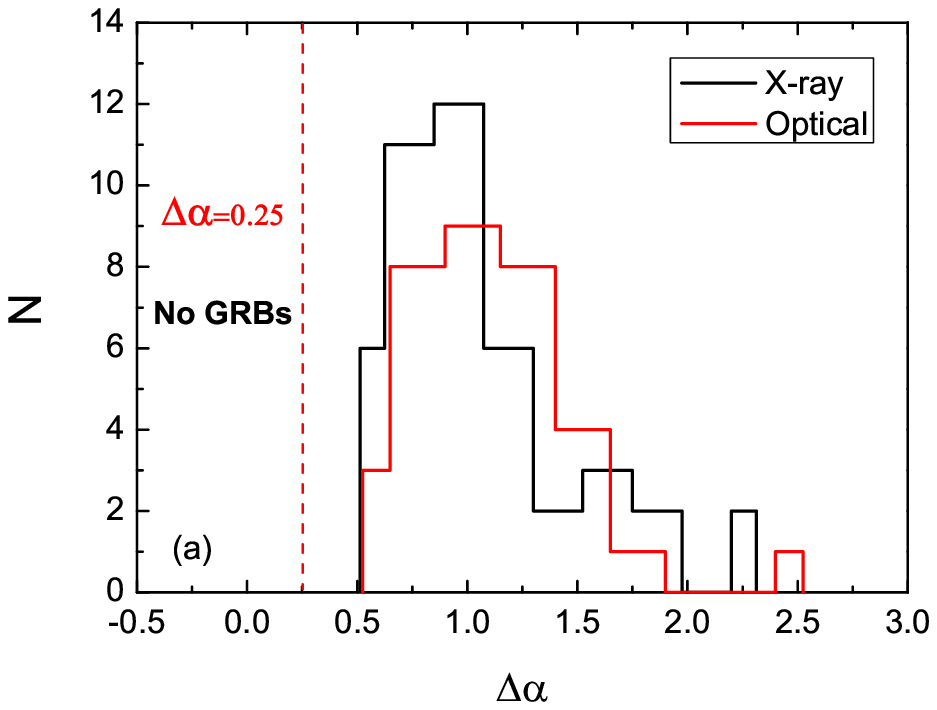}
\includegraphics[angle=0,scale=0.350,width=0.3\textwidth,height=0.25\textheight]{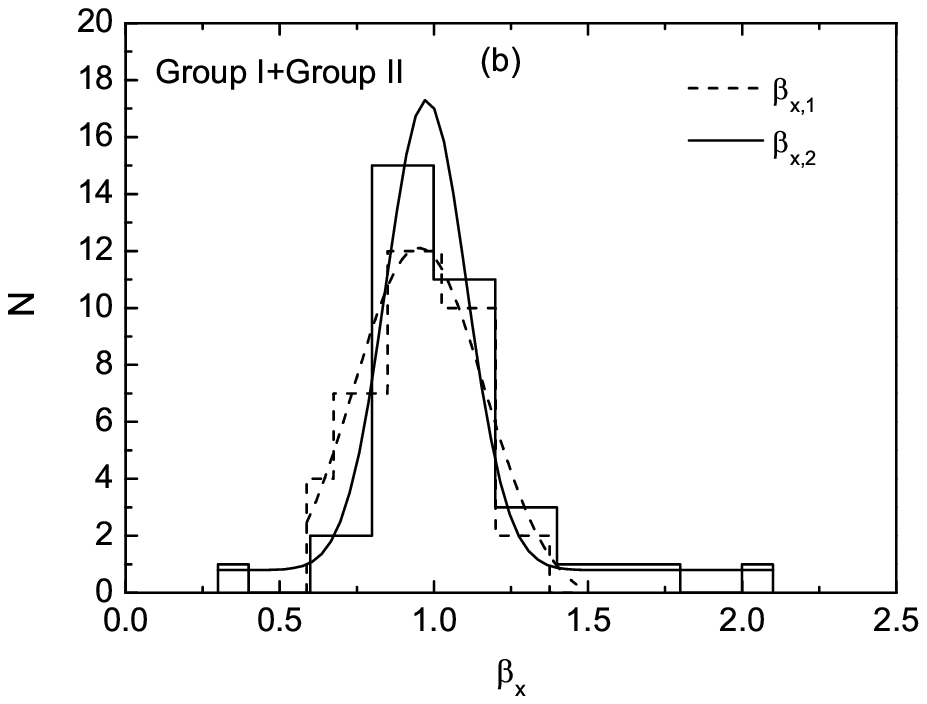}
\includegraphics[angle=0,scale=0.350,width=0.3\textwidth,height=0.25\textheight]{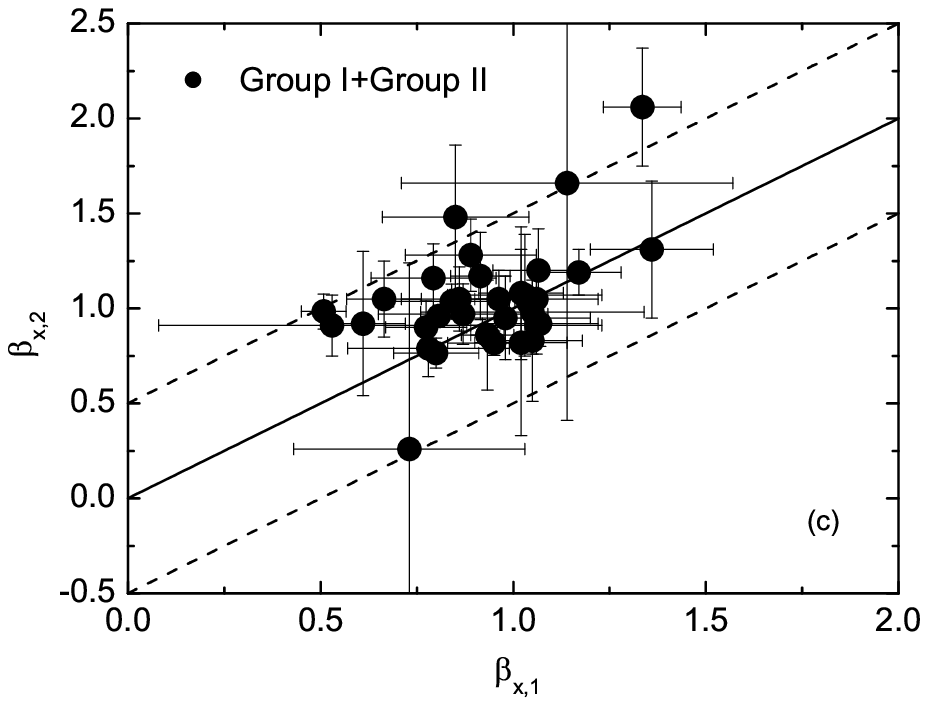}
\caption{The cooling frequency crossing the X-ray and optical band. The change in temporal slope $\Delta\alpha=\alpha_{2}-\alpha_{1}$ for both the X-ray (Group I+II) and optical (Group I+ III) break cases and the spectral index $\Delta\beta_{\rm x}=\beta_{\rm x,2}-\beta_{\rm x,1}$ for the X-ray break cases (Group I+Group II), and the best Gaussian function fitting with the center value $\beta_{\rm x,1(c)}$=0.95$\pm$ 0.03 and $\beta_{\rm x,2(c)}$=0.98$\pm$ 0.01, respectively. In panel (c) shows the relation of spectral slopes of pre and post break segments for X-ray break cases, the black solid line indicates where are the pre and post spectral indices are equal and the black dashed line indicate where $\beta_{\rm x,2}=\beta_{\rm x,1}\pm0.5$}\label{Cooling}
\end{figure*}


\begin{thebibliography}{}
\bibitem[Akerlof et al.(2003)]{2003PASP..115..132A} Akerlof, C.~W., Kehoe, R.~L., McKay, T.~A., et al.\ 2003, \pasp, 115, 132
\bibitem[Antonelli et al.(2006)]{2006A&A...456..509A} Antonelli, L.~A., Testa, V., Romano, P., et al.\ 2006, \aap, 456, 509
\bibitem[Asfandyarov et al.(2006)]{2006GCN..5434....1A} Asfandyarov, I., Pozanenko, A., \& Ibrahimov, M.\ 2006, GRB Coordinates Network, 5434, 1
\bibitem[Barthelmy et al.(2010)]{2010GCN..11023...1B} Barthelmy, S.~D., Baumgartner, W.~H., Cummings, J.~R., et al.\ 2010, GRB Coordinates Network, 11023, 1
\bibitem[Bellm et al.(2008)]{2008ApJ...688..491B} Bellm, E.~C., Hurley, K., Pal'shin, V., et al.\ 2008, \apj, 688, 491
\bibitem[Burrows et al.(2005)]{2005Sci...309.1833B} Burrows, D.~N., Romano, P., Falcone, A., et al.\ 2005, Science, 309, 1833
\bibitem[Burrows et al.(2006)]{2006ApJ...653..468B} Burrows, D.~N., Grupe, D., Capalbi, M., et al.\ 2006, \apj, 653, 468
\bibitem[Butler et al.(2010)]{2010ApJ...711..495B} Butler, N.~R., Bloom, J.~S., \& Poznanski, D.\ 2010, \apj, 711, 495
\bibitem[Butler et al.(2007)]{2007ApJ...671..656B} Butler, N.~R., Kocevski, D., Bloom, J.~S., \& Curtis, J.~L.\ 2007, \apj, 671, 656
\bibitem[Cenko et al.(2011)]{2011ApJ...732...29C} Cenko, S.~B., Frail, D.~A., Harrison, F.~A., et al.\ 2011, \apj, 732, 29
\bibitem[Cenko et al.(2006)]{2006ApJ...652..490C} Cenko, S.~B., Kasliwal, M., Harrison, F.~A., et al.\ 2006, \apj, 652, 490
\bibitem[Chester et al.(2008)]{2008AIPC.1000..421C} Chester, M.~M., Wang, X.~Y., Cummings, J.~R., et al.\ 2008, American Institute of Physics Conference Series, 1000, 421
\bibitem[Chevalier \& Li(2000)]{2000ApJ...536..195C} Chevalier, R.~A., \& Li, Z.-Y.\ 2000, \apj, 536, 195
\bibitem[Chevalier \& Li(1999)]{1999ApJ...520L..29C} Chevalier, R.~A., \& Li, Z.-Y.\ 1999, \apjl, 520, L29
\bibitem[Covino et al.(2010)]{2010A&A...521A..53C} Covino, S., Campana, S., Conciatore, M.~L., et al.\ 2010, \aap, 521, AA53
\bibitem[Covino et al.(2008)]{2008MNRAS.388..347C} Covino, S., D'Avanzo, P., Klotz, A., et al.\ 2008, \mnras, 388, 347
\bibitem[Cucchiara et al.(2011)]{2011ApJ...743..154C} Cucchiara, A., Cenko, S.~B., Bloom, J.~S., et al.\ 2011, \apj, 743, 154
\bibitem[Cusumano et al.(2006)]{2006ApJ...639..316C} Cusumano, G., Mangano, V., Angelini, L., et al.\ 2006, \apj, 639, 316
\bibitem[Dado \& Dar(2010)]{2010ApJ...708L.112D} Dado, S., \& Dar, A.\ 2010, \apjl, 708, L112
\bibitem[Dado \& Dar(2008)]{2008arXiv0812.3340D} Dado, S., \& Dar, A.\ 2008, arXiv:0812.3340
\bibitem[Dai et al.(2007)]{2007ApJ...658..509D} Dai, X., Halpern, J.~P., Morgan, N.~D., et al.\ 2007, \apj, 658, 509
\bibitem[Dai \& Lu(2002)]{2002ApJ...565L..87D} Dai, Z.~G., \& Lu, T.\ 2002, \apjl, 565, L87
\bibitem[Dai \& Lu(1998)]{1998A&A...333L..87D} Dai, Z.~G., \& Lu, T.\ 1998, \aap, 333, L87
\bibitem[Dai et al.(2006)]{2006Sci...311.1127D} Dai, Z.~G., Wang, X.~Y., Wu, X.~F., \& Zhang, B.\ 2006, Science, 311, 1127
\bibitem[Daigne \& Mochkovitch(1998)]{1998MNRAS.296..275D} Daigne, F., \& Mochkovitch, R.\ 1998, \mnras, 296, 275
\bibitem[De Pasquale et al.(2010)]{2010ApJ...709L.146D} De Pasquale, M., Schady, P., Kuin, N.~P.~M., et al.\ 2010, \apjl, 709, L146
\bibitem[De Pasquale et al.(2006)]{2006MNRAS.365.1031D} De Pasquale, M., Beardmore, A.~P., Barthelmy, S.~D., et al.\ 2006, \mnras, 365, 1031
\bibitem[de Ugarte Postigo et al.(2007)]{2007A&A...462L..57D} de Ugarte Postigo, A., Fatkhullin, T.~A., J{\'o}hannesson, G., et al.\ 2007, \aap, 462, L57
\bibitem[Deng et al.(2009)]{2009arXiv0912.5435D} Deng, J., Zheng, W., Zhai, M., et al.\ 2009, arXiv:0912.5435
\bibitem[Drenkhahn \& Spruit(2004)]{2004ASPC..312..357D} Drenkhahn, G., \& Spruit, H.~C.\ 2004, Gamma-Ray Bursts in the Afterglow Era, 312, 357
\bibitem[Duffell \& MacFadyen(2014)]{2014arXiv1407.8250D} Duffell, P.~C., \& MacFadyen, A.~I.\ 2014, arXiv:1407.8250
\bibitem[Elliott et al.(2013)]{2013A&A...556A..23E} Elliott, J., Kr{\"u}hler, T., Greiner, J., et al.\ 2013, \aap, 556, AA23
\bibitem[Elliott et al.(2014)]{2014A&A...562A.100E} Elliott, J., Yu, H.-F., Schmidl, S., et al.\ 2014, \aap, 562, AA100
\bibitem[Evans et al.(2009)]{2009MNRAS.397.1177E} Evans, P.~A., Beardmore, A.~P., Page, K.~L., et al.\ 2009, \mnras, 397, 1177
\bibitem[Evans et al.(2007)]{2007A&A...469..379E} Evans, P.~A., Beardmore, A.~P., Page, K.~L., et al.\ 2007, \aap, 469, 379
\bibitem[Fan \& Wei(2005)]{2005MNRAS.364L..42F} Fan, Y.~Z., \& Wei, D.~M.\ 2005, \mnras, 364, L42
\bibitem[Ferrero et al.(2008)]{2008AIPC.1000..257F} Ferrero, P., Kann, D.~A., Klose, S., et al.\ 2008, American Institute of Physics Conference Series, 1000, 257
\bibitem[Filgas et al.(2012)]{2012A&A...546A.101F} Filgas, R., Greiner, J., Schady, P., et al.\ 2012, \aap, 546, AA101
\bibitem[Filgas et al.(2011)]{2011A&A...526A.113F} Filgas, R., Kr{\"u}hler, T., Greiner, J., et al.\ 2011, \aap, 526, AA113
\bibitem[Fynbo et al.(2009)]{2009ApJS..185..526F} Fynbo, J.~P.~U., Jakobsson, P., Prochaska, J.~X., et al.\ 2009, \apjs, 185, 526
\bibitem[Gao et al.(2013)]{2013NewAR..57..141G} Gao, H., Lei, W.-H., Zou, Y.-C., Wu, X.-F., \& Zhang, B.\ 2013, \nar, 57, 141
\bibitem[Gehrels et al.(2004)]{2004ApJ...611.1005G} Gehrels, N., Chincarini, G., Giommi, P., et al.\ 2004, \apj, 611, 1005
\bibitem[Gendre et al.(2010)]{2010MNRAS.405.2372G} Gendre, B., Klotz, A., Palazzi, E., et al.\ 2010, \mnras, 405, 2372
\bibitem[Gendre et al.(2011)]{2011A&A...530A..74G} Gendre, B., Stratta, G., Laas-Bourez, M., et al.\ 2011, \aap, 530, AA74
\bibitem[Giannios \& Spruit(2006)]{2006A&A...450..887G} Giannios, D., \& Spruit, H.~C.\ 2006, \aap, 450, 887
\bibitem[Gomboc et al.(2008)]{2008ApJ...687..443G} Gomboc, A., Kobayashi, S., Guidorzi, C., et al.\ 2008, \apj, 687, 443
\bibitem[Gorbovskoy et al.(2012)]{2012MNRAS.421.1874G} Gorbovskoy, E.~S., Lipunova, G.~V., Lipunov, V.~M., et al.\ 2012, \mnras, 421, 1874
\bibitem[Greiner et al.(2008)]{2008PASP..120..405G} Greiner, J., Bornemann, W., Clemens, C., et al.\ 2008, \pasp, 120, 405
\bibitem[Greiner et al.(2013)]{2013A&A...560A..70G} Greiner, J., Kr{\"u}hler, T., Nardini, M., et al.\ 2013, \aap, 560, AA70
\bibitem[Greiner et al.(2007)]{2007Msngr.130...12G} Greiner, J., Bornemann, W., Clemens, C., et al.\ 2007, The Messenger, 130, 12
\bibitem[Grupe et al.(2006)]{2006ApJ...645..464G} Grupe, D., Brown, P.~J., Cummings, J., et al.\ 2006, \apj, 645, 464
\bibitem[Grupe et al.(2007)]{2007ApJ...662..443G} Grupe, D., Gronwall, C., Wang, X.-Y., et al.\ 2007, \apj, 662, 443
\bibitem[Guetta et al.(2011)]{2011A&A...525A..53G} Guetta, D., Pian, E., \& Waxman, E.\ 2011, \aap, 525, AA53
\bibitem[Guidorzi et al.(2014)]{2014MNRAS.438..752G} Guidorzi, C., Mundell, C.~G., Harrison, R., et al.\ 2014, \mnras, 438, 752
\bibitem[Guidorzi et al.(2007)]{2007A&A...474..793G} Guidorzi, C., Vergani, S.~D., Sazonov, S., et al.\ 2007, \aap, 474, 793
\bibitem[Hartoog et al.(2013)]{2013MNRAS.430.2739H} Hartoog, O.~E., Wiersema, K., Vreeswijk, P.~M., et al.\ 2013, \mnras, 430, 2739
\bibitem[Huang et al.(2012)]{2012ApJ...748...44H} Huang, K.~Y., Urata, Y., Tung, Y.~H., et al.\ 2012, \apj, 748, 44
\bibitem[Huang et al.(2009)]{2009AIPC.1133..212H} Huang, K.~Y., Wang, S.~Y., \& Urata, Y.\ 2009, American Institute of Physics Conference Series, 1133, 212
\bibitem[Huang et al.(2000)]{2000ApJ...543...90H} Huang, Y.~F., Gou, L.~J., Dai, Z.~G., \& Lu, T.\ 2000, \apj, 543, 90
\bibitem[Jakobsson et al.(2007)]{2007GCN..7076....1J} Jakobsson, P., Fynbo, J.~P.~U., Vreeswijk, P.~M., Malesani, D., \& Sollerman, J.\ 2007, GRB Coordinates Network, 7076, 1
\bibitem[Jakobsson et al.(2006)]{2006GCN..5319....1J} Jakobsson, P., Vreeswijk, P., Ellison, S., et al.\ 2006, GRB Coordinates Network, 5319, 1
\bibitem[Jin et al.(2009)]{2009MNRAS.400.1829J} Jin, Z.~P., Xu, D., Covino, S., et al.\ 2009, \mnras, 400, 1829
\bibitem[Kann et al.(2010)]{2010ApJ...720.1513K} Kann, D.~A., Klose, S., Zhang, B., et al.\ 2010, \apj, 720, 1513
\bibitem[Kelly et al.(2013)]{2013ApJ...775L...5K} Kelly, P.~L., Filippenko, A.~V., Fox, O.~D., Zheng, W., \& Clubb, K.~I.\ 2013, \apjl, 775, LL5
\bibitem[Klotz et al.(2009)]{2009AJ....137.4100K} Klotz, A., Bo{\"e}r, M., Atteia, J.~L., \& Gendre, B.\ 2009, \aj, 137, 4100
\bibitem[Klotz et al.(2005)]{2005A&A...439L..35K} Klotz, A., Bo{\"e}r, M., Atteia, J.~L., et al.\ 2005, \aap, 439, L35
\bibitem[Klotz et al.(2009)]{2009ApJ...697L..18K} Klotz, A., Gendre, B., Atteia, J.~L., et al.\ 2009, \apjl, 697, L18
\bibitem[Klotz et al.(2008)]{2008A&A...483..847K} Klotz, A., Gendre, B., Stratta, G., et al.\ 2008, \aap, 483, 847
\bibitem[Kobayashi et al.(2004)]{2004ApJ...601L..13K} Kobayashi, S., M{\'e}sz{\'a}ros, P., \& Zhang, B.\ 2004, \apjl, 601, L13
\bibitem[Kobayashi et al.(1997)]{1997ApJ...490...92K} Kobayashi, S., Piran, T., \& Sari, R.\ 1997, \apj, 490, 92
\bibitem[Kobayashi \& Zhang(2003)]{2003ApJ...597..455K} Kobayashi, S., \& Zhang, B.\ 2003, \apj, 597, 455
\bibitem[Kopa{\v c} et al.(2013)]{2013ApJ...772...73K} Kopa{\v c}, D., Kobayashi, S., Gomboc, A., et al.\ 2013, \apj, 772, 73
\bibitem[Kr{\"u}hler et al.(2009)]{2009A&A...508..593K} Kr{\"u}hler, T., Greiner, J., Afonso, P., et al.\ 2009, \aap, 508, 593
\bibitem[Kr{\"u}hler et al.(2009)]{2009ApJ...697..758K} Kr{\"u}hler, T., Greiner, J., McBreen, S., et al.\ 2009, \apj, 697, 758
\bibitem[Kr{\"u}hler et al.(2013)]{2013A&A...557A..18K} Kr{\"u}hler, T., Ledoux, C., Fynbo, J.~P.~U., et al.\ 2013, \aap, 557, AA18
\bibitem[Krimm et al.(2007)]{2007GCN..7081....1K} Krimm, H., Barbier, L., Barthelmy, S., et al.\ 2007, GRB Coordinates Network, 7081, 1
\bibitem[Krimm et al.(2009)]{2009ApJ...704.1405K} Krimm, H.~A., Yamaoka, K., Sugita, S., et al.\ 2009, \apj, 704, 1405
\bibitem[Kruehler et al.(2013)]{2013yCat..35579018K} Kruehler, T., Ledoux, C., Fynbo, J.~P.~U., et al.\ 2013, VizieR Online Data Catalog, 355, 79018
\bibitem[Kruehler et al.(2008)]{2008GCN..8075....1K} Kruehler, T., Schrey, F., Greiner, J., et al.\ 2008, GRB Coordinates Network, 8075, 1
\bibitem[Kuin et al.(2009)]{2009MNRAS.395L..21K} Kuin, N.~P.~M., Landsman, W., Page, M.~J., et al.\ 2009, \mnras, 395, L21
\bibitem[Li et al.(2012)]{2012ApJ...758...27L} Li, L., Liang, E.-W., Tang, Q.-W., et al.\ 2012, \apj, 758, 27
\bibitem[Liang et al.(2013)]{2013ApJ...774...13L} Liang, E.-W., Li, L., Gao, H., et al.\ 2013, \apj, 774, 13
\bibitem[Liang et al.(2008)]{2008ApJ...675..528L} Liang, E.-W., Racusin, J.~L., Zhang, B., Zhang, B.-B., \& Burrows, D.~N.\ 2008, \apj, 675, 528
\bibitem[Liang et al.(2007)]{2007ApJ...670..565L} Liang, E.-W., Zhang, B.-B., \& Zhang, B.\ 2007, \apj, 670, 565
\bibitem[Littlejohns et al.(2012)]{2012MNRAS.421.2692L} Littlejohns, O.~M., Willingale, R., O'Brien, P.~T., et al.\ 2012, \mnras, 421, 2692
\bibitem[M{\'e}sz{\'a}ros \& Rees(1997)]{1997ApJ...476..232M} M{\'e}sz{\'a}ros, P., \& Rees, M.~J.\ 1997, \apj, 476, 232
\bibitem[M{\'e}sz{\'a}ros et al.(1998)]{1998ApJ...499..301M} M{\'e}sz{\'a}ros, P., Rees, M.~J., \& Wijers, R.~A.~M.~J.\ 1998, \apj, 499, 301
\bibitem[Mangano et al.(2007)]{2007A&A...470..105M} Mangano, V., Holland, S.~T., Malesani, D., et al.\ 2007, \aap, 470, 105
\bibitem[Markwardt(2009)]{2009ASPC..411..251M} Markwardt, C.~B.\ 2009, Astronomical Data Analysis Software and Systems XVIII, 411, 251
\bibitem[Marshall et al.(2011)]{2011ApJ...727..132M} Marshall, F.~E., Antonelli, L.~A., Burrows, D.~N., et al.\ 2011, \apj, 727, 132
\bibitem[Maselli et al.(2014)]{2014Sci...343...48M} Maselli, A., Melandri, A., Nava, L., et al.\ 2014, Science, 343, 48
\bibitem[Meszaros \& Rees(1993)]{1993ApJ...418L..59M} Meszaros, P., \& Rees, M.~J.\ 1993, \apjl, 418, L59
\bibitem[Molinari et al.(2007)]{2007A&A...469L..13M} Molinari, E., Vergani, S.~D., Malesani, D., et al.\ 2007, \aap, 469, L13
\bibitem[Morgan et al.(2014)]{2014MNRAS.440.1810M} Morgan, A.~N., Perley, D.~A., Cenko, S.~B., et al.\ 2014, \mnras, 440, 1810
\bibitem[Nardini et al.(2014)]{2014A&A...562A..29N} Nardini, M., Elliott, J., Filgas, R., et al.\ 2014, \aap, 562, AA29
\bibitem[Nardini et al.(2006)]{2006A&A...451..821N} Nardini, M., Ghisellini, G., Ghirlanda, G., et al.\ 2006, \aap, 451, 821
\bibitem[Nardini et al.(2011)]{2011A&A...531A..39N} Nardini, M., Greiner, J., Kr{\"u}hler, T., et al.\ 2011, \aap, 531, AA39
\bibitem[Nava et al.(2008)]{2008MNRAS.391..639N} Nava, L., Ghirlanda, G., Ghisellini, G., \& Firmani, C.\ 2008, \mnras, 391, 639
\bibitem[Nicuesa Guelbenzu et al.(2011)]{2011A&A...531L...6N} Nicuesa Guelbenzu, A., Klose, S., Rossi, A., et al.\ 2011, \aap, 531, LL6
\bibitem[Nousek et al.(2006)]{2006ApJ...642..389N} Nousek, J.~A., Kouveliotou, C., Grupe, D., et al.\ 2006, \apj, 642, 389
\bibitem[Nysewander et al.(2009)]{2009ApJ...701..824N} Nysewander, M., Fruchter, A.~S., \& Pe'er, A.\ 2009, \apj, 701, 824
\bibitem[Oates et al.(2011)]{2011MNRAS.412..561O} Oates, S.~R., Page, M.~J., Schady, P., et al.\ 2011, \mnras, 412, 561
\bibitem[Oates et al.(2009)]{2009MNRAS.395..490O} Oates, S.~R., Page, M.~J., Schady, P., et al.\ 2009, \mnras, 395, 490
\bibitem[Paczynski \& Rhoads(1993)]{1993ApJ...418L...5P} Paczynski, B., \& Rhoads, J.~E.\ 1993, \apjl, 418, L5
\bibitem[Page et al.(2009)]{2009MNRAS.400..134P} Page, K.~L., Willingale, R., Bissaldi, E., et al.\ 2009, \mnras, 400, 134
\bibitem[Page et al.(2007)]{2007ApJ...663.1125P} Page, K.~L., Willingale, R., Osborne, J.~P., et al.\ 2007, \apj, 663, 1125
\bibitem[Panaitescu(2005)]{2005MNRAS.363.1409P} Panaitescu, A.\ 2005, \mnras, 363, 1409
\bibitem[Panaitescu(2005)]{2005MNRAS.362..921P} Panaitescu, A.\ 2005, \mnras, 362, 921
\bibitem[Panaitescu \& Kumar(2001)]{2001ApJ...554..667P} Panaitescu, A., \& Kumar, P.\ 2001, \apj, 554, 667
\bibitem[Panaitescu \& Kumar(2000)]{2000ApJ...543...66P} Panaitescu, A., \& Kumar, P.\ 2000, \apj, 543, 66
\bibitem[Panaitescu et al.(2006)]{2006MNRAS.369.2059P} Panaitescu, A., M{\'e}sz{\'a}ros, P., Burrows, D., et al.\ 2006, \mnras, 369, 2059
\bibitem[Panaitescu et al.(1998)]{1998ApJ...503..314P} Panaitescu, A., M{\'e}sz{\'a}ros, P., \& Rees, M.~J.\ 1998, \apj, 503, 314
\bibitem[Pandey et al.(2006)]{2006A&A...460..415P} Pandey, S.~B., Castro-Tirado, A.~J., McBreen, S., et al.\ 2006, \aap, 460, 415
\bibitem[Pandey et al.(2010)]{2010ApJ...714..799P} Pandey, S.~B., Swenson, C.~A., Perley, D.~A., et al.\ 2010, \apj, 714, 799
\bibitem[Pelassa \& Ohno(2010)]{2010arXiv1002.2863P} Pelassa, V., \& Ohno, M.\ 2010, arXiv:1002.2863
\bibitem[Perley et al.(2008)]{2008ApJ...672..449P} Perley, D.~A., Bloom, J.~S., Butler, N.~R., et al.\ 2008, \apj, 672, 449
\bibitem[Perley et al.(2010)]{2010GCN..11024...1P} Perley, D.~A., Li, W., \& Filippenko, A.~V.\ 2010, GRB Coordinates Network, 11024, 1
\bibitem[Perley et al.(2010)]{2010MNRAS.406.2473P} Perley, D.~A., Bloom, J.~S., Klein, C.~R., et al.\ 2010, \mnras, 406, 2473
\bibitem[Perna et al.(2006)]{2006ApJ...636L..29P} Perna, R., Armitage, P.~J., \& Zhang, B.\ 2006, \apjl, 636, L29
\bibitem[Pisani et al.(2013)]{2013A&A...552L...5P} Pisani, G.~B., Izzo, L., Ruffini, R., et al.\ 2013, \aap, 552, LL5
\bibitem[Proga \& Zhang(2006)]{2006MNRAS.370L..61P} Proga, D., \& Zhang, B.\ 2006, \mnras, 370, L61
\bibitem[Rana et al.(2009)]{2009AAS...21361003R} Rana, V., Cenko, B., Harrison, F., Fox, D., \& Kelemen, J.\ 2009, American Astronomical Society Meeting Abstracts \#213, 213, \#610.03
\bibitem[Rau et al.(2010)]{2010ApJ...720..862R} Rau, A., Savaglio, S., Kr{\"u}hler, T., et al.\ 2010, \apj, 720, 862
\bibitem[Rees \& M{\'e}sz{\'a}ros(1998)]{1998ApJ...496L...1R} Rees, M.~J., \& M{\'e}sz{\'a}ros, P.\ 1998, \apjl, 496, L1
\bibitem[Rees \& Meszaros(1994)]{1994ApJ...430L..93R} Rees, M.~J., \& Meszaros, P.\ 1994, \apjl, 430, L93
\bibitem[Rees \& Meszaros(1992)]{1992MNRAS.258P..41R} Rees, M.~J., \& Meszaros, P.\ 1992, \mnras, 258, 41P
\bibitem[Rhoads(1999)]{1999A&AS..138..539R} Rhoads, J.~E.\ 1999, \aaps, 138, 539
\bibitem[Romano et al.(2006)]{2006A&A...456..917R} Romano, P., Campana, S., Chincarini, G., et al.\ 2006, \aap, 456, 917
\bibitem[Rossi et al.(2011)]{2011A&A...529A.142R} Rossi, A., Schulze, S., Klose, S., et al.\ 2011, \aap, 529, AA142
\bibitem[Ruffini et al.(2014)]{2014arXiv1405.5723R} Ruffini, R., Wang, Y., Kovacevic, M., et al.\ 2014, arXiv:1405.5723
\bibitem[Ruiz-Velasco et al.(2007)]{2007ApJ...669....1R} Ruiz-Velasco, A.~E., Swan, H., Troja, E., et al.\ 2007, \apj, 669, 1
\bibitem[Rykoff et al.(2009)]{2009ApJ...702..489R} Rykoff, E.~S., Aharonian, F., Akerlof, C.~W., et al.\ 2009, \apj, 702, 489
\bibitem[Rykoff et al.(2006)]{2006ApJ...638L...5R} Rykoff, E.~S., Mangano, V., Yost, S.~A., et al.\ 2006, \apjl, 638, L5
\bibitem[Sari \& M{\'e}sz{\'a}ros(2000)]{2000ApJ...535L..33S} Sari, R., \& M{\'e}sz{\'a}ros, P.\ 2000, \apjl, 535, L33
\bibitem[Sari et al.(1999)]{1999ApJ...519L..17S} Sari, R., Piran, T., \& Halpern, J.~P.\ 1999, \apjl, 519, L17
\bibitem[Sari et al.(1998)]{1998ApJ...497L..17S} Sari, R., Piran, T., \& Narayan, R.\ 1998, \apjl, 497, L17
\bibitem[Schlegel et al.(1998)]{1998ApJ...500..525S} Schlegel, D.~J., Finkbeiner, D.~P., \& Davis, M.\ 1998, \apj, 500, 525
\bibitem[Siegel et al.(2010)]{2010GCN..10645...1S} Siegel, M.~H., Marshall, F.~E., Holland, S.~T., et al.\ 2010, GRB Coordinates Network, 10645, 1
\bibitem[Singer et al.(2013)]{2013ApJ...776L..34S} Singer, L.~P., Cenko, S.~B., Kasliwal, M.~M., et al.\ 2013, \apjl, 776, LL34
\bibitem[Soderberg et al.(2007)]{2007ApJ...661..982S} Soderberg, A.~M., Nakar, E., Cenko, S.~B., et al.\ 2007, \apj, 661, 982
\bibitem[Sollerman et al.(2007)]{2007A&A...466..839S} Sollerman, J., Fynbo, J.~P.~U., Gorosabel, J., et al.\ 2007, \aap, 466, 839
\bibitem[Stanek et al.(2007)]{2007ApJ...654L..21S} Stanek, K.~Z., Dai, X., Prieto, J.~L., et al.\ 2007, \apjl, 654, L21
\bibitem[Starling et al.(2009)]{2009MNRAS.400...90S} Starling, R.~L.~C., Rol, E., van der Horst, A.~J., et al.\ 2009, \mnras, 400, 90
\bibitem[Stratta et al.(2009)]{2009A&A...503..783S} Stratta, G., Pozanenko, A., Atteia, J.-L., et al.\ 2009, \aap, 503, 783
\bibitem[Th{\"o}ne et al.(2010)]{2010A&A...523A..70T} Th{\"o}ne, C.~C., Kann, D.~A., J{\'o}hannesson, G., et al.\ 2010, \aap, 523, AA70
\bibitem[Thompson(1994)]{1994MNRAS.270..480T} Thompson, C.\ 1994, \mnras, 270, 480
\bibitem[Troja et al.(2007)]{2007ApJ...665..599T} Troja, E., Cusumano, G., O'Brien, P.~T., et al.\ 2007, \apj, 665, 599
\bibitem[Ukwatta et al.(2010)]{2010GCN..10875...1U} Ukwatta, T.~N., Barthelmy, S.~D., Baumgartner, W.~H., et al.\ 2010, GRB Coordinates Network, 10875, 1
\bibitem[Ukwatta et al.(2010)]{2010ApJ...711.1073U} Ukwatta, T.~N., Stamatikos, M., Dhuga, K.~S., et al.\ 2010, \apj, 711, 1073
\bibitem[Updike et al.(2008)]{2008ApJ...685..361U} Updike, A.~C., Haislip, J.~B., Nysewander, M.~C., et al.\ 2008, \apj, 685, 361
\bibitem[Usov(1992)]{1992Natur.357..472U} Usov, V.~V.\ 1992, \nat, 357, 472
\bibitem[van Eerten \& MacFadyen(2013)]{2013ApJ...767..141V} van Eerten, H., \& MacFadyen, A.\ 2013, \apj, 767, 141
\bibitem[Vergani et al.(2011)]{2011A&A...535A.127V} Vergani, S.~D., Flores, H., Covino, S., et al.\ 2011, \aap, 535, AA127
\bibitem[Vestrand et al.(2004)]{2004AN....325..549V} Vestrand, W.~T., Borozdin, K., Casperson, D.~J., et al.\ 2004, Astronomische Nachrichten, 325, 549
\bibitem[Vestrand et al.(2014)]{2014Sci...343...38V} Vestrand, W.~T., Wren, J.~A., Panaitescu, A., et al.\ 2014, Science, 343, 38
\bibitem[Wiersema et al.(2012)]{2012MNRAS.426....2W} Wiersema, K., Curran, P.~A., Kr{\"u}hler, T., et al.\ 2012, \mnras, 426, 2
\bibitem[Wijers et al.(1997)]{1997MNRAS.288L..51W} Wijers, R.~A.~M.~J., Rees, M.~J., \& Meszaros, P.\ 1997, \mnras, 288, L51
\bibitem[Willingale et al.(2007)]{2007ApJ...662.1093W} Willingale, R., O'Brien, P.~T., Osborne, J.~P., et al.\ 2007, \apj, 662, 1093
\bibitem[Wo{\'z}niak et al.(2005)]{2005ApJ...627L..13W} Wo{\'z}niak, P.~R., Vestrand, W.~T., Wren, J.~A., et al.\ 2005, \apjl, 627, L13
\bibitem[Wu et al.(2003)]{2003MNRAS.342.1131W} Wu, X.~F., Dai, Z.~G., Huang, Y.~F., \& Lu, T.\ 2003, \mnras, 342, 1131
\bibitem[Wu et al.(2004)]{2004ApJ...615..359W} Wu, X.~F., Dai, Z.~G., \& Liang, E.~W.\ 2004, \apj, 615, 359
\bibitem[Xin et al.(2010)]{2010MNRAS.401.2005X} Xin, L.~P., Zheng, W.~K., Wang, J., et al.\ 2010, \mnras, 401, 2005
\bibitem[Yost et al.(2006)]{2006ApJ...636..959Y} Yost, S.~A., Alatalo, K., Rykoff, E.~S., et al.\ 2006, \apj, 636, 959
\bibitem[Yost et al.(2007)]{2007ApJ...657..925Y} Yost, S.~A., Swan, H.~F., Rykoff, E.~S., et al.\ 2007, \apj, 657, 925
\bibitem[Yuan et al.(2008)]{2008AIPC.1065..103Y} Yuan, F., Rykoff, E.~S., Schaefer, B.~E., et al.\ 2008, American Institute of Physics Conference Series, 1065, 103
\bibitem[Yuan et al.(2010)]{2010ApJ...711..870Y} Yuan, F., Schady, P., Racusin, J.~L., et al.\ 2010, \apj, 711, 870
\bibitem[Zaninoni et al.(2013)]{2013A&A...557A..12Z} Zaninoni, E., Bernardini, M.~G., Margutti, R., Oates, S., \& Chincarini, G.\ 2013, \aap, 557, AA12
\bibitem[Zerbi et al.(2001)]{2001grba.conf..434Z} Zerbi, F.~M., Chincarini, G., Rodon{\'o}, M., et al.\ 2001, Gamma-ray Bursts in the Afterglow Era, 434
\bibitem[Zhang et al.(2007)]{2007ApJ...666.1002Z} Zhang, B.-B., Liang, E.-W., \& Zhang, B.\ 2007, \apj, 666, 1002
\bibitem[Zhang(2014)]{2014ApJ...780L..21Z} Zhang, B.\ 2014, \apjl, 780, LL21
\bibitem[Zhang(2011)]{2011CRPhy..12..206Z} Zhang, B.\ 2011, Comptes Rendus Physique, 12, 206
\bibitem[Zhang(2007)]{2007ChJAA...7....1Z} Zhang, B.\ 2007, \cjaa, 7, 1
\bibitem[Zhang et al.(2006)]{2006ApJ...642..354Z} Zhang, B., Fan, Y.~Z., Dyks, J., et al.\ 2006, \apj, 642, 354
\bibitem[Zhang \& M{\'e}sz{\'a}ros(2004)]{2004IJMPA..19.2385Z} Zhang, B., \& M{\'e}sz{\'a}ros, P.\ 2004, International Journal of Modern Physics A, 19, 2385
\bibitem[Zhang \& M{\'e}sz{\'a}ros(2001)]{2001ApJ...552L..35Z} Zhang, B., \& M{\'e}sz{\'a}ros, P.\ 2001, \apjl, 552, L35
\bibitem[Zhang \& Pe'er(2009)]{2009ApJ...700L..65Z} Zhang, B., \& Pe'er, A.\ 2009, \apjl, 700, L65
\bibitem[Zhang \& Yan(2011)]{2011ApJ...726...90Z} Zhang, B., \& Yan, H.\ 2011, \apj, 726, 90
\end{thebibliography}
\end{document}